\RequirePackage{fix-cm}
\documentclass[10pt,letterpaper,oneside,openany]{book}
\usepackage[margin=1in,footskip=0.2in]{geometry}
\usepackage[utf8]{inputenc}

\usepackage{amsthm, amsmath, amsfonts, amssymb,booktabs,graphicx, bm}
\usepackage[table]{xcolor}

\usepackage{mathtools}

\usepackage{enumerate}
\usepackage{hyperref}
\usepackage[nameinlink]{cleveref}
\usepackage{multirow}
\usepackage[colorinlistoftodos,prependcaption]{todonotes}
\usepackage{xfrac}

\usepackage{thmtools} 
\usepackage{thm-restate}

\usepackage[
    style=alphabetic,
    backref=true,
    maxalphanames=99,
    minalphanames=99,
    maxbibnames=99
]{biblatex}

\DeclareLabelalphaNameTemplate{%
  \namepart[use=true, pre=true, strwidth=1, compound=true]{prefix}%
  \namepart[strwidth=1, compound=true]{family}%
}

\DeclareLabelalphaTemplate{%
  \labelelement{%
    \field[strwidth=1,strside=left]{labelname}% ← only *one* letter per name
  }%
  \labelelement{%
    \field[strwidth=2,strside=right]{year}%   ← last two digits of year
  }%
}

\hypersetup{
    hypertexnames=false,
    colorlinks,
    linkcolor=teal,
    citecolor=blue
}
\usepackage[nottoc,notlot,notlof]{tocbibind}

\usepackage{algorithm}
\usepackage[noend]{algpseudocode}

\newtheorem{theorem}{Theorem}[chapter]

\newtheorem{lemma}[theorem]{Lemma}
\newtheorem{claim}[theorem]{Claim}
\newtheorem{corollary}[theorem]{Corollary}

\newtheorem*{question*}{Question}
\newtheorem{fact}[theorem]{Fact}

\newtheorem{definition}[theorem]{Definition}
\newtheorem{problem}[theorem]{Problem}

\newtheorem*{definition*}{Definition}

\theoremstyle{remark}
\newtheorem*{remark}{Remark}

\numberwithin{equation}{chapter}



\newcommand{\cA}{\mathcal{A}}

\newcommand{\cC}{\mathcal{C}}
\newcommand{\cD}{\mathcal{D}}
\newcommand{\cE}{\mathcal{E}}

\newcommand{\cG}{\mathcal{G}}

\newcommand{\cL}{\mathcal{L}}
\newcommand{\cM}{\mathcal{M}}
\newcommand{\cN}{\mathcal{N}}
\newcommand{\cO}{\mathcal{O}}

\newcommand{\E}{\operatornamewithlimits{\mathbb{E}}}

\newcommand{\N}{\mathbb{N}}
\newcommand{\bO}{\mathbb{O}}

\newcommand{\R}{\mathbb{R}}
\newcommand{\bS}{\mathbb{S}}

\newcommand{\Z}{\mathbb{Z}}

\newcommand{\mA}{\mathbf{A}}
\newcommand{\mB}{\mathbf{B}}

\newcommand{\mD}{\mathbf{D}}
\newcommand{\mI}{\mathbf{I}}
\newcommand{\mJ}{\mathbf{J}}

\newcommand{\mM}{\mathbf{M}}

\newcommand{\mQ}{\mathbf{Q}}
\newcommand{\mU}{\mathbf{U}}
\newcommand{\mV}{\mathbf{V}}
\newcommand{\mW}{\mathbf{W}}

\newcommand{\mZ}{\mathbf{Z}}

\newcommand{\va}{\mathbf{a}}
\newcommand{\vb}{\mathbf{b}}
\newcommand{\vc}{\mathbf{c}}

\newcommand{\ve}{\mathbf{e}}
\newcommand{\vf}{\mathbf{f}}
\newcommand{\vg}{\mathbf{g}}

\newcommand{\vm}{\mathbf{m}}

\newcommand{\vq}{\mathbf{q}}
\newcommand{\vr}{\mathbf{r}}

\newcommand{\vu}{\mathbf{u}}
\newcommand{\vv}{\mathbf{v}}
\newcommand{\vw}{\mathbf{w}}
\newcommand{\vx}{\mathbf{x}}
\newcommand{\vy}{\mathbf{y}}
\newcommand{\vz}{\mathbf{z}}

\newcommand{\mSigma}{\mathbf{\Sigma}}
\newcommand{\Diag}{\mathbf{Diag}}

\newcommand{\sL}{\mathcal{L}}

\newcommand{\p}[1]{\left(#1\right)}
\newcommand{\pb}[1]{\left[#1\right]}
\newcommand{\pc}[1]{\left\{#1\right\}}
\newcommand{\pa}[1]{\left|#1\right|}
\newcommand{\pg}[1]{\left\langle#1\right\rangle}
\newcommand{\pgo}[1]{\langle#1\rangle}

\newcommand{\pn}[1]{\left\|#1\right\|}

\newcommand{\conv}{\mathrm{conv}}
\newcommand{\vspan}{\mathrm{span}}
\newcommand{\vol}{\mathrm{vol}}
\newcommand{\aff}{\mathrm{aff}}
\newcommand{\diam}{\mathrm{diam}}
\newcommand{\eps}{\varepsilon}

\newcommand{\nnz}{\mathsf{nnz}}
\newcommand{\kappaol}{\kappa^{\mathsf{OL}}}
\newcommand{\defeq}{\stackrel{\textnormal{def}}{=}}

\newcommand{\norm}[1]{\ensuremath{\left\lVert #1 \right\rVert}}
\newcommand{\gnorm}[2]{\norm{#1}_{\cG_{#2}}}
\newcommand{\gnorml}[2]{\norm{#1}_{\cG_{#2}(\vlambda)}}
\newcommand{\inparen}[1]{\left(#1\right)}
\newcommand{\insquare}[1]{\left[#1\right]}
\newcommand{\inbraces}[1]{\left\{#1\right\}}           %\inbrace{x+y}  is {x+y}
 
\newcommand{\mtilde}{\widetilde{m}}

\newcommand{\vrho}{\bm{\rho}}
\newcommand{\logv}[1]{\log\inparen{#1}}
\newcommand{\abs}[1]{\ensuremath{\left\lvert #1 \right\rvert}}
\newcommand{\vbrho}{\eps^{-2}\inparen{\log d}^2\logv{\sfrac{d}{\eps}}}
\newcommand{\suchthat}{{\;\; : \;\;}}
\newcommand{\rank}[1]{\mathsf{rank}\inparen{#1}}
\newcommand{\dtwo}{\mathsf{dist}}
\newcommand{\myfunc}[3]{#1\colon#2\to#3}
\newcommand{\polynorm}[1]{\norm{#1}_{\cG,\vrho,\infty, S}}
\newcommand{\exvv}[2]{\underset{#1}{\E}\insquare{#2}}
\newcommand{\vlambda}{\bm{\lambda}}
\newcommand{\mLambda}{\mathbf{\Lambda}}

\newcommand{\Fstar}{F^{\star}}
\newcommand{\ip}[1]{\left\langle #1 \right\rangle}
\newcommand{\subgnorm}[1]{\norm{#1}_{\psi_2}}
\newcommand{\exv}[1]{\E\insquare{#1}}
\newcommand{\expv}[1]{\mathsf{exp}\inparen{#1}}
\newcommand{\prv}[1]{\Pr\insquare{#1}}
\newcommand{\valpha}{\bm{\alpha}}
\newcommand{\vmu}{\bm{\mu}}

\newcommand{\mR}{\mathbf{R}}
\newcommand{\zstar}{\vz^{\star}}
\newcommand{\mF}{\mathbf{F}}
\newcommand{\xhat}{\widehat{\vx}}
\newcommand{\lnv}[1]{\ln\inparen{#1}}
\newcommand{\opt}{\mathsf{OPT}}
\newcommand{\argmin}[1]{\underset{#1}{\arg\min}}
\newcommand{\xstar}{\vx^{\star}}
\newcommand{\prvv}[2]{\underset{#1}{\Pr}\insquare{#2}}
\newcommand{\onev}{\mathbf{1}}
\newcommand{\ceil}[1]{\left\lceil #1 \right\rceil}
\newcommand{\ones}{\mathbf{1}}

\newcommand{\regret}{\mathrm{Regret}}

\newcommand{\thesistitle}{Computation and Applications of Euclidean and Normed Representations of Massive Data}
\newcommand{\thesisauthor}{Max Ovsiankin}
\newcommand{\thesisdate}{September 2026}

\title{\thesistitle}
\author{\thesisauthor}
\date{\thesisdate}

\begin{document}

\frontmatter
\pagestyle{plain}

\begin{titlepage}
\centering
\vspace*{0.25in}

{\Large\bfseries \MakeUppercase{\thesistitle}\par}

\vspace{0.65in}
{\normalsize BY\par}
\vspace{0.15in}
{\large \MakeUppercase{\thesisauthor}\par}

\vfill

A thesis submitted\par
in partial fulfillment of the requirements for\par
the degree of\par

\vspace{0.35in}
Doctor of Philosophy in Computer Science\par

\vfill

at the\par

\vspace{0.35in}
{\large \MakeUppercase{Toyota Technological Institute at Chicago}\par}
Chicago, Illinois\par

\vspace{0.35in}
\thesisdate\par

\vfill

Thesis Committee:\par
Yury Makarychev (Thesis Advisor)\par
Haotian Jiang\par
Lorenzo Orecchia\par
Madhur Tulsiani\par
\end{titlepage}
\setcounter{page}{2}

\thispagestyle{plain}

\begin{center}
{\Large \textbf{Abstract}}
\end{center}

This thesis investigates Euclidean-space and \(\ell_p\)-norm representations of different forms of data,
with a focus on efficient algorithms for computing these representations in settings where the amount of data is very large.
The applications of such representations are also discussed:
they may be used to summarize the data in a more compact form for downstream tasks
while preserving its most salient properties; and further, they may also be used to extract insights about the data that may not be apparent in its original form.
This thesis presents novel algorithms for computing Euclidean representations in the case where the data comes with linear structure, as well as in the case where the data comes only with metric,
or distance structure.
In addition, we give algorithms for computing \(\ell_p\)-norm representations in linear structured cases.
The analyses of these algorithms use tools from geometry, probability, and optimization, and these tools are used to
illuminate other algorithms for similar problems.

\chapter*{Acknowledgements}
\addcontentsline{toc}{chapter}{Acknowledgements}

I am grateful beyond words for those with whom I have shared this PhD, academic,
and life journey.
I will use these acknowledgements to celebrate some of these incredible people.

Thank you to my PhD advisor, Yury Makarychev.
Yury helped me grow as a researcher in every possible sense;
from learning to pick the right questions, to exploring and
developing results, to communicating and sharing my work.
Yury has been the chief facilitator of what I consider to be an 
extraordinarily special privilege: spending a few years learning about, working on,
and solving
beautiful mathematical problems.
Beyond working on papers, I also thank Yury for his consistent feedback,
patience, care, and wisdom.

Thank you to the other members of my thesis committee, Haotian Jiang, Madhur Tulsiani,
and Lorenzo Orecchia.
I am grateful for all of your warmth, support, and extremely helpful
advice through my PhD.
Thanks Haotian for the fascinating class and discussions on discrepancy.
Thanks Madhur for the lovely teaching experience.
Thanks Lorenzo, for the wonderful research discussions on optimization
and related topics, and for the non-research discussions.
I greatly appreciated your everlasting readiness to entertain my research ideas,
as well as your kindness and friendliness.

I would like to thank all of my other collaborators
and co-authors:
Sara Ahmadian,
Antares Chen,
Romain Cosson,
Anupam Gupta,
Vishesh Jain,
Kostas Kollias,
Debmalya Panigrahi,
Marco Pirazzini,
Aaron Schild,
and Erasmo Tani.
It has been a pleasure to work with you;
whether that was in weekly meetings, impromptu late night study sessions,
or something in between.

There are countless people I could thank for my academic and mathematical
journey.
I will start with Susan Holtzapple, who helped me see back in middle school
the beauty and complexity of math.
Thanks to Mary D. Hill and others at the Middle College and College Now programs
at FUHSD and at De Anza College for the invaluable opportunity to 
pursue and deepen my physics and math studies (among other subjects!)
in high school.

I have learned a great deal from some incredible internships
and mentors in industry.
Thanks to Eishay Smith at Kifi;
Leah McGuire,
Shubha Nabar,
and Matthew Tovbin at Salesforce Einstein;
and Eugene Burmako and Brad Scheppler at Twitter.
Thank you to Eugene for an excellent first research internship
and all the support and guidance on my very first conference and research talks.

I had the privilege of a fantastic environment at Berkeley to start doing teaching
and research.
Thank you to Vyassa Bharatham, Alessandro Chiesa, Christopher Hunn,
Prasad Raghavendra, Satish Rao, Nicholas Spooner, and Luca Trevisan.
Thank you, Nick, for your guidance in my very first theory research project.
Thanks Satish, for the many fascinating research conversations and very helpful
academic advice.

I am also very grateful for everyone at Google who made it such a warm
and welcoming place to do research.
It was lovely to work with the GeoAI team, including
Sara Ahmadian,
Sreenivas Gollapudi,
Dee Guo,
Tarun Kathuria,
Kostas Kollias,
and Aaron Schild.
Also thanks to the many others I met around Google Research for the
fascinating and friendly conversations.

Thank you to Matthew Turk for your exceptional leadership
at TTIC while I was there.
Thanks to
Adam Bohlander, Erica Cocom,
Chrissy Coleman, Mary Marre, Amy Minick,
and the other incredible TTIC staff for your hard work
in making the PhD such a delightful, smooth, and pleasant experience.

TTIC has been such a lovely tight-knit community
to be a part of, and I am thankful to have been in the company
of such wonderful friends in my fellow PhD students.
Thank you especially to
Sudarshan Babu,
Gene Li, 
Naren Sarayu Manoj,
Omar Montasser,
Kumar Kshitij Patel,
Kavya Ravichandran,
Pushkar Shukla,
and Shashank Srivastava.
I also thank other friends I met during the PhD,
whether it was in UChicago, the broader academic community,
or elsewhere,
including
Emma Corless,
Agastya Vibhuti Jha,
Anmol Kabra,
Frederic Koehler,
Sheng Long,
Kunal Marwaha,
Tushant Mittal,
Marco Pirazzini,
Aniruddh Raghu,
Shannon Sequeira,
Vaidehi Srinivas,
and Hanwen Zheng.
And thank you to the friends I have known
during undergrad and before, including
William Bassett,
Antares Chen,
Gillian Chu,
Dee Guo,
Mudit Gupta,
Emaan Hariri,
Catherine Huang,
James Hulett,
Vishnu Iyer,
Lakshya Jain,
Marissa Jasso,
Tarun Kathuria,
Vinay Koshy,
Neha Kunjal,
Harry Lee,
Jonathan Liu,
Sean Luchen,
Carlos Ortega,
Rishi Veerapenani,
Daniel Zhang,
and David Zhang.
Your support has been invaluable
during the easy times, and even more important during the hard ones.

Thank you Safta Ira, my brother, and my mother and father. I cannot
express in words what you mean to me. I love you.

\tableofcontents
\listoftables
\listoffigures

\mainmatter

\chapter{Introduction}

Motivated by applications across statistics, physics, signal processing, computer science, among many other fields,
modern data science has increasingly demanded procedures for analyzing and extracting insights from massive datasets
(characterized by high dimensionality, a large number of observations, or both).
A basic and important paradigm developed to tackle this challenge is preprocessing, summarization, or dimension reduction: finding a representation of the data
that is more economical, ideally significantly so, while preserving its most salient properties, ideally as closely as possible.
These representations can then be used by practitioners in downstream tasks, with their economy enabling the use of a wide variety of methods that may not scale
well with the size of the data.
Even further, these representations can illuminate the structure of the data in ways that are not apparent from its original form.
Euclidean space and \(\ell_p\) space
turn out to be a natural choice for data representation in many cases, as they are often easier to optimize over,
and for data analysis they allow for the use of intuitive geometric techniques.

In this thesis, we focus on the problem of efficiently computing normed-space representations of different forms of data.
For Euclidean space, we consider the case where the data comes equipped with linear structure, as well as cases where the data comes only with metric, or distance structure.
In both cases, we give novel efficient algorithms for computing such representations, and we analyze their theoretical guarantees.

We now discuss these two areas in more detail.
We call Euclidean space \(\ell_2\) as is done throughout functional analysis, owing to the fact that its
norm is \(\|\vx\| = \p{\sum_i \vx_i^p}^{1/p}\) for \(p=2\).

\section{\texorpdfstring{\(\ell_2\)-Representations of Convex Polytopes}{ell2-Representations of Convex Polytopes}}
Say that one is attempting to find a feasible point of some convex set \(K \subseteq \R^d\),
which is nontrivial if e.g. \(K\) is defined as the intersection of a collection of convex sets.
A basic principle of convex optimization is that adding structural assumptions on \(K\) usually allows for more efficient specialized optimization methods.
If we can find such a simpler set that is geometrically close to \(K\), then the answer to the optimization problem will not change too much,
while optimization over the simpler set can be easier and faster.
A good choice for a simpler set is some linear transformation of the unit ball, i.e. an ellipsoid.
This basic idea of approximating a set by a simpler one appears throughout convex optimization; not least in the foundational ellipsoid method,
where a sequence of ellipsoids with fast-decreasing volume is used to approximate parts of the feasible set \cite{grotschel2012geometric}.
Indeed, we will use this idea for one of the algorithms later in this thesis.

The natural notion of `geometric closeness' that is especially useful in this context is the following.
For \(\alpha \ge 1\), we say a set \(L\) is an \(\alpha\)-approximation to \(K\) if for some point \(\vc \in \R^d\), we have \(\frac{1}{\alpha} (L - \vc) + \vc \subseteq K \subseteq L\).
This condition says that \(L\) must contain \(K\), and also \(K\) must contain a version of \(L\) that has been scaled down around some point \(\vc\) by a factor of \(\alpha\).
One of the most fundamental results regarding \(\alpha\)-approximations in convex geometry is the following.

\begin{theorem}[\cite{john1948extremum}]
    Let \(K \subseteq \R^d\) be a convex body. Then some ellipsoid is a \(d\)-approximation to \(K\).
    Further, if \(K\) is symmetric about the origin, then some ellipsoid is a \(\sqrt{d}\)-approximation.
    Both these bounds are tight.
\end{theorem}
A convex body is just a convex set that is closed, bounded, and with nonempty interior, 
all of which are usually light conditions in the context of convex optimization.
Saying that \(K\) is symmetric about the origin simply means that \(K = -K\).

Now, we elaborate the sense in which ellipsoidal approximations can be thought of as a way to represent data with linear structure by \(\ell_2\).
Every normed finite-dimensional vector space (a vector space equipped with a norm) has an associated unit ball consisting of all vectors with norm at most one, which is necessarily a symmetric convex body.
Correspondingly, every symmetric convex body is the unit ball of some normed space.
One important topic of study in functional analysis is the `smoothness' of maps between normed spaces,
and in particular a key quantity is the `condition number' of a map between spaces of the same dimension (the product of the Lipschitz constants of the map and its inverse).
This condition number serves as a measure of how well the spaces geometrically correspond under the map. 
The smallest condition number of any linear map between two fixed finite-dimensional normed spaces is called the `Banach-Mazur distance', 
and the log of this quantity is in fact a distance measure on the `space' of all possible normed spaces of a fixed dimension
\cite{bourgain1988banach}.

The Banach-Mazur distance is closely related to the notion of convex set approximation introduced earlier.
If we fix two bases for both normed spaces, then the images of both spaces' unit balls under the canonical map to \(\R^d\) are convex sets. The smallest \(\alpha\)-approximation achievable between these two sets, fixing \(\vc = 0\), across all possible choices of bases, then equals the Banach-Mazur distance between the two spaces.
And in particular, the Banach-Mazur distance between some arbitrary normed space to \(\ell_2\) is equal to the best approximation of the unit ball of that space achievable by an ellipsoid.
In other words, that ellipsoid represents a linear isomorphism with which we approximately represent the original normed space by \(\ell_2\). 

Thus, John's theorem gives us some idea of how well we can represent arbitrary data with \(\ell_2\), although it does not immediately give ideas on how to compute this representation. John's theorem is proved by showing that an ellipsoid containing \(K\) of minimal volume achieves the desired approximation; and as the problem of finding that ellipsoid is convex, this idea has led to fast algorithms for computing that optimal ellipsoid \cite{cohen2019near}.
We develop algorithms for approximating \(K\) with an ellipsoid with a similar fast runtime in the case that \(K\) is defined by a massive dataset; in particular, when it is the convex hull of \(n\) points in \(\R^d\) where \(n\) is very large.
Any convex hull of a finite set of points, as in the case we study, is called a convex polytope.
Our algorithms are streaming algorithms, requiring no more space than needed to represent the answer (\(d^2\) real numbers), and only requiring a single pass over the points.
The approximation factor we achieve is nearly the best possible, with some logarithmic dependence on an `aspect ratio' (much like a condition number) for the convex body.

Further, we elaborate on some applications of this result and generalize it in new ways.
We give the first streaming algorithm that works when \(K\) is an arbitrary convex polytope, and not just symmetric about the origin; this algorithm has similar approximation guarantees to the symmetric special case.
We generalize the algorithm to be fully online (with no dependence on initialization),
only with a dependence on an online version of the aspect ratio.
For an application, we show how to use this algorithm to compute a `coreset' of a convex polytope;
meaning selecting some subset of its vertices so that the convex hull of this subset is an approximation of the original subset. This application can be thought of a convex-geometric form of sparsification, a notion that was applied before e.g. to approximate sums of rank-one matrices by their subsets.
Finally, we show that for a natural class of similar algorithms to the one we propose,
the dependence we have on the geometric quantity of aspect ratio is the best possible.

\section{\texorpdfstring{Approximating \(\ell_p\)-Matrix Block Norms}{Approximating lp-Matrix Block Norms}}

In the previous section, we discussed work showing how `coresets', or subsets of the data approximating the original data
in some geometric sense, could be constructed with an algorithmic version of John's theorem; and further that this worked in a streaming setting.
Different notions of `approximation' are most useful for different downstream applications: for applications such as \(\ell_p\) regression (see e.g. \cite{mmwy21}), a rich literature has developed studying \(\ell_p\) row-sampling; and for applications such as preserving hypergraph \(\ell_2\) energy, comparable sampling strategies have also been developed (see e.g. \cite{jambulapati2023chaining}).

In the case of \(\ell_p\) row-sampling, we are given some data in a matrix \(A \in \R^{n \times d}\) and
would like to find an \(A'\) with a few rescaled rows of \(A\) such that \(\| A' x\|_{p} \approx \|A x \|_{p}\) for all \(x \in \R^d\).
A foundational result here comes from \cite{blm89}, who essentially showed that such a sparse weighted approximation could be constructed by sampling from a scaling found by Lewis and now called Lewis weights, although this result was originally described as an \(\ell_p\)-subspace embedding in the language of functional analysis. Motivated by an increasing number of applications in statistics and data analysis, these types of results were eventually made algorithmic, see e.g. \cite{cohen2015lp}.

We study a simultaneous generalization of the \(\ell_p\)-row sampling and hypergraph \(\ell_2\) energy preservation problems
called \(\ell_p\)-block norm sampling. Here, the rows of \(A\), which we can think of as the individuals in our dataset,
come in a group structure, specifically a partition of the rows into sets \(S_1, \ldots, S_m\).
We use \(A_{S_i}\) as notation to denote the selection of rows of \(A\) that are in group \(S_i\).
Our reweighting is forced to preserve this group structure; we can only select and scale an entire group at once.
In particular, given \(p, p_1, \ldots, p_m \ge 1\), we wish to find sparse weights \(b_1, \ldots, b_m \ge 0\) so that for all \(x \in \R^d\), 
\[\sum_{i=1}^m b_i^p \|A_{S_i} x\|_{p_i}^p \approx \sum_{i=1}^m \|A_{S_i} x\|_{p_i}^p\]
Observe that the \(\ell_p\) row-sampling case is recovered when all the groups consist of exactly one element.
Similarly, it can be seen that taking \(p = 2\) and all \(p_i\) large recovers the hypergraph \(\ell_2\) energy preservation problem.

We develop a new method to construct this sparse approximation, using the same high-level scheme as is used throughout
row sampling or \(\ell_p\)-subspace embedding.
First, we establish that there exist weights on the groups, such that sampling in accordance with these weights
produces the desired sparse approximation of the original grouped dataset.
Second, we show that in certain regimes of the norm parameters \(p, p_1, \ldots, p_m\),
there exist efficient algorithms for (approximately) computing these weights, thus giving
an end-to-end practical method for finding the sparse approximation.

From a broad point of view, establishing the sampling result requires tightly understanding the geometry of the ball \(\{x \colon \sum_{i=1}^m \|A_{S_i} x\|_{p_i}^p \le 1 \}\) in a metric determined by the weights we choose. The past work of e.g. \cite{blm89} use a `change-of-measure' argument to understand the required geometry, in their case the \(\ell_p\) ball, `through' Euclidean space, by finding a linear transformation of a Euclidean ball that is a good approximation to the \(\ell_p\) ball.
We observe that a natural generalization of Lewis weights, which we call \emph{block Lewis weights},
can be used to perform a similar change of measure, although this argument requires extra care due to the nesting of norms.
For the second part, we show that computing block Lewis weights reduces to iteratively solving a particular sequence of reweighted subproblems (inspired by one definition of Lewis weights as the solution to a determinant maximization problem), which in certain regimes of \(p\) can be done efficiently.

Finally, we establish the applications of this sparse approximation to some downstream tasks in statistics and data science. The group structure translates well to the regression task of minimizing sums of Euclidean norms, where sparsification using our method allows for schemes whose runtime depend on the intrinsic dimensionality of the problem rather than the number of groups. The minimizing sums of Euclidean norms problem is quite general, including as special cases Euclidean facility location and \(\ell_1\) regression.

\section{\texorpdfstring{\(\ell_2\)-Embeddings of Metric Spaces and \(\ell_2\)-Configuration}{ell2-Embeddings of Metric Spaces and ell2-Configuration}}
The following problem has appeared across statistics, computer science, and mathematics:
given a non-negative matrix representing pairwise `dissimilarities' between individuals, assign points in Euclidean space to these individuals so that the Euclidean distances between the points match the respective dissimilarities as closely as possible.
In statistics and psychometrics, this problem is called multidimensional scaling (MDS), where the input is often a matrix of similarities or dissimilarities rather than a full distance metric \cite{borg2005modern}.
Classical MDS gives a spectral solution when the data are treated as squared Euclidean distances, while nonmetric MDS and stress-based MDS solve fitting problems designed for general similarity or dissimilarity data.
In mathematics and computer science, it is usually assumed that the matrix of dissimilarities has the structure of a metric space, and so the dissimilarities are called distances. A common objective is then to minimize a quantity called `distortion', in which case the map to Euclidean space is called a `low-distortion embedding' (\cite{naor2015metric}, \cite{matousek2013lectures}); this is the objective we focus on in this thesis.

The quantity called `distortion' can be thought of as a generalization of the notion of condition number of maps between normed spaces.
Once again, the distortion of a map between metric spaces is defined as the product of the Lipschitz constants of the map and its inverse.
One may interpret this quantity as follows: if a map has distortion \(D\),
then the distance between the images of any two points under the map approximates the distance between the original points up to a \(D\) factor (allowing for some global scaling of the distances).
A fundamental result in this area is again regarding how well these spaces can be approximated by \(\ell_2\). Bourgain showed that every \(n\)-point metric space has an embedding into \(\ell_2\) of distortion \(O(\log n)\) \cite{bourgain1985lipschitz}.
In theoretical computer science, low-distortion embeddings are algorithmic primitives rather than merely visualization tools.
The work of Linial, London, and Rabinovich made explicit the connection between metric embeddings, cuts, and multicommodity flow, and embeddings into Euclidean or negative-type geometries play a central role in SDP relaxations and rounding procedures for cut problems, including sparsest cut
\cite{linial1995geometry,arora2005euclidean}.

Bourgain's theorem does actually provide a construction of this embedding (and the construction can be made efficiently); however, in many cases we are not only interested in the embedding that is guaranteed for any metric space, but the best embedding into \(\ell_2\) possible for one \textit{particular} metric space.
This problem can be solved as a generic semidefinite programming problem, and we develop the first algorithm using first-order techniques that can approximate the lowest-distortion embedding of a metric space into \(\ell_2\) with a geometry-independent runtime.
We also show our solver's guarantee for a problem we call graph-constrained \(\ell_2\) configuration,
which generalizes finding an embedding of low distortion by allowing the problem instance
to specify distance constraints for a certain group of pairs of points, with no constraints on the allowable embedding distances outside this group.
% For example, if we are concerned with preserving distance between \(m\) pairs of points up to 30\% accuracy, then our algorithm requires a time proportional to \(1400 m \sqrt{n}\).
This problem brings the theory of embeddings closer to the applications of multi-dimensional scaling, where often dissimilarities are given that are not metrics, and where some entries of the dissimilarity matrix are empty.

In addition to the applications in data analysis and statistics, solving a metric embedding into \(\ell_2\) problem is a core piece of SDP-based approaches to generalized or nonuniform sparsest cut, a fundamental problem in graph partitioning \cite{arora2005euclidean}.
In short, this approach solves two semidefinite programs in sequence: one `relaxation' of the sparsest cut problem from optimizing over cuts to optimizing over a particular class of metric spaces, and then another program that `rounds' the answer to the first program to \(\ell_2\) in order to convert it back into a cut.
Our new solver targets this second embedding or rounding program, while the first SDP remains a separate task.
Very fast algorithms are known for important special cases of sparsest cut, such as uniform sparsest cut \cite{sherman2009breaking};
the contribution here is an \(m\)-sensitive first-order routine for the Euclidean embedding subproblem that appears in the SDP-based picture.

A closely related line of work is Euclidean distance geometry and Euclidean distance matrix completion, where one seeks a point configuration realizing a partially observed set of distances \cite{liberti2014euclidean}.
These problems arise in applications such as molecular conformation, sensor network localization, and statics, and have been studied using SDP relaxations and low-rank matrix-completion methods
\cite{ding2010sensor,tasissa2018exact}.
This is a close analog of well-known results in matrix completion (e.g. \cite{candes2012exact}), where a low-rank matrix can be exactly recovered with only some uniformly random subsample of its entries.
Sparse distance constraints also arise in graph drawing and manifold learning, where graph distances or neighborhood distances are used as proxies for latent geometric structure.
As mentioned before, it is very useful to be able to visualize metric spaces by embedding them into \(\ell_2\) of very low dimension, e.g. the plane or 3d space. Unfortunately, under the most straightforward measures of quality like distortion or `stress' (as commonly used in statistics), finding an optimal embedding is NP-hard \cite{matouvsek2010inapproximability}, \cite{demaine2021multidimensional}.

Matrix completion can be written as a convex optimization problem, and so is polynomial-time solvable; although as with the other questions in this thesis, we are most interested in how to make algorithms that are efficient enough for massive data sets. In this vein, recent work has been able to establish specialized routines to complete matrices no slower than the time it takes to verify some low-rank factorization (\cite{kelner2023matrix}, \cite{kelner2024semi}). We think that using some ideas from this recent work will allow for similarly efficient algorithms for Euclidean distance matrix recovery, where to our knowledge not much is known beyond standard solvers for convex programs.

\chapter[Streaming Ellipsoidal Rounding]{Streaming Algorithms for Ellipsoidal Approximation of Convex Polytopes}

\section{Introduction}

Let $X$ be a centrally symmetric convex body in $\mathbb{R}^d$. We say that an ellipsoid $\cE$ is an \textit{$\alpha$-ellipsoidal approximation for $X$} if $\sfrac{\cE}{\alpha} \subseteq X \subseteq \cE$ (where $\alpha \geq 1$). Calculating ellipsoidal approximations has applications to problems in machine learning and data science, including sampling and volume estimation (see, e.g., \cite{cousins2015bypassing} and \cite{jia2021reducing}), obstacle collision detection in robotics (see \cite{rimon1997obstacle}), differential privacy (see \cite{nikolov2013geometry}), and online learning (see \cite{li2019stochastic}). Further, the ellipsoid
$\cE$ provides a very succinct approximate representation of $X$ -- $\cE$ can be stored using only
$\binom{d+1}{2} = O(d^2)$ floats, while the exact representation of $X$ may be arbitrarily large.

John's theorem \cite{john1948extremum} states that the minimum-volume outer ellipsoid, called \emph{John's Ellipsoid of $X$}, is a $\sqrt{d}$-ellipsoidal approximation when $X$ is symmetric -- that is, when $X = -X$. Similarly, John's theorem implies that the maximum-volume inner ellipsoid also yields a $\sqrt{d}$-ellipsoidal approximation when $X$ is symmetric. Furthermore, the approximation factor of $\sqrt{d}$ given by John's Ellipsoid cannot be improved in the worst case (e.g. for the hypercube or cross-polytope).
John's result can be made algorithmic: Cohen, Cousins, Lee, and Yang designed an highly efficient algorithm for computing the maximum volume interior ellipsoid when $X$ is a symmetric polytope defined by linear constraints, giving a $\sim \sqrt{d}$ approximation  in the offline setting \cite{cohen2019near}.

\noindent\textbf{Streaming Algorithm.} A natural follow-up question is whether a similar approximation guarantee exists for convex polytopes given in the \textit{streaming setting}. Specifically, suppose that we are given points or constraints defining a symmetric convex polytope one-at-a-time. Our goal is to design an algorithm that computes an $\alpha$-ellipsoidal approximation to the polytope and uses as little memory as possible. Such an algorithm will be useful in a memory-constrained environment. For instance, consider a streaming data summarization task in which a user wishes to obtain an approximation of a dataset that is too large to fit in memory. By computing a good ellipsoidal approximation, the user can summarize the dataset in only $\binom{d+1}{2}$ floating point numbers, whereas to store all vertices of the polytope $X$ we may need $nd$ floats. To our knowledge, existing solutions (such as that of \cite{cohen2019near}) require the entire dataset to be stored in memory and therefore cannot be applied to a streaming setting.

\paragraph{Related Works and Applications} The problem of calculating an ellipsoidal approximation to a convex body has been well-studied; see \cite{todd2016minimum} for an overview of the area. The recent paper of \cite{cohen2019near} presents an $\widetilde{O}\p{nd^2}$-time algorithm for computing a $\sim \sqrt{d}$-approximation for $X$ when $X$ is specified by symmetric linear constraints.

The problem of approximating a \textit{non-symmetric} convex hull with an ellipsoid was introduced in \cite{mukhopadhyay2010approximate}.
The authors present a greedy algorithm for this problem and show that its approximation factor is unbounded for every $d \geq 2$. However, they do not provide any upper bounds on the approximation ratio.

Efficiently calculating ellipsoidal approximations has implications to the problem of estimating the volume of a convex body. For instance, it is known (see \cite{cousins2015bypassing}) that if a convex body $X$ satisfies $B_2^d \subseteq X \subseteq R \cdot B_2^d$, then its volume can be approximated in time $\widetilde{O}\p{\max\pc{d^2R^2, d^3}}$ using a procedure known as \emph{Gaussian cooling}. Observe that computing an $\alpha$-ellipsoidal approximation yields a linear transformation $T$ that transforms a convex body $X$ into a position such that $B_2^d \subseteq TX \subseteq \alpha \cdot B_2^d$. Hence, an efficient algorithm to compute an $\alpha$-ellipsoidal approximation for small $\alpha$ yields an efficient algorithm for estimating the volume of a convex polytope.

Ellipsoidal approximations have also been used as \emph{exploration bases} in some online optimization problems. For instance, the work of \cite{li2019stochastic} considers a stochastic linear optimization setting with adversarial corruptions. Here, a learner observes noisy evaluations of a linear function with the goal of maximizing this function over a convex constraint set. The algorithm used in \cite{li2019stochastic} uses an ellipsoidal rounding of the constraint set to construct an exploration basis. The learner then uses this exploration basis to sample actions during its exploration phases. The approximation factor of the rounding plays a role in the expected regret of the algorithm. Thus, a good, efficiently computable ellipsoidal approximation can be used as a black box to obtain a more efficient, lower-regret algorithm for this setting.

It is highly desirable to give a streaming algorithm for problems in compute-constrained scenarios. For instance, the work of \cite{rimon1997obstacle} studies a problem in which a robot must estimate the distance between a collection of obstacles and itself. At a high level, their workflow involves computing an ellipsoidal approximation of the convex hull of the set of obstacles. The robot then calculates the distance between itself and this ellipsoidal approximation. Now, the robot could be operating using a microcontroller or some other device with limited computing capabilities and at the same time, the amount of data may be very large. Therefore, we need a memory-efficient algorithm for this setting.

Finally, a streaming algorithm to compute ellipsoidal approximations would yield an algorithm that can adapt to certain changes in a dataset over time while maintaining a consistent approximation guarantee. For instance, suppose that the robot in the setting of \cite{rimon1997obstacle} has only a limited sight distance. As the robot moves, it acquires knowledge of new obstacles. A streaming algorithm for ellipsoidal approximations would allow the robot to quickly update its summary of the set of obstacles as it moves.

\subsection{Main Result}

In this paper, we study the following formalization of the \emph{Ellipsoidal Approximation Problem}, stated below.
\begin{problem}[Ellipsoidal Approximation Problem]
\label{problem:ellipsoidal_approx_general}
Given a symmetric convex body $X$, find an ellipsoid $\cE$ so that $\sfrac{\cE}{\alpha} \subseteq X \subseteq \cE$ for some $\alpha \geq 1$. The goal is to find an approximation with a small value of $\alpha$. We say that $\cE$ is an $\alpha$-approximation to $X$.
\end{problem}

We study the Ellipsoidal Approximation Problem in the following streaming model. We assume that points $x_1, \dots, x_n \in X$ arrive one-by-one. After the algorithm receives point $x_t$, it must output an ellipsoid $\cE_t$ centered at the origin such that all points $x_1,\dots, x_t$ lie in $\cE_t$. The \emph{approximation factor} of the algorithm is the smallest $\alpha$ such that
$\sfrac{\cE_n}{\alpha} \subseteq \conv\p{\pc{\pm x_1, \dots, \pm x_n}} \subseteq \cE_n$ for every input sequence. The value of $n$ might not be known to the algorithm beforehand.

\begin{problem}[Formal Problem Statement]
\label{problem:main_problem}
We observe a stream of points $x_1, \dots, x_n$ where points arrive one-by-one and $n$ might not be known beforehand. Upon receiving point $x_t$, find an ellipsoid $\cE_t$ such that:
\begin{itemize}
  \item For all $t \in \{1,\dots,n\}$, we have $\conv\p{\pc{\pm x_1, \dots, \pm x_t}} \subseteq \cE_t$;
  \item At the end of the stream, we have $\sfrac{\cE_n}{\alpha} \subseteq \conv\p{\pc{\pm x_1, \dots, \pm x_n}} \subseteq \cE_n$.
\end{itemize}
\end{problem}

The primary results of our work are algorithms presented in Theorem~\ref{thm:main_result_informal}.

\begin{theorem}[Main Result]
  \label{thm:main_result_informal}
  Let $X =\conv\p{\p{\pm x_1, \dots, \pm x_n}}$. Assume that $X$ contains a ball of radius $r$ and is contained in a ball of radius $R$.
  \begin{enumerate}
    \item There is a streaming algorithm that given $r$ and a stream of points $x_1,\dots,x_n$ provides a solution for Problem~\ref{problem:main_problem} with $\alpha = O(\sqrt{d\log\p{\sfrac{R}{r}+1}})$.
    \item There is a streaming algorithm that given $\sfrac{R}{r}$ (but not $r$ and $R$) and a stream of points $x_1,\dots,x_n$ provides a solution for Problem~\ref{problem:main_problem} with $\alpha = O(\sqrt{d\log\p{\sfrac{R}{r}+1}})$.
  \end{enumerate}
  The algorithms run in time $\widetilde{O}\p{nd^2}$ and store $O(d^2)$ floating-point numbers.
\end{theorem}

The ratio \(\sfrac{R}{r}\) gives an upper bound to the \textit{aspect ratio} \(\kappa(X)\), a quantity we formally define in Definition~\ref{def:aspect_ratio}.
We present and analyze the algorithm for item 1 in Theorem~\ref{thm:simple_alg_arb_ellipsoid} and the one for item 2 in  Theorem~\ref{thm:greedy_approx}.
Our results provide a nearly optimal approximation, since $\sqrt{d\log\p{\sfrac{R}{r} +1}}$ is worse than $\sqrt{d}$ (the best possible factor for the offline setting) by only a factor of $O\p{\sqrt{\log\p{\sfrac{R}{r} +1}}}$. Our algorithms store only $O(d^2)$ floats -- this amount of memory is necessary to represent an ellipsoid in $\mathbb{R}^d$.
We note that the algorithm from item 1 is similar to the algorithm for the non-symmetric case of the problem presented in~\cite{mukhopadhyay2010approximate}.

\textbf{Hyperplane representation of a polytope.} Our algorithms also work in the setting where instead of receiving points $x_i$, we receive hyperplanes (or, more precisely, slabs) $\{y:|\langle x_i, y\rangle| \leq 1\}$ defining body $X$. We describe this setting in Appendix~\ref{section:mvie}.

\textbf{Applications to volume estimation.} Our algorithms directly yield concrete results for the problem of estimating the volume of a symmetric convex polytope. In particular, composing our ellipsoidal approximation procedures with the Gaussian cooling algorithm of \cite{cousins2015bypassing} yields an $\widetilde{O}\p{nd^2+d^3\log\p{\sfrac{R}{r}} +d^3}$-time algorithm for approximating the volume of a symmetric convex polytope. This guarantee is comparable to that obtained by using the algorithm of \cite{cohen2019near} as a preprocessing step prior to Gaussian cooling. Furthermore, this improves upon the best guarantee for this problem suggested by \cite{jia2021reducing}, which is $\widetilde{O}\p{nd^{3.2}\mathrm{polylog}\p{\sfrac{R}{r}}}$ time (\cite{jia2021reducing} assumes the KLS hyperplane conjecture but does not require that $X$ be symmetric).

\textbf{Limitations to approximating John's ellipsoid.} We note that one could try to obtain a good ellipsoidal approximation $\cE$ for $X$ in the streaming model by finding a $\beta$-approximation for John's ellipsoid. Such an ellipsoid $\cE$ would provide a $\beta \sqrt{d}$ approximation for $X$. However, we give a lower bound on the approximability of John's ellipsoid for a natural class of streaming algorithms. We show that none of the algorithms in this class can find a better than $\sqrt{d}$-approximation for John's ellipsoid. Thus, the approach of approximating $X$ via approximating John's ellipsoid potentially may only yield an $O(\sqrt{d} \cdot \sqrt{d}) = O(d)$ approximation.

\iffalse
  \edits{\textbf{Converting vertex approximations to ellipsoidal approximations. } In an independent and concurrent work, the authors of \cite{woodruff2022high} study the closely related problem of finding an online core-set for a convex body given as a stream. Specifically, \cite{woodruff2022high} gives an algorithm that, given a stream of vertices, constructs an online core-set $S \subseteq [n]$ with the property that, for $X\vert_S \coloneqq \conv\p{\pc{\pm x_i \colon i \in S}}$, we have $X\vert_S \subseteq X \subseteq \widetilde{O}\p{\sqrt{d\log\p{\sfrac{R}{r}+1}}}\cdot X\vert_S$. They then use the fact that John's Ellipsoid $\cE^{*}_S$ for $X\vert_S$ gives a $\sqrt{d}$-approximation to $X\vert_S$. Finally, they combine this with the $\widetilde{O}\p{\sqrt{d\log\p{\sfrac{R}{r}+1}}}$-approximation for $X\vert_S$ to $X$ to prove that $\cE^{*}_S$ is an $\widetilde{O}\p{d\sqrt{\log\p{\sfrac{R}{r}+1}}}$-approximation to $X$. While this strategy yields a streaming ellipsoidal approximation algorithm in the setting of Problem \ref{problem:main_problem}, our results give an improvement of $\widetilde{\Omega}(\sqrt{d})$ over those of \cite{woodruff2022high}.}
\fi

\paragraph{Independent and concurrent work on convex hull approximation.}
% Independent of and concurrent to our work,
% Woodruff and Yasuda~\cite{woodruff2022high} consider a problem of maintaining a coreset for $\ell_\infty$ subspace embedding: given a stream of points, choose a small subset of them so that the symmetric convex hull of this subset is an $\widetilde{O}(\sqrt{d})$-approximation to the convex hull of the original points (the hidden constant in $\widetilde{O}$ depends on the online condition number $\kappa^{\mathrm{OL}}$ of the dataset; see~\cite{woodruff2022high} for details).
% This problem and the one we study are closely related. In both of them, we need to maintain an approximation for the symmetric convex hull of a stream of points. However, Woodruff and Yasuda approximate the convex hull with a convex hull of a small coreset, while we approximate the convex hull with an ellipsoid. The approximation guarantees for both problems have $O(\sqrt{d})$ dependence on $d$ and additionally depend on $\sqrt{\log \kappa^{\mathrm{OL}}}$ and $\sqrt{\log \kappa}$, respectively, which are related but different parameters.
% Also, by applying John's Theorem, the authors of \cite{woodruff2022high} get an $\widetilde{O}\p{d}$ ellipsoidal approximation to the convex hull (which is a weaker result than the one we get).
Independent of and concurrent to our work,
Woodruff and Yasuda~\cite{woodruff2022high} consider a problem of maintaining a coreset for $\ell_\infty$ subspace embedding: given a stream of points, choose a small subset of them so that the symmetric convex hull of this subset is a good approximation to the convex hull of the original points.
This problem and the one we study are closely related. In both of them, we need to maintain an approximation for the symmetric convex hull of a stream of points. However, Woodruff and Yasuda approximate the convex hull with a convex hull of a small coreset, while we approximate the convex hull with an ellipsoid. The approximation guarantee of \cite{woodruff2022high} is $O(\sqrt{d\log\p{n\kappaol(X)}}$, where $\kappaol(X)$ is the online condition number.
It has the same dependence on $d$ as our guarantee; $\kappaol(X)$ is closely related to but different from the parameter $\kappa(X)$  we use.
Also, by applying John's Theorem, the authors of \cite{woodruff2022high} get an $O(d\sqrt{\log\p{n\kappaol(X)}})$ ellipsoidal approximation to the convex hull. Our approximation guarantee is better than this one by a factor of $\sim \sqrt{d}$ and does not depend on $n$.

\paragraph{A Proof Sketch of the Main Result.} We now give a proof outline of Theorem \ref{thm:main_result_informal}.

Our first algorithm, Algorithm~\ref{alg:greedy_maxev}, maintains an ellipsoid $\cE_t$ covering $\conv\p{\pc{\pm x_1, \dots, \pm x_t}}$. Initially, $\cE_0$ is simply a ball of radius $r$. Upon receiving point $x_{t+1}$, the algorithm updates $\cE_t$ to $\cE_{t+1}$ by computing the minimal volume ellipsoid containing $\cE_t$ and $\pm x_{t+1}$.

We need to prove that this algorithm achieves an $\alpha = O(\sqrt{d\log\p{\sfrac{R}{r} +1}})$ approximation. By construction, $\cE_n$ contains $X$. Now we need to prove that $\cE_n \subseteq \alpha \cdot X$. We first prove a  different statement. We compare ellipsoid $\cE_n$ not to $X$ but rather to an ellipsoid $\cE^{*}$ that contains $X$. We show that $\cE_n \subseteq \alpha' \cE^{*}$ for every ellipsoid $\cE^{*}$ such that (i) $\cE^{*}$ contains $X$ and (ii) $\cE^{*}$ has aspect ratio (the ratio of its longest to shortest semi-axis) at most $\sfrac{R}{r}$.
To this end, we define a potential function $\Phi$ (see Definition~\ref{def:potential}) with the following properties:
\begin{itemize}
  \item $\cE_n \subseteq \alpha' \cE^{*}$ for $\alpha'=O(\sqrt{\Phi})$ (see Lemma~\ref{lemma:phi_control_max_sv})
  \item Initially, $\Phi$ is $O(d\log\p{\sfrac{R}{r} +1})$ (see Lemma~\ref{lemma:simple_alg_initial}).
  \item The value of the potential function is non-increasing over time (see Lemma~\ref{lemma:invariants_basic}).
\end{itemize}
These properties imply that
$\cE_n \subseteq \alpha' \cE^{*}$ for $\alpha'=O(\sqrt{d\log\p{\sfrac{R}{r} +1}})$.
Then we prove in Theorem~\ref{thm:simul_ellipsoid_approx}, that this implies that $\cE_n\subseteq \alpha X$ with $\alpha = \sqrt{2}\alpha'$, as required.

For this algorithm to perform well, it must know $r$ or a reasonable estimate for $r$.
However, if we do not have any estimate on $r$, the performance of the algorithm may be arbitrarily bad.
On the technical level, the challenge is that the initial value of potential $\Phi$ may be arbitrarily large.

Algorithm~\ref{alg:general_greedy_maxev} does not need to know $r$ but instead needs to have an estimate $\xi$ for the aspect ratio of $X$. Algorithm~\ref{alg:general_greedy_maxev} updates $\cE_i$ in two steps: first it performs the update step from Algorithm~\ref{alg:greedy_maxev} and then ensures that the aspect ratio of the obtained ellipsoid is roughly at most $\xi$ (if it is more than that, it expands the semiaxes of the ellipsoid appropriately). The analysis of Algorithm~\ref{alg:general_greedy_maxev} is based on that of Algorithm~\ref{alg:greedy_maxev} but is substantially more complex. We use a pair of potential function $S$ and $P$ and keep track of their evolutions over time.

\paragraph{Outline} The rest of our paper is organized as follows. In Section~\ref{section:ellipsoid_preliminaries_notation}, we present definitions and notation used in this paper. In Section~\ref{section:natural_attempt}, we describe the first algorithm from Theorem~\ref{alg:greedy_maxev}. The algorithm itself is very simple but its analysis is insightful and  captures the core technical ideas used later. In Section \ref{section:revised_algorithm}, we present the second algorithm from Theorem~\ref{alg:general_greedy_maxev}.
The analyses of Theorems~\ref{alg:greedy_maxev} and \ref{alg:general_greedy_maxev} relies on the fact that every centrally symmetric convex body is well approximated by an intersection of ellipsoids with bounded aspect ratio. We prove this fact in Appendix \ref{section:ellipsoidal_approximations}. In Appendix~\ref{section:tracking_mvoe}, we present our lower bound on the approximability of John's ellipsoid. Finally, in Appendix~\ref{section:mvie}, we discuss the equivalence between the problem we study and an alternate formulation wherein we receive linear constraints one-at-a-time instead of points.

\section{Preliminaries and Notation}
\label{section:ellipsoid_preliminaries_notation}

\paragraph{Notation} Consider a sequence of points $\pc{x_1,\dots, x_n} \subset {\mathbb R}^d$. We denote the symmetric convex hull of the first $t$ points by $X_t=\conv\p{\pc{\pm x_1, \dots, \pm x_t}}$
and the symmetric convex hull of all points $\{\pm x_i\}$ by $X = X_n$. 
We denote the standard Euclidean norm of a vector $v$ by $\|v\|$ and the Frobenius norm of a matrix $A$ by $\|A\|_F^2$. We denote the singular values of a matrix $A \in \R^{d\times d}$ by $\sigma_1(A),\dots, \sigma_d(A)$.
Let $\sigma_{\max}(A)$ and $\sigma_{\min}(A)$ be the largest and smallest  singular values of $A$, respectively. We say that $X$ is centrally symmetric if $X=-X$.

Denote the $\ell_p$-unit ball by $B_p^d = \{x\in \R^d: \|x\|_p \leq 1\}$.   Given a set $S \subseteq \mathbb{R}^d$, its polar is $S^\circ \coloneqq \{ y \in \mathbb{R}^d \colon \sup_{x \in S} | \langle x, y \rangle | \leq 1 \}$. We use natural logarithms unless otherwise specified. 

In this paper, we will work extensively with ellipsoids. We will always assume that all ellipsoids and balls we consider are centered at the origin; we will not explicitly state that. We use the following representation of ellipsoids.
For a non-singular matrix $A\in\R^{d\times d}$, let $\cE_A \coloneqq \pc{x \colon \| \mA \vx \| \le 1}$. In other words, matrix $A$ defines a bijective map  of $\cE_A$ to the unit ball $B_2^d$.
Every ellipsoid (centered at the origin) has such a representation.
We note that this representation is not unique as matrices $A$ and $MA$ define the same ellipsoid if matrix $M$ is orthogonal (since $\|Av\| =\|MAv\|$ for every vector $v$). Now consider the singular value decomposition of  $A$: $A = U\Sigma^{-1} V^T$ (it will be convenient for us to write $\Sigma^{-1}$ instead of standard $\Sigma$ in the decomposition). The diagonal entries of $\Sigma$ are exactly the semi-axes of $\cE_A$. As mentioned above, matrices $U\Sigma^{-1} V^T$ and $U'\Sigma^{-1} V^T$ define the same ellipsoid for any orthogonal \(U' \in \mathbb{R}^{d \times d}\); in particular, every ellipsoid can be represented by a matrix of the form $A=\Sigma^{-1} V^T$.

Our goal is to design an algorithm for Problem \ref{problem:main_problem} that achieves a good approximation $\alpha$ and at the same time uses as little memory as possible. To understand what value of $\alpha$ is achievable in the offline case, recall John's Theorem.

{\begin{theorem}[John's Theorem, \cite{john1948extremum}]
\label{thm:johns}
I. Let $X$ be a centrally symmetric convex body. Consider the minimum volume ellipsoid $\cE$ containing $X$. Then we have $\sfrac{\cE}{\sqrt{d}} \subseteq X \subseteq \cE$.

II. There exists a centrally symmetric body $X$ in ${\mathbb R}^d$ (e.g. hypercube  $B_\infty^d$ and cross-polytope $B_1^d$) such that
there is no ellipsoid $\cE$ that approximates $X$ within a factor 
of $\alpha < \sqrt{d}$: $\sfrac{\cE}{\sqrt{d}} \subseteq X \subseteq \cE$.
\end{theorem}}

%%we already say this in the Intro
%Theorem~\ref{thm:johns} suggests an optimal \textit{offline} algorithm for Problem~\ref{problem:main_problem}: an algorithm to estimate John's Ellipsoid (e.g. that given in \cite{pmlr-v99-cohen19a}) achieves an $\alpha=\sqrt{d}$ approximation
%and no better approximation is possible in the worst case.
%Hence, as we design a streaming algorithm for Problem~\ref{problem:main_problem}, we would like $\alpha$ to be close to $\sqrt{d}$.

In this work, we consider a natural class of one-pass streaming algorithms that we call \emph{monotonic algorithms}.

\begin{definition}[Monotonic Algorithm]
We call an algorithm for Problem \ref{problem:main_problem} \emph{monotonic} if it produces a sequence of ellipsoids $\cE_t$ satisfying $X_t \subseteq \cE_t$ and $\cE_{t} \subseteq \cE_{t+1}$ for all timestamps $t$. 
\end{definition}

Monotonic algorithms have the advantage that once they decide
that a certain point $x$ belongs to ellipsoid $\cE_t$, they commit to this decision: all consecutive ellipsoids $\cE_{t+1}, \cE_{t+2}, \dots$ also contain point $x$.
The approximation factor of our algorithms depend sublogarithmically on the \textit{aspect ratio} of the convex body $X$. 

\begin{definition}[Aspect Ratio]\label{def:aspect_ratio}
Consider a centrally symmetric convex body $X$. Let $r$ be the radius of the largest ball $r\cdot B_2^d$ contained in $X$ and $R$ be the radius of the smallest ball $R\cdot B_2^d$ that contains $X$.
Then the aspect ratio of $X$ is written as $\kappa(X) = \sfrac{R}{r}$.
\end{definition}

Logarithmic dependences on the aspect ratio have previously appeared for algorithms on convex bodies; for example, the algorithms in \cite{jia2021reducing} for rounding and computing the volume of a convex body have a runtime that depends on \(\log \kappa(X)\).
We also recall the condition number of a matrix:

\begin{definition}[Condition Number of a Matrix]
The condition number $\kappa(A)$ of a symmetric nonsingular matrix $A$ is the ratio of its largest to smallest singular values:
$\kappa(A) = \sfrac{\sigma_{\max}(A)}{\sigma_{\min}(A)}$.
\end{definition}
The notions of aspect ratio of a convex body and condition number of a matrix are closely related. It is immediate that the aspect ratio of an ellipsoid $\cE$ equals the ratio of its longest to shortest semi-axes. Consequently, $\kappa(\cE_A) = \kappa(A)$.

\section{Scale-Dependent Algorithm for Ellipsoid Approximation}
\label{section:natural_attempt}

In this section, we present and analyze a simple algorithm for Problem \Cref{problem:main_problem}; see \Cref{alg:greedy_maxev}. This algorithm must be given a radius $r$ such that $r \cdot B_2^d \subseteq X$.
The approximation guarantee of the algorithm linearly depends on 
$\sqrt{\log\p{\sfrac{R}{r}}}$ where $R = \max_t \|x_t\|$ is the radius of the smallest ball that contains $X$.
\begin{algorithm}\label{alg:greedy_maxev}
\caption{Streaming Ellipsoidal Approximation -- Scale-Dependent Algorithm}
\begin{algorithmic}[1]
    \Require A stream of points $x_1, \dots, x_n$ and a value $r$ such that: \begin{align*}
        r \cdot B_2^d \subset X = \conv \pc{\pm x_1, \dots, \pm x_n}
    \end{align*}
    \Ensure Ellipsoid $\cE_n$ that covers $X$.
    \State Initialize $\cE_0 = r \cdot B_2^d$.
    \For{$t = 1, \ldots, n$}
        \State Read point $x_t$ from the stream
        \State Let $\cE_t$ be the ellipsoid of smallest volume (centered at $0$) that contains both $\cE_{t-1}$ and $x_t$
        \label{line:simple_update_rule}
    \EndFor
    \State \Return $\cE_n$
\end{algorithmic}
\end{algorithm}

The key line in the algorithm is Line \ref{line:simple_update_rule} which updates \(\cE_t\) to contain \(x_t\); we  refer to it as the ``update rule.''
Claim \ref{claim:minimum_volume_ellipsoid_update}, which we prove in Appendix~\ref{sec:rank-one-update}, shows how to compute the update.
{\begin{claim}
\label{claim:minimum_volume_ellipsoid_update}
Given a matrix $A_{t-1}$ for $\cE_{t-1}$, the updated matrix 
$A_{t}$ for $\cE_{t}$ can be updated using the following formula: $A_t = \widehat{A} A_{t-1}$, where
\begin{equation}
\label{eq:A-hat}
\widehat{A} = \p{I - \p{1 - \frac{1}{\|A_{t-1}x_t\|}}\p{\frac{\p{A_{t-1}x_t}\p{A_{t-1}x_t}^T}{\|A_{t-1}x_t\|^2}}}.
\end{equation}
\end{claim}}

We first show that ${\cal E}_n$ provides a good approximation to every ellipsoid \({\cal E}^{*}\) with aspect ratio $\sfrac{R}{r}$ containing $X$.
{\begin{theorem}
\label{thm:simple_alg_arb_ellipsoid}
Let ${\cal E}^{*} \supseteq X$ be an ellipsoid containing $X$ with aspect ratio at most $\sfrac{R}{r}$.
Then the output of Algorithm \ref{alg:greedy_maxev} $\cE_n$ satisfies:
\begin{align*}
    {\cal E}_n \subseteq \alpha \cdot \cE^{*}
\end{align*}
where $\alpha = O\p{\sqrt{d\p{\log\p{\sfrac{R}{r}}+1}}}$. Algorithm \ref{alg:greedy_maxev} runs in time $O(nd^2)$ and stores at most $O(d^2)$ floating point numbers.
\end{theorem}}

As stated, \Cref{thm:simple_alg_arb_ellipsoid} does not say that $\cE_n$ provides an
$\alpha$ approximation for $X$. However, the statement of \Cref{thm:simple_alg_arb_ellipsoid} holds simultaneously for all ellipsoids 
${\cal E}^{*}$ whose aspect ratio is at most that of \(X\).
As the following theorem shows, this is sufficient to get the desired result that $\cE_n \subseteq \alpha\sqrt{2}\cdot X$.
% $\cE_n \subseteq (\sqrt{2} \alpha) {\cE^{*}}$.

{\begin{theorem}
\label{thm:simul_ellipsoid_approx}
Consider a centrally symmetric convex body $X$ and an ellipsoid $\cE$. 
Assume that every ellipsoid $\cE^{*}$ that satisfies properties (i) and (ii)
\begin{center}
(i) $\cE^{*}$ contains $X$ \quad and \quad (ii) $\cE^{*}$ has aspect ratio of at most $\kappa(X)$
\end{center}
also contains $\cE$. Then $\cE \subseteq \sqrt{2} \cdot X$.
\end{theorem}}
We prove \Cref{thm:simul_ellipsoid_approx} in \Cref{section:ellipsoidal_approximations}.
Combining these theorems, we have \Cref{thm:simple_alg}.

{\begin{theorem}
\label{thm:simple_alg}
Algorithm~\ref{alg:greedy_maxev} gets an $\alpha = O\p{\sqrt{d\log\p{\sfrac{R}{r} + 1}}}$ approximation:
$
    \sfrac{\cE_n}{\alpha} \subseteq X \subseteq \cE_n
$.
\end{theorem}}

We focus now on proving \Cref{thm:simple_alg_arb_ellipsoid}.
\begin{proof}[Proof of \Cref{thm:simple_alg_arb_ellipsoid}]
We first analyze the runtime and memory complexity, then prove correctness.
% TODO Fix
\paragraph{Runtime and Memory Complexity}
The key observation is that Algorithm \ref{alg:greedy_maxev} only stores the matrix $A_t$ representing the ellipsoid $\cE_t$ between iterations and does not require additional memory within an iteration. Hence, the memory complexity of Algorithm \ref{alg:greedy_maxev} is $O(d^2)$. Next, observe that computing the update rule as per Claim \ref{claim:minimum_volume_ellipsoid_update} requires only three matrix-vector products, which takes time $O(d^2)$. It immediately follows that the runtime of Algorithm \ref{alg:greedy_maxev} is $O(nd^2)$, as desired.

% TODO Fix
\paragraph{Correctness}
We assume without loss of generality that for all $t\geq 1$, $\|A_{t-1}x_t\| > 1$, since Algorithm \ref{alg:greedy_maxev} ignores all points $x_t$ with $\|A_{t-1}x_t\| \leq 1$. In particular, when Algorithm \ref{alg:greedy_maxev} encounters such a point, it simply lets $A_t = A_{t-1}$.
Additionally, we assume that the shortest semi-axis of $\cE^{*}$ is at most $R$.
If not, we prove the statement for $\cE^{**} = R\cdot B_2^d$ (which is contained in $\cE^{*}$ by our assumption)  and get $\cE_n \subseteq  \alpha \cE^{**} \subseteq \alpha \cE^{*}$, as required.

Let $\mJ^*$ be a matrix that defines the ellipsoid $\cE^{*}$: $\cE^{*} = \pb{x \colon \pn{\mJ^* x} \le 1}$. Since all points $x_t$ lie in $X \subseteq \cE^{*}$, we have $\pn{\mJ^* x_t} = \pn{\mJ^* (-x_t)} \le 1$.
Since the shortest semi-axis of $\cE^{*}$ is at most $R$, we have \(\sigma_{\max}(\mJ^*) \geq \sfrac{1}{R}\).
Further, as \(r \cdot B_2^d \subseteq X \subseteq \cE^{*}\), we have \(\sigma_{\max}(\mJ^*) \leq \sfrac{1}{r}\).

Now we  define a potential function $\Phi_{\mJ^*}(\cE_t)$. We will show that the value of this function does not increase over time. 
We will then upper bound the approximation factor $\alpha$ in terms of $\Phi_{\mJ^*}(\cE_n)$.

\begin{definition}[Potential Function $\Phi_{\mA^*}\p{\cdot}$]
\label{def:potential}
We define the potential function $\Phi_{\mJ^*}(A_t)$ as:
\begin{align*}
S_t &\defeq \|\mJ^* \mA_t^{-1}\|_F^2 = \sum_{i=1}^d \sigma_i(\mJ^* A_t)^2\\
P_t &\defeq 2\log\det\p{\mJ^*\cdot A_{t}^{-1}} =\log \prod_{i=1}^d \sigma_i(\mJ^* A_t)^2 \\
\Phi_{\mJ^*}(A_t) &\defeq S_t - P_t = \|\mJ^* \cdot A_t^{-1}\|_F^2 - 2\log \det\p{\mJ^* \cdot A_t^{-1}}
\end{align*}
\end{definition}

%If we let $P_t \coloneqq 2\log\det\p{\mJ^*\cdot A_{t}^{-1}}$ and $S_t \coloneqq \|\\|_F^2star \cdot A_{t}^{-1}}^2$, we have \(\Phi_{\mJ^*}(A_t) = S_t - P_t\).
%\(P_t\) has a convenient interpretation as the the volume of \(\cE_t\) in the coordinate system where \(\cE^{*}\) is the unit ball, while \(S_t\) is the sum of the squares of the axes of \(\cE_t\) in this coordinate system.

Let us see how we use $\Phi_{\mJ^*}(\cdot)$ to upper bound the singular values of $\mJ^* A_n^{-1}$ and the approximation factor $\alpha$. 
{\begin{lemma}
\label{lemma:phi_control_max_sv}
\begin{enumerate}
    \item $\sigma_{\mathrm{max}}(\mJ^* \cdot A_{n}^{-1})^2 \le \frac{e}{e-1}\cdot \Phi_{\mJ^*}(A_n)$ 
    \item $\cE_n \subseteq \alpha \cE^{*}$, where $\alpha = \sigma_{\mathrm{max}}(\mJ^* \cdot A_{n}^{-1})$.
\end{enumerate}    
\end{lemma}}
\begin{proof}
1. Let $M=\mJ^* \cdot A_n^{-1}$. Note that $f(t):=t^2 - \log t^2 \geq \frac{e-1}{e}\cdot t^2 > 0$ for $t>0$.
Therefore,
\begin{align*}
    \Phi_{\mJ^*}(A_n) &=
    \sum_{i=1}^d \p{\sigma_{i}^2(M) - \log \sigma_{i}^2(M)} 
    = \sum_{i=1}^d f(\sigma_{i}(M))
    \geq f(\sigma_{\mathrm{max}}(M))
    \geq \frac{e-1}{e}\cdot\sigma_{\mathrm{max}}^2(M)
\end{align*}
as desired.

2. Consider a point $x\in \cE_n$. Then $\|A_n x\|\leq 1$. We get
$$\|\mJ^* x\| = \|\mJ^* A_{n}^{-1} \cdot A_nx\|
\le \sigma_{\max}\p{\mJ^* \cdot A_{n}^{-1}} \cdot \|A_n x\| 
\leq \alpha.$$
We conclude that $x\in \alpha \cE^{*}$. Thus, $\cE_n \subseteq \alpha \cE^{*}$.
\end{proof}

Let us describe the plan for the rest of the proof.
\begin{itemize}
\item In \Cref{lemma:invariants_basic}, we will prove that $\Phi_{\mJ^*}(A_t) \le \Phi_{\mJ^*}(A_{t-1})$ for all $t\geq 1$; that is, values $\Phi_{\mJ^*}(A_{t-1})$ are non-increasing.
\item In Lemma~\ref{lemma:simple_alg_initial}, we give an upper bound  
$\Phi_{\mJ^*}(A_{0}) \leq O(d \log (\sfrac{R}{r} +1))$. 
\end{itemize}
From Lemma \ref{lemma:phi_control_max_sv}, we get that $\cE_n \subseteq \alpha \cdot \cE^{*}$ with
$$
\alpha \leq O(\sqrt{\Phi_{\mJ^*}(A_{n})}) \leq O(\sqrt{\Phi_{\mJ^*}(A_{0})}) \leq
O(\sqrt{d \log (\sfrac{R}{r} +1)}).
$$ 
It remains to prove Lemmas~\ref{lemma:invariants_basic} and~\ref{lemma:simple_alg_initial}
mentioned above. We start with Lemma~\ref{lemma:invariants_basic}.

{\begin{lemma}
\label{lemma:invariants_basic}
Under the update rule for Algorithm \ref{alg:greedy_maxev}, we have $\Phi_{\mJ^*}(A_t) \le \Phi_{\mJ^*}(A_{t-1})$ for all $t\geq 1$.
\end{lemma}}
\begin{proof}
We first derive formulas that express $P_t$ and $S_t$ in terms of $P_{t-1}$ and $S_{t-1}$.
{\begin{lemma}
\label{lemma:invariants_precorrection}
We have,
\begin{align*}
    P_t &= 2\log\pn{A_{t-1}x_t} + P_{t-1}\\
    S_t &= S_{t-1} + \p{1 - \frac{1}{\pn{A_{t-1}x_t}^2}}\pn{\mJ^*\cdot x_t}^2
\end{align*}
% \begin{align*}
%     P_t &= \pn{A_{t-1}x_t} \cdot P_{t-1}\\
%     S_t &= S_{t-1} + \p{1 - \frac{1}{\pn{A_{t-1}x_t}^2}}\pn{\mJ^*\cdot x_t}^2
% \end{align*}
\end{lemma}}
\begin{proof}
By Claim~\ref{claim:minimum_volume_ellipsoid_update}, $A_t = \widehat{A} A_{t-1}$ where $\hat A$ is given by (\ref{eq:A-hat}).

\paragraph{Determinant Update}
We start by calculating $\det\p{\widehat{A}}$. Note that 
$\p{1 - \frac{1}{\|{A_{t-1}x_t}\|}\p{\frac{\p{A_{t-1}x_t}\p{A_{t-1}x_t}^T}{\pn{A_{t-1}x_t}^2}}}$ is a symmetric rank-1 matrix, whose only non-zero
eigenvalue equals $\p{1 - \frac{1}{\|A_{t-1}x_t\|}}$.
Therefore, the spectrum of $\widehat{A}$ consists of $1$ with multiplicity \(d-1\) and $\pn{A_{t-1}x_t}^{-1}$ with multiplicity \(1\). Thus, $\det\p{\widehat{A}} = \pn{A_{t-1}x_t}^{-1}$ and we have:
\begin{align*}
    \det\p{\mJ^* A_{t}^{-1}}
    = \det\p{\mJ^* A_{t-1}^{-1}\widehat{A}^{-1}}
    = \det\p{\mJ^* A_{t-1}^{-1}} \cdot \det\p{\widehat{A}^{-1}}
\end{align*}% \textcolor{red}{nsm: Fix this typo}
% We use Fact \ref{fact:frobenius_orthonormal}\ynote{we wanted to find a textbook reference for this}, which says that for any linear transformation $A$ and orthonormal matrix $V$ (with $v_1, \dots, v_d$ as its columns):
\paragraph{Frobenius Norm Update} It is well-known (see, e.g., \cite{horn1994topics}) that for any matrix $A$ and orthonormal matrix $\mV$ (with $v_1, \dots, v_d$ as its columns):
\begin{align*}
    \|A\|_F^2 = \|A \mV \|_F^2= \sum_{i=1}^d \pn{Av_i}^2
\end{align*}
Let $V$ be a matrix consisting of the eigenvectors of $\widehat{A}$. Observe that one of these vectors must be $v_1 \coloneqq \sfrac{A_{t-1}x_t}{\pn{A_{t-1}x_t}}$; let $v_2, \dots, v_d$ denote the remaining eigenvectors, all of which have an associated eigenvalue of $1$. Now we calculate:
\begin{align*}
    S_t &= \|\mJ^* \cdot A_{t}^{-1}\|_F^2 = \|\mJ^* \cdot A_{t-1}^{-1}\widehat{A}^{-1}\|_F^2 
%    &= \|\\|_F^2star\cdot A_{t-1}^{-1}\widehat{A}^{-1}V}^2 \\
    = \sum_{i=1}^d \pn{\mJ^*\cdot A_{t-1}^{-1}\widehat{A}^{-1}v_i}^2 \\
    &= \pn{A_{t-1}x_t}^2 \pn{\mJ^*\cdot A_{t-1}^{-1}v_1}^2 + \sum_{i=2}^d \pn{\mJ^*\cdot A_{t-1}^{-1}v_i}^2 \\
    &= \p{\pn{A_{t-1}x_t}^2-1} \pn{\mJ^*\cdot A_{t-1}^{-1}v_1}^2 + \sum_{i=1}^d \pn{\mJ^*\cdot A_{t-1}^{-1}v_i}^2 \\
%    &= \p{\pn{A_{t-1}x_t}^2-1} \pn{\mJ^*\cdot A_{t-1}^{-1}v_1}^2 + \|\\|_F^2star\cdot A_{t-1}^{-1}}^2 \\
    &= \p{\pn{A_{t-1}x_t}^2-1} \pn{\mJ^*\cdot A_{t-1}^{-1} \cdot \frac{A_{t-1}x_t}{\pn{A_{t-1}x_t}}}^2 + \|\mJ^* \cdot A_{t-1}^{-1}\|_F^2 \\
    &= S_{t-1} + \p{1 - \frac{1}{\pn{A_{t-1}x_t}^2}}\pn{\mJ^*\cdot x_t}^2
\end{align*}
\end{proof}

Now are ready to prove Lemma~\ref{lemma:invariant_basic_stronger}, which implies
Lemma~\ref{lemma:invariants_basic}.
In fact, it is stronger than what we need to prove Lemma \ref{lemma:invariants_basic}; however, it will be necessary in the sequel.

{\begin{lemma}
\label{lemma:invariant_basic_stronger}
For all $t\in\{1,\dots, n\}$, we have:
\begin{align*}
    S_t - S_{t-1} &\le \p{P_t - P_{t-1}}\pn{\mJ^* \cdot x_t}^2
\end{align*}
\end{lemma}}
\begin{proof}
From Lemma \ref{lemma:invariants_precorrection}, we have:
\begin{align*}
    \frac{S_t - S_{t-1}}{P_t - P_{t-1}} &= \frac{\p{1-\sfrac{1}{\pn{A_{t-1}x_t}^2}}\pn{\mJ^* \cdot x_t}^2}{2\log\p{\pn{A_{t-1}x_t}}} \le \pn{\mJ^* x_t}^2\quad\quad\\
\end{align*}
where the inequality follows from the fact that \(\frac{\p{1-\sfrac{1}{y^2}}}{2\log y} < 1\) for all \(y > 1\).
We now multiply both sides by $P_t - P_{t-1}$ to obtain \(S_t - S_{t-1} \leq (P_t - P_{t-1}) \cdot \| J^{*} x_t\|^2\).
\end{proof}
As $x_t \in \cE^{*}$, we have $\pn{\mJ^* x_t} \le 1$. Using Lemma \ref{lemma:invariants_basic} and rearranging, we find: \[\Phi_{\mJ^*}(A_t) = S_t - P_t \leq S_{t-1} - P_{t-1} = \Phi_{\mJ^*}(A_{t-1})\]
This concludes the proof of Lemma~\ref{lemma:invariants_basic}.
\end{proof}

{\begin{lemma}
\label{lemma:simple_alg_initial}
We have $\Phi_{\mJ^*}(A_0) \leq d
\p{1 + 4\log\p{\sfrac{R}{r}}}$.
\end{lemma}}
\begin{proof}
Recall that $R = \max_{x \in X} \pn{x}$.

Let $f(t):=t^2 - \log t^2$ as in the proof of 
Lemma~\ref{lemma:phi_control_max_sv}.
By the definition of the potential function, we have
$$
    \Phi_{\mJ^*}(A_0) = \|\mJ^* \cdot A_0^{-1}\|_F^2 - 2\log \det\p{\mJ^* \cdot A_0^{-1}} 
    = \sum_{i=1}^d f(\sigma_i(\mJ^* \cdot A_0^{-1}))
$$
Now we bound the range of $\sigma_i(\mJ^* \cdot A_0^{-1})$. Observe that we have $\cE_0 = r\cdot B_2^d$ and, accordingly, $A_0 = \sfrac{1}{r} \cdot I$. Hence, $\mJ^* \cdot A_0^{-1} =  r \mJ^*$ and $\sigma_i(\mJ^* \cdot A_0^{-1}) = r \sigma_i(\mJ^*)$.
Now recall that by assumption \(\sfrac{1}{R} \leq \sigma_{\max}(\mJ^*) \leq \sfrac{1}{r}\), and the condition number of \(\mJ^*\) is at most \(\sfrac{R}{r}\).
Thus \(\sigma_{\min}(\mJ^*) \geq \frac{\sigma_{\max}(\mJ^*)}{\sfrac{R}{r}} \geq \sfrac{r}{R^2}\), and consequently \(\sfrac{r^2}{R^2} \leq \sigma_i(\mJ^* \cdot A_0^{-1}) \leq 1\).
Now we bound $f(\sigma_i(\mJ^* \cdot A_0^{-1}))$. Since \(f\) is convex,
\begin{align*}
    f(t) \leq \max\pb{f\p{ \sfrac{r^2}{R^2}},f(1)} \leq 1 + 2\log\p{ \sfrac{R^2}{r^2}} = 1 + 4 \log\p{\sfrac{R}{r}} \quad\text{for } t\in \pb{ \sfrac{r^2}{R^2},1}
\end{align*}
We conclude that
\begin{align*}
    \Phi_{\mJ^*}(A_0) = \sum_{i=1}^d f(\sigma_i(\mJ^* \cdot A_0^{-1})) \leq d\p{1 + 4\log\p{\sfrac{R}{r}}}.
\end{align*}
\vspace{-\baselineskip} %\qedhere
\end{proof}
We have proven Lemmas~\ref{lemma:invariants_basic} and~\ref{lemma:simple_alg_initial}. This concludes the proof of Theorem~\ref{thm:simple_alg_arb_ellipsoid}.
\end{proof}

\section{Scale-Independent Algorithm}
\label{section:revised_algorithm}

To use Algorithm~\ref{alg:greedy_maxev}, we need to know some lower bound $r$ on the radius of the largest ball contained in $X$. The approximation guarantee of the algorithm linearly depends on $\sqrt{\log\p{\sfrac{R}{r}}}$, so as long as we have a reasonable estimate on $r$, we can use Algorithm~\ref{alg:greedy_maxev}. However, if we have no prior information about the scale of $X$ and do not have any reasonable estimate $r$, we cannot use Algorithm~\ref{alg:greedy_maxev}.
In this section, we present Algorithm~\ref{alg:general_greedy_maxev} that only requires an upper bound $\xi \geq \kappa(X)$ on the aspect ratio of $X$.
\begin{algorithm}[ht]
\caption{Streaming Ellipsoidal Approximation\label{alg:general_greedy_maxev}}
\begin{algorithmic}[1]
    \Require A stream of points $x_1, \dots, x_n$, and an aspect ratio estimate \(\xi\)
    % , a factor-$q \ge 1$ estimate of $\kappa(X)$.
    \Ensure Matrix $A_n$ that defines ellipsoid $\cE_n = \pc{x \colon \pn{A_nx} \le 1}$.
    %where $X = \conv\p{\pc{\pm x_1,\dots,\pm x_n}} \subseteq \cE_n$.
    \State Receive point $x_1$.
    \State Set $V_1$ to be an orthonormal matrix satisfying $V_1^T x_1 = \pn{x_1} \cdot e_1$.
    \State $U_1 = I$, $\Sigma_1 = \Diag{\pn{x_1}, \sfrac{\pn{x_1}}{\xi}, \dots, \sfrac{\pn{x_1}}{\xi}}$
    \State  \(A_1 = U_1 \Sigma_1^{-1} V_1^T\) % should we remove lines that explicitly compute A_i? no, I addressed that. sounds good
    \For{$t = 2, \ldots, n$}
        \State Receive point $x_t$
        \If{$\pn{A_{t-1} x_t} > 1$}
            \State $a = \sfrac{A_{t-1} x_t}{\pn{A_{t-1} x_t}} \text{ and } b = A_{t-1}^T a$.
            \State $M_t = \max(M_{t-1}, \|x_t\|)$ \quad \textit{(that is, $M_t = \max_{1\leq i \leq t} \|x_i\|$)}
%            \State Compute $U_t, \Sigma_t', V_t^T$ as the SVD of $A_{t-1} - \p{1 - \sfrac{1}{\pn{A_{t-1} x_t}}}ab^T$, where $U_t$ and $V_t^T$ are orthonormal matrices in $\R^{d\times d}$ and $\Sigma_t'$ is a diagonal matrix.
            \State $(U_t, (\Sigma_t')^{-1}, V_t) = \textsc{SVDRankOneUpdate}((U_{t-1}, \Sigma_{t-1}^{-1}, V_{t-1}),  - \p{1 - \sfrac{1}{\pn{A_{t-1} x_t}}}a, b)$
            \label{line:svd_update}
            \State \(\Sigma_t = \Diag{\tau_{1, t}, \ldots, \tau_{d, t}}\), where $\tau_{i,t} = \max([\Sigma'_{t}]_{i i}, \sfrac{M_t}{\xi})$ for every $i \in [d]$
             \label{line:sv_correction}
        \Else
            \State $U_t = U_{t-1}, V_t = V_{t-1}, \Sigma_t = \Sigma_{t-1}$
        \EndIf
        \State  \(A_{t} = U_t \Sigma_t^{-1} V_t^T\)
%%% too slow
    \EndFor
    \State \textbf{Output}: $\cE_n = \pc{x \colon \pn{A_n x}\le 1}$
\end{algorithmic}
\end{algorithm}
We present Algorithm~\ref{alg:general_greedy_maxev} with its full implementation details.
However, conceptually the only difference between Algorithms~\ref{alg:greedy_maxev} and~\ref{alg:general_greedy_maxev} is a new ``singular value correction step'', presented on line~\ref{line:sv_correction}. As in Algorithm~\ref{alg:greedy_maxev}, we perform the basic update rule (line~\ref{line:svd_update}) -- compute the minimum volume ellipsoid $\cE'_t$ containing the current ellipsoid $\cE_{t-1}$ and the new point $x_t$.
Then, we increase the semi-axes of $\cE'_t$ (if necessary) so that all of them are at least $M_t/\xi$, where $M_t = \max(\|x_1\|,\dots,\|x_t\|)$ is the length of the longest vector we have received so far. This ensures that the matrix $A_t$ is well-conditioned. 

Formally, we compute matrix $A_t'=\widehat{A} A_{t-1}$ where $\widehat A$ is given by (\ref{eq:A-hat}), and update its singular values.
To implement this efficiently, we use a procedure 
$$(U', \Sigma', V') = \textsc{SVDRankOneUpdate}((U, \Sigma, V), y, z),$$
which gives \(U', \Sigma', V'\) as the SVD of the matrix \(U \Sigma V^T + y z^T\).
Per \cite{stange2008efficient}, this can be implemented in time $O(d^2\log^2 d)$.
As noted in Section \ref{section:ellipsoid_preliminaries_notation}, 
the \(U_t\)s have no effect on the definition of \(\mathcal{E}_t\)s and thus will not factor into our analysis. In fact, the algorithm may discard the value of $U_t$ after it computes the SVD for $A_t$ and later use the identity matrix instead of $U_t$. However, we keep them in the algorithm so that the exposition is closer to that of Algorithm~\ref{alg:greedy_maxev}.
%Observe the effect of Line \ref{line:sv_correction} in Algorithm \ref{alg:general_greedy_maxev} is to correct the singular values of $A_t$ in order to keep it well-conditioned. We will refer to this this process as the ``singular value correction step.''

The remainder of this section is devoted to proving Theorem~\ref{thm:greedy_approx}.% \ynote{I would just assume that we are given $K\geq \kappa(X)$; that is, $q\geq 1$ and get rid of $q$.}

{\begin{theorem}
\label{thm:greedy_approx}
% For all ellipsoids $\cE^{*}$ containing $X$ with aspect ratio at most $\xi$, Algorithm \ref{alg:general_greedy_maxev} outputs an ellipsoid $\cE_n$ satisfying:
% \begin{align*}
%     \cE_n \subseteq  \sqrt{\magicnumber{\xi}} \cdot\cE^{*}
% \end{align*}
Algorithm \ref{alg:general_greedy_maxev} outputs an ellipsoid $\cE_n$ satisfying:
\begin{align*}
    \frac{\cE_n}{\sqrt{6 + 28d \log \xi + 16d}
    %\sqrt{2}\cdot\sqrt{\magicnumber{\xi}}
    } \subseteq X \subseteq \cE_n
\end{align*}
assuming $\xi \ge \kappa(X)$. 
Moreover, Algorithm \ref{alg:general_greedy_maxev} runs in time $O(nd^2\log^2d)$ and stores at most $O(d^2)$ floats.
\end{theorem}}
\begin{proof}
We first compute the running time and then prove the correctness of the algorithm.

\paragraph{Runtime and Memory Complexity}
It is easy to see that Algorithm \ref{alg:general_greedy_maxev} keeps track of three matrices $U_t, \Sigma_t, V_t$ at every iteration, in addition to the new point $x_t$. The algorithm does not explicitly store the matrices $A_t = U_t \Sigma_t^{-1}V_t$. Instead, when 
it is asked to compute $A_t w$ or $A_t^Tw$, it simply consecutively performs 3 matrix-vector multiplications.
Algorithm \ref{alg:general_greedy_maxev} does not store anything else in between iterations, so the memory complexity is $O(d^2)$ floating-point numbers, as desired.

We now analyze the running time. 
%Observe that for all $w\in\R^d$, $U_{t-1}\Sigma_{t-1}^{-1}V_{t-1}^Tw$ can be computed in time $O(d^2)$ as three matrix-vector products in succession. 
The matrix-vector multiplications require time $O(d^2)$. With the rank-one update, each iteration takes time $O(d^2\log^2d)$. 
%Finally, observe that Algorithm~\ref{alg:general_greedy_maxev} never needs to explicitly calculate $A_t$. 
Hence, Algorithm \ref{alg:general_greedy_maxev} has time complexity $O(nd^2\log^2d) = \widetilde{O}\p{nd^2}$ as desired.  

\paragraph{Correctness}

As in the analysis of Algorithm~\ref{alg:greedy_maxev}, it will in fact be enough to show that for all ellipsoids $\cE^{*}$ that cover $X$ and have aspect ratio at most $\xi$, we have:
\begin{align*}
    \cE_n \subseteq \sqrt{3+14d\log\xi + 8d} \cdot \cE^{*}
\end{align*} 
Then, by Theorem \ref{thm:simul_ellipsoid_approx}, we will have that the intersection of all ellipsoids $\cE^{*}$ covering $X$ with aspect ratio at most $\xi$ is itself a $\sqrt{2}$-approximation to $X$. Hence, $\cE_n$ will be a $\sqrt{2}\cdot\sqrt{3+14d\log\xi + 8d}$ to $X$, as required.
Similarly, as in the analysis of Algorithm~\ref{alg:greedy_maxev}, we assume without loss of generality that for all $t$, $\|A_{t-1}x_t\| > 1$. 

\iffalse
Observe that the singular value correction step does not change the convex body that the algorithm is covering -- in particular, we have that if $z_i \notin \cE_t'$ (and therefore adding $z_i$ would cause $\cE_t'$ to change), then $z_i \in X$. To see this, let $r$ and $R$ be the largest and smallest values, respectively, such that $r \cdot B_2^d \subseteq X \subseteq R \cdot B_2^d$. Then:
\begin{align*}
    \pn{z_i} = \frac{M_t}{\xi} \le \frac{M_t}{\sfrac{R}{r}} = \frac{M_t}{R} \cdot r \le r
\end{align*}
as required.
\fi

Observe that Lemma \ref{lemma:phi_control_max_sv} implies that it is sufficient to analyze $\sigma_{\max}\p{\mJ^* \cdot A_{n}^{-1}}$ for our choice of $\mJ^*$ defining the ellipsoid $\cE^{*} = \pc{x\colon\pn{\mJ^* x}\le 1}$ satisfying $\kappa(\cE^{*})\le\xi$. Specifically, our goal is now to show that $\sigma_{\max}\p{\mJ^* \cdot A_{n}^{-1}}^2 \le 3+14d\log\xi + 8d$.

Define $Q_t = \max_{1\le i \le t} \pn{\mJ^* x_i}^2$. Clearly, $Q_1\leq  \dots \leq Q_n$. Now we prove a version of Lemma \ref{lemma:invariant_basic_stronger} that applies to Algorithm \ref{alg:general_greedy_maxev}. We give the proofs of all the lemmas stated below in Appendix~\ref{sec:moved_proofs}.
\begin{lemma}
\label{lemma:invariants}
Under the update rule for Algorithm \ref{alg:general_greedy_maxev}, we have the following

\begin{itemize}
\item For any two timestamps $u$, $t$, such that $1 \leq u \leq t \leq n$:
\begin{equation}
\label{eqn:multi_step_p_inv}
    S_t \leq S_u + Q_t \cdot (P_t - P_u)
\end{equation}
\item For any timestamp \(2 \leq t \leq n\),
\begin{equation}
\label{eqn:single_step_inv}
    S_t \leq S_{t-1} + d \cdot Q_t
\end{equation}
\end{itemize}
\end{lemma}

Let $\sigma_{i,j}$ be the $i$th singular value of $\mJ^* A_{j}^{-1}$ and $\sigma_{\max, j} = \max_i \sigma_{i,j}$.
By Lemma~\ref{lemma:phi_control_max_sv} (part 2), $\alpha \leq \sigma_{\max, n} \leq \sqrt{\sum_{i=1}^d \sigma_{i,n}^2}$, hence upper bounding \(S_n = \sum_{i=1}^d \sigma_{i,n}^2\) is sufficient to finish the analysis.
We do this by bounding $S_1$ and then applying Lemma~\ref{lemma:invariants}, as described below.

\begin{lemma}
\label{lemma:s_initial_value}
We have $S_1 \le d$.
\end{lemma}

By applying the equations from Lemma \ref{lemma:invariants} in sequence, we can relate \(S_n\) to \(S_1\). 
Only applying (\ref{eqn:single_step_inv}) repeatedly is insufficient, as then the resulting upper bound on \(S_n - S_1\) grows linearly with \(n\).
Even though (\ref{eqn:multi_step_p_inv}) is efficient for large stretches where \(t\) is much later than \(u\), it also does not give a good bound when  \(Q_t\) is more than a constant factor larger than \(Q_u\) (this phenomenon may not be apparent from the equation itself, but becomes clear from the upcoming analysis).
In order to obtain the desired bound on $S_n$, we split the \(Q_t\)s into contiguous groups such that \(Q_t\) does not increase significantly within any group, and differs by at least \(e\) outside the groups.

\begin{lemma}\label{lem:bound-on-S_n}
After running Algorithm \ref{alg:general_greedy_maxev} for $n$ steps, we have
\begin{align*}
    S_n = \sum_{i=1}^d \sigma_{i,n}^2 \leq 3+14d\log\xi + 8d
\end{align*}
\end{lemma}

Finally, we write $\sigma_{\max, n}^2 \le \sum_{i=1}^d \sigma^2_{i,n} \le 3+14d\log\xi + 8d$.
Recall that this is sufficient to conclude the proof of \Cref{thm:greedy_approx} -- specifically, since we have $\sigma_{\max, n} = \sigma_{\max, n}\p{\mJ^* \cdot A_n^{-1}} \le \sqrt{3+14d\log\xi + 8d}$, we can invoke part 2 of Lemma~\ref{lemma:phi_control_max_sv} to arrive at $\cE_n \subseteq \sqrt{3+14d \log\xi + 8d}\cdot\cE^{*}$. 
\end{proof}

% \appendix

\section[Proof of minimum-volume ellipsoid update]{Proof of \Cref{claim:minimum_volume_ellipsoid_update}}
\label{sec:rank-one-update}
Note that the volume of the ellipsoid determined by $A_t$ is proportional to $\det\p{A_t^{-1}}$. Therefore, $A_t$ is the solution to the following optimization problem, where we use that the volume of the ellipsoid determined by $A_t$ is proportional to $\det\p{A_t^{-1}}$.
\begin{align*}
    \max \det\p{A_t} \text{ such that } A_t \preceq A_{t-1} \text{ and } \pn{A_tx_t} \le 1
\end{align*}
Additionally, since $\det\p{AB} = \det\p{A}\cdot\det\p{B}$, we have that this objective is invariant under linear transformations. It thus follows that our objective can be rewritten as:
\begin{align*}
    &\max \det\p{A_t} &\text{ such that } &A_t \preceq A_{t-1} \text{ and } \pn{A_tx_t} \le 1 \\
    \equiv &\max \det\p{A_t \cdot A_{t-1}^{-1}} &\text{ such that } &A_t \cdot A_{t-1}^{-1} \preceq I \text{ and } \pn{\p{A_t\cdot A_{t-1}^{-1}}A_{t-1}x_t} \le 1 \\
    \equiv &\max \det\p{\widehat{A}} &\text{ such that } &\widehat{A} \preceq I \text{ and } \pn{\widehat{A}A_{t-1}x_{t}} \le 1
\end{align*}
where the last line follows from using the intermediate variable $\widehat{A} = A_t\cdot A_{t-1}^{-1}$.

In other words, after the transformation, the problem is equivalent to finding the minimum volume ellipsoid that contains (i) the unit ball and (ii) point $A_{t-1}x_t$. Geometrically, it is clear what the optimal ellipsoid for this problem is: one of its semi-axes is $A_{t-1}x_t$; all others are orthogonal to $A_{t-1}x_t$ and have length 1 (this can be formally proved using symmetrization). However, we do not use this observation and derive a formula for $\widehat{A}$ using linear algebra.

We first give an upper bound on the objective value of the above optimization problem. Since $\widehat{A} \preceq I$, we have that all its singular values must be at most $1$. Additionally, since $1 \ge \pn{\widehat{A}A_{t-1}x_t} \ge \sigma_{\min}\p{\widehat{A}} \cdot \pn{A_{t-1}x_t}$, we have that at least one singular value of $\widehat{A}$ must be $\le \sfrac{1}{\pn{A_{t-1}x_t}}$. Putting everything together and using the fact that the determinant is the product of the singular values gives $\det\p{\widehat{A}} \le \sfrac{1}{\pn{A_{t-1}x_t}}$.

We now show that there exists a setting of $\widehat{A}$ that achieves this upper bound. Let $v_1$ be a unit vector in the direction of $A_{t-1}x_t$ and $v_2, \dots, v_d$ complete the orthonormal basis for $\R^d$ from $v_1$, and write \(
    \widehat{A} = \frac{1}{\pn{A_{t-1}x_t}}v_1v_1^T + \sum_{i=2}^d v_iv_i^T
\). We will show that $\widehat{A}$ satisfies the constraints imposed by the optimization problem. Since we have $\pn{A_{t-1}x_t} \ge 1$ (as we impose that $x_t \notin \cE_{A_{t-1}}$), the fact that $\widehat{A} \preceq I$ follows immediately. For the second constraint, we write:
\begin{align*}
    \pn{\widehat{A}A_{t-1}x} = \pn{\p{\frac{1}{\pn{A_{t-1}x_t}}v_1v_1^T + \sum_{i=2}^d v_iv_i^T}A_{t-1}x_t} = \pn{\frac{A_{t-1}x_t}{\pn{A_{t-1}x_t}}} = 1
\end{align*}
Furthermore, it is easy to see that $\det\p{\widehat{A}} = \sfrac{1}{\pn{A_{t-1}x_t}}$, which achieves our upper bound.

Finally, recall that we wrote $\widehat{A} = A_t\cdot A_{t-1}^{-1}$; rearranging this gives us what we want.

\section[Proof of ellipsoidal approximation theorem]{Proof of \Cref{thm:simul_ellipsoid_approx}: Approximating Convex Polytopes with Ellipsoids}
\label{section:ellipsoidal_approximations}

In this section, we prove \Cref{thm:simul_ellipsoid_approx}.
To this end, we show that every centrally symmetric convex polytope $X$ is well-approximated by the intersection of all ellipsoids $\cE$ containing $X$ with comparable aspect ratio.
This will immediately imply Theorem~\ref{thm:simul_ellipsoid_approx}.
Let $q \geq \sfrac{1}{\kappa(X)}$ and $\delta = \sqrt{1+\sfrac{1}{q^2}} - 1$. 
Define 
$$
E_{q} = \pc{\cE \colon X \subseteq \cE \text{ and } \kappa(\cE) \le q\cdot\kappa(X)}
\quad \text{and}\quad
\cA_q = \bigcap_{\cE \in E_q}\cE 
$$
We show that $\cA_q$ provides a good approximation for $X$.
\begin{lemma}
\label{thm:ellipsoid_intersections_general}
We have \(
    \frac{1}{1 + \delta}\cdot\cA_q \subseteq X \subseteq \cA_q
\).
\end{lemma}
\begin{proof}
The first inclusion $X \subseteq \cA_q$ is trivial, since all ellipsoids $\cE$ in $E_q$ contain $X$. We now prove the the first inclusion: 
\begin{equation}\label{eq:AqisinX}
cA_q\subseteq (1+\delta) X.
\end{equation}
Consider the set of all slabs of the form $\{x: |\langle a,x\rangle| \leq 1\}$ that contain $X$.
Since $X$ is a centrally symmetric convex body,
the intersection of all the slabs in $S$ equals $X$.
Further, let set $S'$  consist of the slabs in $S$ expanded by a factor of $(1+\delta)$: for every slab $|\langle a,x\rangle| \leq 1$ in $S$, there is a slab $|\langle a,x\rangle| \leq (1+\delta)$ in $S'$.
Then the intersection of  slabs in $S'$ equals $(1+\delta)X$.
Thus, to prove inclusion~(\ref{eq:AqisinX}), it is sufficient to prove that $|\langle a,x\rangle| \leq (1+\delta)$ in $S$.
To this end, we construct an ellipsoid $\cE$ in $E_q$ that 
lies in the slab $|\langle a,x\rangle| \leq (1+\delta)$
(and contains $\cA_q$, by the definition of $\cA_q$).

We complement vector $\sfrac{a}{\|a\|}$ to an orthonormal basis for $\R^d$:
$w_1 = \sfrac{a}{\|a\|},w_2,\dots,w_d$. 
Recall that the condition number $\kappa(X)$ equals the ratio of the radius $r$ of the largest inscribed ball to the radius $R$ of the smallest circumscribed ball. Since the width of slab $|\langle a,x\rangle| \leq 1$ containing $X$ is $\sfrac{2}{\|a\|}$, we have $r \leq \sfrac{1}{\|a\|}$. Accordingly, $R\leq \sfrac{\kappa(X)}{\|a\|}$.
Therefore, for $x\in X\subseteq R \cdot B_2^d$, 
$\sum_{i=1}^d \langle w_i, x\rangle^2 = \|x\|^2 \leq R^2 \leq 
\p{\sfrac{\kappa(X)}{\|a\|}}^2$. 
Also note that 
$\langle w_1, x\rangle^2 = \sfrac{\pgo{a,x}^2}{\pn{a}^2} \leq \sfrac{1}{\pn{a}^2}$. 
We are ready to define ellipsoid $\cE$:
$$\cE = \left\{x: \langle w_1,x\rangle^2 + \sum_{i=2}^d \frac{\pgo{w_i,x}^2}{q^2\kappa(X)^2} \le \frac{1}{\|a\|^2}\Bigl(1 + \frac{1}{q^2}\Bigr)\right\}.$$

Now we verify that (i) $\cE\in E_q$ and (ii) all points in $\cE$ satisfy
$|\langle a,x\rangle| \leq (1+\delta)$.
First, note that the aspect ratio of $\cal E$ is $q\kappa(X)$.
Using the bounds we derived above, we get that for all $x\in X$
$$\langle w_1,x\rangle^2 + \sum_{i=2}^d \frac{\pgo{w_i,x}^2}{q^2\kappa(X)^2}
\leq  \frac{1}{\|a\|^2}  + 
\frac{\kappa(X)^2}{\|a\|^2} \cdot \frac{1}{q^2\kappa(X)^2}
= \frac{1}{\|a\|^2}\Bigl(1 + \frac{1}{q^2}\Bigr).
$$
Therefore, all points from $X$ lie in $\cE$; that is, $X\subseteq \cE$. We conclude that $\cE \in E_q$, as required.

Finally, if $x\in \cE$, then 
$$\langle x, a\rangle = \sqrt{\|a\|^2 \langle w_1,x\rangle^2} \leq \sqrt{1 + \sfrac{1}{q^2}} = 1 + \delta.$$
This concludes the proof.
\end{proof}

To get \Cref{thm:simul_ellipsoid_approx}, we apply \Cref{thm:ellipsoid_intersections_general} with $q = 1$.
Clearly, an ellipsoid \(\cE\) that is contained in every \(\cE^{*} \in E_1\) is also contained in \(\cA \subseteq \sqrt{2} \cdot X\).

\section{Proofs from Section~\ref{section:revised_algorithm}}
\label{sec:moved_proofs}
For the purposes of our analysis, we will ``simulate'' the singular value correction step using the following procedure. Let $w_{1}, \dots, w_{d}$ be the the $i$-th column of $V_t$ (note that the $w_i$-s
are unit vectors that are the directions of the semi-axes of $\cE_t$).
%semi-axes of our ellipsoid $\cE_t$ at the current timestamp $t$; $w_i$ is directed in the same direction as the $i$-th column of $V_t$.
Let $\cE_t'$ be the ellipsoid obtained prior to Line \ref{line:sv_correction}.
i.e., $\cE_t' = \pc{x\colon\pn{(\Sigma_t')^{-1}V_t^Tx}\leq 1}$.
We create ``ghost'' points $z_{1}, \dots, z_{d}$: $z_{i} = \tau_{i,t} w_i = \max([\Sigma'_{t}]_{i i}, 
\sfrac{M_t}{\xi}) w_i $ (see line \ref{line:sv_correction} in Algorithm \ref{alg:general_greedy_maxev}).
Note that $X$ contains the ball of radius $M_t/\xi$ centered at $0$, since the aspect ratio of $X$ is at most $\xi$. Thus, each point $z_i$ either lies in 
$X$ (if $\|z_i\| = \sfrac{M_t}{\xi}$) or 
in $\cE_t'$ (if $\|z_i\| = [\Sigma'_{t}]_{i i}$).
We finally start with matrix $A_{t-1}$ and consecutively apply the update rule from Algorithm~\ref{alg:greedy_maxev} for each of the points $x_t, z_{1}, \dots, z_{d}$. 
%(The update step will not perform any updates for points $z_i$ already in $\cE_t'$ and will update the ellipsoid only when $z_i \in X$.)

Next, we show Lemma~\ref{lemma:sv_update_step_simulation}, which states that this simulation of the singular value correction step yields $A_t$, the same matrix that we obtain when we perform the update rule from Algorithm~\ref{alg:general_greedy_maxev}.

\begin{lemma}
\label{lemma:sv_update_step_simulation}
The process described above yields matrix $A_t$.
\end{lemma}
\begin{proof}
Consider the executions of Algorithms \ref{alg:greedy_maxev} and \ref{alg:general_greedy_maxev}.
After Algorithm~\ref{alg:greedy_maxev} processes point $x_t$ and Algorithm \ref{alg:general_greedy_maxev} executes Line \ref{line:svd_update}, both algorithms are in the same state. Namely they store matrix $A_{t}'$ given by Claim~\ref{claim:minimum_volume_ellipsoid_update}, 
\begin{align*}
    A_t' = A_{t-1} - \p{1 - \frac{1}{\pn{A_{t-1}x_t}}}\p{\frac{\p{A_{t-1}x_t}\p{A_{t-1}x_t}^T}{\pn{A_{t-1}x_t}^2}}A_{t-1}
\end{align*}
It remains to show that executing Line \ref{line:sv_correction} in Algorithm \ref{alg:general_greedy_maxev} is equivalent to injecting these ``ghost'' points $z_1,\dots,z_d$ into Algorithm \ref{alg:greedy_maxev}.

It is sufficient to consider the effect of injecting one point $z_i$. Assume we started with matrix $A$ and obtained matrix $A'$ by injecting $z_i$. Observe that if $\sigma_i(A) \leq \sfrac{1}{\|z_i\|}$ (that is, $z_i$ lies in the ellipsoid defined by $A$; in particular, if $z_i \in \cE_t'$), then $A'=A$.
We prove that if $\sigma_i(A) > \sfrac{1}{\|z_i\|}$, then $A' = U{\Sigma'}^{-1}V^T$ where $A = U\Sigma^{-1}V^T$, entry $\Sigma_{ii}' = \|z_i\|$, and all other entries of $\Sigma'$ are equal to the corresponding entries of $\Sigma$. We have,
$$
    A' = \widehat{A} A = \left(I - \p{1 - \frac{1}{\pn{A z_i}}}\p{\frac{\p{Az_i}\p{Az_i}^T}{\pn{Az_i}^2}}\right)A 
$$
Since $z_i = \tau_{i,t} w_{i}$, we have $V^T z_i = \|z_i\| e_i = \tau_{i,t} e_i$. Accordingly, 
$Az_i = U\Sigma^{-1} \tau_{i,t} e_i =  \frac{\tau_{i,t}}{\Sigma_{ii}} Ue_i$.
Thus, $\widehat{A} = I - (1 - \frac{\Sigma_{ii}}{\tau_{i,t}}) Ue_i e_i^T U^T$.
Since $U$ is an orthogonal matrix, $UU^T = I$ and thus 
$$A'=\left(I - \p{1 - \frac{\Sigma_{ii}}{\tau_{i,t}}} Ue_i e_i^T U^T\right)U\Sigma^{-1}V^T = 
U \biggl(\underbrace{\Sigma^{-1} \left(I-\p{1 - \frac{\Sigma_{ii}}{\tau_{i,t}}} e_ie_i^T\right)}_{\Sigma'^{-1}}\biggr)V^T.$$
Note that $I-\bigl(1 - \frac{\Sigma_{ii}}{\tau_{i,t}}\bigr) e_ie_i^T$ is a diagonal matrix; all of its diagonal entries are equal to 1 except for the $i$-th diagonal entry, which is $\sfrac{\Sigma_{ii}}{\tau_{i,t}}$.
We get that $\Sigma'^{-1}$ differs from $\Sigma^{-1}$ only in the $i$-th diagonal entry: $(\Sigma'^{-1})_{ii} = \sfrac{1}{\tau_{it}}$, as required.
\end{proof}

\begin{proof}[Proof of Lemma~\ref{lemma:invariants}]
By Lemma~\ref{lemma:sv_update_step_simulation}, we can break the evolution of our potential functions into two main steps: that after Algorithm~\ref{alg:greedy_maxev} gets $x_t$ and that after Algorithm~\ref{alg:greedy_maxev} gets the ghost points.
Let $S_t'$, $P_t'$, $Q_t$ represent $\pn{\mJ^*\cdot (A_t')^{-1}}_F^2$, $2\log\det\p{(A_t')^{-1}}$, and $\max_{i\in[t]} \pn{\mJ^* x_i}^2$, respectively, where $A_t'$ is the matrix defined in Lemma~\ref{lemma:sv_update_step_simulation}.

By Lemma \ref{lemma:invariant_basic_stronger}, for all $t$, we have $S_t' - S_{t-1} \le (P_t' - P_{t-1}) \cdot Q_t$. Similarly, as \(\|\mJ^* x_t\| \leq 1\) and \(\|A_{t-1} x_t\| > 1\), we have:
\begin{align*}
    S_t' - S_{t-1} = \p{1 - \frac{1}{\pn{A_{t-1}x_t}^2}}\pn{\mJ^* x_t}^2 \le \p{1 - \frac{1}{\pn{A_{t-1}x_t}^2}}Q_t \le Q_t
\end{align*}
We now analyze the singular value correction step. By Lemma~\ref{lemma:sv_update_step_simulation},it can be simulated by adding the ghost points $z_1,\dots,z_d$.
%as defined in the proof of Lemma \ref{lemma:sv_update_step_simulation}. 
First, observe that we only need to analyze points $z_i$ with  $z_i = \sfrac{M_t}{\xi} w_{i}$ (because other points are in $\cE_t'$ and do not cause any update). We now show that $\pn{\mJ^* z_i} \le \max_{i \in [t]} \pn{\mJ^* x_t} = \sqrt{Q_t}$. We have:
\begin{align*}
    \pn{\mJ^* z_i} = \frac{M_t}{\xi} \cdot \pn{\mJ^* w_i} \le M_t \cdot \frac{\sigma_{\max}\p{\mJ^*}}{\xi} \le \sigma_{\min}\p{\mJ^*} \cdot M_t
\end{align*}
Let $p \coloneqq \arg\max_{i \in [t]} \pn{x_i}$. Then:
\begin{align*}
    \pn{\mJ^* z_i} \le \sigma_{\min}\p{\mJ^*} \cdot M_t = \sigma_{\min}\p{\mJ^*} \cdot \pn{x_p} \le \pn{\mJ^* x_p} \le \max_{i \in [t]} \pn{\mJ^* x_t} = \sqrt{Q_t}.
\end{align*}
Consider the state of Algorithm~1 in the simulation after it gets points $z_1,\dots,z_i$.
Let $A_{t,i}'$ be the resulting state matrix, and $S_{t,i}'$ and $P_{t,i}'$ be the values of functions $S$ and $P$. Observe that $S_{t,0}' = S_{t}'$ and $P_{t,0}' = P_{t}'$ and that $S_t = S_{t,d}'$ and $P_t = P_{t,d}'$. We now invoke Lemma~\ref{lemma:invariant_basic_stronger} repeatedly:
\begin{align*}
    S_{t,0}' - S_{t-1} &\le (P_{t,0}' - P_{t-1}) \cdot Q_t \\
    S_{t,1}' - S_{t,0}' &\le (P_{t,1}' - P_{t,0}') \cdot Q_t \\
    &\vdots \\
    S_{t,i}' - S_{t,i-1}' &\le \p{P_{t,i}' - P_{t,i-1}'} \cdot Q_t \\
    &\vdots \\
    S_{t} - S_{t,d-1}' &\le \p{P_{t} - P_{t,d-1}'} \cdot Q_t
\end{align*}
Adding all these inequalities yields $S_{t}-S_{t-1} \le \p{P_t-P_{t-1}}\cdot Q_t$. Since numbers $Q_t$ are non-decreasing, this implies that for all $u \le t$, we have $S_t \leq S_u + Q_t \cdot (P_t - P_u)$.

Similarly, we repeatedly write:
\begin{align*}
    S_{t,0}' - S_{t-1} &\le Q_t \\
    S_{t,1}' - S_{t,0}' &\le Q_t \\
    &\vdots \\
    S_{t,i}' - S_{t,i-1}' &\le Q_t \\
    &\vdots \\
    S_{t} - S_{t,d-1}' &\le Q_t
\end{align*}
Note that at least one of the semi-axes of $\cE_t'$ has length at least $M_t$
(since all points $x_1,\dots, x_t$ are in $\cE_t'$). That is, $[\Sigma']_{ii} \geq M_t \geq M_t/\xi_i$ for some point $z_i$.
This means that we do not perform any updates for $z_i$ and $S_{t,i}' = S_{t,i-1}'$. Summing up the inequalities above and taking into account that $S_{t,i}' = S_{t,i-1}'$ for at least one $i$ yields $ S_t \leq S_{t-1} + d \cdot Q_t$, as required.
\end{proof}

\begin{proof}[Proof of Lemma~\ref{lemma:s_initial_value}]
Consider ellipsoid $\cE_1$. It has semi-axes $\tilde w_{1}=x_1$ and some $\tilde w_{2},\dots, \tilde w_{d}$. Since $\|\tilde w_{i}\| = \sfrac{\|x_1\|}{\xi}$ for $i \geq 2$ (see step 5 of Algorithm~\ref{alg:general_greedy_maxev}), all points $\tilde w_{i}$ lie inside $X\subset \cE^{*}$. In particular, $\|\mJ^* \tilde w_{i}\| \leq 1$. Finally, note that $A_1 \tilde w_{i} = e_i$.
We have:
\begin{align*}
   S_1 = \|\mJ^* A_1^{-1}\|_F = \sum_{i=1}^d \|\mJ^* A_1^{-1} e_i\|^2 = 
\sum_{i=1}^d \|\mJ^* \tilde w_i\|^2 \leq d 
\end{align*}
\end{proof}

\begin{proof}[Proof of Lemma~\ref{lem:bound-on-S_n}]
Set $t_0 = n$, and define $t_i \in \{1, \ldots, n\}$ for $i \geq 1$ recursively as follows:
\begin{enumerate}
    \item Set \(t_i = \max \{j \in \{1, \ldots, n\} \colon Q_j < \sfrac{Q_{t_{i-1}}}{e} \}\)
    \item If there is no such \(j\), then finish.
\end{enumerate}
Let \(t_0 > \ldots > t_m\) be the indices defined by this process. Observe that for each \(0 \leq i \leq m\), we have \(Q_{t_i} \leq e^{-i}\). Further, for each \(0 \leq i \leq m -1\), we have \(\sfrac{Q_{t_i}}{Q_{t_{i+1} +1}} \leq e\), and \(\sfrac{Q_{t_m}}{Q_1} \leq e\).

Now we apply bound (\ref{eqn:multi_step_p_inv}). For every \(0 \leq i \leq m - 1\), we have
$$S_{t_i} \leq S_{t_{i+1} + 1} + Q_{t_i} (P_{t_i} - P_{t_{i+1} + 1})$$
From bound (\ref{eqn:single_step_inv}), we get 
$$S_{t_{i+1} + 1} \leq S_{t_{i+1}} + d \cdot Q_{t_{i+1} + 1}.$$
Combining these equations, we obtain for each \(0 \leq i \leq m - 1\):
\begin{equation}
\label{eqn:combined_inv}
S_{t_i} - S_{t_{i+1}} \leq d \cdot Q_{t_{i+1} + 1} + Q_{t_i} \cdot (P_{t_i} - P_{t_{i+1} + 1})
\end{equation}
Now we add up inequalities (\ref{eqn:combined_inv}) for all \(0 \leq i \leq m-1\) and get:
\[S_{t_m} - S_{t_0} = \sum_{i=0}^{m-1} \left(S_{t_i} - S_{t_{i+1}}\right) \leq d \cdot \sum_{i=0}^{m-1} Q_{t_{i+1} + 1} + \sum_{i=0}^{m-1} Q_{t_{i}} \cdot P_{t_i} + \sum_{i=0}^{m-1} Q_{t_{i}} \cdot \left(- P_{t_{i+1} + 1}\right)\]
%The left hand side is a telescoping sum that simplifies to \(S_{t_0} - S_{t_m}\).
By (\ref{eqn:multi_step_p_inv}) we have \(S_{t_m} - S_1 \leq Q_{t_m} \cdot (P_{t_m} - P_1)\). Combining this with the previous inequality, we get
\begin{equation}
\label{eqn:final_bound}
S_{n} - S_1 \leq d \cdot \underbrace{\sum_{i=0}^{m-1} Q_{t_{i+1} + 1}}_{\text{I}} + \underbrace{\sum_{i=0}^{m} Q_{t_{i}} \cdot P_{t_i}}_{\text{II}} + \underbrace{\left(-Q_{t_m} \cdot P_1\right) + \sum_{i=0}^{m-1} Q_{t_{i}} \cdot \left(- P_{t_{i+1} + 1}\right)}_{\text{III}}
\end{equation}
Now, we bound each individual term in (\ref{eqn:final_bound}).
For I, observe that as \(t_{i+1} + 1 \leq t_i\), we have \(Q_{t_{i+1}+1} \leq Q_{t_i} \leq e^{-i}\). Thus
\[\sum_{i=0}^{m-1} Q_{t_{i+1} + 1} \leq \sum_{i=0}^{m-1} e^{-i} \leq \sum_{i=0}^\infty e^{-i} = \frac{e}{e-1}\]
For II, we have \(P_{t_i} \leq P_{t_0}\) for all \(i\), therefore
\[\sum_{i=0}^m Q_{t_i} P_{t_i} \leq P_{t_0} \cdot \sum_{i=0}^m Q_{t_i} \leq P_{t_0} \cdot \frac{e}{e-1}\]

Now we bound III. First, we prove the following lemma to bound terms of the form \(- Q_t \cdot P_u\):

\begin{lemma}
\label{lemma:prod_qp_bound}
Consider indices \(u \leq t\). If \(\sfrac{Q_{t}}{Q_u} \leq e\), then
\[- Q_t \cdot P_u \leq Q_t \cdot \left(d + 4d \log \xi + d \log\left(\frac{1}{Q_t}\right)\right)\]
\end{lemma}

\begin{proof}
By definition, we have \(-P_u = - \sum_{i=1}^d \log(\sigma^2_{i, u})\). Next, notice that for all $t\in\{1,\dots,u\}$:
\begin{align*}
    \sigma_{\min}\p{\mJ^* \cdot A_{u}^{-1}} &\ge \sigma_{\min}\p{\mJ^*} \cdot \sigma_{\min}\p{A_{u}^{-1}} \\
    &\ge \sigma_{\min}\p{\mJ^*} \cdot \frac{\pn{x_t}}{\xi}\quad\quad\text{since the correction step ensures that } \sigma_{\min}\p{A_u^{-1}} \ge \frac{\pn{x_t}}{\xi} \\
    &\ge \frac{\sigma_{\max}\p{\mJ^*}}{\xi}\cdot \frac{\pn{x_t}}{\xi} \\
    &\ge \frac{\pn{\mJ^* x_t}}{\xi^2}
\end{align*}
Since this is true for all $t \in [u]$, we can maximize the RHS over $t \in [u]$, and we obtain \(\sigma_{i, u}^2 \geq \sfrac{Q_u}{\xi^4}\). Hence, \(-P_u \leq d \log\p{\sfrac{1}{Q_u}} + 4d \log \xi\). As \(\sfrac{1}{Q_u} \leq \sfrac{e}{Q_t}\), we get \(-P_u \leq d + 4d \log \xi + d \log\p{\sfrac{1}{Q_t}}\). Multiplying by \(Q_t\) gives the claim.
\end{proof}

Applying Lemma \ref{lemma:prod_qp_bound} to each term in III, we obtain
\[-Q_{t_m} \cdot P_1 + \sum_{i=0}^{m-1} Q_{t_{i}} \cdot - P_{t_{i+1} + 1} \leq (d + 4d \log \xi) \cdot \sum_{i=0}^m Q_{t_i} + d \cdot \sum_{i=0}^m Q_{t_i} \log\left(\frac{1}{Q_{t_i}}\right)\]
As before, we can bound the first term with \(\sum_{i=0}^m Q_{t_i} \leq \frac{e}{e-1}\).
For the second term, observe that \(y \mapsto y \log(\sfrac{1}{y})\) is increasing on \(\pb{0,\sfrac{1}{e}}\). Thus, \(Q_{t_i} \log(\sfrac{1}{Q_{t_i}}) \leq e^{-i} \log(\sfrac{1}{e^{-i}}) = e^{-i} \cdot i\) for \(i \geq 1\). The maximum of \(y \log\p{\sfrac{1}{y}})\) is \(\sfrac{1}{e}\); therefore \(Q_{t_0} \log(\sfrac{1}{Q_{t_0}}) \leq \sfrac{1}{e}\).
Thus we can bound the second term by 
$$d \cdot \left(\frac{1}{e} + \sum_{i=1}^{m} i \cdot e^{-i}\right) \leq d \cdot \left(\frac{1}{e} + \sum_{i=1}^{\infty} i \cdot e^{-i}\right) = d \cdot \p{\frac{1}{e} + \frac{e}{(1-e)^2}}.$$
To summarize, we bound term III by
\[d \cdot \left(\frac{1}{e} + \frac{e}{e-1} + \frac{e}{(1-e)^2}\right) + 4d \log \xi \cdot \frac{e}{e-1}\]

Combining the bounds on the terms of (\ref{eqn:final_bound}) results in:
\[
    S_n - S_1 \leq d \cdot \left(\frac{1}{e} + \frac{2 \cdot e}{e-1} + \frac{e}{(1-e)^2}\right) + P_{t_0} \cdot \frac{e}{e-1} + 2d \log \xi \cdot \frac{e}{e-1}
\]
% Rearranging, we obtain:
% \[
%     S_n -  P_{n} \cdot \frac{e}{e-1} \leq S_1 + d \cdot \left(\frac{1}{e} + \frac{2 \cdot e}{e-1} + \frac{e}{(1-e)^2}\right) + 2d \log K \cdot \frac{e}{e-1}
% \]
Rearranging and applying \Cref{lemma:s_initial_value}, we obtain
\begin{equation}
\label{eqn:final_sn_pn}
    S_n -  P_{n} \cdot \frac{e}{e-1} \leq  d \cdot \left(1 + \frac{1}{e} + \frac{2 \cdot e}{e-1} + \frac{e}{(1-e)^2}\right) + 2d \log \xi \cdot \frac{e}{e-1}
\end{equation}
Observe that
\(S_n - P_n \cdot \frac{e}{e-1} = \sum_{i=1}^d (\sigma^2_{i, n} - \frac{e}{e-1} \log(\sigma^2_{i, n}))\). As \(y - \frac{e}{e-1}\log y \geq \frac{e-2}{e-1}\, y\) for all \(y > 0\), we get
\[
\sum_{i=1}^d \sigma^2_{i, n} \leq \frac{e-1}{e-2} \left(S_n - P_n \cdot \frac{e}{e-1}\right)
\]

Using (\ref{eqn:final_sn_pn}) and replacing constants with their integer ceilings, we finish with
\[\sum_{i=1}^d \sigma^2_{i,n} \leq 3+14d\log \xi + 8d\]
\end{proof}

\section{Tracking the Minimum-Volume Outer Ellipsoid}
\label{section:tracking_mvoe}

Observe that the guarantee of Theorem \ref{thm:greedy_approx} gives a guarantee similar to that given by John's Theorem for centrally-symmetric convex bodies. Therefore, a natural question is, ``how closely can any one-pass monotonic algorithm approximate the minimum-volume outer ellipsoid for a centrally-symmetric convex body?'' We formalize this notion below.

\begin{definition}[Approximation to Minimum Volume Outer Ellipsoid]
We say a streaming algorithm $A$ $\alpha$-approximates the minimum volume outer ellipsoid if $A$ outputs an ellipsoid $\cE_n$ satisfying $\cE_n \subseteq \alpha \cdot J(X)$, where $J(X)$ is the minimum volume outer ellipsoid for $X$.
\end{definition}

Theorem \ref{thm:greedy_approx_tight} asserts that for a natural class of streaming algorithms, it is not possible to approximate the minimum volume outer ellipsoid up to factor $< \sqrt{d}$ in the worst case.

\begin{theorem}
\label{thm:greedy_approx_tight}
Every one-pass monotonic deterministic streaming algorithm for Problem \ref{problem:main_problem} has approximation factor to the minimum volume outer ellipsoid of at least $\sqrt{d}$, for infinitely many $d$.
\end{theorem}
\begin{proof}
Before we begin, recall that a Hadamard basis is a set of vectors $v_1, \dots, v_d$ such that:
\begin{itemize}
    \item $\pn{v_i} = 1$, for all $i \in [d]$;
    \item For all $i \neq j$, $\pgo{v_i,v_j}=0$;
    \item Every entry of $v_i$ is in $\pc{\pm \sfrac{1}{\sqrt{d}}}$.
\end{itemize}

Our family of hard instances proceeds in two phases.

\paragraph{Phase 1} Let $d$ be such that there exists a Hadamard basis for $\R^d$. Consider a corresponding Hadamard basis $v_1, \dots, v_d$. The adversary gives the algorithm the points $v_1, \dots, v_d$.

\paragraph{Phase 2} The adversary selects $i \in [d]$ arbitrarily and $\eps \in (0, d-1)$ arbitrarily. They then define the vectors $w_i \defeq e_i \cdot \sfrac{1}{\sqrt{d-\eps}}$ and $w_j \defeq e_j \cdot \sqrt{\sfrac{d-1}{\eps}}$ for all $j \neq i$. The adversary gives the algorithm the points $w_1, \dots, w_d$. Call the outcome here ``Outcome (i).''

It is easy to see that at the end of Phase 1, the minimum volume outer ellipsoid is simply $B_2^d$. Furthermore, the algorithm's solution $\widehat{\cE}$ contains $\conv\p{\pm v_1, \dots, \pm v_d}$. On the other hand, consider the following claim.
\begin{lemma}
The following ellipsoid is the minimum-volume outer ellipsoid for Outcome (i):
\begin{align*}
    \cE_{OPT(i)} = \pc{x \colon 1 \ge \frac{x_i^2}{\p{\sfrac{1}{\sqrt{d-\eps}}}^2} + \sum_{j \neq i}^d \frac{x_j^2}{\p{\sqrt{\sfrac{d-1}{\eps}}}^2}}
\end{align*}
\end{lemma}
\begin{proof}
Notice that all the points $w_j$ are orthogonal. Thus, the minimum-volume outer ellipsoid containing all the $w_j$ must be the one whose axes are along the directions of $w_j$ and whose poles are located on $w_j$. Observe that $\cE_{OPT(i)}$ satisfies this, so it must be the minimum-volume outer ellipsoid for the convex body whose vertices are determined by the $w_j$.

It now remains to show that every Hadamard basis vector is on the surface of $\cE_{OPT(i)}$:
\begin{align*}
    \frac{\p{\sfrac{1}{\sqrt{d}}}^2}{\p{\sfrac{1}{\sqrt{d-\eps}}}^2} + \sum_{j\neq i}^d \frac{\p{\sfrac{1}{\sqrt{d}}}^2}{\p{\sqrt{\sfrac{d-1}{\eps}}}^2} = \frac{1}{d}\p{(d-\eps) + (d-1) \cdot \frac{\eps}{d-1}} = 1
\end{align*}
Since the minimum volume ellipsoid containing all the $w_j$ also contains the Hadamard basis vectors, it (i.e., $\cE_{OPT(i)}$) must be the minimum-volume outer ellipsoid for Outcome (i).
\end{proof}

We will now show that any that outputs an ellipsoid $\widehat{\cE}$ at the end of Phase 1 must have an approximation factor of at least $\sqrt{d-\eps}$ on at least one of Outcomes ($1, \dots, i$). Suppose that in each of Outcome (i), we obtain an ellipsoid $\widehat{\cE_i}$ that satisfies $C \cdot \cE_{OPT(i)} \supseteq \widehat{\cE_i}$. We now have:
\begin{align*}
    \conv\p{\pc{\pm v_1, \dots, v_d}} \subseteq \widehat{\cE} \subseteq \bigcap_{i=1}^d \widehat{\cE_{i}} \subseteq C \cdot \bigcap_{i = 1}^d \cE_{OPT(i)}
\end{align*}
We therefore want to argue about $\widehat{\cE}$ given that it must contain $\conv\p{\pc{\pm v_1, \dots, v_d}}$ and be contained by $C \cdot \bigcap_{i = 1}^d \cE_{OPT(i)}$. Let $A$ be a matrix mapping $\widehat{\cE}$ to the unit ball. Then, notice that we can write for all $i \in [d]$:
\begin{align*}
    \pn{Av_i} &\le 1 & \pn{A \cdot \frac{Ce_i}{\sqrt{d-\eps}}} &\ge 1
\end{align*}
In particular, the rightmost exclusion follows from the fact that $\sfrac{Ce_i}{\sqrt{d-\eps}}$ lies on the boundary of $C \cdot \bigcap_{i=1}^d \cE_{OPT(b.i)}$. Now, recall the well-known fact that for any unitary matrix $W$, we have $\pn{AW}_F^2 = \pn{A}_F^2$ (see, e.g., \cite{horn1994topics}), and observe that we have:
\begin{align*}
    d \ge \sum_{i=1}^d \pn{Av_i}^2 = \pn{AV}_F^2 = \pn{A}_F^2 = \pn{AI}_F^2 = \sum_{i=1}^d \pn{Ae_i}^2 \ge \frac{d(d-\eps)}{C^2}
\end{align*}
Rearranging gives $C \ge \sqrt{d-\eps}$, as desired.
\end{proof}

\section{Dual Problem -- Inner Ellipsoidal Approximation}
\label{section:mvie}

We design our algorithms for Problem \ref{problem:main_problem} in the setting where we receive points $x_1,\dots, x_n$ defining $X=\conv\p{\{\pm x_1,\dots, \pm x_n\}}$. Alternatively, we may define a centrally symmetric convex polytope is by providing a set of its faces, or more generally, a set of slabs of the form $\{x:|a^Tx| \leq 1\}$. Accordingly, we may consider a different online model where 
inequalities $\{x:|a^Tx| \leq 1\}$ arrive one-by-one and the resulting polytope is their intersection. Using the notion of a polar set, we show that this model is essentially equivalent to the model we study in this paper. All our results equally apply to it. 

Thus, another possible formulation for Problem \ref{problem:main_problem} involves the algorithm receiving the linear constraints one-at-a-time instead of points from the body.

In this section, we show that this choice of formulation does not matter. Specifically, an algorithm for one of these variants yields an algorithm for the other. In fact, as written in Table \ref{table:primal_vs_dual}, these problems are dual to one another.

\begin{table}[H]
\centering
\begin{tabular}{|l|l|}
\hline
\textbf{Primal}                                                                                                                                                               & \textbf{Dual}                                                                                                                                                                                                   \\ \hline
\begin{tabular}[c]{@{}l@{}}Find $\cE$ (outer ellipsoid) such that:\\  $\sfrac{\cE}{\alpha} \subseteq X \subseteq \cE$\\ where $X = \conv\p{\pc{\pm x_1,\dots,\pm x_n}}$\end{tabular} & \begin{tabular}[c]{@{}l@{}}Find $\cE$ (inner ellipsoid) such that:\\  $\cE \subseteq Y \subseteq \alpha\cdot\cE$\\ where $Y = \pc{y \colon \pa{\pgo{x_i,y}}\le 1\text{ for all } i\in[n]}$\end{tabular} \\ \hline
\end{tabular}
\caption{The primal and dual version of the ellipsoidal approximation problem.\label{table:primal_vs_dual}}
\end{table}

We first address the duality between the two problems in Table \ref{table:primal_vs_dual}. To do so, observe the following useful facts regarding convex polars.
\begin{itemize}
    \item If $A$ and $B$ are convex bodies, and if $A \subseteq B$, then $B^{\circ} \subseteq A^{\circ}$ \cite[Proposition 7.16(iv)]{bauschke2011convex}.
    \item In Table \ref{table:primal_vs_dual}, $X^{\circ} = Y$, and $Y^{\circ} = X$.
    \item If an ellipsoid $\cE = \pc{x \colon \pn{Ax}\le 1}$, then $\cE^{\circ} = \pc{x \colon \pn{A^{-T}x} \le 1}$ \cite[Definition 2.17]{vishnoi2021algorithms}.
\end{itemize}
The first and third are well-known, and the second follows from the definition of the polar and that if $X$ is closed, convex, and contains the origin, then $(X^{\circ})^{\circ} = X$ \cite[Corollary 7.19(i)]{bauschke2011convex}.

We now put these facts together to show that a solution to the primal problem can be converted to one for the dual problem. First, notice that we have $\cE$ and $\alpha$ such that $\sfrac{\cE}{\alpha} \subseteq X \subseteq \cE$. Using the first fact, we have $\cE^{\circ} \subseteq X^{\circ} \subseteq \p{\sfrac{\cE}{\alpha}}^{\circ}$. Using the second fact, we have $\cE^{\circ} \subseteq Y \subseteq \p{\sfrac{\cE}{\alpha}}^{\circ}$. Finally, using the third fact, we have $\p{\sfrac{\cE}{\alpha}}^{\circ} = \alpha \cdot \cE^{\circ}$, which yields $\cE^{\circ} \subseteq Y \subseteq \alpha\cdot\cE^{\circ}$. A similar argument shows that a solution to the dual yields a solution to the primal.

We now address how a solution to the streaming variant of the primal problem can be converted to a solution to the streaming variant of the dual problem. Specifically, suppose we are in the dual setting, wherein we receive linear constraints one-at-a-time. 
Our task is to find an $\alpha$-ellipsoidal approximation to $Y_t = \pc{y \colon \pa{\pgo{x_i,y}}\le 1 \text{ for all } i \in [t]}$. 
Observe that every incoming linear constraint $\pc{y \colon \pa{\pgo{x_i, y}} \le 1}$ can be treated as an incoming point $\pm x_i$ in the primal space. 
This means that we can apply our algorithm in the primal setting to obtain a solution in the primal space, which for all $t$ gives an ellipsoid such that $X_t \subseteq \cE_t$. We then compute the polar of the outer ellipsoid we obtain in the primal space (i.e., $\cE_t$) to obtain an inner ellipsoid in the dual space (i.e., $\cE_t^{\circ}$), which yields $\cE_t^{\circ} \subseteq Y_t$. As per our previous argument, this preserves the approximation factor -- at the end of the stream, we have $\cE_n^{\circ} \subseteq Y \subseteq \alpha \cdot \cE_n^{\circ}$, as desired.

\chapter[Streaming Ellipsoidal Rounding for General Convex Polytopes]{Near-Optimal Streaming Ellipsoidal Rounding for General Convex Polytopes}
\newcommand{\Gbar}{\overline{\mG}}
\newcommand{\cbar}{\overline{\vc}}

%\mnote{Todo -- check when we're mentioning previous work it doesn't reveal our identities}

\section{Introduction}

We consider the problem of approximating convex polytopes in $\R^d$ with ``simpler'' convex bodies.
Consider a convex polytope $Z \subset \R^d$.
Our goal is to find a convex body $\widehat{Z}\subset \R^d$ from a given family of convex bodies, a translation vector $\vc \in \R^d$, and a scaling factor $\alpha \in (0, 1]$ such that
\begin{align}
    \vc + \alpha \cdot \widehat{Z} \subseteq Z \subseteq \vc + \widehat{Z}\label{eq:intro_approx_general}.
\end{align}
We say that $\widehat{Z}$ is a $\sfrac{1}{\alpha}$-approximation to $Z$; an algorithm that computes $\widehat{Z}$ is a $\sfrac{1}{\alpha}$-approximation algorithm. In this paper, we will be interested in approximating $Z$ with (a) ellipsoids and (b) polytopes defined by small number of vertices.

This problem has many applications in computational geometry, graphics, robotics, data analysis, and other fields (see \cite{agarwal2005geometric} for an overview of some applications).
It is particularly relevant when we are in the big-data regime and  storing polytope $Z$ requires too much memory. In this case, instead of storing $Z$, we find a reasonable approximation $\widehat{Z}$ with a succinct representation and then use it as a proxy for $Z$. In this setting, it is crucial that we use a \textit{low-memory} approximation algorithm to find $\widehat{Z}$.
%extending existing algorithms relying on approximations of convex polytopes to big-data scenarios. memory-efficient %low-memory algorithms 

In this paper, we study the problem of approximating convex polytopes in the streaming model. The streaming model is a canonical big-data setting that conveniently lends itself to the study of low-memory algorithms. 
We assume that $Z$ is the convex hull of points $\vz_1,\dots, \vz_n$: $Z = \conv\p{\{\vz_1,\dots,\vz_n\}}$; the stream of points $\{\vz_1,\dots,\vz_n\}$ contains all the vertices of $Z$ and additionally may contain other points from polytope $Z$.
In our streaming model, points $\vz_1,\dots, \vz_n$ arrive one at a time. At every timestep $t$, we must maintain an approximating body $\widehat{Z}_t$ and translate $\vc_t$ such that
% \begin{align}
%     \vp_t + \alpha_t \cdot X_t' \subseteq \conv\p{\pc{\vx_1,\dots,\vx_t}} \subseteq \vp_t + X_t'.\label{eq:intro_approx_streaming}
% \end{align}
\begin{align}
    \conv\p{\pc{\vz_1,\dots,\vz_t}} \subseteq \vc_t + \widehat{Z}_t.\label{eq:intro_approx_streaming}
\end{align}
Once a new point $\vz_{t+1}$ arrives, the algorithm must compute a new approximating body $\widehat{Z}_{t+1}$ and translation $\vc_{t+1}$ such that the guarantee \eqref{eq:intro_approx_streaming} holds for timestep $t+1$. Finally, after the algorithm has seen all $n$ points, we must have 
\begin{align}
    \vc_n + \alpha \cdot \widehat{Z}_n \subseteq \underbrace{\conv\p{\pc{\vz_1,\dots,\vz_n}}}_{Z} \subseteq \vc_n + \widehat{Z}_n\label{eq:intro_approx_final}
\end{align}
for some $0 < \alpha \le 1$ (where $\sfrac{1}{\alpha}$ is the approximation factor).
Note that the algorithm may not know the value of $n$ beforehand. We consider two types of approximation. 

\paragraph{Ellipsoidal roundings.} In one thrust, we aim to calculate an \textit{ellipsoidal rounding} of $Z$ -- we are looking for ellipsoidal approximation $\widehat Z=\cE$. Formally, we would like to output an origin-centered ellipsoid $\cE$, a center/translate $\vc \in \R^d$, and a scaling parameter $0 < \alpha \le 1$ such that
\begin{align*}
    \vc + \alpha \cdot \cE \subseteq Z \subseteq \vc + \cE.
\end{align*}
Ellipsoidal roundings are convenient representations of convex sets. They have applications to preconditioning convex sets for efficient sampling and volume estimation \cite{jia2021reducing}, algorithms for convex programming \cite{nesterov2008rounding}, robotics \cite{rimon1997obstacle}, and other areas. They also require the storage of at most $\sim d^2$ floating point numbers, as every ellipsoid can be represented with a center $\vc$ and semiaxes $\vv_1,\dots,\vv_{d'}$ for $d' \le d$.

We note that by John's theorem \cite{john1948extremum}, the minimum-volume outer ellipsoid for $Z$ achieves approximation $\sfrac{1}{\alpha} \le d$. Moreover, the upper bound of $d$ is tight, which is witnessed when $Z$ is a $d$-dimensional simplex (that is, the convex hull of $d+1$ points in general position).

We now formally state the streaming ellipsoidal rounding problem.

\begin{problem}[Streaming ellipsoidal rounding]\label{prob:main}
Let \(Z = \conv\p{\pc{\vz_1, \ldots, \vz_n}} \subseteq \R^d\). 
%Assume that the points \(\vz_1, \ldots, \vz_n\) are received one at a time in a stream.
A streaming algorithm \(\cA\)  receives points \(\vz_1, \ldots, \vz_n\) one at a time  and produces a sequence of ellipsoids \(\vc_t + \cE_t\) and scalings \(\alpha_t\).
The algorithm must satisfy the following guarantee at the end of the stream.
\[\vc_n + \alpha_n \cdot \cE_n \subseteq Z \subseteq c_n + \cE_n\]
We say that \(\vc_n + \cE_n\) is an ellipsoidal rounding of \(Z\) with approximation factor \(\sfrac{1}{\alpha_n}\).
\end{problem}

We note that in the special case where $Z$ is centrally symmetric (i.e., $Z = -Z$), there are algorithms with nearly optimal approximation factors  $O(\sqrt{d\log\p{n\kappa^{\mathsf{OL}}}})$ and $O(\sqrt{d\log \kappa})$ due to \cite{woodruff2022high} and \cite{makarychev2022streaming}, respectively (here, $\kappa^{\mathsf{OL}}$ is the online condition number and $\kappa$ is the aspect ratio of the dataset). The running times of these algorithms nearly match those of the best-known offline solutions. However, these algorithms do not work with non-symmetric polytopes and we are not aware of any way to adapt them so that they do. We defer a more detailed discussion of the algorithms for the symmetric case to Section~\ref{sec:rw}. 

\paragraph{Convex hull approximation.} In another thrust, we want to find a translate $\vc \in \R^d$, subset $S \subseteq [n]$, and scale $\alpha$ such that
\begin{align*}
\conv\p{\pc{\vz_i: i \in S}} \subseteq \conv\p{\pc{\vz_1,\dots,\vz_n}} \subseteq  \vc + \frac{1}{\alpha} \cdot \conv\p{\pc{\vz_i - \vc : i \in S}}.
\end{align*}
Note that $\vc + \sfrac{1}{\alpha} \cdot \conv\p{\pc{\vz_i - \vc : i \in S}}$ is a $\sfrac{1}{\alpha}$-scaled copy of $\conv\p{\pc{\vz_i : i \in S}}$. In other words, we desire to find a \textit{coreset} $\pc{\vz_i : i \in S}$ that approximates $Z$. This approach has the advantage of yielding an interpretable solution -- one can think of a coreset as consisting of the most ``important'' datapoints of the input dataset.

We formally state the streaming convex hull approximation problem we study in Problem \ref{prob:coreset}.

\begin{problem}[Streaming convex hull approximation]
\label{prob:coreset}
Let \(Z = \conv\p{\vz_1, \ldots, \vz_n} \subseteq \R^d\).
%be a convex set where the points \(\vz_1, \ldots, \vz_n\) are received one-by-one in a stream.
A streaming algorithm \(\cA\) receives points \(\vz_1, \ldots, \vz_n\) one at a time and produces a sequence of scalings \(\alpha_t\), centers \(\vc_t\), subsets \(S_t \subseteq [n]\) such that \(S_t \subseteq S_{t+1}\).
The algorithm must satisfy the following guarantee at the end of the stream.
\[\conv\p{\pc{\vz_i : i \in S_n}} \subseteq \conv\p{\pc{\vz_1,\dots,\vz_n}} \subseteq \vc_n + \frac{1}{\alpha} \cdot \conv\p{\pc{\vz_i - \vc_n : i \in S_n}}\]
We say that $\pc{\vz_i : i \in S_n}$ is a coreset of \(Z\) with approximation factor \(\sfrac{1}{\alpha_n}\). We will also call \(S_n\) a coreset.
\end{problem}
Note that the model considered in Problem \ref{prob:coreset} is essentially the same as the \textit{online coreset model} studied by \cite{woodruff2022high}.
Similar to Problem \ref{prob:main}, Problem \ref{prob:coreset} has been studied in the case where $Z$ is centrally symmetric. In particular, \cite{woodruff2022high} obtain approximation factor $O(\sqrt{d\log\p{n\kappa^{\mathsf{OL}}}})$ (where $\kappa^{\mathsf{OL}}$ is the same online condition number mentioned earlier).
However, whether analogous results for asymmetric polytopes hold was an important unresolved question.

\subsection{Our contributions}

In this section, we present our results for Problems \ref{prob:main} and \ref{prob:coreset}.

\subsubsection{Algorithmic results}

We start with defining several quantities that we need to state the results and describe their proofs.

\paragraph{Notation.} We will denote the linear span of a set of points $A$ by $\vspan\p{A}$. That is, $\vspan\p{A}$ is the minimal linear subspace that contains $A$. We denote the affine hull of $A$ by $\aff{A}$. That is, $\aff{A}$ is the minimal affine subspace that contains $A$. Note that $\aff{A} = \va + \vspan\p{A-\va}$ if $\va \in A$. Finally, we denote the unit ball centered at the origin by $B_2^d$.

\begin{definition}[Inradius]
\label{def:inradius}
Let $K \subset \R^d$ be a convex body. The \textit{inradius} $r(K)$ of $K$ is the largest $r$ such that there exists a point $\vc_I$ (called the \textit{incenter}) for which $\vc_I + r \cdot \p{B_2^d \cap \vspan\p{K - \vc_I}} \subseteq K$. 
\end{definition}

\begin{definition}[Circumradius]
\label{def:circumradius}
Let $K \subset \R^d$ be a convex body.  The \textit{circumradius} $R(K)$ of $K$ is the smallest $R$ such that there exists a point $\vc_C$ (called the \textit{circumcenter}) for which $K \subseteq \vc_C + R \cdot B_2^d$.
\end{definition}

\begin{definition}[Aspect Ratio]
\label{def:aspect-ratio}
Let $K \subset \R^d$ be a convex body. We say that $\kappa(K) \coloneqq \sfrac{R(K)}{r(K)}$ is the \textit{aspect ratio} of $K$.
\end{definition}

We now state Theorem \ref{thm:main_one}, which provides an algorithm for Problem \ref{prob:main}. In addition to the data stream of $z_1, \dots, z_n$, this algorithm needs a suitable initialization: a ball $\vc_0 + r_0 \cdot B_2^d$ inside $Z$.
%; if the radius $r_0$ of this ball is close to the inradius $r(Z)$ of $Z$, the algorithm gives an $O(d\log \kappa(Z))$ approximation.

\begin{theorem}
\label{thm:main_one}
Consider the setting of Problem \ref{prob:main}. Suppose the algorithm is given an initial center $\vc_0$ and radius $r_0$ for which it is guaranteed that $\vc_0 + r_0 \cdot B_2^d \subseteq \conv\p{\pc{\vz_1,\dots,\vz_n}}$.
There exists an algorithm (Algorithm \ref{alg:ellipsoid_main}) that, for every timestep $t$, maintains an origin-centered ellipsoid $\cE_t$, center $\vc_t$, and scaling factor $\alpha_t$ such that at every timestep $t$: $\conv\p{\pc{\vz_1,\dots,\vz_t}} \subseteq \vc_t + \cE_t$ 
and at timestep $n$: $\vc_n + \alpha_n \cdot \cE_n \subseteq Z \subseteq c_n + \cE_n$, where  \[\sfrac{1}{\alpha_n} = O\p{\min\p{ \sfrac{R(Z)}{r_0}, d\log\p{\sfrac{R(Z)}{r_0}}}}\] The algorithm has runtime $\widetilde{O}(nd^2)$ and stores $O(d^2)$ floating point numbers.
\end{theorem}

Note that 
%Theorem \ref{thm:main_one} requires the algorithm to know some center and ball initialization $(\vc_0, r_0)$, and 
the final approximation factor depends on the quality of the initialization $(\vc_0, r_0)$. If the radius $r_0$ of this ball is reasonably close to the inradius $r(Z)$ of $Z$, the algorithm gives an $O(\min (\kappa(Z),d\log \kappa(Z)))$ approximation.
In Theorem \ref{thm:main_two}, we adapt the algorithm form Theorem~\ref{thm:main_one} to the setting where the algorithm does not have the initialization information.
%this approach to obtain an algorithm that yields a solution for Problem \ref{prob:main} when the algorithm does not have this information. 
Note that the approximation guarantee of $O(\min (\kappa(Z),d\log \kappa(Z)))$ is a natural analogue of the bounds by~\cite{makarychev2022streaming} and \cite{woodruff2022high} for the symmetric case (see Section~\ref{sec:rw}).

%See Theorem \ref{thm:main_two}.

\begin{theorem}
\label{thm:main_two}
Consider the setting of Problem \ref{prob:main}. There exists an algorithm (Algorithm \ref{alg:fully_online_approx}) that, for every timestep $t$, maintains an ellipsoid $\cE_t$, center $\vc_t$, and approximation factor $\alpha_t$ such that
\begin{align*}
    \vc_t + \alpha_t \cdot \cE_t \subseteq \conv\p{\pc{\vz_1,\dots,\vz_t}} \subseteq \vc_t + \cE_t.
\end{align*}
Additionally, let $r_t$ and $R_t$ be the largest and smallest parameters, respectively, for which there exists $\vc^{\star}_t$ such that
\begin{align*}
    \vc^{\star}_t + r_t \cdot \p{B_2^d \cap \vspan\p{\vz_1-\vc^{\star}_t,\dots,\vz_t-\vc^{\star}_t}} \subseteq \conv\p{\pc{\vz_1,\dots,\vz_t}} \subseteq \vc^{\star}_t + R_t \cdot B_2^d
\end{align*}
and $d_t \coloneqq \mathsf{dim}\p{\vspan\p{\vz_1,\dots,\vz_t}}$.
Then, for all timesteps $t$, we have
\begin{align*}
    \sfrac{1}{\alpha_t} = O\p{d_t\log\p{d_t \cdot \max_{t' \le t} \frac{R_t}{r_{t'}}}}.
\end{align*}
The algorithm runs in time $\widetilde{O}(nd^2)$ and stores $O(d^2)$ floating point numbers.
\end{theorem}
Let us now quickly compare the guarantees of Theorem~\ref{thm:main_one} and~\ref{thm:main_two}. Notice that the algorithm in Theorem~\ref{thm:main_two} does not require an initialization pair $(\vc_0, r_0)$. Additionally, the algorithm in Theorem~\ref{thm:main_two} outputs a per-timestep approximation as opposed to just an approximation at the end of the stream. However, these advantages come at a cost -- it is easy to check that the aspect ratio term seen in Theorem \ref{thm:main_two} can be larger than that in Theorem \ref{thm:main_one}, e.g., it is possible to have $\sfrac{R(Z)}{r_0} \le \max_{t' \le n} \sfrac{R_n}{r_{t'}}$.

However, when we impose the additional constraint that the points $\vz_t$ have coordinates that are integers in the range $[-N,N]$, we can improve over the guarantee in \Cref{thm:main_two} and obtain results that are independent of the aspect ratio. This is similar in spirit to the condition number-independent bound that \cite{woodruff2022high} obtain for the sums of online leverage scores. However, a key difference is that our results still remain independent of the length of the stream. See \Cref{thm:main_two_ints}.

\begin{theorem}
\label{thm:main_two_ints}
Consider the setting of Problem \ref{prob:main}, where in addition, the points $\vz_1,\dots,\vz_n$ are such that their coordinates are integers in $\pc{-N, -N+1, \dots, N-1, N}$. There exists an algorithm (Algorithm \ref{alg:fully_online_approx}) that, for every timestep $t$, maintains an ellipsoid $\cE_t$, center $\vc_t$, and approximation factor $\alpha_t$ such that
\begin{align*}
    \vc_t + \alpha_t \cdot \cE_t \subseteq \conv\p{\pc{\vz_1,\dots,\vz_t}} \subseteq \vc_t + \cE_t.
\end{align*}
Let $d_t \coloneqq \mathsf{dim}\p{\vspan\p{\vz_1,\dots,\vz_t}}$. Then, for all timesteps $t$, we have
\begin{align*}
    \sfrac{1}{\alpha_t} = O\p{d_t\log\p{d N}}.
\end{align*}
The algorithm runs in time $\widetilde{O}(nd^2)$ and stores $O(d^2)$ floating point numbers.
\end{theorem}

We prove Theorems \ref{thm:main_one}, \ref{thm:main_two}, and \ref{thm:main_two_ints} in Section \ref{sec:main_alg_top}.
With Theorems \ref{thm:main_two} and \ref{thm:main_two_ints} in hand, obtaining results for Problem \ref{prob:coreset} becomes straightforward. We use the algorithm guaranteed by Theorem \ref{thm:main_two} along with a simple subset selection criterion to arrive at our result for Problem \ref{prob:coreset}. %See Theorem \ref{thm:main_three}.

\begin{theorem}
\label{thm:main_three}
Consider $Z = \conv\p{\pc{\vz_1,\dots,\vz_n}}$. For a subset $S \subseteq [n]$, let $Z\vert_S=\conv\p{\pc{\vz_i : i \in S}}$.
Consider the setting of Problem \ref{prob:coreset}. There exists a streaming algorithm (Algorithm \ref{alg:ellipse_to_coreset}) that, for every timestep $t$, maintains a subset $S_t$, center $\vc_t$, and scaling factor $\alpha_t$ such that
\begin{align*}
    Z\vert_{S_t} \subseteq \conv\p{\pc{\vz_1,\dots,\vz_t}} \subseteq \vc_t + \frac{1}{\alpha_t} \cdot \p{Z\vert_{S_t} - \vc_t}.
\end{align*}
Additionally, for $d_t$, $r_t$ and $R_t$ as defined in Theorem \ref{thm:main_two}, we have for all $t$ that
\begin{align*}
    \frac{1}{\alpha_t} &= O\p{d_t\log\p{d_t \cdot \max_{t' \le t} \frac{R_t}{r_{t'}}}} & \text{ and }& & \pa{S_t} &= O\p{d_t\log\p{\max_{t' \le t} \frac{R_t}{r_{t'}}}},
\end{align*}
and, if the $\vz_t$ have integer coordinates ranging in $\pb{-N, N}$, then
\begin{align*}
    \frac{1}{\alpha_t} &= O\p{d_t\log\p{dN}} & \text{ and }& & \pa{S_t} &= O\p{d_t\log\p{dN}}.
\end{align*}
Each $S_t$ is either $S_{t-1}$ or $S_{t-1} \cup \{t\}$ (where $t\geq 1$ and $S_0 = \varnothing$).
The algorithm runs in time $\widetilde{O}(nd^2)$ and stores at most $O(d^2)$ floating point numbers.
\end{theorem}

We prove Theorem \ref{thm:main_three} in Section \ref{sec:coreset}.

\subsubsection{Approximability lower bound}

Observe that the approximation factors obtained in Theorems \ref{thm:main_one}, \ref{thm:main_two}, and \ref{thm:main_three} all incur a mild dependence on (variants of) the aspect ratio of the dataset. A natural question is whether this dependence is necessary. In Theorem \ref{thm:main_four}, we conclude that the approximation factor from Theorem~\ref{thm:main_one} is in fact nearly optimal for a wide class of \textit{monotone} algorithms. We defer the discussion of the notion of a monotone algorithm to Section~\ref{sec:monotone}. Loosely speaking, a monotone algorithm commits to the choices it makes; namely, the outer ellipsoid may only increase over time $\vc_{t} + \cE_{t} \supseteq \vc_{t-1} + \cE_{t-1}$ and the inner ellipsoid $\vc_t + \alpha_t \cE_t$ satisfies a related but more technical condition $\vc_t + \alpha_t \cE_t \subseteq \conv\p{(\vc_{t-1} + \alpha_{t-1} \cdot \cE_{t-1}) \cup \{\vz_t\}}$.
%Now we are ready to state our lower bound.

\begin{theorem}
\label{thm:main_four}
Consider the setting of Problem \ref{prob:main}. Let \(\cA\) be any monotone algorithm (see  Definition \ref{def:alg-invariant} in Section~\ref{sec:monotone}) that solves \Cref{prob:main} with approximation factor $\sfrac{1}{\alpha_n}$.
For every \(d \ge 2\), there exists a sequence of points \(\pc{\vz_1, \ldots, \vz_n} \subset \R^d\) 
such that algorithm \(\cA\) gets  an approximation factor of \(\sfrac{1}{\alpha_n} \geq \Omega\p{\frac{d \log\p{\kappa(Z)}}{\log d}}\) on \(Z = \conv\p{\pc{z_1,\dots,z_n}}\).
\end{theorem}

We prove \Cref{thm:main_four} in \Cref{sec:lb}.

\subsection{Related work and open questions}
\label{sec:rw}

\paragraph{Streaming asymmetric ellipsoidal roundings.}

To our knowledge, the first paper to study ellipsoidal roundings in the streaming model is that of \cite{mukhopadhyayapproximate}. The authors consider the case where $d=2$ and prove that the approximation factor of the greedy algorithm (that which updates the ellipsoid to be the minimum volume ellipsoid containing the new point and the previous iterate) can be unbounded. Subsequent work by  \cite{mukhopadhyay2010approximate} generalizes this result to all $d \ge 2$.

\paragraph{Nearly-optimal streaming symmetric ellipsoidal roundings.} 
Recently, \cite{makarychev2022streaming}, and \cite{woodruff2022high} gave the first positive results for streaming ellipsoidal roundings. Both \cite{makarychev2022streaming} and \cite{woodruff2022high} considered the problem only in the \textit{symmetric setting} -- when the goal is to approximate the polytope $\conv\p{\pc{\pm \vz_1,\dots, \pm \vz_n}}$.  \cite{makarychev2022streaming} and \cite{woodruff2022high} obtained $O(\sqrt{d\log \kappa(Z)})$ and $O(\sqrt{d\log n\kappa^{\mathsf{OL}}})$-approximations, respectively (here, $\kappa^{\mathsf{OL}}$ is the online condition number; see  \cite{woodruff2022high} for details). Their algorithms use only $\widetilde{O}(\mathsf{poly}(d))$ space, where the $\widetilde{O}$ suppresses $\log d$, $\log n$, and aspect ratio-like terms. 
Note that by John's theorem, the $\Omega(\sqrt{d})$ dependence is required in the symmetric setting even for offline algorithms.

A natural question is whether the techniques of \cite{makarychev2022streaming} or \cite{woodruff2022high} extend to Problems \ref{prob:main} and~\ref{prob:coreset}. The update rule used in \cite{makarychev2022streaming} essentially updates $\cE_{t+1}$ to be the minimum volume ellipsoid covering both $\cE_t$ and points $\pm\vz_{t+1}$. In the non-symmetric case, it would be natural to consider the minimum volume ellipsoid covering $\cE_t$ and point $\vz_{t+1}$. However, this approach does not give an $\tilde O(d)$ approximation.
%Although this update has a simple closed form in the symmetric case, the form of the asymmetric update is far more challenging to work with.
The algorithm in \cite{woodruff2022high} maintains a quadratic form that consists of sums of outer products of ``important points'' (technically speaking, those with a constant online leverage score). Unfortunately, this approach does not suggest how to move the previous center $\vc_{t-1}$ to a new center $\vc_{t}$ in a way that allows the algorithm to maintain a good approximation factor. It is not hard to see that there exist example streams for which the center $\vc_{t-1}$ must be shifted in each iteration to maintain even a bounded approximation factor. This means that any nontrivial solution to Problems \ref{prob:main} and \ref{prob:coreset} must overcome this difficulty.

\paragraph{Offline ellipsoidal roundings for general convex polytopes.} \cite{nesterov2008rounding} gives an efficient \textit{offline} $O(d)$-approximation algorithm for the ellipsoidal rounding problem, with a runtime of $\widetilde{O}(nd^2)$. Observe that this is essentially the same runtime as those achieved by the algorithms we give (see Theorems \ref{thm:main_one} and \ref{thm:main_two}).

\paragraph{Streaming convex hull approximations.}  \cite{agarwal2010streaming} studied related problems of computing \textit{extent measures} of a convex hull in the streaming model, in particular finding coresets for the minimum enclosing ball, and obtained both positive and negative results.
% Note that these imply similar inapproximability results for Problem \ref{prob:coreset} for any sufficiently small approximation factor. 
% In contrast, our algorithms obtain a \(O(d)\) approximation rather than \((1+\eps)\).  Similarly,
\cite{blum2017approximate} showed that one cannot maintain an \textit{$\eps$-hull} in space proportional to the number of vertices belonging to the offline optimal solution (where a body $\widehat{Z}$ is an $\eps$-hull for $Z$ if every point in $\widehat{Z}$ is distance at most $\eps$ away from $Z$).

\paragraph{Offline convex hull approximations.} The problem of approximating a convex body with the convex hull of a small number of points belonging to the body has been well-studied. Existentially, \cite{barvinok2014thrifty} shows that if the input convex set is sufficiently symmetric, then one can choose $(d/\eps)^{d/2}$ points to obtain a $1+\eps$ approximation. Moreover, \cite{lu2020note} shows that one can obtain a $d+2$ approximation with $d+1$ points, which is witnessed by choosing the $d+1$ points to be the maximum volume simplex contained within the convex body (for this reason, this construction is called ``John's Theorem for simplices''; see \cite{leme2020costly} for more details). However, none of these works study a streaming or online setting, as we do here.

\paragraph{Coresets for the minimum volume enclosing ellipsoid problem (MVEE).} 
Let $\mathsf{MVEE}(K)$ denote the minimum volume enclosing ellipsoid for a convex body $K \subset \R^d$.
%A related, though not equivalent, problem to \Cref{prob:coreset}, is the problem of finding a coreset for the MVEE problem. 
We say that a subset $S \subseteq [n]$ is an $\eps$-coreset for the MVEE problem if we have
\begin{align}
    \vol\p{\mathsf{MVEE}(Z)} \le \p{1+\eps}^d\vol\p{\mathsf{MVEE}(Z\vert_S)}\label{eq:mvee_coreset}.
\end{align}
There is extensive literature on coresets for the MVEE problem, and we refer the reader to papers by \cite{kumar2005minimum}, 
\cite{todd2007minimum}, \cite{clarkson2010coresets}, \cite{bhaskara2023tight}, and the book by \cite{todd2016minimum}.

Importantly, $\mathsf{MVEE}(Z\vert_S)$ may not be a good approximation for $\mathsf{MVEE}(Z)$ (for that reason, 
some authors refer to coresets satisfying~(\ref{eq:mvee_coreset}) as weak coresets for MVEE). 
Therefore, even though $\mathsf{MVEE}(Z)$ provides a good ellipsoidal rounding for $Z$, $\mathsf{MVEE}(Z\vert_S)$ generally speaking does not. 
See \cite[page 2]{todd2007minimum} and \cite[Section 2.1]{bhaskara2023tight} for an extended discussion.

\iffalse
Although at a first glance our coreset result (\Cref{thm:main_three}) appears to apply to a problem that is equivalent to that of finding an $\eps$-coreset for MVEE, this is not the case. This is because a coreset satisfying \eqref{eq:mvee_coreset} may be a \textit{weak} coreset, whereas \Cref{prob:coreset} is essentially asking for a \textit{strong} coreset (see the discussion in Section 2.1 of \cite{bmv23}). In other words, observe that there exist subsets $S$ for which $S$ satisfies \eqref{eq:mvee_coreset} but $Z\vert_S$ is not a convex hull approximation to $Z$, even when $d=2$. 

Additionally, to our knowledge, all prior work on constructing coresets for MVEE applies in the offline setting. Since our , it follows that the setting and problem we study is totally incomparable to those of the aforementioned prior work.
\fi

% \paragraph{Open questions.}

% % \maxcmt{to turn into a real thing} Wy algorithm gives nnz time.
% % Connection between suggests this may be possible for the general case.
% Although the runtime of our algorithms nearly matches those of the best-known offline algorithms in the worst case, it would be  interesting to design algorithms that run in input-sparsity time. In turn, this would be useful for more structured inputs.
% \mnote{thanks for writing this, but I would like to add some more to this paragraph after I finish appendix B}

\section{Summary of Techniques}

In this section, we give an overview of the technical methods behind our results.

\subsection{Monotone algorithms}
\label{sec:monotone}
The algorithm we give in \Cref{thm:main_one} belongs to a class we term \textit{monotone algorithms}, which we now define.

\begin{figure}[ht]
\centering
\includegraphics[height=6cm]{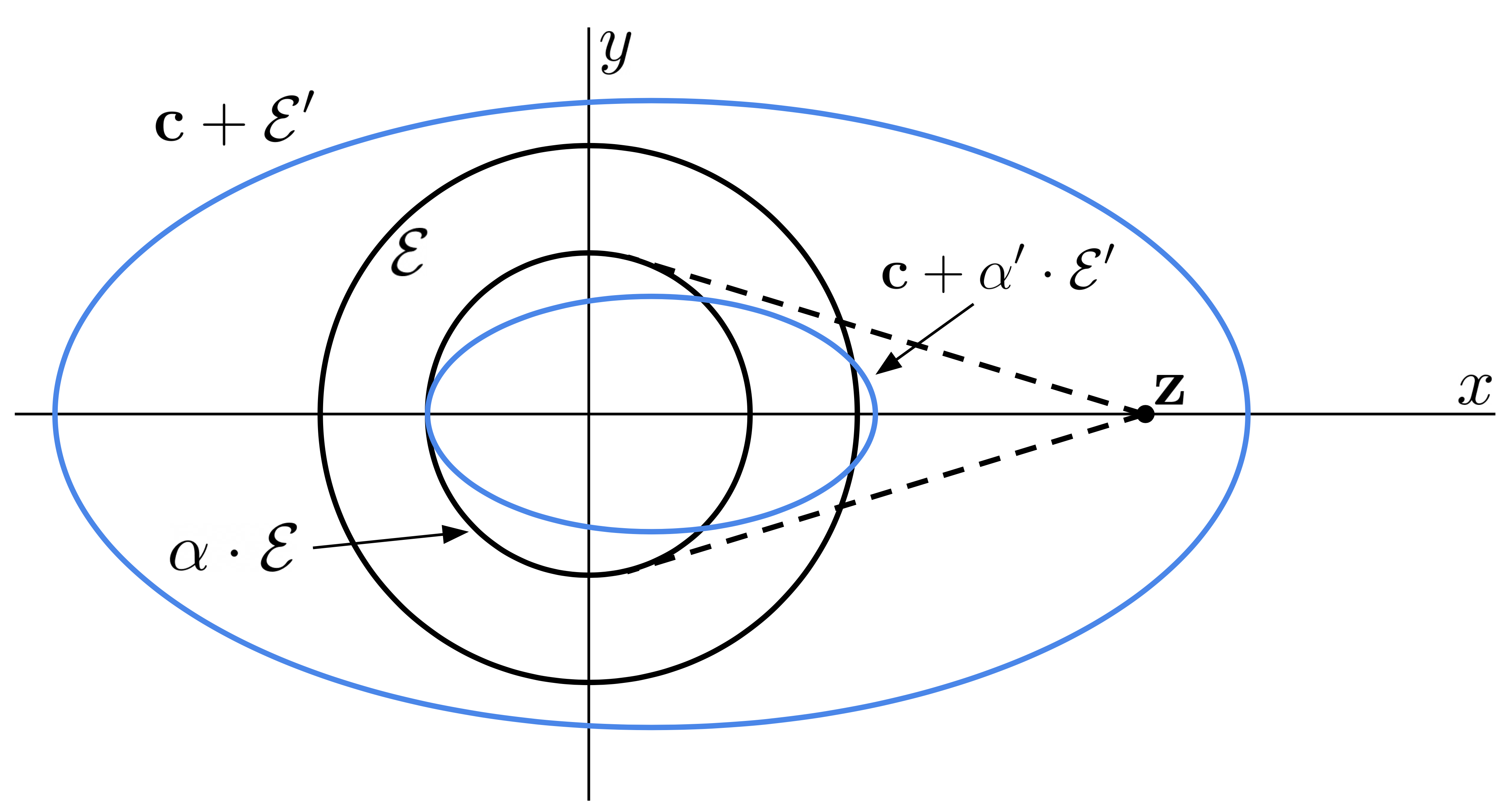}
\caption[A monotone update step.]{A monotone update step. For brevity, we refer to \(\cE\) and \(\alpha \cdot \cE\) as the previous ellipsoids \(\cE_{t-1}, \alpha \cE_{t-1}\), and \(\cE'\) and \(\alpha' \cdot \cE'\) as the next ellipsoids \(\cE_{t}, \alpha_t \cdot \cE_t\). \(\cE\) and \(\alpha \cE\) are, respectively, the larger and smaller black circles. 
\(c + \cE'\) and \(c + \alpha' \cE'\) are the larger and smaller blue ellipses. The dotted lines show \(\partial(\conv\p{\alpha \cE \cup \{\vz\}}) \setminus \partial(\alpha \cE)\), i.e. the the boundary of \(\conv\p{\alpha \cdot \cE \cup \{\vz\}}\) minus the boundary of \(\alpha \cE\). }
\label{fig:update_step}
\end{figure}

\begin{definition}[Monotone algorithm]
\label{def:alg-invariant}
Consider the setting of Problem \ref{prob:main}. Note the following invariants for every timestep \(t\).
\begin{align}
    \vc_t + \cE_t 
     &\supseteq 
     \conv\p{(\vc_{t-1} + \cE_{t-1}) \cup \{\vz_t\}}     
\label{item:def-invariant-1} \\
    \vc_t + \alpha_t \cE_t &\subseteq \conv\p{(\vc_{t-1} + \alpha_{t-1} \cdot \cE_{t-1}) \cup \{\vz_t\}} \label{item:def-invariant-2}
\end{align}

We say that an algorithm \(\cA\) is \textnormal{monotone} if for any initial \((\vc_0 + \cE_0, \alpha_0)\) and sequence of data points \(\vz_1, \ldots, \vz_n\), the resulting sequence \(\{(\vc_0 + \cE_0, \alpha_0), (\vc_1 + \cE_1, \alpha_1), \ldots, (\vc_n + \cE_n, \alpha_n)\}\) arising from applying \(\cA\) to the stream satisfies the two invariants \eqref{item:def-invariant-1} and \eqref{item:def-invariant-2}. Refer to Figure \ref{fig:update_step}.

We will sometimes consider how a monotone algorithm \(\cA\) makes a single update upon seeing a new point $\vx$. In this setting, we will call \(\cA\) a \textit{monotone update rule}.
\end{definition}

% Link to edit figures: https://drive.google.com/drive/folders/1i6EOC9rYHTdqFn0iK4SrGEW2UYLTMktq?usp=share_link

Here we will refer to \(\vc_t + \cE_t, \vc + \alpha_t \cE_t\) as the ``next'' ellipsoids and 
to \(\vc_{t-1} + \cE_{t-1}, \vc + \alpha_{t-1} \cE_{t-1}\) as the ``previous'' ellispoids.
The first condition we require is that 
\begin{equation}\vc_{t} + \cE_{t} \supseteq \vc_{t-1}+\cE_{t-1}.\label{item:def-invariant-1-alt}
\tag{\ref*{item:def-invariant-1}a}
\end{equation}
It ensures that each successive outer ellipsoid contains the previous outer ellipsoid. Thus once the algorithm decides  that some $\vz \in \vc_t + \cE_t$, it makes a commitment that $\vz \in \vc_{t'} + \cE_{t'}$ for all $t'\geq t$.
Note that (\ref{item:def-invariant-1-alt}) implies (\ref{item:def-invariant-1}), since $\vz_t$ must be in $\vc_t + \cE_t$ and $\vc_t + \cE_t$ is convex.
The second condition (\ref{item:def-invariant-2}) looks more complex but is also very natural. Assume that the algorithm only knows that (a) $\vc_{t-1} + \alpha_{t-1}\cE_{t-1}\subseteq Z$ (this is true from induction) and (b) $\vz_t\in Z$ (this is true by the definition of $Z$). Then, we must have that $A = \conv\p{(\vc_{t-1} + \alpha_{t-1} \cdot \cE_{t-1}) \cup \{\vz_t\}}$ lies in $Z$; as far as the algorithm is concerned, any point outside of $A$ may also be outside of $Z$.
Since the algorithm must ensure that $\vc_t + \alpha_t {\cE}_t \subseteq Z$, it will also ensure that $\vc_t + \alpha_t {\cE}_t\subseteq A$ and thus satisfy (\ref{item:def-invariant-2}).

\subsection{Streaming ellipsoidal rounding (Theorems \ref{thm:main_one}, \ref{thm:main_two}, and \ref{thm:main_two_ints})}

Now we describe the algorithm from Theorem~\ref{thm:main_one} in more detail. 
Our algorithm keeps track of the current ellipsoid $\cE_t$, center $\vc_t$, and scaling parameter $\alpha_t$. Initially, $\vc_0 + \cE_0$ is the ball of radius $r_0$ around $\vc_0$ ($r_0$ and $\vc_0$ are given to the algorithm),  and $\alpha_0=1$. Each time the algorithm gets a new point $\vz_t$, it updates $\cE_{t-1}$, $\vc_{t-1}$, $\alpha_{t-1}$ using a \textit{monotone} update rule (as defined in Definition~\ref{def:alg-invariant}) and obtains $\cE_{t}$, $\vc_{t}$, $\alpha_{t}$. 
The monotonicity condition is sufficient to guarantee that the algorithm gets a $1/\alpha_n$ approximation to $Z$.
Indeed, first using condition~(\ref{item:def-invariant-1}), we get
$$\vc_n + \cE_n \supseteq (\vc_{n-1} + \cE_{n-1}) \cup\{\vz_n\} \supseteq (\vc_{n-2} + \cE_{n-2}) \cup\{\vz_{n-1},\vz_n\}\supseteq \dots \supseteq \{\vz_1,\dots,\vz_n\}.$$
Thus, $\vc_n + \cE_n \supseteq Z$.
Then, using condition~(\ref{item:def-invariant-2}), we get
\begin{align*}
\vc_n + \alpha_n \cE_n &\subseteq
\conv\p{(\vc_{n-1} + \alpha_{n-1} \cE_{n-1}} \cup \{\vz_n\}) \subseteq \conv\p{(\vc_{n-2} + \alpha_{n-2} \cE_{n-2}} \cup \{\vz_{n-1},\vz_n\}) \\
&\subseteq \dots \subseteq \conv\p{(\vc_0 + \alpha_0 \cE_0) \cup \{\vz_1,\dots,\vz_n\}}.
\end{align*}
The initial ellipsoid $\vc_0 + \alpha_0 \cE_0 = \vc_0 + r_0 B_2^d$ is in $Z$ and therefore $\vc_n + \alpha_n \cE_n\subseteq \conv\p{\vz_1,\dots, \vz_n} = Z$.
We verified that the algorithm finds a $\sfrac{1}{\alpha_n}$ approximation for $Z$.

Now, the main challenge is to design an update rule that ensures that $1/\alpha_n$ is small (as in the statement Theorem~\ref{thm:main_one}) and prove that the rule satisfies the monotonicity conditions/invariants from Definition~\ref{def:alg-invariant}.
We proceed as follows.
%There are three main components to the proofs of Theorem \ref{thm:main_one} and \ref{thm:main_two}. 

First, we design a monotone update rule that satisfies a particular evolution condition. This condition upper bounds the increase of the approximation factor $\sfrac{1}{\alpha_t} - \sfrac{1}{\alpha_{t-1}}$. Second, we prove that any monotone update rule satisfying the evolution condition yields the approximation we desire. These two parts imply Theorem \ref{thm:main_one}. Finally, we remove the initialization requirement from Theorem \ref{thm:main_one} and obtain Theorem \ref{thm:main_two}. 
%The second component is straightforward, so we focus on the first and third components.

\paragraph{Designing a monotone update rule.} Suppose that at the end of timestep $t-1$ our solution consists of a center $\vc_{t-1}$, ellipsoid $\cE_{t-1}$, and scaling parameter $\alpha_{t-1}$ for which the invariants in Definition \ref{def:alg-invariant} hold. We give a procedure that, given the next point $\vz_{t}$, computes $\vc_{t}, \cE_{t}, \alpha_{t}$ that still satisfy the invariants of Definition \ref{def:alg-invariant}. Further, we prove that the resulting update satisfies an evolution condition \eqref{eq:overview_evolution}
\begin{align}
    \frac{\sfrac{1}{\alpha_{t}} - \sfrac{1}{\alpha_{t-1}}}{\log\mathsf{vol}(\cE_{t}) - \log\mathsf{vol}(\cE_{t-1})} \le C,\label{eq:overview_evolution}
\end{align}
where $C$ is an absolute constant and $\vol{\cE}$ denotes the volume of the ellipsoid $\cE$. 
%It is possible to find \textit{some} monotone update via convex optimization techniques. However it is unclear how to show that such an update satisfies the evolution condition \eqref{eq:overview_evolution}. In particular, the increase in approximation factor in this iteration may be large compared to the increase of the volume of the ellipsoids.
While it is possible to find the optimal update using convex optimization (the update that satisfies the invariants and minimizes the ratio on the left of (\ref{eq:overview_evolution})), we instead provide an explicit formula for an update that readily satisfies (\ref{eq:overview_evolution}) and as we show is monotone. 

We now describe how we get the formula for the update rule.
By applying an affine transformation, we may assume that ${\cE}_{t-1}$ is a unit ball and $\vc_{t-1} = 0$. Further, we may assume that $\vz_t$ is colinear with $\ve_1$ (the first basis vector): $\vz_t = \|\vz_t\| \ve_1$. Importantly, affine transformations preserve (a) the invariants in Definition~\ref{def:alg-invariant} (if they hold for the original ellipsoids and points, then they also do for the transformed ones and vice versa) and (b) the value of the ratio in (\ref{eq:overview_evolution}), since they preserve the value of $\vol({\cE_t})/\vol({\cE_{t-1}})$.

% \todo{mention Steiner symmetrization instead?} 
Now consider the group $G = \bO(d)_{\ve_1}\cong \bO(d-1)$ of orthogonal transformations that map $\ve_1$ to itself: all of them map the unit ball ${\cE}_{t-1}$ to itself and $\vz_{t}$ to itself. Thus, it is natural to search for an update $(\vc_t, {\cE}_t)$ that is symmetric with respect to all these transformations.
%\footnote{In fact, when we prove our lower bound, we show that the optimal update is  symmetric with respect to $G$.} 
It is easy to see that in this case ${\cE}_t$ is defined by equation 
$(x_1/a)^2 + \sum_{i=2}^d (x_i/b)^2 =1$ where $a$ and $b$ are some parameters (equal to the semiaxes of ${\cE}_{t}$) and $\vc_t = c \ve_1$ for some $c$. 
Since all ellipsoids and points appearing in the invariant conditions are symmetric with respect to $G$, it is sufficient now to restrict our attention to their sections in the $2$d-plane $\vspan\p{\ve_1,\ve_2}$ and prove that the invariants hold in this plane. Hence, the problem reduces to a statement in two-dimensional Euclidean geometry (however, when we analyze~(\ref{eq:overview_evolution}), we still use that the volume of ${\cE}_t$ is proportional to $ab^{d-1}$ and not $ab$).

Let us denote the coordinates corresponding to basis vectors $\ve_1$ and $\ve_2$ by $x$ and $y$. For brevity, let ${\cE} = {\cE}_{t-1}$, $\vz = \vz_t$, ${\cE}' = {\cE}_{t}$, $\vc = \vc_t =c\ve_1$, $\alpha = \alpha_{t-1}$, and $\alpha' = \alpha_t$.
We now need to choose parameters $a$, $b$, and $c$ so that invariants from Definition \ref{def:alg-invariant} and (\ref{eq:overview_evolution}) hold. See \Cref{fig:update_step}.
%We will later see that this statement is just the restriction of the monotone update condition to the special case where the previous ellipses are unit balls. The newly received point will be on the \(x\)-axis as depicted in \Cref{fig:update_step}. 
As shown in that figure, the new outer ellipse \(\vc + \cE'\) must contain the previous outer ellipse \(\cE\) and the newly received point \(\vz\). The new inner ellipse \(\vc + \alpha' \cE'\) must be contained within the convex hull of the previous inner ellipse \(\alpha \cE\) and \(\vz\).

It is instructive to consider what happens when point $\vz$ is at infinitesimal distance $\Delta$ from $\cE$: $\|\vz\| = 1 +\Delta$. We consider a minimal axis-parallel outer ellipse $\cE'$ that contains $\cE$ and $\vz$. It must go through $\vz = (1+\Delta,0)$ and touch $\cE$ at two points symmetric w.r.t.\ the $x$-axis, say, $(-\sin \varphi, \pm \cos \varphi)$. Angle $\varphi$ uniquely determines $\cE'$. Now we want to find the largest value of the scaling parameter $\alpha'$ so that $\alpha' \cE'$ fits inside the convex hull of $\cE$ and $\vz$. When $\Delta$ is infinitesimal, this condition splits into two lower bounds on $\alpha'$ --  loosely speaking, they say that $\cE$ does not extend out beyond the convex hull in the horizontal (one bound) and vertical directions (the other). The former bound becomes stronger (gives a smaller upper bound on $\alpha'$) when $\varphi$ increases, and the latter becomes stronger when $\varphi$ decreases. When $\varphi = \alpha/2 \pm O(\alpha^2)$, then all terms linear in $\alpha$ vanish in both bounds and then $\alpha' = \alpha - \Theta(\alpha^2\Delta)$ satisfies both of them; for other choices of $\varphi$, we have $\alpha' \leq \alpha - \Omega(\alpha\Delta)$. 
So we let $\varphi=\alpha/2$ and from the formula for $\alpha'$ get $1/\alpha' = 1/\alpha + O(\Delta)$. On the other hand, $\vol(\cE') \geq (1 + \Delta/2)\vol(\cE)$, since $\cE'$ covers $\vz = (1 + \Delta,0)$. It is easy to see now that the evolution condition (\ref{eq:overview_evolution}) holds: the numerator is $O(\Delta)$ and the denominator is $\Omega(\Delta)$ in (\ref{eq:overview_evolution}).

We remark that letting $\vc + {\cE}'$ be the minimum volume ellipsoid that contains $\cE$ and $\vz$ is a highly suboptimal choice (it corresponds to setting $\varphi=\Theta(1/d)$). To derive our specific update formulas for arbitrary $\vz$, we, loosely speaking,  represent an arbitrary update as a series of infinitesimal updates, get a differential equation on $a$, $b$, $c$, and $\alpha'$, solve it, and then simplify the solution (remove non-essential terms, etc).  We get the following.

Our updates come from a family parameterized by \(\gamma \geq 0\). %We now describe this family. 
 Define \(\alpha'\) by \(\sfrac{1}{\alpha'} = \sfrac{1}{\alpha} + 2\gamma\). With this choice of $\alpha'$, define the new ellipses to be
% \begin{gather*}
% c + \cE' \defeq \pc{(x, y) \in \R^2 \middle| \frac{1}{\exp\p{\gamma}^2}\p{x+ \p{\alpha - \frac{1}{\frac{1}{\alpha} + 2 \gamma} \cdot \exp(\gamma)}}^2 + \frac{1}{\p{1 + \frac{\alpha - \frac{1}{\frac{1}{\alpha} + 2 \gamma}}{2}}^2} y^2 \leq 1} \\
% c + \alpha' \cE' \defeq \pc{(x, y) \in \R^2 \middle| \frac{1}{\exp\p{\gamma}^2}\p{x+ \p{\alpha - \frac{1}{\frac{1}{\alpha} + 2 \gamma} \cdot \exp(\gamma)}}^2 + \frac{1}{\p{1 + \frac{\alpha - \frac{1}{\frac{1}{\alpha} + 2 \gamma}}{2}}^2} y^2 \leq \frac{1}{\p{\frac{1}{\alpha} + 2\gamma}^2}}
% \end{gather*}
% where \(\alpha' = \frac{1}{\frac{1}{\alpha} + 2 \gamma}\).
\[\underbrace{\frac{1}{a^2}(x-c)^2 + \frac{1}{b^2} y^2 = 1}_{\vc + \cE'}, \qquad \underbrace{\frac{1}{a^2}(x-c)^2 + \frac{1}{b^2} y^2 = \alpha'^2}_{\vc + \alpha' \cE'}\]
where we use parameters
\begin{equation*}
% \label{eqn:update_params}
\left.\begin{aligned}
a &= \exp\p{\gamma} \\
b &=1 + \frac{\alpha - \alpha'}{2} \\
c &= - \alpha + \alpha' \cdot a
\end{aligned}\qquad\right\}.
\end{equation*}
Choose $\gamma\approx \ln \|\vz\|$ so that $\vc + \cE'$ covers point $\vz$.
We use two-dimensional geometry to prove that $\cE'$, $\vc$, and $\alpha'$ satisfy the invariants (see Figure~\ref{fig:update_step}).
Now to prove the evolution condition, we observe two key properties: (1) the increase in the approximation factor is given by \(\frac{1}{\alpha'} - \frac{1}{\alpha} = 2 \gamma\) and (2) the length of the horizontal semiaxis of the new outer ellipse is \(\exp(\gamma)\). The length of the vertical semiaxis is at least $1$, so by the second property we have \(\log \vol(\cE') - \log \vol(\cE) \geq \gamma\). We combine this with the first property to prove that this update satisfies the evolution condition \eqref{eq:overview_evolution}.

Finally, we obtain an upper bound on $1/\alpha_n$ from the evolution equation. We have
\begin{align*}
    \sfrac{1}{\alpha_{n}} = \sfrac{1}{\alpha_0} + \sum_{t=1}^n
\left(\sfrac{1}{\alpha_{t}} - \sfrac{1}{\alpha_{t-1}}\right) \stackrel{\eqref{eq:overview_evolution}}{\leq} 1+ C\sum_{t=1}^n(\log\mathsf{vol}(\cE_{t}) - \log\mathsf{vol}(\cE_{t-1})) = 1 + C\log \frac{\vol{\cE_n}}{\vol{\cE_0}}.
\end{align*}
It remains to get an upper bound on $\vol(\cE_n)$. We know that $\cE_n$ approximates $Z$, and $Z$, in turn, is contained in the ball of radius $R(Z)$. Loosely speaking, we get $\vol(\cE_n) \approx \vol(Z) \leq R(Z)^d \vol(B_2^d)$. Since $\cE_0$ is the ball of radius $r$, $\vol{\cE_0}= r^d \vol(B_2^d)$. We conclude that 
the approximation factor is at most $\sfrac{1}{\alpha_{n}} \lessapprox 1 + C\log \frac{R(Z)^d}{r^d} = 1 + O(d\log \frac{R(Z)}{r})$, as desired.

\paragraph{Removing the initialization assumption.} Once we have a monotone update rule and  guarantee on its approximation factor, we have to convert this to a guarantee where the algorithm does not have access to the initialization. 

One natural approach is as follows. Let $d' \le d$ be the largest timestep for which points $\vz_1, \dots,\vz_{d'+1}$ are in general position. We can compute the John ellipsoid for $\conv\p{\pc{\vz_1,\dots,\vz_{d'+1}}}$ and after that apply the monotone update rule guaranteed by Theorem \ref{thm:main_one} to obtain the rounding for every $t \ge d'+2$, so long as for every such timestep we have $\vz_t  \in \vspan\p{\vz_1,\dots,\vz_{t-1}}$.

\begin{figure}[ht]
\centering
\includegraphics[height=6cm]{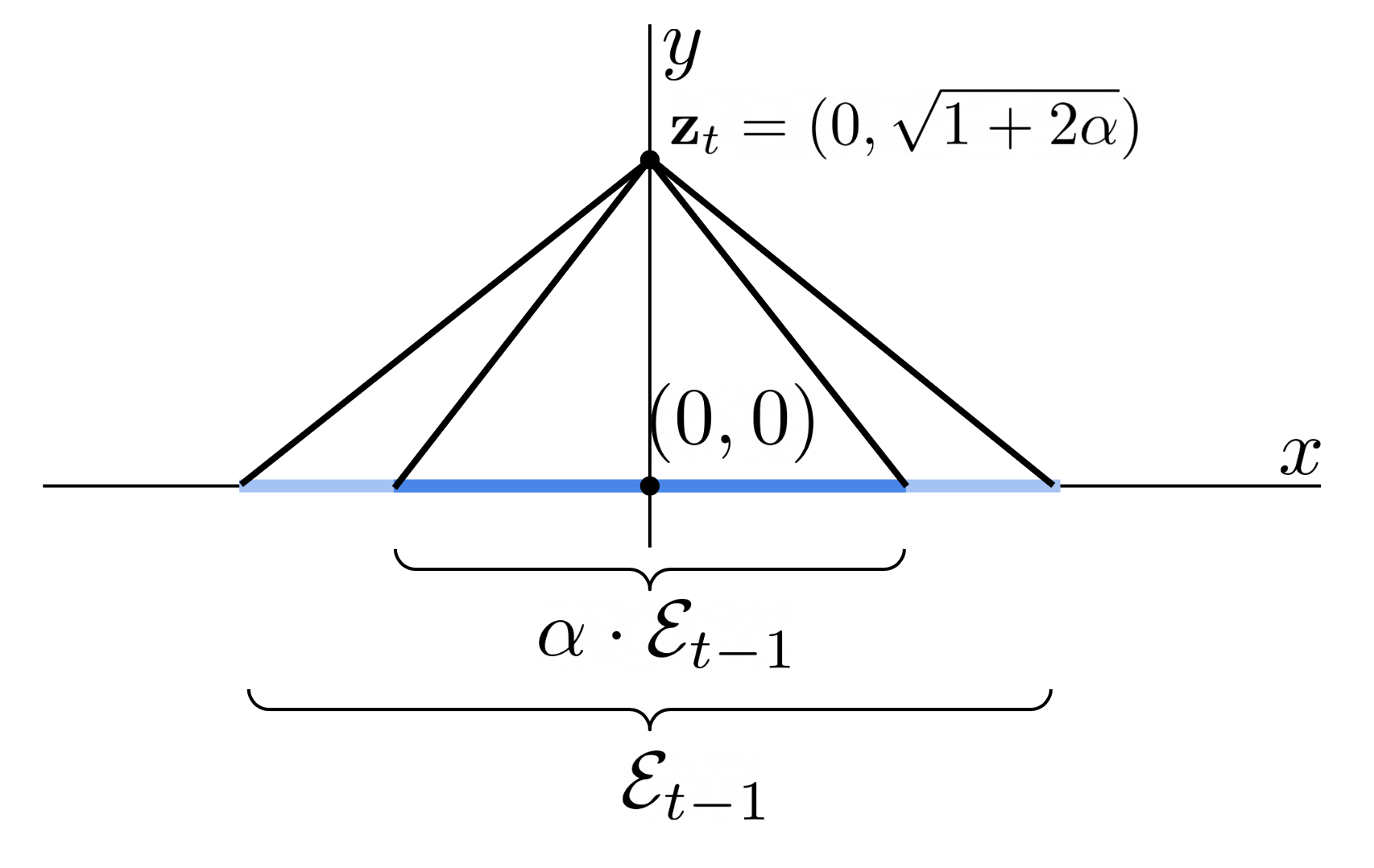}
\caption[Irregular update step.]{Irregular update step. \(\cE_{t-1}\) and \(\alpha \cdot \cE_{t-1}\) are, respectively, the light blue strip on the \(x\)-axis and the dark blue strip on the \(x\)-axis. \(\vz_t = (0, \sqrt{1+2 \alpha})\) is the newly received point.
}
\label{fig:update_step_irreg}
\end{figure}

The principal difficulty in this approach is designing an \textit{irregular update step} that will handle points $\vz_t$ outside of $\vspan\p{\vz_1,\dots,\vz_{t-1}}$; when we add these points the dimensionality of the affine hull increases by 1. We consider the special case where the new point $\vz_t$ is conveniently located with respect to our previous ellipsoid $\cE_{t-1}$ (see \Cref{fig:update_step_irreg} for a $2$d-picture). Specifically, $\cE_{t-1}$ is the unit ball in $\vspan\p{\ve_1,\dots,\ve_{d'}}$, and the new point $\vz_t = (0,\dots, 0, \sqrt{1+2\alpha}),0,\dots)$. In $\vz_t$, only coordinate $d'+1$ is nonzero. We show that we can design an irregular update step for this special case that makes the new approximation factor $\sfrac{1}{\alpha_{t}}$ satisfy $\sfrac{1}{\alpha_t} = \sfrac{1}{\alpha_{t-1}}+1$.

% Link to edit figures: https://drive.google.com/drive/folders/1i6EOC9rYHTdqFn0iK4SrGEW2UYLTMktq?usp=share_link

It turns out that it is sufficient to consider only this special case. To see this, note that we can choose an affine transformation that maps any new point $\vz_{t}$ and previous ellipsoid $\cE_{t-1}$ to the setting shown in \Cref{fig:update_step_irreg}.
%; recall that set containments are maintained under affine transformations. 
Next, observe that there are at most $d-1$ irregular update steps. This means that the irregular update steps contribute at most an additive $d-1$ to the final approximation factor.

Finally, observe that the inradius of $\conv\p{\pc{\vz_1,\dots,\vz_t}}$ is not monotone in $t$. In particular, it can decrease after each irregular update step. Nonetheless, we can still give a bound on the radius of a ball that our convex body $\conv\p{\vz_1,\dots,\vz_t}$ contains for all $t$. This will give us everything we need to apply \Cref{thm:main_one} to this setting, and \Cref{thm:main_two} follows.

\paragraph{Improved bounds on lattices.} 
% \nnote{what's a good way to write this?} 
Finally, we briefly discuss how to remove the aspect ratio dependence in the setting where the input points $\vz_t$ have coordinates in $[-N,N]$. At a high level, this improvement follows from carefully tracking how the approximation factors of our solutions change after an irregular update step. Following \eqref{eq:overview_evolution}, recall that our goal is to analyze (where we write $\alpha_0 = 1$)
\begin{align*}
    \sum_{t \ge 1} \frac{1}{\alpha_t}-\frac{1}{\alpha_{t-1}}.
\end{align*}
By \eqref{eq:overview_evolution}, we see that for all ``regular'' updates, we have
\begin{align*}
    \frac{1}{\alpha_t}-\frac{1}{\alpha_{t-1}} \lesssim \log\p{\frac{\vol_{d_t}\p{\cE_t}}{\vol_{d_t}\p{\cE_{t-1}}}},
\end{align*}
where $d_t = \mathsf{dim}\p{\vspan\p{\vz_1,\dots,\vz_t}}$. Furthermore, as previously mentioned, in our irregular update step, we get
\begin{align*}
    \frac{1}{\alpha_t}-\frac{1}{\alpha_{t-1}} = 1.
\end{align*}
In order to control the sum of the $\sfrac{1}{\alpha_{t}}-\sfrac{1}{\alpha_{t-1}}$, it remains to bound $\sfrac{\vol_{d_t}(\cE_t)}{\vol_{d_{t-1}}(\cE_{t-1})}$ for an irregular update step $t$. We will then get a telescoping upper bound whose last term is the ratio of the volume of the final ellipsoid to the Euclidean ball in the same affine span.

Similarly to the improvements of \cite{woodruff2022high} in the integer-valued case, it will turn out that we will be interested in the total product of these volume changes. By carefully tracking these, we will get that this product can be expressed as the determinant of a particular integer-valued matrix. Then, since this matrix has integer entries, the magnitude of its determinant must be at least $1$. We then observe that the volume of $\cE_n$ after normalizing by the volume of $\vol(B_2^{d_n})$ must be at most $(N\sqrt{d})^{d_n}$, since the length of any vector in this lattice is at most $N\sqrt{d}$. The desired result then follows.

\subsection{Coresets for convex hull \texorpdfstring{(\Cref{thm:main_three})}{(Theorem~\ref{thm:main_three})}}
\label{sec:overview_coreset}

We now outline our proof strategy for \Cref{thm:main_three}.
Our main task is to design an appropriate selection criterion for every new point -- in other words, we must check whether a new point $\vz_t$ is ``important enough'' to be added to our previous set of points $S_{t-1}$. We then have to show that this selection criterion yields the approximation guarantee promised by Theorem \ref{thm:main_three}.

To design the selection criterion, we run an instance of the algorithm in Theorem \ref{thm:main_two} on the stream. For every new point $\vz_t$, we ask two questions -- ``Does $\vz_t$ result in an irregular update step? Does it cause $\mathsf{vol}(\cE_{t})$ to be much larger than $\mathsf{vol}(\cE_{t-1})$?'' If the answer to any of these questions is affirmative, we add $\vz_t$ to the coreset. The first question is necessary to obtain even a bounded approximation factor (for example, imagine that the final point $\vz_n$ results in an irregular update step -- then, we must add it). The second question is quite natural, as it ensures that the algorithm adds ``important points'' -- those that necessitate a significant update.

%-- in particular, note that if $\vz_{t}$ is ``close enough'' to $\cE_{t-1}$, a constant-factor rescaling of $\cE_{t-1}$ will contain $\vz_t$ while only losing a constant in the final approximation factor.

We now observe that at every irregular update step $t_{d'}$ for $d' \le d$ and subsequent timestep $t \ge t_{d'}$ for which there are no irregular update steps in between $t_{d'}$ and $t$, there exists a translation $\vc_{d'}$ (which is the center for $\cE_{d'}$ that the algorithm maintains) and a value $r_{d'}$ for which we know
\begin{align*}
    \vc_{d'} + r_{d'} \cdot \p{B_2^d \cap \vspan\p{\vz_1-\vc_{d'},\dots,\vz_{d'}-\vc_{d'}}} \subseteq \conv\p{\vz_1,\dots,\vz_{t}} \subseteq \vc_C + R_{t} \cdot B_2^d,
\end{align*}
where $\vc_C$ is the circumcenter of $\conv\p{\pc{\vz_1,\dots,\vz_t}}$. The resulting bound on $\pa{S_t}$ follows easily from the above observation and a simple volume argument.

Finally, we obtain the approximation guarantee from noting that for all \(t\), the output of the algorithm from Theorem \ref{thm:main_two} given the first \(t\) points is the same as running it only on the points selected by \(S_t\).

% CMT: LOWER BOUND
\subsection{Lower bound (Theorem \ref{thm:main_four})}

Whereas in the upper bound we demonstrated a particular algorithm that satisfies the evolution condition \eqref{eq:overview_evolution}, 
for the lower bound it suffices to show that for any monotone algorithm, there exists an instance of the problem (a sequence of $\vz_1$,\dots, $\vz_n$) where the algorithm must satisfy the ``reverse evolution condition'', i.e.
\begin{align}
    \frac{\sfrac{1}{\alpha_{t}} - \sfrac{1}{\alpha_{t-1}}}{\log\mathsf{vol}(\cE_{t}) - \log\mathsf{vol}(\cE_{t-1})} \ge C\label{eq:lb_overview_evolution}
\end{align}
for some $C > 0$.
In analogy to the argument of the upper bound, showing this reverse evolution condition yields a lower bound of the form \(\frac{1}{\alpha_n} \geq \widetilde{\Omega}\p{d \log(\kappa)}\).
Given any monotone algorithm \(\cA\), the instance we use is produced by an adversary that repeatedly feeds \(\cA\) a point that is a constant factor away from the previous ellipsoid.

In order to simplify showing this reverse evolution condition, we use a symmetrization argument.
Specifically, by a particular sequence of Steiner symmetrizations, we see that the optimal response of \(\cA\) can be completely described in two dimensions. Thus, it is sufficient to only show this reverse evolution condition in the two-dimensional case where the previous outer ellipsoid is the unit ball.

% show that suffices to only show that this reverse evolution condition holds in the two-dimensional case where the previous outer ellipsoid is the unit ball. 

This transformed two-dimensional setting is significantly simpler to analyze. Specifically, we can assume that the point given by the adversary is always \(2 \ve_1\). The rest of the argument proceeds by cases, again using two-dimensional Euclidean geometry. On a high level, the constraints placed on the new outer and inner ellipsoid by the monotonicity condition force the update of \(\cA\) to satisfy the reverse evolution condition.

\section{Preliminaries}
\label{section:asymmetric_preliminaries}

\subsection{Notation} 

We denote the standard Euclidean norm of a vector $\vv$ by $\pn{\vv}$ and the Frobenius norm of a matrix $\mA$ by $\pn{\mA}_F$. We denote the singular values of a matrix $\mA \in \R^{d\times d}$ by $\sigma_1(\mA),\dots, \sigma_d(\mA)$. Let $\sigma_{\max}(\mA)$ and $\sigma_{\min}(\mA)$ be the largest and smallest singular values of $\mA$, respectively.
We write \(\Diag{a_1, \ldots, a_d}\) to mean the \(d \times d\) diagonal matrix whose diagonal entries are \(a_1, \ldots, a_d\).
We use \(\bS_{++}^d\) to denote the set of \(d \times d\) positive definite matrices.
We use \(\ve_1, \ldots, \ve_d\) for the standard basis in \(\R^d\).

Denote the $\ell_2$-unit ball by $B_2^d = \pc{\vx \in \R^d \colon \pn{\vx} \le 1}$,
and \(\bS^{d-1} = \pc{\vx \in \R^d \colon \|\vx\| = 1}\) the unit Euclidean sphere. We use \(\partial S\) for the boundary of an arbitrary set \(S\). We use natural logarithms unless otherwise specified. 

In this paper, we will work extensively with ellipsoids. We will always assume that all ellipsoids and balls we consider are centered at the origin. We use the following representation of ellipsoids. For a non-singular matrix $\mA\in\R^{d\times d}$, let $\cE_\mA \coloneqq \pc{\vx \colon \pn{\mA\vx}\le 1}$. In other words, the matrix $\mA$ defines an bijective linear map satisfying $\mA \cE_\mA = B_2^d$. Every full-dimensional ellipsoid (centered at the origin) has such a representation. We note that this representation is not unique as matrices $\mA$ and $\mM\mA$ define the same ellipsoid if matrix $\mM$ is orthogonal (since $\pn{\mA\vv} = \pn{\mM\mA\vv}$ for every vector $\vv$). Sometimes, we will have to consider lower-dimensional ellipsoids within an ambient space of higher dimension; in this case, we will use the notation $\cE \cap H$ where $H$ is some linear or affine subspace -- note that $\cE \cap H$ is also an ellipsoid. 

Now consider the singular value decomposition of  $\mA$: $\mA = \mU\Sigma^{-1} \mV^T$ (it will be convenient for us to write $\Sigma^{-1}$ instead of standard $\Sigma$ in the decomposition). The diagonal entries of $\Sigma$ are exactly the semi-axes of $\cE_\mA$. As mentioned above, matrices $\mU\Sigma^{-1} \mV^T$ and $\mU'\Sigma^{-1} \mV^T$ define the same ellipsoid for any orthogonal \(\mU' \in \mathbb{R}^{d \times d}\); in particular, every ellipsoid can be represented by a matrix of the form $\mA=\Sigma^{-1} \mV^T$.

\subsection{Geometry}
We restate the well-known result that five points determine an ellipse.
This is usually phrased for conics, but for nondegenerate ellipses the usual condition that no three of the five points are collinear is vacuously true.
\begin{claim}[Five points determine an ellipse]\label{claim:conic_5pt}
Let \(\vc_1 + \partial \cE_1, \vc_2 + \partial \cE_2\) be two ellipses in \(\R^2\).
If they intersect at five distinct points, then \(\vc_1 + \partial \cE_1\) and \(\vc_2 + \partial \cE_2\) are the same.
\end{claim}
% \ynote{This is well known. We don't need this paragraph.}\mnote{Citation?}
% For an arbitrary conic, this claim comes with the condition that no three of the five points of interesection are collinear.
% However, no three points of any nondegenerate ellipse in \(\R^2\) can be collinear, hence for ellipses this condition is vacuously true and we get the claim above.

The following claim, that every full-rank ellipsoid (i.e. an ellipsoid whose span has full dimension) can be represented by a positive definite matrix, follows from looking at the singular value decomposition of $\mA$.
\begin{claim}\label{claim:prelim_ellips_sym}
    Let \(\cE \subseteq \R^d\) be a full-rank ellipsoid.
    Then there exists \(\mA \succ 0\) such that \(\cE = \cE_{\mA}\).
\end{claim}

We also have the standard result relating volume and determinants, which follows from observing \(\mA \cE_{\mA} = B_2^d\).
\begin{claim}\label{claim:prelim_vol}
Let \(\mA \succ 0\). Then 
\(\vol(\cE_{\mA}) = \det(\mA^{-1}) \vol(B_2^d)\).
\end{claim}

In order to give the reduction in the lower bound from the general case to the two-dimensional case, we use the technique of Steiner symmetrization (see e.g. \cite[Section 1.1.7]{artstein2015asymptotic}).
Given some unit vector \(\vu \in \R^d\) and convex body \(K \subseteq \R^d\), we write \(S_{\vu}(K)\) for the \textit{Steiner symmetrization} in the direction of \(\vu\).
Recall that the Steiner symmetrization is defined so that for any \(\vx \perp \vu\),
\[\vol((\vx + \R \vu) \cap K) = \vol((\vx + \R \vu) \cap S_{\vu}(K)),\]
and so that \((\vx + \R \vu) \cap S_{\vu}(K)\) is an interval centered at \(\vx\).
Note that we will overload notation slightly, as we will allow \(\vu\) to be a vector of any non-zero length while Steiner symmetrization is usually
defined with \(\vu\) being a unit vector, but we will simply take \(S_{\vu} = S_{\frac{\vu}{\|\vu\|_2}}\).

Importantly, Steiner symmetrization will preserve important properties of the update.
We have the key facts that \(\vol(S_{\vu}(K)) = \vol(K)\), \(S_{\vu}(K') \subseteq S_{\vu}(K)\) if \(K \subseteq K'\), and further the Steiner symmetrization preserves \(K\) being an ellipsoid:
\begin{claim}[\protect{\cite[Lemma 2]{bourgain2006estimates}}]\label{claim:steiner_ellipsoid}
    If \(c + \cE \subseteq \R^d\) is an ellipsoid, \(S_{\vu}(c + \cE)\) is still an ellipsoid.
\end{claim}

Further, if we apply Steiner symmetrization to a body that is a body of revolution about an axis,
it does not change the body if \(\vu\) is perpendicular to the axis of revolution.
\begin{claim}\label{claim:steiner_sym}
    Let \(K \subseteq \R^d\) be a body of revolution about the \(\ve_1\)-axis.
    Then if \(\vu \perp \ve_1\), \(S_{\vu}(K) = K\).
\end{claim}

\section{Streaming Ellipsoidal Rounding}
\label{sec:main_alg_top}

Our goal in this section is to prove Theorems \ref{thm:main_one} and \ref{thm:main_two}.

\subsection{Monotone algorithms solve \texorpdfstring{\Cref{prob:main}}{Theorem \ref{prob:main}}}

To design algorithms to solve the streaming ellipsoidal rounding problem,
we first show that any monotone algorithm gives a valid solution.
We let \(\vc_0 \in \R^d\) and \(r_0 \geq 0\) be given so that
\(\vc_0 + r_0 \cdot B_2^d \subseteq Z\),
and denote the initial ellipsoid as \(\cE_0 = r_0 \cdot B_2^d\).
Note that \(r_0\) need not be the inradius, although it is upper bounded by the inradius.

If we had for each intermediate step \(t\)
that \(\vc_t + \alpha_t \cdot \cE_t \subseteq \conv\p{\vz_1, \ldots \vz_t} \subseteq \vc_t + \cE_t\),
then clearly any algorithm that satisfies this would give a valid final solution as well.
However, in intermediate steps it is not clear that 
\(\vc_t + \alpha_t \cdot \cE_t \subseteq \conv\p{\vz_1, \ldots \vz_t}\),
due to the initialization of \(\vc_0 + \cE_0\) in our monotone algorithm framework.
Instead, we relax this invariant to \(\vc_t + \alpha_t \cdot \cE_t \subseteq \conv\p{\{\vz_1, \ldots \vz_t\} \cup (\vc_0 + \cE_0)}\),
which still suffices to produce a valid final solution.
\begin{claim}\label{claim:alg_invariant}
To solve \Cref{prob:main}, it suffices for %\(\vc_0 + \alpha_0 \cE_0 \subseteq \vc_0 + r B_2^d\), and for
the sequence of ellipsoids \(\vc_i + \cE_i\) and scalings \(\alpha_i\) to satisfy the invariants of \Cref{def:alg-invariant}.
\end{claim}

\begin{proof}
First, we argue that \(\conv\p{\vz_1, \ldots, \vz_n} \subseteq \vc_n + \cE_n\).
As \(\cE_n\) is an ellipsoid and therefore a convex set, it suffices to show \(\{\vz_1, \ldots, \vz_n\} \subseteq \vc_n + \cE_n\).
We actually argue by induction that \(\{\vz_1, \ldots, \vz_t\} \subseteq \vc_t + \cE_t\) for all \(0 \leq t \leq n\).
This is vacuously true for \(t = 0\). At each step \(t > 0\) the inductive hypothesis gives \(\{\vz_1, \ldots, \vz_{t-1}\} \subseteq \vc_{t-1} + \cE_{t-1}\), and thus by (\ref{item:def-invariant-1}) we have \(\{\vz_1, \ldots, \vz_{t}\} \subseteq \vc_t + \cE_t\).

Now, we argue that \(\vc_n + \alpha_n \cdot \cE_n \subseteq \conv\p{\vz_1, \ldots, \vz_n}\).
We show by induction that \(\vc_t + \alpha_t \cdot \cE_t \subseteq \conv\p{\{\vz_1, \ldots, \vz_t\} \cup (\vc_0 + \cE_0)}\)
for all \(0 \leq t \leq n\).
This is sufficient as \(\conv\p{\{\vz_1, \ldots, \vz_n\} \cup (\vc_0 + \cE_0)} = Z\).
The case for \(t = 0\) is trivial.
For \(t > 0\), the inductive hypothesis gives \(\vc_{t-1} + \alpha_{t-1} \cdot \cE_{t-1} \subseteq \conv\p{\{\vz_1, \ldots, \vz_{t-1}\} \cup (\vc_0+\cE_0)}\),
and by (\ref{item:def-invariant-2}) we have
\[c_t + \alpha_t \cdot \cE_t \subseteq \conv\p{(c_{t-1} + \alpha_{t-1} \cdot \cE_{t-1}) \cup \{\vz_i\}} \subseteq \conv\p{\{\vz_1, \ldots, \vz_t\} \cup (\vc_0 + \cE_0)},\]
as desired.
\end{proof}

\subsection{Special case}\label{sec:two_dim_update}

In light of \Cref{claim:alg_invariant}, our strategy is to design an algorithm that preserves the invariants given in \Cref{def:alg-invariant}. This algorithm can be thought of as an \textit{update rule} that,
given the previous outer and inner ellipsoids \(\vc_{t-1} + \cE_{t-1}, \vc_{t-1} + \alpha_{t-1} \cE_{t-1}\) and next point \(\vz_t\), produces the next outer and inner ellipsoids \(\vc_t + \cE_t, \vc_t + \alpha_t \cE_t\).

It is in fact sufficient to consider the simplified case where the previous outer ellipsoid is the unit ball, and the previous inner ellipsoid is some scaling of the unit ball;
we will show this in \Cref{sec:gen_high_dim}. 
% (essentially, by an appropriate choice of coordinates where the previous inner ellipsoid ).
We can further specialize by considering only the two-dimensional case \(d=2\).
We will later show that the high-dimensional case is not much different, as all the relevant sets \(\vc_{t_1} + \cE_{t-1}, \vc_t + \cE_{t}\) and \(\conv\p{\alpha \cdot \cE_{t-1} \cup \{\vz_t\}}\) form bodies of revolution about the axis through \(\vc_{t-1}\) and \(\vz_t\).

We now describe our two-dimensional update rule.
In order to simplify notation, we will let \(\alpha\) be the previous scaling \(\alpha_{t-1}\), and \(\alpha'\)
be the next scaling \(\alpha_t\). 
We will assume that \(\alpha \leq \sfrac{1}{2}\) to simplify the analysis
of our update rule; this will not affect the quality of our final approximation as this update rule
will only be used in the ``large approximation factor'' regime.
We will also overload notation by writing \(c + \cE\) even when \(c\)
is a scalar to mean \((c, 0) + \cE\).
% The previous outer ellipsoid is the unit ball, and the previous inner ellipsoid is \(\alpha\) times the unit ball so that the previous scaling is \(\alpha\).
% Further, let \(y^2 = x_2^2 + \ldots + x_d^2\) refer to the coordinates except \(x_1\).
% We will also overload notation and use \(x\) to refer to \(x_1\).
We can describe the previous outer ellipsoid \(\cE\) with the equation \(x^2 + y^2 \leq 1\), and the previous inner ellipsoid \(\alpha \cE\) with \(x^2 + y^2 \leq \alpha^2\).
We define the next outer and inner ellipsoids \(c + \cE'\), \(c + \alpha' \cE'\) as
\[\underbrace{\frac{1}{a^2}(x-c)^2 + \frac{1}{b^2} y^2 \leq 1}_{c + \cE'}, \qquad \underbrace{\frac{1}{a^2}(x-c)^2 + \frac{1}{b^2} y^2 \leq \alpha'^2}_{c + \alpha' \cE'}\]
where we use parameters
\begin{equation}\label{eqn:update_params}
\left.\begin{aligned}
a &= \exp\p{\gamma} \\
b &=1 + \frac{\alpha - \alpha'}{2} \\
c &= - \alpha + \alpha' \cdot a \\
\alpha'& = \frac{1}{\frac{1}{\alpha} + 2\gamma} \\
\end{aligned}\qquad\right\}
\end{equation}

We will let \(\vz\) be the rightmost point of \(c + \cE'\), so that \(\vz = (c+ a, 0)\). Eventually, we will choose \(\gamma\) so that \(\vz\) coincides with \(\vz_{t}\), the point received in the next iteration.
In \Cref{sec:gen_alg}, these parameters \(a(\gamma), b(\gamma), c(\gamma), \alpha'(\gamma)\)
%, as well as the resulting ellipsoids \(c(t) + \cE'(t), c(t) + \alpha' (t)\cE'(t)\)
will be used as functions of the parameter \(\gamma \geq 0\). However, we will not yet explicitly specify \(\gamma\), so in this section these parameters can be thought of as constants for some fixed \(\gamma\). 
This update rule is pictured in \Cref{fig:update_step}.

We first collect a few straightforward properties of this update rule.
\begin{claim}\label{claim:update_params}
The parameters in the setup (\ref{eqn:update_params}) satisfy the following.
\begin{enumerate}
    \item\label{item:claim_update_params_1} \(\frac{1}{\alpha'} = \frac{1}{\alpha} + 2\gamma\)
    \item\label{item:claim_update_params_2} \(b \geq 1\)
    \item\label{item:claim_update_params_3} \(c \geq 0\)
    \item\label{item:claim_update_params_4} \(c + \alpha' \cdot a \geq \alpha\)
\end{enumerate}
\end{claim}
Before proving these properties, we provide geometric interpretations.
Intuitively, (\ref{item:claim_update_params_1}) means that \(\gamma\) is proportional to the increase in the approximation factor at this step, a fact that we will use when analyzing the general-case algorithm.
(\ref{item:claim_update_params_2}) means that the outer ellipsoid grows on every axis;
and (\ref{item:claim_update_params_3}) means that the centers of the next ellipsoids are to the right of the \(y\)-axis, i.e. the centers of the next ellipsoids are further towards \(\vv\) than those of the previous ellipsoids.
The rightmost point of \(c + \alpha' \cE'\) is \(c + \alpha' \cdot a\), so (\ref{item:claim_update_params_4}) shows that this point is to the right of the rightmost point of \(\alpha \cdot \cE\).

We now prove \Cref{claim:update_params}.

\begin{proof}[Proof of \Cref{claim:update_params}]
(\ref{item:claim_update_params_1}) is clear from rearranging the definition of \(\alpha'\).
From (\ref{item:claim_update_params_1}) we also have \(\alpha' \leq \alpha\), so that (\ref{item:claim_update_params_2}) follows immediately.

For (\ref{item:claim_update_params_3}), observe that \(\frac{\alpha}{\alpha'} = 1 + 2 \gamma \alpha\).
When \(\alpha \leq \sfrac{1}{2}\), this means \begin{equation}\label{eqn:pup} % ceqn: property update params
\frac{\alpha}{\alpha'} \leq 1 + \gamma \leq \exp\p{\gamma} = a
\end{equation}
using \(1 + x \leq e^x\), \Cref{claim:wk_e}-(\ref{item:wk_1}).
By definition of \(c\), \(\alpha/\alpha' \leq a\) is equivalent to \(c \geq 0\).

To show (\ref{item:claim_update_params_4}), by definition we have that \(c + \alpha' \cdot a = - \alpha + 2 \alpha' a\).
Thus showing \(c + \alpha' \cdot a \geq \alpha\) is equivalent to showing that \(\alpha' a \geq \alpha\), which is equivalent to the inequality in (\ref{eqn:pup}).
\end{proof}

As \Cref{fig:update_step} depicts, the update step we defined satisfies the invariants in \Cref{def:alg-invariant} and so is monotone;
in the rest of this section we make this picture formal.
To start, we consider the invariant concerning outer ellipsoids; we will show that \(\cE \subseteq c + \cE'\).
For now we can think of \(\vz\) as replacing \(\vz_t\)
%(the point received in the next iteration)
, and clearly \(\vz \in c + \cE'\), so if we show that \(\cE \subseteq c + \cE'\), then \(\conv\p{\cE \cup \{\vz\}} \subseteq c + \cE'\) as well since \(c + \cE'\) is convex.

\begin{claim}\label{claim:update_step_outer}
    We have \(\cE \subseteq c + \cE'\).
\end{claim}
\begin{proof}
First, observe that \(\cE \subseteq \cE'\) because both axes of \(\cE'\) have greater length than those of \(\cE\): \(a \geq 1\) by definition, and \(b \geq 1\) from \Cref{claim:update_params}-(\ref{item:claim_update_params_2}).
Now, we translate \(\cE'\) to the right until it touches \(\cE\) at two points. 
We call this translated ellipse \(c_r + \cE'\), as shown in \Cref{fig:update_step_outer}.
Observe that as long as \(c \leq c_r\), we have \(\cE \subseteq c + \cE'\).
We now determine \(c_r\).

\begin{figure}[ht]
\centering
\includegraphics[height=6cm]{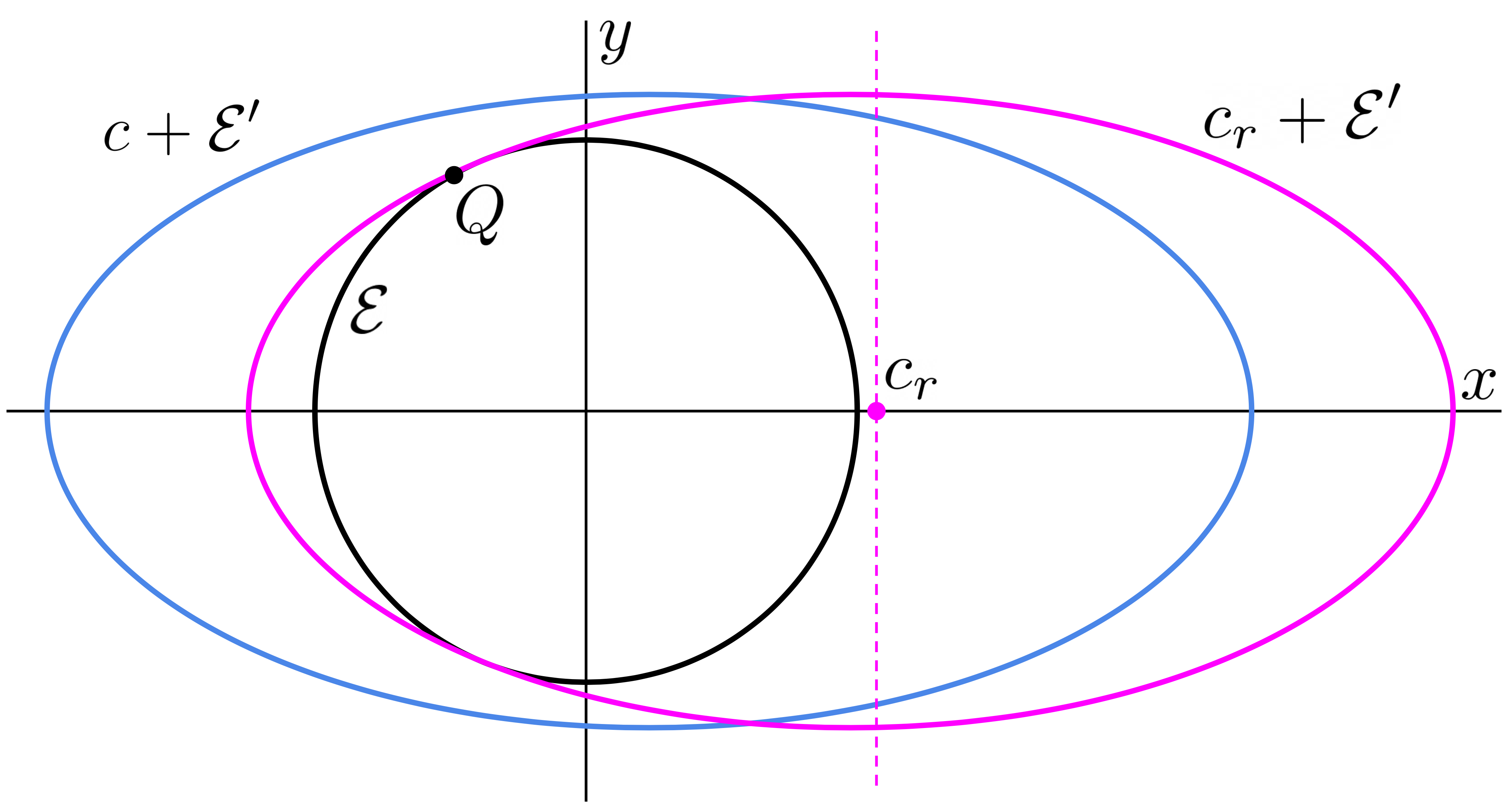}
\caption[Outer ellipses of the update step.]{Outer ellipses of the update step. As before, \(\cE\) is the black circle and \(c + \cE'\) is the blue ellipse.
\(c_r + \cE'\) is the magenta ellipse, with its center at \(c_r\) and the dotted magenta line showing the position of \(c_r\) along the \(x\)-axis.
\(c_r\) is defined so \(c_r + \cE'\) and \(\cE\) are tangent at two points. \(Q\) is one of these two tangent points.}
\label{fig:update_step_outer}
\end{figure}

First, note points on the boundary of \(c_r + \mathcal{E}'\) are described by the equation \begin{equation}\label{eqn:def_cr}
\frac{(x - c_r)^2}{a^2} + \frac{y^2}{b^2} = 1
\end{equation}
Let \(Q = (x', y')\) be the point of intersection between \(\cE\) and \(c_r + \cE'\) where \(y' > 0\).
Since \(Q\) is on the boundary of both ellipses, the vectors \(\left(\frac{2(x'-c_r)}{a^2}, \frac{2 y'}{b^2}\right)\) and \(\left(2x', 2y'\right)\), which are the normal vectors at \(Q\) of \(c_r + \cE'\) and \(\cE\) respectively, must be parallel.
Thus \(\frac{4 (x' - c_r)}{a^2} \cdot y' = \frac{4 y' x'}{b^2}\), which simplifies to \begin{equation}\label{eqn:x_cr}
x' = \frac{c_r}{1 - \frac{a^2}{b^2}}.
\end{equation}
At this point we have a system of three equations relating \((x', y')\) and \(c_r\): (\ref{eqn:x_cr}), \(Q\) lying on \(\cE\), and \(Q\) satisfying (\ref{eqn:def_cr}).
We now solve this system to find \(c_r\).
To start, we expand (\ref{eqn:def_cr}) into
\(x'^2 - 2 x' c_r + c_r^2 + y'^2 \frac{a^2}{b^2} = a^2\),
which we rewrite into \(x'^2 \frac{a^2}{b^2} + x'^2 \left(1 - \frac{a^2}{b^2}\right) - 2 x' c_r + c_r^2 + y'^2 \frac{a^2}{b^2} = a^2\).
As \(Q\) lies on \(\cE\), this becomes \(x'^2 \left(1 - \frac{a^2}{b^2}\right) - 2 x' c_r + c_r^2 +  \frac{a^2}{b^2} = a^2\).
Substituting in (\ref{eqn:x_cr}), we get
\[\frac{c_r^2}{1 - \frac{a^2}{b^2}} - 2 \frac{c_r^2}{1 - \frac{a^2}{b^2}} + c_r^2 + \frac{a^2}{b^2} = a^2.\]
Simplifying, we have \(c_r^2 \left(1 - \frac{b^2}{b^2 - a^2}\right) = a^2 \left(1 - \frac{1}{b^2}\right)\), i.e.
\[c_r^2 = \frac{b^2 - 1}{b^2} (a^2 - b^2).\]
To complete the proof of \cref{claim:update_step_outer}, it suffices to show \(c^2 \leq \frac{b^2 - 1}{b^2} (a^2 - b^2)\). This will follow from \Cref{claim:pty_outer}.
\end{proof}

Now, we move on to the inner ellipsoid invariant of \Cref{def:alg-invariant}.
In particular, we will argue that \(c + \alpha' \cE' \subseteq \conv\p{\alpha\cE \cup \{\vz\}}\).
On a high level, we show this by arguing that the boundary of \(c + \alpha' \cE'\) does not intersect the boundary of \(\conv\p{\alpha\cE \cup \{\vz\}}\), except at points of tangency.

\begin{figure}[ht]
\centering
\includegraphics[height=6cm]{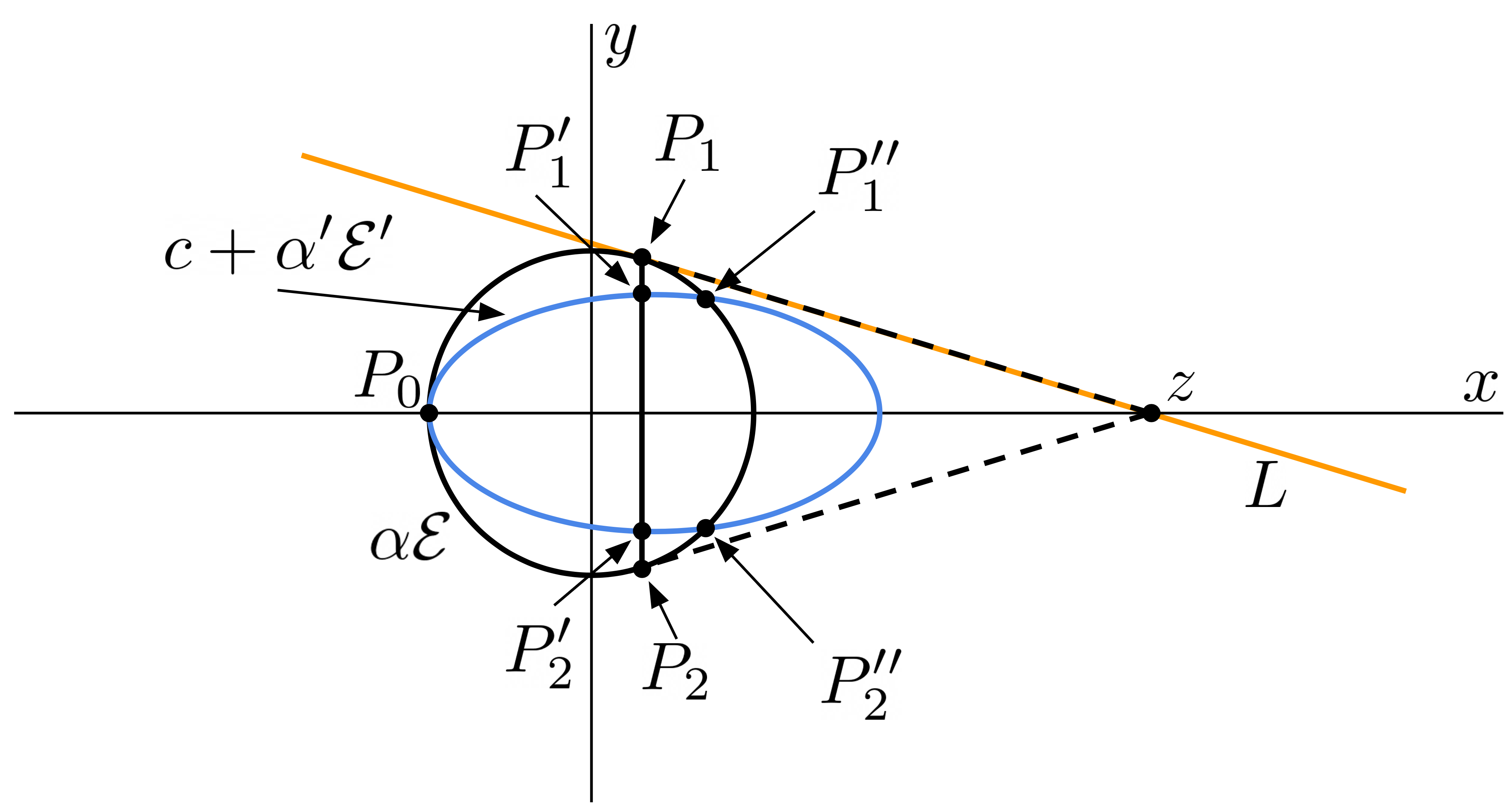}
\caption[Inner ellipses of the update step.]{Inner ellipses of the update step. As before, \(\alpha \cE\) is the black circle and \(c + \alpha' \cE\) is the blue ellipse. 
\(P_0\) is the shared leftmost point of \(\alpha \cE\) and \(c + \alpha' \cE'\).
There are two lines through \(\vv\) that are tangent to \(\alpha \cE\), one of which we call \(L\) and pictured in orange. We call the tangent points \(P_1\) and \(P_2\).
The line segments \(\overline{P_1 \vz}, \overline{P_2 \vz}\) are the dotted black lines.
\(P_1'\) and \(P_2'\) are the two points of intersection between \(c + \alpha' \cE\) and the line segment \(\overline{P_1 P_2}\). \(P_1''\) and \(P_2''\)
are the two points of intersection between \(\partial (c + \alpha' \cE')\) and \(\partial \alpha \cE\) to the right of the \(y\)-axis. Note that \(P_2, P_2', P_2''\) are the reflections of \(P_1, P_1', P_1''\) across the \(x\)-axis.}
\label{fig:update_step_inner}
\end{figure}

We can split the boundary of \(\conv\p{\alpha\cE \cup \{\vz\}}\) into two pieces: the part that intersects with the boundary of \(\alpha \cE\), which is an arc of the boundary of \(\alpha \cE\); and the remainder, which can described as two line segments connecting \(\vz\) to that arc. In particular, there are two lines that go through \(\vz\) and are tangent to \(\alpha \cE\), one of which we call line \(L\), and the other line is the reflection of \(L\) across the \(x\)-axis.
We define \(P_1\) and \(P_2\) as the tangent points of these lines to \(\alpha \cE\). 
Then, the boundary of \(\conv\p{\alpha\cE \cup \{\vz\}}\) consists of an arc \(P_1 P_2\) and the line segments \(\overline{P_1 \vz}, \overline{P_2 \vz}\). This is illustrated in \Cref{fig:update_step_inner}.
Note that at this point it is possible a priori for the arc \(P_1 P_2\) that coincides with the boundary of \(\conv\p{\alpha\cE \cup \{\vz\}}\)
to be either the major or minor arc; we will later show it must be the major arc.
We will take \(L\) to be the line whose tangent point to \(\alpha \cE\), \(P_1\), is above the \(x\)-axis, though this choice is arbitrary due to symmetry across the \(x\)-axis.

We first show that \(c + \alpha' \cE'\) does not intersect \(\overline{P_1 \vz}\) and \(\overline{P_2 \vz}\), except possibly at points of tangency.
In fact, we show a slightly stronger statement, in similar fashion to \Cref{claim:update_step_outer}.

\begin{claim}\label{claim:translate_ellipse_angle}
 % Ellipse 
 \(c + \alpha' \cE'\) lies inside the angle \(\angle P_1 \vz P_2\).
\end{claim}
% When we say that \(c + \alpha' \cE'\) lies inside the angle \(\angle P_1 v P_2\), we mean that it is contained in \(\mathsf{conv}(\overrightarrow{v P_1}, \overrightarrow{v P_2})\).
\begin{proof}
We translate \(c + \alpha' \cE'\) to the right until it touches \(L\) (and, by symmetry, \(\overline{P_2 \vz}\)).
We call this translated ellipse \(c_+ + \alpha' \cE'\), as shown in \Cref{fig:update_step_inner_translate}. 
(Formally, the center \(c_+\) can be described not as a translation from some other ellipse,
but as \(c_+\) such that \(c_+ + \alpha' \cE'\) intersects \(L\) at one point).
Observe that 
if \(c \leq c_{+}\), then \(c + \alpha' \cE'\) lies inside the angle \(\angle P_1 \vz P_2\).
We now determine \(c_{+}\).

\begin{figure}[ht]
\centering
\includegraphics[height=6cm]{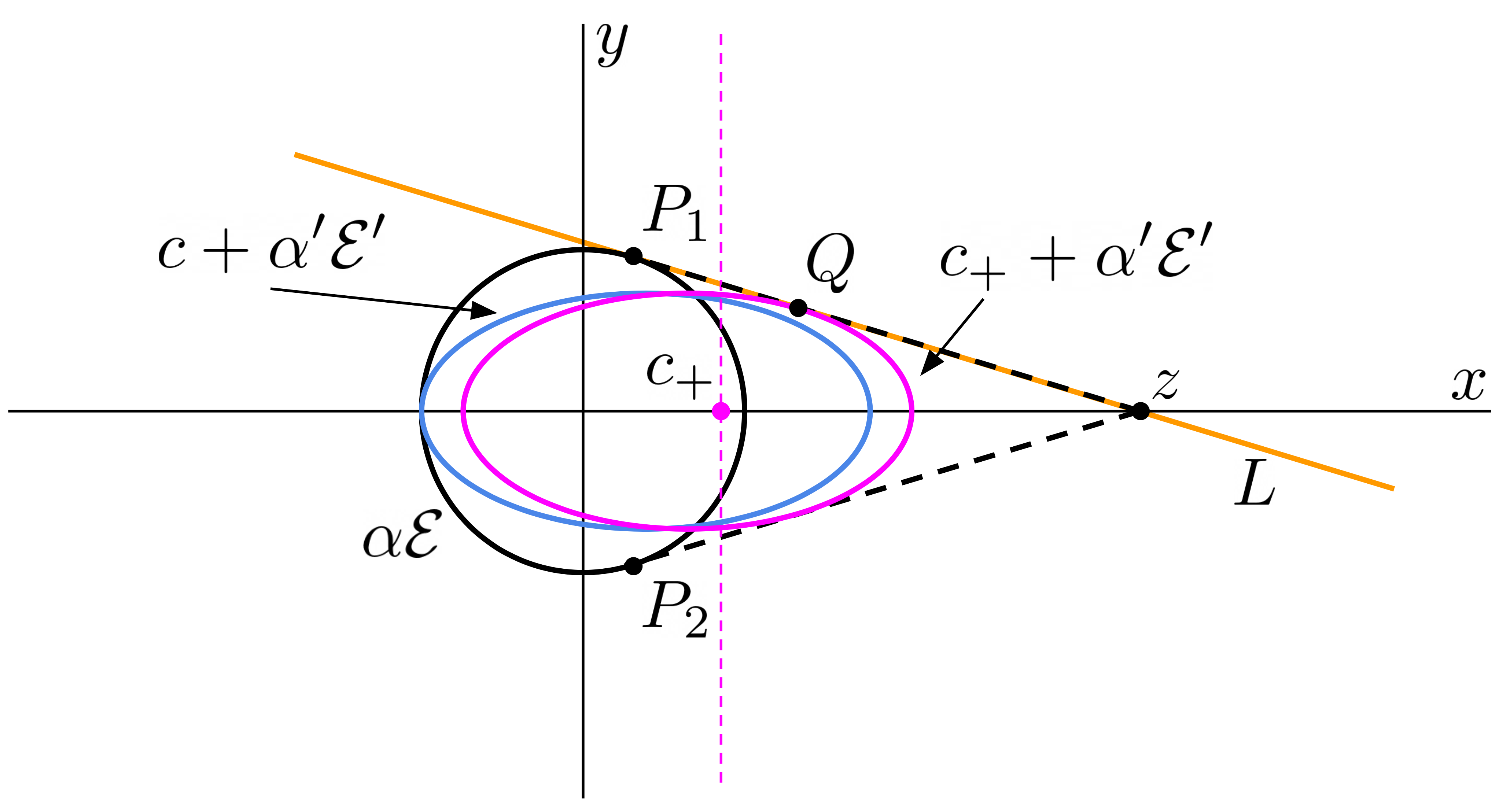}
\caption[Inner ellipses of the update step.]{Inner ellipses of the update step. As before, \(\alpha \cE\) is the black circle, \(c + \alpha' \cE\) is the blue ellipse, \(L\) is the orange line through \(\vz\) and tangent to \(\alpha \cE\), \(P_1\) and \(P_2\) are the tangent points on the lines through \(\vz\) tangent to \(\alpha \cE\), and \(\overline{P_1 \vz}, \overline{P_2 \vz}\) are the dotted black lines.
\(c_{+} + \alpha' \cE'\) is the magenta ellipse, with its center at \(c_{+}\) and magenta dotted line showing its position on the \(x\)-axis.
\(c_{+}\) is defined so that \(c_{+} + \alpha' \cE'\) is tangent to \(\overline{P_1 \vz}\) and \(\overline{P_2 \vz}\), with \(Q\) as the tangent point of \(c_{+} + \alpha' \cE'\) and \(\overline{P_1 \vz}\).
}
\label{fig:update_step_inner_translate}
\end{figure}

The equation of \(L\) is \[\underbrace{\frac{1}{c+a}}_{\ell_1} \cdot x + \underbrace{\sqrt{\frac{1}{\alpha^2} - \frac{1}{(c+a)^2}}}_{\ell_2} \cdot y = 1\]
where we define \(\ell_1, \ell_2\) as the coefficents for \(x\) and \(y\).
Observe that \(\vz\) is on \(L\), and \(L\) is tangent to \(\alpha \cE\) at \(P_1\), which has coordinates \begin{equation}\label{eqn:usit_p1} % equation: update step inner translate p1
P_1 = \left(\frac{\alpha^2}{c+a}, \alpha^2 \sqrt{\frac{1}{\alpha^2} - \frac{1}{(c+a)^2}} \right).
\end{equation}
Tangency can be confirmed by checking that \(P_1\) is parallel to \((\ell_1, \ell_2)\), the normal vector definining \(L\).

Let \(Q = (x', y')\) be the point of intersection of \(L\) and \(c_{+} + \alpha' \cE\), there are three properties that define \(Q\). First it lies on the boundary of \(c_{+} + \alpha' \cE\), so it satisfies
\begin{equation}\label{eqn:pty_inner_oe} % equation: inner property on ellipsoid
    \frac{(x'-c_{+})^2}{a^2} + \frac{y'^2}{b^2} = \alpha'^2.
\end{equation}

Second, at \(Q\) the normal vectors for the equations defining \(c_{+} + \alpha' \cE\) and \(L\) are parallel, i.e.
\((\frac{2(x-c_{+})}{a^2},\frac{2y}{b^2})\) is parallel to \((\ell_1, \ell_2)\). So
\begin{equation}\label{eqn:pty_inner_gm} % equation: inner property gradient matching
    \frac{(x'-c_{+})}{a^2} \ell_2 = \frac{y'}{b^2} \ell_1.
\end{equation}
Finally, \(Q\) lies on \(L\), so we have \(\ell_1 x' + \ell_2 y' = 1\).
Solving this for \(y'\), we get
\begin{equation}\label{eqn:pty_inner_ol} % equation: inner property on line
    y' = \frac{1-\ell_1 x'}{\ell_2}.
\end{equation}
These three equations form a system for \(x', y'\) and \(c_{+}\), which we now solve to find \(c_{+}\).
Taking the square of (\ref{eqn:pty_inner_gm}) and rearranging gives \(\frac{y'^2}{b^2} = \frac{b^2 (x'-c_{+})^2 \ell_2^2}{a^4 \ell_1^2} \).
Substituting this into (\ref{eqn:pty_inner_oe}), we get
\(\frac{(x'-c_{+})^2}{a^2} + \frac{b^2 (x'-c_{+})^2 \ell_2^2}{a^4 \ell_1^2} = \alpha'^2\).
Now, defining \(r \defeq\frac{a^2 \ell_1^2}{b^2 \ell_2^2}\), we group the terms of this equation into the form
% \[(x'-c_t)^2 \cdot \frac{1}{a^2} \left(1 + \frac{b^2 \ell_2^2}{a^2 \ell_1^2}\right) = \alpha'^2\]
% We, as we will see this quantity again.
% Thus we get
\begin{equation}\label{eqn:pty_inner_1}
    (x'-c_{+})^2 \cdot \frac{1}{a^2} \left(1 + \frac{1}{r}\right) = \alpha'^2.
\end{equation}
We substitute (\ref{eqn:pty_inner_ol}) into (\ref{eqn:pty_inner_gm})
to get \(\frac{x'-c_{+}}{a^2} \ell_2 = \frac{\ell_1}{b^2} \frac{1 - x' \ell_1}{\ell_2}\).
%Expanding gives \(\frac{x'-c_t}{a^2} \ell_2 = \frac{\ell_1}{b^2 \ell_2} - \frac{x' \ell_1^2}{b^2 \ell_2}\),
%Grouping for \(x'\) yields %\(x' \left(1 + \frac{a^2\ell_1^2}{b^2 \ell_2}^2\right) = c_t + \frac{a^2 \ell_1}{b^2 \ell_2}\), i.e. 
%\[
%x' (1+r) = c_t + \frac{r}{\ell_1}
%\]
%Thus \(x' = \frac{c_t + \frac{r}{\ell_1}}{1+r}\), and so \(x'-c_t = \frac{c_t+\frac{r}{\ell_1} - c_t(1+r)}{1 + r} = \frac{\frac{r}{\ell_1} - c_t r}{1+r}\), and we get
Grouping for \(x'\) and rearranging yields
\begin{equation}\label{eqn:pty_inner_2}
    x' - c_{+} = \frac{r}{1+r} \left(\frac{1}{\ell_1} - c_{+}\right).
\end{equation}
%Next, we substitute (\ref{eqn:pty_inner_2}) into (\ref{eqn:pty_inner_1}) and get
% \[ \left(\frac{r}{1+r}\right)^2 \left(\frac{1}{\ell_1} - c_t\right)^2 \cdot \frac{1}{a^2} \left(\frac{r+1}{r}\right) = \alpha'^2\]
%After some cancellation, we have
Next, we substitute (\ref{eqn:pty_inner_2}) into (\ref{eqn:pty_inner_1}), and get after some cancellation
\[ \left(\frac{1}{\ell_1} - c_{+}\right)^2 = \alpha'^2 a^2 \cdot \frac{1+r}{r}.\]
Observe on the left hand side that \(\frac{1}{\ell_1} - c_{+} = c+a-c_{+}\).
Clearly the center \(c_{+}\) must be to the left of \(\vz\), so this must be non-negative.
%Hence after taking the positive square root, we obtain \(\frac{1}{\ell_1} - c_t = \alpha' \cdot a \sqrt{\frac{1+r}{r}}\), i.e. 
%\[c_t = \frac{1}{\ell_1} - \alpha' \cdot a \sqrt{\frac{1+r}{r}} = c + a - \alpha' \cdot a \sqrt{\frac{1+r}{r}}\]
Hence after taking the positive square root, we obtain
\[c_{+} = c + a - \alpha' \cdot a \sqrt{\frac{1+r}{r}}\]
It remains to show that \(c \leq c_{+}\), or equivalently that
\[a - \alpha' \cdot a \sqrt{\frac{1+r}{r}} \geq 0,\]
which we do in \Cref{claim:pty_inner}. This completes the proof of \Cref{claim:translate_ellipse_angle}. 
\end{proof}

Now, we build on the previous claim to show the inner ellipsoid invariant.
\begin{claim}\label{claim:update_step_inner}
    We have \(c + \alpha' \cdot \cE' \subseteq \conv\p{\alpha\cdot\cE \cup \{\vz\}}\).
\end{claim}
\begin{proof}
We will argue that the boundary of \(c + \alpha' \cE'\) does not intersect the boundary of \(\conv\p{\alpha\cE \cup \{\vz\}}\), except at points of tangency.
This is sufficient to establish the claim, as \Cref{claim:translate_ellipse_angle} shows that \(c + \alpha' \cE'\) is internal to \(\angle P_1 \vz P_2\), and so if \(c + \alpha' \cE'\) does not intersect the boundary of \(\conv\p{\alpha\cE \cup \{\vz\}}\), \(c + \alpha' \cE'\) must lie inside of, or be disjoint from \(\conv\p{\alpha\cE \cup \{\vz\}}\).
Since the leftmost points of \(\alpha \cE\) and \(c + \alpha' \cE'\) coincide, \(c + \alpha' \cE'\) must then lie inside of  \(\conv\p{\alpha\cE \cup \{\vz\}}\).
Recall that the boundary of \(\conv\p{\alpha\cE \cup \{\vz\}}\) consists of the arc \(P_1 P_2\) and the line segments \(\overline{P_1 \vz}, \overline{P_2 \vz}\).
\Cref{claim:translate_ellipse_angle} already shows that the boundary of \(c + \alpha' \cE'\) does not intersect \(\overline{P_1 \vz}\) and \( \overline{P_2 \vz}\), so we only need to show that the boundary of \(c + \alpha' \cE'\) does not intersect the arc \(P_1 P_2\).

To do this, we start by enumerating the points of intersection of \(\partial \alpha \cE\) and \(\partial(c + \alpha' \cE')\), recalling that \(P_1 P_2\) is an arc of \(\partial \alpha \cE\).
Observe that the leftmost points of \(\alpha \cE\) and \(c + \alpha' \cE'\) coincide, as the leftmost point of \(c + \alpha' \cE'\) is \(c - \alpha' \cdot a = -\alpha\) by definition; we call this point \(P_0\). \(P_0\) is a point of tangency and hence has intersection multiplicity 2, because the centers of \(\alpha \cE\) and \(c + \alpha' \cdot \cE'\) both lie on the \(x\)-axis.

Next, we argue for the existence of two more distinct intersection points \(P_1'', P_2''\) as depicted in \Cref{fig:update_step_inner}.
The leftmost point of \(c + \alpha' \cE'\) is \((- \alpha, 0)\), and the rightmost point is \(c + \alpha'\), which by \Cref{claim:update_params}-(\ref{item:claim_update_params_4}) is to the right of \((\alpha, 0)\), the rightmost point of \(\alpha \cE\).
Thus, by lying on \(\partial \alpha \cE\), \(P_1, P_2\) lie between the leftmost and rightmost points of \(c + \alpha' \cE'\), and so \(c + \alpha' \cE'\) intersects the line through \(P_1\) and \(P_2\).
Further, by \Cref{claim:translate_ellipse_angle}, as \(c + \alpha' \cE'\) lies in the angle \(\angle P_1 \vv P_2\), \(c + \alpha' \cE'\) actually intersects the line segment \(\overline{P_1 P_2}\).
Observe that this intersection happens at two distinct points, which we call \(P_1'\) and \(P_2'\).
Both points are inside of \(\alpha \cE\), yet \(\partial (c + \alpha' \cE')\) is a continuous path that connects both to
the rightmost point of \(c + \alpha' \cE'\), which is outside of \(\alpha \cE\).
Thus \(\partial (c + \alpha' \cE')\) intersects \(\partial \alpha \cE\) at two more distinct points, which we call \(P_1''\) and \(P_2''\).

Now, we argue that \(P_1''\) and \(P_2''\) lie on the minor arc \(P_1 P_2\).
First, observe that the arc \(P_1 P_2\) containing \(P_0\) is the major arc. This is because \(P_1\) lies to the right of the \(y\)-axis, as determined in (\ref{eqn:usit_p1}); and by symmetry so does \(P_2\).
This also implies that major arc \(P_1 P_2\) is the arc with which the boundary of \(\conv\p{\alpha \cE \cup \{\vz\}}\) coincides.
\(P_1'\) and \(P_2'\) are collinear with \(P_1\) and \(P_2\), and as \(P_1''\) and \(P_2''\) are to the right of \(P_1'\) and \(P_2'\), this implies that they must lie on the minor arc \(P_1 P_2\).

Counting all the intersection points of \(\partial \alpha \cE\) and \(\partial(c + \alpha' \cE')\), we have \(P_0\) (with multiplicity \(2\))
and \(P_1''\) and \(P_2''\) (both with multiplicity 1); with total multiplicity 4.
Using \Cref{claim:conic_5pt}, it is impossible for them to have another intersection point without both ellipses being the same.
Thus \(\partial(c + \alpha' \cE')\) cannot intersect the major arc \(P_1 P_2\) except at \(P_0\), and so except at points of tangency the boundary of \(c + \alpha' \cE'\) does not intersect the boundary of \(\conv\p{\alpha\cE \cup \{\vz\}}\).
\end{proof}

\subsection{Generalizing to high dimension and arbitrary previous ellipsoids}
\label{sec:gen_high_dim}
Now that we have demonstrated the invariants of \Cref{def:alg-invariant} for the special two-dimensional case where the previous ellipsoid is the unit ball, we generalize slightly to higher dimensions. However, we first still assume the previous ellipsoid is the unit ball.

Using the parameters as defined in (\ref{eqn:update_params}), we will let \(\cE = B_2^d\), and define the boundary of \(\cE'\) as
\[\frac{1}{a^2} (\vx_1 - c)^2 + \frac{1}{b^2} \vx_2^2 + \ldots + \frac{1}{b^2} \vx_d^2 = 1.\]
Observe that we can also write \(\cE' = \cE_{\mD}\) where \(\mD = \Diag{\frac{1}{a^2}, \frac{1}{b^2}, \ldots, \frac{1}{b^2}}\).
Similarly to before, we let \(\vz = (c+a, 0, 0, \ldots, 0) \in \R^d\),
the furthest point of \(c + \cE'\) in the positive direction of the \(x_1\)-axis.
% As before, we overload notation so that \(c + \cE\) when \(c\) is a scalar means \((c, 0, \ldots, 0) + \cE\).

Now, we argue that the invariants of \cref{def:alg-invariant}
still hold in this setting.
\begin{claim}\label{claim:i_gen_dim1}
The inner and outer ellipsoid invariants hold in this setting:
\begin{enumerate}
    \item \(\cE \subseteq c \cdot \ve_1 + \cE'\)
    \item \(c \cdot \ve_1 + \alpha' \cE' \subseteq \conv\p{\alpha\cE \cup \{\vz\}}\)
\end{enumerate}
\end{claim}
\begin{proof}
Observe that \(\cE\), \(c \cdot \ve_1 + \cE'\), \(c \cdot \ve_1 + \alpha' \cE'\), and \(\conv\p{\alpha \cE \cup \{\vz\}}\) are all bodies of revolution about the \(x_1\)-axis,
with their cross-sections given by their counterparts in \Cref{sec:two_dim_update}.
As \Cref{claim:update_step_outer} and \Cref{claim:update_step_inner}
hold for these cross sections, the set containments hold for
the bodies of revolution as well.
\end{proof}

We further generalize to the case where the previous ellipsoid is arbitrary.
In particular, let \(\vc^{\circ} + \cE\) be the previous ellipsoid, with a vector \(\vc^{\circ} \in \R^d\) and \(\cE = \cE_{\mA}\) for non-singular matrix \(\mA \in \R^{d \times d}\).
Let \(\vz^{\circ} \in \R^d\) be an arbitrary vector, representing the next point received.
We let \(\vu = \mA(\vz^{\circ} - \vc^{\circ})\), and \(\mW \in \R^{d \times d}\) be an orthogonal matrix with \(\vw = \frac{\vu}{\|\vu\|}\) as its first column (e.g. by using as its columns an orthonormal basis containing \(\vw\)).
We define the next outer ellipsoid as \(\vc^{\circ}+c \mA^{-1} \vw + \cE' \) for \(\cE' = \cE_{\mW \mD \mW^\top \mA}\), with \(\mD = \Diag{\frac{1}{a^2}, \frac{1}{b^2}, \ldots, \frac{1}{b^2}}\) as before.
Observe that \(\vz = \vc^{\circ}+(c+a) \mA^{-1} \vw\) is the furthest point of \(\vc^{\circ}+c \mA^{-1} \vw + \cE'\) from the previous center \(\vc^{\circ}\) towards \(\vz^{\circ}\).

This setup works to preserve the key invariants, as we see in the next claim.
\begin{claim}\label{claim:i_gen_us}
   The inner and outer ellipsoid invariants hold in this setting:
    \begin{enumerate}
        \item \(\vc^{\circ} + \cE \subseteq \vc^{\circ}+c \mA^{-1} \vw + \cE' \)
        \item \(\vc^{\circ}+c \mA^{-1} \vw + \alpha' \cE' \subseteq \conv\p{(\vc^{\circ} + \alpha\cE) \cup \{\vz\}}\)
    \end{enumerate}
\end{claim}
\begin{proof}
    We translate both set inclusions by \(- \vc^{\circ}\), then apply the nonsingular linear transformation \(\mW^\top \mA\).
    Observe that the set inclusions we wish to prove hold if and only if the transformed ones do.
    Noting that \(\mW^\top \mA \cE' = \cE_{\mW \mD}\), the transformed set inclusions are
    \(\cE_{\mW} \subseteq c \cdot \ve_1 + \cE_{\mW \mD}\)
    and \(c \cdot \ve_1 + \alpha' \cE_{\mW \mD} \subseteq \conv\p{\alpha \cE_{\mW} \cup \{(c + a) \cdot \ve_1\}}\).
    However, since \(\mW\) is an orthogonal matrix,
    \(\cE_{\mW} = B_2^d\) and \(\cE_{\mW \mD} = \cE_{\mD}\),
    and so the inclusions are exactly those shown in \Cref{claim:i_gen_dim1}.
\end{proof}

Choosing \(\gamma\) correctly in (\ref{eqn:update_params})
ensures that \(\vz \in \vc^{\circ}+c \mA^{-1} \vw + \cE'\) coincides with \(\vz^{\circ}\),
as stated in the upcoming claim. This can be seen by looking at the definition of \(\vz\).
\begin{claim}\label{claim:i_gen_v}
    If \(\gamma\) is chosen so that \(c + a = \|\vu\|\), then \(\vz = \vz^{\circ}\).
\end{claim}

\subsection{General algorithm}
\label{sec:gen_alg}
The goal of this section is to give and analyze a full algorithm that solves the streaming ellipsoid approximation problem, building on the analysis of the update rule from the previous sections.

Before we describe the complete algorithm, we give pseudocode in \Cref{alg:full_update} for its primary primitive. It is an update step like the one we analyzed in the previous section, \Cref{sec:gen_high_dim}.

% old: full algorithm with just the old update step
% \begin{algorithm}[H]
% \caption{Streaming ellipsoid approximation}\label{alg:main}

% \textbf{input}: \(\vc_0+r B_2^d \subseteq Z\) \\
% \textbf{output}: \(\vc_n + \cE_n, \vc_n + \alpha_n \cdot \cE_n\)
% %represented as \(\{x \in \R^d \colon \|A_n (x - c_n)\| \leq 1\}\), \(\{x \in \R^d \colon \|A_n (x - c_n)\| \leq \alpha_n\}\)
% \begin{algorithmic}[1]
% \State Initialize \(\mA_0 = \frac{1}{r} \mI_d, \vc_0 = \vc_0, \alpha_0 = \frac{1}{2}\)
% \For{\(i = 1, \ldots, n\)}
%     \State Receive point \(\vz_i\)
%     \State Let \(\vu = \mA_{i-1}(\vz_i - \vc_{i-1})\), \(\vw = \frac{\vu}{\|\vu\|}\)
%     \If{\(\|\vu\| > 1\)}
%         \State Let \(\gamma^{\star}_t\) be such that \(a(\gamma^{\star}_t) + c(\gamma^{\star}_t) = \|\vu\|\)\label{alg_line:compute_update_t}
%         \State \(\hat{\mA} = \frac{1}{b(\gamma^{\star}_t)} \mI_d + \left(\frac{1}{a(\gamma^{\star}_t)} - \frac{1}{b(\gamma^{\star}_t)}\right) \vw \vw^\top\)\label{alg_line:ellipsoid_update}
%         \State \(\mA_i = \hat{\mA} \cdot \mA_{i-1}, \ \vc_i = \vc_{i-1} + c(t^*) \mA_{i-1}^{-1} \vw, \ \alpha_i = \alpha'(\gamma^{\star}_t)\) \label{alg_line:update_mat}
%     \Else
%         \State \(\mA_i = \mA_{i-1}, \vc_i = \vc_{i-1}, \alpha_i = \alpha_{i-1}\)
%         \label{alg_line:no_update}
%     \EndIf
% \EndFor
% \State \Return \(\mA_n, \vc_n, \alpha_n\)
% \end{algorithmic}
% \end{algorithm}

\begin{algorithm}[ht]
\caption{Full update step \(\cA^{\mathsf{full}}\)}\label{alg:full_update}

\textbf{input}: \(\mA_{t-1} \in \R^{d \times d}, \vc_{t-1} \in \R^d, \alpha_{t-1} \in [0, \frac{1}{2}], \vz_t \in \R^d\) \\
\textbf{output}: \(\mA_{t} \in \R^{d \times d}, \vc_{t} \in \R^{d}, \alpha_{t} \in [0, \alpha_{t-1}]\)
%represented as \(\{x \in \R^d \colon \|A_n (x - c_n)\| \leq 1\}\), \(\{x \in \R^d \colon \|A_n (x - c_n)\| \leq \alpha_n\}\)
\begin{algorithmic}[1]
% \For{\(t = j+1, \ldots, n\)}
%     \State Receive point \(\vz_i\)
    \State Let \(\vu = \mA_{t-1}(\vz_t - \vc_{t-1})\), \(\vw = \frac{\vu}{\|\vu\|}\)
    \If{\(\|\vu\| > 1\)}
        \State Let \(\gamma^{\star}_t\) be such that \(a(\gamma^{\star}_t) + c(\gamma^{\star}_t) = \|\vu\|\)\label{alg_line:compute_update_t}
        \State \(\hat{\mA} = \frac{1}{b(\gamma^{\star}_t)} \mI_d + \left(\frac{1}{a(\gamma^{\star}_t)} - \frac{1}{b(\gamma^{\star}_t)}\right) \vw \vw^\top\)\label{alg_line:ellipsoid_update}
        \State \Return \(\mA_t = \hat{\mA} \cdot \mA_{t-1}, \ \vc_t = \vc_{t-1} + c(\gamma^{\star}_t) \mA_{t-1}^{-1} \vw, \ \alpha_t = \alpha'(\gamma^{\star}_t)\) \label{alg_line:update_mat}
    \Else
        \State \Return \(\mA_t = \mA_{t-1}, \vc_t = \vc_{t-1}, \alpha_i = \alpha_{t-1}\)
        \label{alg_line:no_update}
    \EndIf
% \EndFor
% \State \Return \(\mA_n, \vc_n, \alpha_n\)
\end{algorithmic}
\end{algorithm}

In Lines \ref{alg_line:compute_update_t}, \ref{alg_line:ellipsoid_update} and \ref{alg_line:update_mat}, we use the definition of \(a(\gamma), b(\gamma), c(\gamma), \alpha'(\gamma)\) from (\ref{eqn:update_params}),
substituting \(\alpha_{t-1}\) for \(\alpha\).
Although the update step does not explicitly mention ellipsoids,
we use \(\cE_t = \cE_{\mA_t}\) so that at iteration \(t\) the next outer and inner ellipsoids are \(\vc_t + \cE_{\mA_t}\)
and \(\vc_t + \alpha_t \cE_{\mA_t}\), respectively.
If at this iteration \(\|\vu\| \leq 1\), we will refer to this as the case where the ellipsoids are not updated, as is clear from Line \ref{alg_line:no_update}.

Observe also that if in iteration \(t\) we let \(\mW \in \R^{d \times d}\) be an orthogonal matrix with \(\vw\) as its first column,
we can write
\begin{equation}\label{eqn:alg_mat_update}
\hat{\mA} = \mW \cdot \Diag\p{\frac{1}{a(\gamma^{\star}_t)}, \frac{1}{b(\gamma^{\star}_t)}, \cdots, \frac{1}{b(\gamma^{\star}_t)}} \cdot \mW^\top
\end{equation}

% Now, we analyze the approximation factor \(\frac{1}{\alpha_n}\) given by the algorithm.
%Let \(\gamma^{\star}_t\) be \(t^*\) in Line \ref{alg_line:compute_update_t} at the \(t\)th iteration of the algorithm. 
Now, we argue that this algorithm satisfies the invariants defined in \Cref{def:alg-invariant}.
This argument is essentially the observation that the update step in the algorithm is the one analyzed in \Cref{claim:i_gen_us}.
\begin{claim}\label{claim:alg_correctness}
\Cref{alg:full_update} is a monotone update; i.e., it satisfies the invariants in \Cref{def:alg-invariant}.
\end{claim}
\begin{proof}
If \(\|\vu\| \leq 1\), then \(\vz_i \in \vc_n + \cE_n\) and the inner and outer ellipsoids are not updated, so the invariants clearly hold.
Otherwise, we apply \Cref{claim:i_gen_us} and \Cref{claim:i_gen_v} setting \(\mA = \mA_{t-1}, \vc^{\circ} = \vc_{t-1}, \vz^{\circ} = \vz_{t}, \alpha=\alpha_{t-1}\).
Using (\ref{eqn:alg_mat_update}), \(\cE_{\mA_t}\) is the same as  \(\cE'\) in \Cref{claim:i_gen_us}; and clearly \(\alpha_{t} = \alpha'\).
This establishes the inner ellipsoid invariant \(\vc_{t} + \alpha_{t} \cE_{t} \subseteq \conv\p{(\vc_{t-1} + \alpha_{t-1} \cE_{t-1}) \cup \{\vz_{t}\}}\) directly.
To show \(\conv\p{(\vc_{t-1} + \cE_{t-1}) \cup \{\vz_{t}\}} \subseteq \vc_{t} + \cE_{t}\),
observe that we have \(\vc_{t-1} + \cE_{t-1} \subseteq \vc_{t} + \cE_{t}\) from \Cref{claim:i_gen_us},
and \(\vz_{t} \in \vc_{t} + \cE_{t}\) from \Cref{claim:i_gen_v}.
Then the outer ellipsoid invariant follows as \(\vc_{t} + \cE_{t}\) is a convex set.
\end{proof}

Finally, we bound the relevant quantities that will be used in the analysis of the full algorithm's approximation factor.
In particular, we show that \(\exp(\gamma^{\star}_t)\) gives a lower bound on the increase in volume at each iteration \(t\).
If \(\|\vu\| \leq 1\), and the ellipsoids are not updated, in that iteration we think of \(\gamma^{\star}_t = 0\).
\begin{claim}\label{claim:vol_t}
For any input given to \Cref{alg:full_update}, we have  \(\vol(\cE_i) \geq \exp(\gamma^{\star}_t) \vol(\cE_{t-1})\).
\end{claim}
\begin{proof}
This formula is clearly true when the ellipsoids are not updated because \(\gamma^{\star}_t = 0\), so we consider the nontrivial case.
Recall the formula \(\vol(\cE_\mA) = \det(\mA^{-1}) \vol(B_2^d)\) from \Cref{claim:prelim_vol}. Then we have
\[\vol(\cE_{\mA_i}) = \det(\mA_i^{-1}) \vol(B_2^d) = \det(\hat{\mA}^{-1}) \cdot \det(\mA_{t-1}^{-1}) \cdot \vol(B_2^d) = \det(\hat{\mA}^{-1}) \vol(\cE_{\mA_{t-1}})\]
where we use the definition of \(\hat{\mA}\) from Line \ref{alg_line:ellipsoid_update} on the \(t\)-th iteration.
Then \begin{align*}
    \det(\hat{\mA}^{-1}) &= a(\gamma^{\star}_t) \cdot b(\gamma^{\star}_t)^{d-1} & \text{using (\ref{eqn:alg_mat_update})}\\
    &\geq a(\gamma^{\star}_t) & \text{by \Cref{claim:update_params}-(\ref{item:claim_update_params_2})} \\
    &= \exp(\gamma^{\star}_t) & \text{by definition of \(a\) in (\ref{eqn:update_params})}
\end{align*}
and using \(\vol(\cE_{\mA_i}) = \det(\hat{\mA}^{-1}) \cdot \vol(\cE_{\mA_{t-1}})\) completes the proof.
\end{proof}

We are now ready to present the complete algorithm in \Cref{alg:ellipsoid_main}.
The algorithm is explicitly given \(\vc_0 + r_0 \cdot B_2^d \subseteq Z\). For simplicity, here, we say \(r = r_0\).
Let \(R = R(Z)\). While the final approximation factor depends on this quantity, the algorithm is not given it.
Note that \(\kappa(Z) \leq \sfrac{R}{r}\), so the quality of the approximation depends not only on \(\kappa(Z)\), but also on how well the given ball \(\vc_0 + r \cdot B_2^d\) is centered within \(Z\).

This algorithm proceeds in two phases. It begins with a ``local'' first phase, where the inner ellipsoid is a ball kept at radius \(r\), and the outer ellipsoid is a ball scaled to contain all the points. For readability, the variables of the algorithm in this phase are annotated with a superscript \(^{(l)}\). The second phase starts if the approximation factor of the first phase ever reaches \(\alpha^{(l)} \leq \frac{1}{d \log d}\),
at which point the algorithm uses the ``full'' update that was just described in \Cref{alg:full_update}.
We use two phases because while the full update reaches a near-optimal approximation factor when \(\sfrac{R}{r} \geq d \log d\), the local phase using balls does better when \(\sfrac{R}{r} \leq d \log d\).
While we cannot tell when to switch phases exactly (this would require knowing \(\sfrac{R}{r}\)), we show that it is enough to approximate the aspect ratio during the first phase up to a constant factor.

\begin{algorithm}[ht]
\caption{Streaming ellipsoid rounding -- complete algorithm}\label{alg:ellipsoid_main}

\textbf{input}: \(\vc_0+r B_2^d \subseteq Z\) \\
\textbf{output}: \(\vc_n + \cE_n, \vc_n + \alpha_n \cdot \cE_n\)
%represented as \(\{x \in \R^d \colon \|A_n (x - c_n)\| \leq 1\}\), \(\{x \in \R^d \colon \|A_n (x - c_n)\| \leq \alpha_n\}\)
\begin{algorithmic}[1]
\State Initialize \(\mA^{(l)}_0 = \frac{1}{r} \mI_d, \vc_0^{(l)} = \vc_0, \alpha_0^{(l)} = 1\)
\State \({t^{(l)}} = 0, R_0 = 0\)
\While{\({t^{(l)}} \leq n\)}
\Comment{Phase I: Local update step that maintains a ball}
    \State Receive point \(\vz_{t^{(l)}}\)
    \If{\(\|\vz_{t^{(l)}} - \vc_0\| \leq r \cdot d \log d\)}\label{alg_line:phaseI_if}
        \If{\(\|\vz_{t^{(l)}} - \vc_0\| > R_{{t^{(l)}}-1}\)}
            \State \(\mA_{t^{(l)}}^{(l)} = \frac{1}{\|\vz_{t^{(l)}} - \vc_0\|} \cdot \mI_d, \vc_{t^{(l)}}^{(l)} = \vc_{{t^{(l)}}-1}^{(l)}, \alpha_{t^{(l)}}^{(l)} = \frac{r}{\|\vz_{t^{(l)}} - \vc_0\|}\)
            \Comment{Grow the ball to contain \(\vz_{t^{(l)}}\)}
            \State \(R_{t^{(l)}} = \frac{\|\vz_{t^{(l)}} - \vc_0\|}{r}\)
        \Else
        \State \(\mA_{t^{(l)}}^{(l)} = \mA_{{t^{(l)}}-1}, \vc_i^{(l)} = \vc_{{t^{(l)}}-1}^{(l)}, \alpha_{t^{(l)}}^{(l)} = \alpha_{{t^{(l)}}-1}\)
        \State \(R_{t^{(l)}} = R_{{t^{(l)}}-1}\)
        \EndIf
    \Else{}
        \State \textbf{break} 
        \Comment{Break the loop and jump to Line \ref{alg_line:end_phase1}}
    \EndIf
    \State \({t^{(l)}} = {t^{(l)}} + 1\)
\EndWhile
\If{\({t^{(l)}} > n\)}\label{alg_line:end_phase1}
    \Comment{If we stayed in Phase I for the entire execution of the algorithm}
    \State \Return \(\mA_n^{(l)}, \vc_n^{(l)}, \alpha_n^{(l)}\)
\EndIf
\State \(t_{s} = {t^{(l)}}\) \label{alg_line:tswitch}
\Comment{Point \(\vz_{t_{s}}\) has not yet been processed}
\State \(\mA_{t_{s}-1} =\frac{1}{r d \log d} \cdot \mI_d, \vc_{t_{s}-1} = \vc_{t_{s}-1}^{(l)}, \alpha_{t_{s}-1} = \frac{1}{d \log d}\)\label{alg_line:phase_trans}
\Comment{Transition: grow the ball to maximium size}
\For{\(t \in \{t_{s}, t_{s}+1, \ldots, n\}\)}
    \Comment{Phase II: full update for the remaining points}
    \State Receive point \(\vz_i\)
    \State \(\mA_i, \vc_i, \alpha_i = \cA^{\mathsf{full}}(\mA_{t-1}, \vc_{t-1}, \alpha_{t-1}, \vz_i)\)\label{alg_line:phaseII_update}
\EndFor
\State \Return \(\mA_n, \vc_n, \alpha_n\)
\end{algorithmic}
\end{algorithm}

Before Line \ref{alg_line:end_phase1}, the algorithm executes the first phase that has the outer and inner ellipsoids as balls.
In Line \ref{alg_line:end_phase1}, we have \({t^{(l)}} > n\) if the algorithm stayed in Phase I for every point, i.e. we had \(\max_{1 \leq {t^{(l)}} \leq n} \| \vz_{t^{(l)}} - \vc_0 \| \leq r \cdot d \log d\). In this case, the algorithm returns the approximation maintained by Phase I.
Otherwise we must have come across a point where \(\|\vz_{t^{(l)}} - \vc_0 \| > r \cdot d \log d\), and the algorithm proceeds with Phase II.
We let \(t_{s}\) in Line \ref{alg_line:tswitch} mark the point received that causes the algorithm to proceed to Phase II.
We then perform a ``transition'' on Line \ref{alg_line:phase_trans} that grows the ball of Phase I to its maximum size. This transition step makes the analysis of the complete algorithm easier, as then the starting approximation for the second phase is exactly \(\alpha_{t_s - 1} = \frac{1}{d \log d}\).
Then the algorithm runs the full update \(\cA^{\mathsf{full}}\) for the rest of the points, including \(\vz_{t_{s}}\).
For simplicity, we write our algorithm so that it `receives' \(\vz_{t_{s}}\) twice, once for each phase.
However, the first phase does not commit to an update for this point, and the ellipsoids in Line \ref{alg_line:phase_trans} are not committed either; the algorithm does not commit to an update for this point until Line \ref{alg_line:phaseII_update}.

Recall the approximation guarantee stated in \Cref{thm:main_one}:
\begin{equation}\label{eq:main_approx_guarantee}
    \frac{1}{\alpha_n} \leq O(\min\p{\sfrac{R}{r}, d \log\p{\sfrac{R}{r}}})
\end{equation}

We can interpret the approximation guarantee \eqref{eq:main_approx_guarantee} by cases depending on if \(\sfrac{R}{r} \geq d \log d\) (i.e. if the algorithm ever enters the second phase):
\begin{claim}
We have for all \(d \geq 2\) that
\[\min\p{\sfrac{R}{r}, d \log\p{\sfrac{R}{r}}} = \Theta\p{\begin{cases}
    d \log\p{\sfrac{R}{r}} & \text{if } \sfrac{R}{r} > d \log d \\
    \sfrac{R}{r} & \text{if } \sfrac{R}{r} \leq d \log d
\end{cases}}.\]
\end{claim}

Now, we claim a straightforward geometric fact -- that the distance of the furthest \(\vz_{t}\) to \(\vc_0\)
approximates the circumradius of \(Z\) up to a constant factor.
We will use this to show that Line \ref{alg_line:phaseI_if} will be able to properly detect when
\(\sfrac{R}{r} > d \log d\) (again, up to a constant factor).
\begin{claim}\label{claim:alg_r_est}
    Let \(\vc_0 + r_0 B_2^d \subseteq Z\), and 
    \(R = R(Z)\). Then,
    \[R \leq \max_{1 \leq t^{(l)} \leq n} \| \vc_0 - \vz_{t^{(l)}} \| \leq 2 \cdot R.\]
\end{claim}
\begin{proof}
    For the left inequality, observe that if we let \(r_{\max} = \max_{1 \leq t^{(l)} \leq n} \| \vc_0 - \vz_{t^{(l)}} \|\),
    then \(Z \subseteq \vc_0 + r_{\max} \cdot B_2^d\).
    For the right inequality, observe that for any containing ball \(\vc' + R' \cdot B_2^d \supseteq Z\),
    its diameter is \(2 R'\).
    But as \(\vc' + R' \cdot B_2^d\) contains \(\vc_0\) and \(\vz_{1}, \ldots, \vz_{n}\), we must have \(\diam \p{\vc' + R' \cdot B_2^d} \geq \diam \p{\{\vc_0\} \cup \{\vz_1, \ldots, \vz_n\}}\) and so \(2R' \geq r_{\max}\).
\end{proof}

Next, we discuss the approximation guarantee that the algorithm achieves, depending on the phase that it terminates with.
We start with if the algorithm only stays in the local phase, in which case we can readily apply the previous claim.
\begin{claim}\label{claim:approx_local}
    If \Cref{alg:ellipsoid_main} never enters Phase II, then its approximation guarantee satisfies
    \(\frac{1}{\alpha_n} \leq \sfrac{2 R}{r}\).
\end{claim}
\begin{proof}
    At the termination of Phase I, the algorithm produces approximation \(\alpha_n^{(l)} = \max_{1 \leq {t^{(l)}} \leq n} \frac{\|\vz_{t^{(l)}} - \vc_0\|}{r}\).
    Using \Cref{claim:alg_r_est} we obtain \[
    \frac{1}{\alpha_n^{(l)}} = \max_{1 \leq {t^{(l)}} \leq n} \frac{\|\vz_{t^{(l)}} - \vc_0\|}{r} \leq \frac{2 R}{r}
    ,\]
    as desired.
\end{proof}

The analysis in the case where the algorithm enters the full phase is more involved.
We use \Cref{claim:vol_t}, which shows that the increase in approximation factor each iteration is not too large compared to the increase in volume, to bound \(\frac{1}{\alpha_n}\).
We know that the volume of the final ellipsoid \(\vc_n + \cE_n\) must be bounded relative to \(R \cdot B_2^d\), as the algorithm produces \(\vc_n + \alpha_n \cdot \cE_n \subseteq Z\). However, this leads to an upper bound that is still a function of \(\frac{1}{\alpha_n}\).
\begin{claim}\label{claim:alpha_rel}
    If \Cref{alg:ellipsoid_main} enters Phase II, the approximation guarantee satisfies
    \[\frac{1}{\alpha_n} \leq 2 \left(d \log\left(\frac{1}{\alpha_n}\right) + d \log\left(\frac{R}{r}\right)\right).\]
\end{claim}
\begin{proof}
The algorithm transitions to Phase II at Line \ref{alg_line:tswitch}, starting at iteration \(t_{s}\).
At each subsequent iteration, we claim that \Cref{alg:full_update} guarantees \(\frac{1}{\alpha_{t}} = \frac{1}{\alpha_{t-1}} + 2 \gamma^{\star}_t\).
By \Cref{claim:update_params}-(\ref{item:claim_update_params_1}), we have for all \(t_{s} \leq t \leq n-1\) where the ellipsoids were updated that \(\frac{1}{\alpha_{t}} = \frac{1}{\alpha_{t-1}} + 2 \gamma^{\star}_t\).
When the ellipsoids are not updated, this still holds, as in that case \(\gamma^{\star}_t = 0\). 

As in Phase II the algorithm begins with \(\alpha_{t_{s} - 1} = \frac{1}{d \log d}\), we have
\begin{equation}\label{eqn:alpha_rel_t}
\frac{1}{\alpha_n} = d \log d + 2 \sum_{t=t_{s}}^{n-1} \gamma^{\star}_t.
\end{equation}
Now applying \Cref{claim:vol_t} for each \(t\), we have \(\vol(\cE_n) \geq \exp\left(\sum_{t=t_{s}}^{n-1} \gamma^{\star}_t\right) \cdot \vol(\cE_{t_{s}})\). Taking logarithms gives
\begin{equation}\label{eq:ub_t_vol}
\log\p{\frac{\vol \cE_n}{\vol \cE_{t_{s} - 1}}} \geq \sum_{t=t_{s}}^{n-1} \gamma^{\star}_t.
\end{equation}

Recall that \(\vc_0 + r \cdot B_2^d \subseteq Z\),
and by \Cref{def:circumradius},
\(Z \subseteq \vc_c + R \cdot B_2^d\) for some center \(\vc_c\).
By \Cref{claim:alg_correctness}, we have \(\vc_n + \alpha_n \cdot \cE_n \subseteq Z\), so
that \(\vol(\cE_n) \leq \frac{1}{\alpha_n^d} \cdot \vol(R \cdot B_2^d)\).
As in Phase II we start with \(\cE_{t_s - 1} = \vc_0 + r d \log d \cdot B_2^d\), this yields
\begin{align*}\sum_{t=t_{s}}^{n-1} \gamma^{\star}_t &\leq \log\p{\frac{\vol(\cE_n)}{\vol (\cE_{t_{s}-1})}} &\text{by \eqref{eqn:alpha_rel_t}} \\
&\leq d \log \p{\frac{1}{\alpha_n}} + \log\p{\frac{\vol(R \cdot B_2^d)}{\vol(r d \log d \cdot B_2^d)}} & \text{by } \vol(\cE_n) \leq \frac{1}{\alpha_n^d} \vol(R \cdot B_2^d)\\
&= d \log\p{\frac{1}{\alpha_n}} + d\log\p{\frac{R}{rd \log d}}\\
&\leq d \log \p{\frac{1}{\alpha_n}} + d \log \p{\frac{R}{r}} - d \log d.
\end{align*}
Plugging into (\ref{eqn:alpha_rel_t}) completes the proof of \Cref{claim:alpha_rel}.
\end{proof}

Intuitively, \(x \leq a + b \cdot \log x\) for some constants \(a, b > 0\) can only be true for bounded \(x\), as \(x = \omega(\log x)\).
As we showed \(1/\alpha_n\) satisfies a relation like this in \Cref{claim:alpha_rel}, we develop this intuition to give a quantitative upper bound on \(1/\alpha_n\).
\begin{claim}\label{claim:approx_full}
If \Cref{alg:ellipsoid_main} enters Phase II, then we have \[\frac{1}{\alpha_n} \leq 8 d(\log d + \log \sfrac{R}{r}).\]
\end{claim}
\begin{proof}
Assume towards contradiction that \(\frac{1}{\alpha_n} > 8 d (\log d + \log \sfrac{R}{r})\).
Observe then that \(\frac{1}{\alpha_n} - \frac{3}{4} \cdot\frac{1}{\alpha_n} > 2 d (\log d + \log \sfrac{R}{r})\). Using \Cref{claim:alpha_rel}, we have
\[2 (d \log \sfrac{1}{\alpha_n} + d \log \sfrac{R}{r}) \geq \frac{1}{\alpha_n} > 2 (d \log d + d \log \sfrac{R}{r}) + \frac{3}{4} \cdot \frac{1}{\alpha_n}\]
Simplifying the above inequality gives \(2d \log \sfrac{1}{d \cdot \alpha_n} > \frac{3}{4} \cdot \frac{1}{\alpha_n}\), i.e. 
\(2\log \sfrac{1}{d \cdot \alpha_n} > \frac{7}{8} \cdot \frac{1}{d \cdot \alpha_n}\).
It is clear that this is impossible by looking at the graph of the function \(x \mapsto 2 \log x - \frac{3}{4} x\), which is concave with a maximum of \(2(\log(\sfrac{8}{3}) - 1) < 0\).
\end{proof}

Now we combine the previous claims to prove the guarantees of \Cref{alg:ellipsoid_main} and obtain \Cref{thm:main_one}.

\begin{proof}[Proof of \Cref{thm:main_one}.] We first discuss the approximation guarantee and correctness, then the memory and runtime complexity of \Cref{alg:ellipsoid_main}.

\paragraph{Approximation guarantee}
We break the analysis of the approximation guarantee by cases, depending on the aspect ratio.
If \(\sfrac{R}{r} \leq \frac{1}{2} d \log d\),
then by \Cref{claim:alg_r_est} we have \(\max_{1 \leq t^{(l)} \leq n} \| \vc_0 - \vz_{t^{(l)}} \| \leq r d \log d\),
and the algorithm never enters Phase II.
By \Cref{claim:approx_local}, the final approximation factor is \(\sfrac{2R}{r}\).
If \(\sfrac{R}{r} > d \log d\), then by \Cref{claim:alg_r_est} we have \(\max_{1 \leq t^{(l)} \leq n} \| \vc_0 - \vz_{t^{(l)}} \| > r d \log d\), and the algorithm must enter Phase II.
Then \Cref{claim:approx_full} applies, and the final approximation factor is \(O(d (\log d + \log \sfrac{R}{r}) = O(d \log \sfrac{R}{r})\).

If \(\frac{1}{2} d \log d < \sfrac{R}{r} \leq d \log d\), then it is possible for the algorithm to never enter Phase II or for it to enter Phase II. Either way, we argue that the final approximation factor is \(\frac{1}{\alpha_n} \leq O\p{\sfrac{R}{r}}\).
If it does not enter Phase II, then by \Cref{claim:approx_local}, the approximation guarantee we get is \(\frac{1}{\alpha_n} \leq O(\sfrac{R}{r})\).
If it does enter Phase II, then by \Cref{claim:approx_full}
we have \[
\frac{1}{\alpha_n} \leq O(d \log d + d \log \sfrac{R}{r})
\]
Due to the assumption that \(\frac{1}{2} d \log d < \sfrac{R}{r} \leq d \log d\), we also have in this case that \(\frac{1}{\alpha_n} \leq O(\sfrac{R}{r})\).

\paragraph{Correctness} By \Cref{claim:alg_invariant}, to argue that the algorithm solves \Cref{prob:main} it is enough to show that it is monotone, i.e. it satisfies the invariants of \Cref{def:alg-invariant}.
It is clear that the local update in Phase I satisfies the invariants, as the outer ellipsoid is a ball of growing radius
and the inner ellipsoid is kept to the ball of radius \(r\). 
It is also clear that after the algorithm transitions to Phase II, all the full updates are monotone by \Cref{claim:alg_correctness} and the fact that the starting approximation factor for this phase is is \(\alpha_{t_s - 1} = \frac{1}{d \log d} \leq \frac{1}{2}\).
As algorithm transitions to Phase II,
observe that on Line \ref{alg_line:phase_trans} the radius of the outer ellipsoid grows again to \(r d \log d\) before applying the full update,
so the first first full update of Phase II is also monotone.

\paragraph{Memory and runtime complexity}
The memory complexity of the algorithm is \(O(d^2)\).
Observe that \Cref{alg:full_update} only stores a constant number of matrices in \(\R^{d \times d}\), vectors in \(\R^d\), or constants,
so its memory complexity is \(O(d^2)\). It is only instantiated once for each point received in Phase II, so the memory complexity in this phase \(O(d^2)\). Finally, the memory complexity in the first phase is also \(O(d^2)\) because it stores the same kind of quantities as \Cref{alg:full_update}.

To show the runtime of the algorithm is \(\widetilde{O}(nd^2)\), we show that the runtime to process each next point is at most \(\widetilde{O}(d^2)\). This is clear in Phase I, and during the transition to Phase II.
For the full update this is less clear, as \Cref{alg:full_update} uses both \(\mA_{t-1}\) and \(\mA_{t-1}^{-1}\)
which naively would require inverting a matrix on each iteration.
However, if we represent \(\mA\) using the SVD (see the next section and \Cref{claim:full_update_eff}), we can implement the update in \(\widetilde{O}(d^2)\) time.
This would require that \(\mA_{t_s - 1}\) be given in SVD form as well for the first full update, but it is already in that form as a scaled identity matrix.

Put together, these complete the proof of \Cref{thm:main_one}.
\end{proof}

\subsubsection{Efficient implementation of the full update step}

In this section, we use a method similar to that in Algorithm 2 from \cite{makarychev2022streaming} to show that the full update step can be implemented in \(\widetilde{O}(d^2)\) time.
In particular, we use the same subroutine \(\textsc{SVDRankOneUpdate}\) with signature
\begin{align}
    (\mU', \mSigma', \mV') = \textsc{SVDRankOneUpdate}((\mU, \mSigma, \mV), \vy_1, \vy_2)\label{eq:stange_update}
\end{align}
where the result \(\mU' \mSigma' (\mV')^\top\) is the SVD of the matrix \(\mU \mSigma \mV^\top + \vy_1 \vy_2^\top\).
\cite{stange2008efficient} shows that this procedure be done in \(O(d^2 \log d)\) time.
We rewrite \Cref{alg:full_update} in \Cref{alg:full_update_eff} to make it clear how to use the SVD representation and the efficient rank-1 update to efficiently implement the full update.
One can readily see that \Cref{alg:full_update_eff} has the exact same behavior as \Cref{alg:full_update}, and so gives the same approximation and correctness guarantees.

\begin{algorithm}[ht]
\caption{Efficient full update step \(\cA^{\mathsf{full}}\)}\label{alg:full_update_eff}

\textbf{input}: \((\mU_{t-1}, \mSigma_{t-1}, \mV_{t-1}) \in \R^{d \times d}, \vc_{t-1} \in \R^{d}, \alpha_{t-1} \in [0, \frac{1}{2}], \vz_{t} \in \R^{d}\) \\
\textbf{output}: \((\mU_{t}, \mSigma_{t}, \mV_{t}) \in \R^{d \times d}, \vc_{t} \in \R^{d}, \alpha_{t} \in [0, \alpha_{t}]\)
%represented as \(\{x \in \R^d \colon \|A_n (x - c_n)\| \leq 1\}\), \(\{x \in \R^d \colon \|A_n (x - c_n)\| \leq \alpha_n\}\)
\begin{algorithmic}[1]
% \For{\(t = j+1, \ldots, n\)}
%     \State Receive point \(\vz_i\)
    \State Let \(\vu = \mU_{t-1} \mSigma_{t-1} \mV_{t-1}^\top (\vz_{t} - \vc_{t-1})\), \(\vw = \frac{\vu}{\|\vu\|}\)
    \If{\(\|\vu\| > 1\)}
        \State Let \(\gamma^{\star}_t\) be such that \(a(\gamma^{\star}_t) + c(\gamma^{\star}_t) = \|\vu\|\)\label{alg_line:compute_update_t_eff}
        \State \(\vy_1 = \left(\frac{1}{a(\gamma^{\star}_t)} - \frac{1}{b(\gamma^{\star}_t)}\right) \vw, \vy_2 = \mV_{t-1} \mSigma_{t-1} \mU_{t-1}^\top \vw\)
        \State \((\mU_{t}, \mSigma_{t}, \mV_{t}) = \textsc{SVDRankOneUpdate}((\mU_{t-1}, \frac{1}{b(\gamma^{\star}_t)} \mSigma_{t-1}, \mV_{t-1}), \vy_1, \vy_2)\)
        \State \Return \((\mU_{t}, \mSigma_{t}, \mV_{t}), \ \vc_{t} = \vc_{t-1} + c(\gamma^{\star}_t) \mV_{t-1} \mSigma_{t-1}^{-1} \mU_{t-1}^\top \vw, \ \alpha_t = \alpha'(\gamma^{\star}_t)\)
    \Else
        \State \Return \((\mU_{t}, \mSigma_{t}, \mV_{t}) = (\mU_{t-1}, \mSigma_{t-1}, \mV_{t-1}), \vc_t = \vc_{t-1}, \alpha_t = \alpha_{t-1}\)
    \EndIf
% \EndFor
% \State \Return \(\mA_n, \vc_n, \alpha_n\)
\end{algorithmic}
\end{algorithm}

\begin{remark}\label{remark:compute_update_t_eff}
    We briefly explain why Line \ref{alg_line:compute_update_t_eff}, finding \(\gamma^{\star}\) such that \(a(\gamma^{\star}) + c(\gamma^{\star}) = \|\vu\|\)  can be implemented efficiently. This is a one-dimensional optimization problem, and \(\gamma \mapsto a(\gamma) + c(\gamma)\) using \(a, c\) as defined in \eqref{eqn:update_params} is monotone increasing, 
    so finding an approximate \(\gamma^{\star}\) can be done efficiently with binary search.
    In particular, we can choose \(\gamma^{\star}\) to be a slight overestimate so the update is still monotone after slightly increasing \(\alpha_t\).
    This does not affect the final approximation guarantee beyond constant factors.
\end{remark}

This algorithm performs a constant number of taking norms of vectors, matrix-vector products, and algebraic operations; as well as one rank-one SVD update. As explained in the preceding remark, finding \(\gamma_i^*\) can also be done in effectively constant time. Thus for our runtime guarantee, we have:
\begin{claim}\label{claim:full_update_eff}
    \Cref{alg:full_update_eff} runs in time \(O(d^2 \log d)\).
\end{claim}

\subsection{Fully-online asymmetric ellipsoidal rounding algorithm}

In this subsection, we prove Theorem \ref{thm:main_two}. See Algorithm \ref{alg:fully_online_approx}.

\begin{algorithm}%[H]
\caption{Fully online asymmetric ellipsoidal rounding}\label{alg:fully_online_approx}
\begin{algorithmic}[1]
\State \textbf{Input:} Stream of points $\vz_t$; 
 monotone update rule $\cA$ (Definition \ref{def:alg-invariant}) that takes as input the previous ellipsoid matrix $\mA$, center $\vc$, approximation factor $\alpha$, and update point $\vz$ and outputs the next ellipsoid matrix $\mA'$, center $\vc'$, and approximation factor $\alpha'$.
\State \textbf{Output:} Ellipsoid $\cE$, center $\vc$, and scale $\alpha \in (0,1)$ such that $\vc + \alpha \cdot \cE \subseteq \mathsf{conv}\p{\pb{\vz_1,\dots,\vz_n}} \subseteq \vc +\cE$.
\State Receive $\vz_1$; set $\mA = \mI_d$, $d_1 = 1$, $\vc_1 = \vz_1$, $\alpha_1 = 1$.
\For{$t = 2,\dots,n$}
    \State Receive $\vz_{t}$.
    \If{$\vz_{t} - \vc_{t-1} \notin \vspan\p{\vz_{1} - \vc_{t-1},\dots,\vz_{t-1}-\vc_{t-1}}$}\Comment{Irregular update step.}\label{line:irregular_step}
        \State Let $\vv_1,\dots,\vv_{d_{t-1}}$ be the singular vectors of $\mA$ corresponding to the semiaxes of $\cE_{t-1}$.
        \State Let $d_t = d_{t-1}+1$.
        \State Let $\vz_{d_{t}}' \coloneqq \frac{\vz_t - \sum_{i=1}^{d_{t-1}} \vv_i\pgo{\vv_i,\vz_t}}{\pn{\vz_t - \sum_{i=1}^{d_{t-1}} \vv_i\pgo{\vv_i,\vz_t}}_2}$.
        \State Let $\mM \coloneqq \mI_d - \frac{1}{\pgo{\vv_{d_{t}}',\vz}}\cdot\p{\vz_t - \sqrt{1+2\alpha_{t-1}} \cdot \vv_{d_{t}}'}(\vv_{d_{t}}')^T$.
        \State Update $\mA_t \gets \mA_{t-1}\mM$.\Comment{Use \eqref{eq:stange_update} of \cite{stange2008efficient} to update $\vv_1,\dots,\vv_d$.}
        \State Update $\vc_t = \frac{\alpha_{t-1}}{1+2\alpha_{t-1}}\cdot\vz_t + \p{1-\frac{\alpha_{t-1}}{1+2\alpha_{t-1}}}\cdot\vc_{t-1}$.
        \State Update $\sfrac{1}{\alpha_t} \gets \sfrac{1}{\alpha_{t-1}}+1$.
    \Else  
        \State $\mA_t, \vc_t, \alpha_t = \cA(\mA_{t-1}, \vc_{t-1}, \alpha_{t-1}, \vz_{t})$
        \State $d_t \gets d_{t-1}$.
    \EndIf
\EndFor
\State \textbf{Output:} $(\vc_n, \cE_n, \alpha_n)$.
\end{algorithmic}
\end{algorithm}

% \begin{theorem}
% \label{thm:fully_online_approx}
% For all $t \le n$, let $r_t$ be the largest parameter such that there exists $\vc_0 \in \R^d$ for which
% \begin{align*}
%     \vc_0 + r_t \cdot \p{B_2^d \cap \vspan\p{\vz_1-\vc_0,\dots,\vz_t-\vc_0}} \subseteq \mathsf{conv}\p{\vz_1,\dots,\vz_t},
% \end{align*}
% let $R$ be the smallest parameter such that there exists $\vc_0 \in \R^d$ (not necessarily the same $\vc_0$ as before) for which
% \begin{align*}
%     \mathsf{conv}\p{\vz_1,\dots,\vz_n} \subseteq \vc_0 + R \cdot \p{B_2^d \cap \vspan\p{\vz_1-\vc_0,\dots,\vz_n-\vc_0}},
% \end{align*}
% and let
% \begin{align*}
%     d_t &= \mathsf{dim}\p{\vspan\p{\vz_1-\vc_t,\dots,\vz_t-\vc_t}}.
% \end{align*}
% For all $t \le n$, Algorithm \ref{alg:fully_online_approx} maintains an ellipsoid $\cE_t$ and a center $\vc_t$ for which
% \begin{align*}
%     \vc_t + \alpha_t \cdot \cE_t \subseteq \mathsf{conv}\p{\vz_1,\dots,\vz_t} \subseteq \vc_t + \cE_t
% \end{align*}
% where
% \begin{align*}
%     \frac{1}{\alpha_t} &= O\p{d_t\log d_t + d_t\log\p{\max_{i \le t} \frac{R}{r_i}}}.
% \end{align*}
% \end{theorem}

To prove Theorem \ref{thm:main_two}, we need to show that our \textit{irregular update step} (a timestep $t$ when we have to update the dimensionality of our ellipsoid $\cE_{t-1}$ -- see Line \ref{line:irregular_step} of Algorithm \ref{alg:fully_online_approx}) still maintains the invariants we desire (Definition \ref{def:alg-invariant}).

Our plan is to first consider the special case of the irregular update where the new point to cover is conveniently located with respect to our current ellipsoids. We will see later that this special case is nearly enough for us to conclude the proof.

\begin{claim}
\label{claim:irregular_update_easy}
Let $Z \subset \R^{d}$ be a convex body where $Z$ lies in $\vspan\p{\vv_1,\dots,\vv_{d'}}$ for $d' < d$. For $0 < \alpha \le 1$, suppose we have
\begin{align*}
    \alpha \cdot \pb{\vz \in \vspan\p{\vv_1,\dots,\vv_{d'}} \colon \pn{\vz} \le 1} \subseteq Z \subseteq \pb{\vz \in \vspan\p{\vv_1,\dots,\vv_{d'}} \colon \pn{\vz} \le 1}.
\end{align*}
Then, for any $\vv_{d'+1}$ such that $\pgo{\vv_i, \vv_{d'+1}} = 0$ for all $i \in [d']$ and for which
\begin{align*}
    \cE' &\coloneqq \pb{\vz \in \vspan\p{\vv_1,\dots,\vv_{d'+1}} \colon \pn{\vz} \le \frac{1+\alpha}{\sqrt{1+2\alpha}}} \\
    \vc &\coloneqq \frac{\alpha}{\sqrt{1+2\alpha}} \cdot \vv_{d'+1}
\end{align*}
we have
\begin{align*}
    \vc + \frac{1}{1 + \sfrac{1}{\alpha}} \cdot \cE' \subseteq \mathsf{conv}\p{Z \cup \pb{\sqrt{1+2\alpha} \cdot \vv_{d'+1}}} \subseteq \vc + \cE'.
\end{align*}
\end{claim}
\begin{proof}[Proof of Claim \ref{claim:irregular_update_easy}]
We will show that the pair of ellipsoids given below satisfy the conditions promised by the statement of Claim \ref{claim:irregular_update_easy}.
\begin{align}
    &\pb{\vz \in \vspan\p{\vv_1,\dots,\vv_{d'+1}} \colon \pn{\vz} \le \frac{1+\alpha}{\sqrt{1+2\alpha}}} +  \frac{\alpha}{\sqrt{1+2\alpha}} \cdot \vv_{d'+1}\label{eq:irregular_outer} \\
    &\pb{\vz \in \vspan\p{\vv_1,\dots,\vv_{d'+1}} \colon \pn{\vz} \le \frac{1+\alpha}{\sqrt{1+2\alpha}}} \cdot \frac{\alpha}{1+\alpha} +  \frac{\alpha}{\sqrt{1+2\alpha}} \cdot \vv_{d'+1}\label{eq:irregular_inner}
\end{align}
Clearly, the two ellipsoids given above are apart by a factor of $\sfrac{1+\alpha}{\alpha} = \sfrac{1}{\alpha}+1$, which means the approximation factor increases by exactly $1$ as a result of this update. It now suffices to show that the ellipsoid described by (\ref{eq:irregular_outer}) contains $\mathsf{conv}\p{B_2^{d'} \cup \pb{\sqrt{1+2\alpha} \cdot \vv_{d'+1}}}$ and that the ellipsoid described by (\ref{eq:irregular_inner}) is contained by the cone whose base is $\alpha \cdot B_2^{d'}$ and whose apex is $\sqrt{1+2\alpha}\cdot\vv_{d'+1}$.

For the first part, it suffices to verify that every point $\vz \in Z$ and $\sqrt{1+2\alpha} \cdot \vv_{d'+1}$ is contained by (\ref{eq:irregular_outer}). We give both the calculations below, from which the result for (\ref{eq:irregular_outer}) follows.
\begin{align*}
    \vz \in Z :&\quad\quad \pn{\vz - \frac{\alpha}{\sqrt{1+2\alpha}} \cdot \vv_{d'+1}} = \sqrt{\pn{\vz}^2 + \frac{\alpha^2}{1+2\alpha}} \le \frac{1+\alpha}{\sqrt{1+2\alpha}} \\
    \vz = \sqrt{1+2\alpha} \cdot \vv_{d'+1} :&\quad\quad \pn{\vz - \frac{\alpha}{\sqrt{1+2\alpha}} \cdot \vv_{d'+1}} = \sqrt{1+2\alpha} - \frac{\alpha}{\sqrt{1+2\alpha}} = \frac{1+\alpha}{\sqrt{1+2\alpha}}
\end{align*}

We now analyze (\ref{eq:irregular_inner}). Our task is to show the below inclusion.
\begin{align*}
    &\pb{\vz \in \vspan\p{\vv_1,\dots,\vv_{d'+1}} \colon \pn{\vz - \frac{\alpha}{\sqrt{1+2\alpha}}\cdot \vv_{d'+1}} \le \frac{\alpha}{\sqrt{1+2\alpha}}}\\
    \subseteq &\mathsf{conv}\p{\alpha \cdot \pb{\vz \in \vspan\p{\vv_1,\dots,\vv_d} \colon \pn{\vz} \le 1} \cup \pb{\sqrt{1+2\alpha} \cdot \vv_{d'+1}}}
\end{align*}
Let $\vw$ be an arbitrarily chosen unit vector in $\vspan\p{\vv_1,\dots,\vv_{d'}}$. Observe that it is enough to show
\begin{align*}
    \pb{\vz \in \vspan\p{\vw, \vv_{d'+1}} \colon \pn{\vz - \frac{\alpha}{\sqrt{1+2\alpha}} \cdot \vv_{d'+1}} \le \frac{\alpha}{\sqrt{1+2\alpha}}} \subseteq \mathsf{conv}\p{\pm \alpha \cdot \vw, \sqrt{1+2\alpha} \cdot \vv_{d'+1}}.
\end{align*}
Since the above is a two-dimensional problem and that $\pgo{\vw, \vv_{d'+1}} = 0$, it is equivalent to show that the inradius of the triangle with vertices $(-\alpha, 0)$, $(\alpha, 0)$, and $(0, \sqrt{1+2\alpha})$ is $\sfrac{\alpha}{\sqrt{1+2\alpha}}$ and that its incenter is $(0, \sfrac{\alpha}{\sqrt{1+2\alpha}})$.

Recall that the inradius of a triangle can be written as $\sfrac{K}{s}$ where $K$ is the area of the triangle (in this case, $\alpha\sqrt{1+2\alpha}$) and $s$ is the semiperimeter of the triangle (in this case, $1+2\alpha$). This implies that the inradius is indeed $\sfrac{\alpha}{\sqrt{1+2\alpha}}$. Finally, since the triangle in question is isosceles with its apex being the $y$-axis, the $x$-coordinate of its incenter must be $0$. These observations imply that the incenter is $(0, \sfrac{\alpha}{\sqrt{1+2\alpha}})$.

This is sufficient for us to conclude the proof of Claim \ref{claim:irregular_update_easy}.
\end{proof}

We will now see that the analysis for the convenient update that we gave in Claim \ref{claim:irregular_update_easy} is nearly enough for us to fully analyze the irregular update step. See Claim \ref{claim:irregular_update_hard}, where we analyze the irregular update step in full generality (up to translating by $\vc_{t-1}$).

\begin{claim}
\label{claim:irregular_update_hard}
Let $Z \subset \R^d$ be a convex body such that $Z$ lies in a subspace $H$ of dimension $d' < d$. Let $\cE$ be an ellipsoid and let $0 < \alpha \le 1$ be such that
\begin{align*}
    \alpha \cdot \cE \subseteq Z \subseteq \cE.
\end{align*}
Let $\vz \notin H$. Then, there exists a center $\vc$ and an ellipsoid $\cE'$ such that
\begin{align*}
    \vc + \frac{1}{1+\sfrac{1}{\alpha}} \cdot \cE' \subseteq \mathsf{conv}\p{Z \cup \pb{\vz}} \subseteq \vc + \cE'.
\end{align*}
\end{claim}
\begin{proof}[Proof of Claim \ref{claim:irregular_update_hard}]
Recall that $\vv_1,\dots,\vv_{d'} \in \R^{d}$ are the unit vectors corresponding to the semiaxes of $\cE$; notice that these form a basis for $H$. Observe that $\vv_{d'+1}$ is a unit vector orthogonal to $\vv_1,\dots,\vv_{d'}$ such that $\vz$ can be expressed as $\sum_{i=1}^{d'+1} \vv_{i}\pgo{\vv_i, \vz}$.

As stated in Algorithm \ref{alg:fully_online_approx}, let
\begin{align*}
    \mM \coloneqq \mI_d - \frac{1}{\pgo{\vv_{d_t}',\vz_t}}\cdot\p{\vz_t - \sqrt{1+2\alpha} \cdot \vv_{d_t}'}(\vv_{d_t}')^T.
\end{align*}
We calculate
\begin{align*}
    \mM\vz_t = \vz_t - \frac{1}{\pgo{\vv_{d_t}',\vz_t}}\cdot\p{\vz_t - \sqrt{1+2\alpha} \cdot \vv_{d_t}'}(\vv_{d_t}')^T\vz_t = \vz_t - \vz_t + \sqrt{1+2\alpha} \cdot \vv_{d_t}'  = \sqrt{1+2\alpha} \cdot \vv_{d_t}'.
\end{align*}
By the definition of $\mA_{t-1}$, we have
\begin{align*}
    \mA_{t-1}\mM\vz_t = \sqrt{1+2\alpha} \cdot \mA_{t-1}\vv_{d_t}' = \sqrt{1+2\alpha} \cdot \vv_{d_t}'.
\end{align*}
Next, for any $\vz \in Z$, we have $\vz \in H_{t-1}$. This means that
\begin{align*}
    \mM\vz = \vz - \frac{1}{\pgo{\vv_{d_t}',\vz}}\cdot\p{\vz - \sqrt{1+2\alpha} \cdot \vv_{d_t}'}(\vv_{d_t}')^T\vz = \vz - 0 = \vz.
\end{align*}
By Claim \ref{claim:irregular_update_easy}, we know for
\begin{align*}
    \mA_{t-1}\mM\vc_t &= \frac{\alpha}{\sqrt{1+2\alpha}}\cdot\vv_{d_t}'\\
    \mA_{t-1}\mM\cE_t &= \pb{\vz \in \vspan\p{\vv_1,\dots,\vv_{d_{t-1}},\vv_{d_t}'} \colon \pn{\vz} \le 1}\\
\end{align*}
that
\begin{align*}
    \mA_{t-1}\mM\vc_t + \frac{1}{1+\sfrac{1}{\alpha_{t-1}}} \cdot \mA_{t-1}\mM\cE_t \subseteq \mathsf{conv}\p{\mA_{t-1}\mM\cdot Z \cup \pb{\mA_{t-1}\mM\vz_t}} \subseteq \mA_{t-1}\mM\vc_t + \mA_{t-1}\mM\cE_t
\end{align*}
and, since $\mA_{t-1}\mM$ is invertible (owing to the invertibility of $\mA_{t-1}$ and $\mM$),
\begin{align*}
    \vc_t + \frac{1}{1+\sfrac{1}{\alpha_{t-1}}} \cdot \cE_t \subseteq \mathsf{conv}\p{Z \cup \pb{\vz_t}} \subseteq \vc_t + \cE_t.
\end{align*}
Finally, note that
\begin{align*}
    \vc_t &= \frac{\alpha}{\sqrt{1+2\alpha}} \cdot \mM^{-1}\mA_{t-1}^{-1}\vv_{d_t}' = \frac{\alpha}{1+2\alpha} \cdot \vz_t \\
    \cE_t &= \pb{\vz \in \vspan\p{\vz_1,\dots,\vz_t} \colon \pn{\mA_{t-1}\mM\vz} \le 1}
\end{align*}
and then translate by $\vc_{t-1}$, which concludes the proof of Claim \ref{claim:irregular_update_hard}.
\end{proof}

We are now ready to prove \Cref{thm:main_two}.

\begin{proof}[Proof of \Cref{thm:main_two}]
Using \Cref{claim:irregular_update_hard}, we have that the ellipsoids maintain our desired invariants (Definition \ref{def:alg-invariant}) throughout the process. Hence, \Cref{alg:fully_online_approx} maintains an ellipsoidal approximation to $\conv\p{\pb{\vz_1,\dots,\vz_t}}$ for all $t$.

It remains to verify the approximation factor $\alpha_t$ of Algorithm \ref{alg:fully_online_approx}.

Consider a timestep $t$. For every $t' \leq t$, let $H_{t'} = \vspan\p{\vz_1,\dots,\vz_{t'}}$,  $r_{t'} = r(Z_{t'})$ be the inradius of $Z_{t'} = \conv\p{\vz_1,\dots, \vz_{t'}}$, and $R_{t'} = R(Z_{t'})$ be the circumradius of $Z_t$. 
Let ${\hat r} = \min_{t'\leq t} r_{t'}$. Consider the $d$-dimensional ellipsoid $T(\cE_{t'})$ which is exactly equal to $\cE_{t'}$ in the space $H_{t'}$ and whose remaining semiaxes orthogonal to $H_{t'}$ are equal and have length $\hat r$. Observe that for a regular update step $t'$ (with $d_{t'} = d_{t'-1}$), we have
$$\frac{\mathsf{vol}_{d_{t'}}(\cE_{t'})}{\mathsf{vol}_{d_t}(\cE_{t'-1})} = \frac{\mathsf{vol}_d(T(\cE_{t'}))}{\mathsf{vol}_d(T(\cE_{t'-1}))}.$$
Now applying the evolution condition (\ref{eq:overview_evolution}) to the update restricted to $H_{t'}$, we get 
$$
\frac{1}{\alpha_{t'}} - \frac{1}{\alpha_{t'-1}} \leq C\log \frac{\mathsf{vol}_{d_{t'}}(\cE_t)}{\mathsf{vol}_{d_{t'}}(\cE_{t'-1})} = C\log \frac{\mathsf{vol}_d(T(\cE_{t'}))}{\mathsf{vol}_d(T(\cE_{t'-1}))}.
$$
We have obtained the following upper bound on the approximation-factor increase:
\begin{align}
    \frac{1}{\alpha_{t'}} - \frac{1}{\alpha_{t'-1}} \leq \begin{cases} 1 & \text{if $t'$ is an irregular update step} \\ C\log\p{\frac{\mathsf{vol}_{d}(T(\cE_{t'}))}{\mathsf{vol}_{d}(T(\cE_{t'-1}))}} & \text{otherwise}\end{cases}\label{eq:fully_online_invariant}
\end{align}
Let $T_{\mathsf{reg}}$ consist of all the timesteps $t' \le t$ where we perform a regular update. Then we have,
\[
    \frac{1}{\alpha_t} - \frac{1}{\alpha_0}= \alpha_0 + \sum_{t'=1}^t  \left(\frac{1}{\alpha_{t'}} - \frac{1}{\alpha_{t'-1}} \right)\leq d_t + C\sum_{t'\in T_{\mathsf{reg}}} \log\p{\frac{\mathsf{vol}_{d}(T(\cE_{t'}))}{\mathsf{vol}_{d}(T(\cE_{t'-1}))}}.
\]
Now we show that $\log\p{\frac{\mathsf{vol}_{d}(T(\cE_{t'}))}{\mathsf{vol}_{d_t}(T(\cE_{t'-1}))}} \geq 0$ for an irregular step: let $\sigma_1\geq\dots\geq\sigma_d$ and $\sigma_1'\geq \dots\geq\sigma_d'$ be the lengths of semi-axes of $T(\cE_{t'})$ and $T(\cE_{t'-1})$, respectively. Then $\sigma_i \geq \sigma'_i$ for $1\leq i \leq d_{t'}-1$, since $\cE_{t'-1} \subset \cE_{t'}$; $\sigma_{d_{t'}} \geq r_{t'} \geq \hat r = \sigma'_{d_{t'}}$; and $\sigma_{i} = \hat r = \sigma'_{i}$ for $i > d_{t'}$. Therefore,
\[\log\p{\frac{\mathsf{vol}_{d}(T(\cE_{t'}))}{\mathsf{vol}_{d_t}(T(\cE_{t'-1}))}}=
\log\p{\frac{\sigma_1\cdot \ldots \cdot \sigma_d}{\sigma_1'\cdot \ldots \cdot \sigma_d'}} \geq \log 1 = 0.
\]
Using this inequality and plugging in $\alpha_0 = 1$, we get
\begin{align*}
    \frac{1}{\alpha_t} &= 1 + d_t + C\sum_{t'\in T_{\mathsf{reg}}} \log\p{\frac{\mathsf{vol}_{d}(T(\cE_{t'}))}{\mathsf{vol}_{d}(T(\cE_{t'-1}))}}
    \leq 1+ d_t + C\sum_{t'=1}^t \log\p{\frac{\mathsf{vol}_{d}(T(\cE_{t'}))}{\mathsf{vol}_{d}(T(\cE_{t'-1}))}} \\
    &\lesssim d_t + \log\p{\frac{\mathsf{vol}_{d}(T(\cE_{t}))}{\mathsf{vol}_{d}(T(\cE_{0}))}} 
    \lesssim d_t + \log\p{\frac{(R_t/\alpha_t)^{d_t}{\hat r}^{d-d_t}}{{\hat r}^d}} \lesssim d_t +  d_t \log\p{\frac{R_t}{\alpha_t\hat r}}.
\end{align*}
We conclude that
\begin{align*}
    \frac{1}{\alpha_t} \lesssim d_t + d_t \log\p{\frac{R_t}{\hat r}} + d_t\log d_t.
\end{align*}
This concludes the proof of \Cref{thm:main_two}.
\end{proof}

\subsection{Aspect ratio-independent bounds and proof of \texorpdfstring{Theorem~\ref{thm:main_two_ints}}{Theorem 3}}

To prove \Cref{thm:main_two_ints}, we first establish \Cref{claim:irregular_volume_increase}.

\begin{claim}
\label{claim:irregular_volume_increase}
Let $t$ be an iteration corresponding to an irregular update step in \Cref{alg:fully_online_approx}. Then,
\begin{align*}
    \frac{\vol_{d_{t-1}}\p{B_2^{d_{t-1}}}}{\vol_{d_{t}}\p{B_2^{d_{t}}}} \cdot\frac{\vol_{d_t}\p{\cE_t}}{\vol_{d_t-1}\p{\cE_{t-1}}} \ge \frac{\pn{\vz_{t}^{\perp}}}{2}
\end{align*}
where $\pn{\vz_{t}^{\perp}}$ is the length of the component of $\vz_t$ in the orthogonal complement of $\vspan\p{\vz_1,\dots,\vz_{t-1}}$.
\end{claim}
\begin{proof}[Proof of \Cref{claim:irregular_volume_increase}]
By affine invariance, we can apply an affine transformation to map $\vz_t$ and $\cE_{t-1}$ to a convenient position. Hence, following the proof of \Cref{claim:irregular_update_easy}, without loss of generality, suppose we have $\cE_{t-1} = B_{2}^{d_{t-1}}$ and $\vz_t = \sqrt{1+2\alpha} \cdot \ve_{d_t}$. By \Cref{claim:irregular_update_easy}, the ellipsoid $\cE_t$ is a ball of radius $\sfrac{(1+\alpha)}{\sqrt{1+2\alpha}}$. Let $z \coloneqq \sqrt{1+2\alpha}$.
We now have
\begin{align*}
    \frac{\vol_{d_{t-1}}\p{B_2^{d_{t-1}}}}{\vol_{d_{t}}\p{B_2^{d_{t}}}} \cdot \frac{\vol_{d_t}\p{\cE_t}}{\vol_{d_{t-1}}\p{\cE_{t-1}}} = \p{\frac{1+\alpha}{\sqrt{1+2\alpha}}}^{d_t} \geq 1 > \frac{\|\vz_t\|}{2}
\end{align*}
since $\|\vz_t\| =\sqrt{1+2\alpha} \leq \sqrt{3} < 2$. This concludes the Proof of \Cref{claim:irregular_volume_increase}.
\end{proof}

We will also need \Cref{claim:det_prod_identity}, which we take from \cite{gover2010determinants}.

\begin{claim}
\label{claim:det_prod_identity}
Let $\mM \in \R^{r \times d}$ have linearly independent rows $\vm_1,\dots,\vm_r$. Then,
\begin{align*}
    \prod_{i=1}^r \pn{\vm_i} = \sqrt{\det\p{\mM\mM^T}}.
\end{align*}
\end{claim}

We are now ready to prove \Cref{thm:main_two_ints}.

\begin{proof}[Proof of \Cref{thm:main_two_ints}]
Our approach is reminiscent of that used in the proof of \cite[Theorem 1.5]{woodruff2022high}.

By applying a translation to all points, we may assume without loss of generality that $\vz_1 = 0$. We will prove the guarantee for the last timestamp $t=n$ to simplify the notation. By replacing $n$ with $n'$, we can get a proof for any time stamp $t=n'$.

Let $S$ be the set of timestamps of irregular update steps excluding the first step. Since the update rule satisfies the evolution condition \eqref{eq:overview_evolution}, we have for all $t \notin S$ (recall that $d_t=d_{t-1}$ for $t\notin S$)
\begin{align*}
    \frac{\vol_{d_t}\p{\cE_{t}}}{\vol_{d_{t-1}}\p{\cE_{t-1}}} \ge \exp\p{\frac{1}{\alpha_t}-\frac{1}{\alpha_{t-1}}}.
\end{align*}
Next, by \Cref{claim:irregular_volume_increase}, we have for every irregular update step $t> 1$ 
%for which $d_t \ge 2$ that
\begin{align*}
    \frac{\vol_{d_{t-1}}\p{B_2^{d_{t-1}}}}{\vol_{d_{t}}\p{B_2^{d_{t}}}} \cdot\frac{\vol_{d_t}\p{\cE_t}}{\vol_{d_{t-1}}\p{\cE_{t-1}}} \ge \frac{\pn{\vz_{t}^{\perp}}}{2}.
\end{align*}
Here, we assume that $\vol_0(\{0\}) = 1$ and define $\pn{\vz_2^{\perp}} = \pn{\vz_2}$.
Inductively combining the above for all $t > 1$ gives
\begin{align} % change the range of the products
    \vol_{d_n}\p{\cE_n} &\ge \prod_{t \notin S} \exp\p{\frac{1}{\alpha_t}-\frac{1}{\alpha_{t-1}}} \cdot \prod_{t \in S} \frac{\pn{\vz_t^{\perp}}}{2}\cdot \prod_{j=1}^{d_n} \frac{\vol_{j}\p{B_2^{j}}}{\vol_{j-1}\p{B_2^{j-1}}}\nonumber \\
    &= \prod_{t \notin S} \exp\p{\frac{1}{\alpha_t}-\frac{1}{\alpha_{t-1}}} \cdot \prod_{t \in S} \frac{\pn{\vz_t^{\perp}}}{2}\cdot \vol_{d_n}\p{B_2^{d_n}}\label{eq:int_potential}
\end{align}
Here we used that $\vol_0(\cE_{0}) = \vol_0(B_2^0) = 1$. Now invoking~\Cref{claim:det_prod_identity}, we get
\begin{align*}
    \prod_{t \in S} \frac{\pn{\vz_t^{\perp}}}{2} \ge 2^{-\pa{S}}\sqrt{\det\p{\mZ\vert_S\mZ\vert_S^T}} \ge 2^{-\pa{S}} = 2^{-d_n},
\end{align*}
where we used that $\det\p{\mZ\vert_S\mZ\vert_S^T} \ge 1$ because all the vectors $\vz_t$ have integer coordinates. Moreover, since all coordinates are at most $N$ in absolute value, all the vectors  $\vz_t$ have length at most $N\sqrt{d}$. Therefore, $\frac{\vol\p{\cE_n}}{\vol\p{B_2^{d_n}}} \le \p{N\sqrt{d}}^{d_n}$. We plug these bounds back into \eqref{eq:int_potential}, rearrange, and take the logarithm of both sides, yielding
\begin{align*}
    \sum_{t \notin S} \frac{1}{\alpha_t}-\frac{1}{\alpha_{t-1}} \lesssim d_n\log\p{d N}.
\end{align*}
Finally, by \eqref{eq:fully_online_invariant}, we have $\frac{1}{\alpha_t} - \frac{1}{\alpha_{t-1}} = 1$ for every $t \in S$. Combining everything gives
\begin{align*}
    \sum_{t \le n} \frac{1}{\alpha_t}-\frac{1}{\alpha_{t-1}} \lesssim d_n\log\p{d N} + |S| \lesssim d_n\log\p{d N},
\end{align*}
thereby concluding the proof of \Cref{thm:main_two_ints}.
\end{proof}

\section{Forming Small Coresets for Convex Bodies (Proof of Theorem~\ref{thm:main_three})}
\label{sec:coreset}

In this section, we prove Theorem \ref{thm:main_three}. See Algorithm \ref{alg:ellipse_to_coreset}.

\begin{algorithm}[H]
\caption{Streaming coreset for convex hull}\label{alg:ellipse_to_coreset}
\begin{algorithmic}[1]
\State \textbf{Input:} Stream of points $\vz_t$; Update rule for Algorithm \ref{alg:fully_online_approx} $\cA$.
\State \textbf{Output:} Set $S \subseteq [n]$.
\For{$t = 1,\dots,n$}
    \State Receive $\vz_{t}$.
    \State Let $\cE_{\mathsf{test}} = \cA(\vc_{t-1}, \cE_{t-1}, \vz_t)$.
    \State Let $d_t = \mathsf{dim}\p{\vspan{\vz_1-\vc_{t-1},\dots,\vz_t-\vc_{t-1}}}$.
    \If{$d_t > d_{t-1}$ or $\frac{\mathsf{Vol}_{d_t}(\cE_{\mathsf{test}})}{\mathsf{Vol}_{d_t}(\cE_{t-1})} \ge e$}
        \State Let $\vc_t, \cE_{t} = \cA(\vc_{t-1},\cE_{t-1},\vz_{t})$.
        \State Update $S_t = S_{t-1} \cup \pb{\vz_{t}}$.
    \Else
        \State Let $\vc_t,\cE_{t}=\vc_{t-1},\cE_{t-1}$.
        \State Let $S_{t-1} = S_t$.
    \EndIf
\EndFor
\State \textbf{Output:} $S_n$
\end{algorithmic}
\end{algorithm}

For a sketch of the intuition and the argument we will use for the proof, see Section \ref{sec:overview_coreset}.

\begin{proof}[Proof of Theorem \ref{thm:main_three}]

We prove two properties of Algorithm \ref{alg:ellipse_to_coreset}. First, we show
\[
\pa{S_t} \le O\p{d_t \cdot \log\p{d_t \cdot \max_{t' \le t} \sfrac{R_t}{r_{t'}}}},
\]
and, further, $\pa{S_t} \le O\p{d_t \cdot \log\p{dN}}$ if points $\vz_t$ have integer coordinates between $-N$ and $N$. Second, we show that
\begin{align*}
\conv\p{Z\vert_{S_t}} &\subseteq \conv\p{Z\vert_{[t]}} \\
&\subseteq O\p{d_t \cdot \log\p{d_t \cdot \max_{t' \le t} \sfrac{R_t}{r_{t'}}}}\cdot \conv\p{Z\vert_{S_t}},
\end{align*}
and
\begin{align*}
\conv\p{Z\vert_{S_t}} &\subseteq \conv\p{Z\vert_{[t]}} \\
&\subseteq O\p{d_t \cdot \log\p{dN}} \cdot \conv\p{Z\vert_{S_t}}.
\end{align*}

\paragraph{Bounding $\pa{S_t}$.}

It is enough to count the number of steps $t$ for which we have $\frac{\mathsf{Vol}_{d_t}(\cE_{\mathsf{test}})}{\mathsf{Vol}_{d_t}(\cE_{t-1})} \ge e$.

It is easy to see that for all $t$, we have  $r(Z\vert_{[t]}) \cdot \p{B_2^d \cap \vspan{\vz_1-\vc_t,\dots,\vz_t-\vc_t}} \subseteq \vc_{t} + \cE_t$. Additionally, by the definition of $R(Z)$, we always have $Z\vert_{[t]} \subseteq R(Z) \cdot \p{B_2^d \cap \vspan{\vz_1-\vc_t,\dots,\vz_t-\vc_t}}$. These are enough to give volume lower and upper bounds in each step. Next, for each step in which we add an element to $S_{t-1}$ to obtain $S_t$, the volume must increase by a factor of $e$. It easily follows that the number of elements in $S_t$ satisfies
\begin{align*}
    \pa{S_t} \le \log\p{\max_{t' \le t} \frac{\prod_{i=1}^{d_t} R(Z\vert_{[t]})}{\prod_{i=1}^{d_t} r(Z\vert_{[t']})}} = d_t\log\p{\max_{t' \le t} \frac{R(Z\vert_{[t]})}{r(Z\vert_{[t']})}}.
\end{align*}
We now give an upper bound for the case when all coordinated of $\vz_t$ are integers not exceeding $N$ in absolute value. It is easy to see that the update rule in \Cref{alg:ellipse_to_coreset} exactly corresponds to the steps where we have
\begin{align*}
    \frac{1}{\alpha_{t}} - \frac{1}{\alpha_{t-1}} \gtrsim 1,
\end{align*}
and in the same way as in the proof of \Cref{thm:main_two_ints}, we have for all $t$ that
\begin{align*}
    \sum_{t \ge 1} \frac{1}{\alpha_{t}} - \frac{1}{\alpha_{t-1}} \lesssim d_t\log\p{dN}.
\end{align*}
It therefore follows that $\pa{S} \lesssim d_t\log\p{dN}$, as desired.

\paragraph{Bounding the distortion of the chosen points.}

Consider some iteration $t' \le t$. Without loss of generality, let $\vc_{t'-1} = 0$. Suppose $\vz_{t'}$ does not result in an update to $S_{t'-1}$. This implies that $\vz_{t'} \in 2e \cdot \cE_{t'-1}$. Next, observe that $0 \in \vc_t + \cE_t$. Putting these together, we have $\vz_{t'} \in \p{\vc_t + \cE_t} + 2e \cdot \cE_{t'-1}$. Since $\cA$ is monotone, we must have $2e \cdot \cE_{t'-1} \subseteq \vc_t + e\cdot\cE_t$; hence, we may write $\vz_{t'} \in \vc_t + \p{2e+1}\cE_t$.

The inner ellipsoid $\vc_t + \alpha_t \cdot \cE_t$ will still be an inner ellipsoid for the points determined by $S_t$. Stitching together all our inclusions, we have
\begin{align}
    \vc_t + \alpha_t \cdot \cE_t \subseteq Z\vert_{S_t} \subseteq Z \subseteq \vc_t + (2e+1)\cE_t \subseteq \frac{2e+1}{\alpha_t} \cdot Z\vert_{S_t}\label{eq:ellipsoid_coreset_inclusions}.
\end{align}
which means that
\begin{align*}
    Z\vert_{S_t} \subseteq Z \subseteq O\p{d_t \cdot \log\p{d_t \cdot \max_{t' \le t} \frac{R(Z\vert_{([t] \cap S_t)})}{r(Z\vert_{([t'] \cap S_t)})}}} \cdot Z\vert_{S_t}.
\end{align*}
Notice that this is nearly what we want, except that the aspect ratio term is in terms of the subset body $Z\vert_{S_t}$. To obtain the final guarantee in terms of the aspect ratio of $Z\vert_{[t]}$, observe that the above guarantee readily implies that
\begin{align*}
    O\p{d_t \cdot \log\p{d_t \cdot \max_{t' \le t} \frac{R(Z\vert_{([t] \cap S_t)})}{r(Z\vert_{([t'] \cap S_t)})}}} \le O\p{d_t \cdot \log\p{d_t \cdot \max_{t' \le t} \frac{R(Z\vert_{[t]})}{r(Z\vert_{[t']})}}}.
\end{align*}
We now give the corresponding improvement when the $\vz_t$ are integer-valued. As before, \eqref{eq:ellipsoid_coreset_inclusions} holds. From this, we get
\begin{align*}
    Z\vert_{S_t} \subseteq Z \subseteq O\p{d_t \cdot \log\p{dN}} \cdot Z\vert_{S_t},
\end{align*}
as desired. This concludes the proof of Theorem \ref{thm:main_three}.
\end{proof}

% \subsection{Improved Distortion for Symmetric Convex Bodies}
% \input{main_select_vertices_symmetric}

\section{Lower Bound}
\label{sec:lb}

In this section, we show \Cref{thm:main_four}.

\subsection{Lower bound adversary}

Our proof of \Cref{thm:main_four} constructs an adversary, which given a monotone algorithm \(\cA\) and \(\kappa \geq 1\),
constructs a sequence of points \(\vz_1, \ldots, \vz_n\) satisfying \(\kappa(\conv\p{\vz_1, \ldots, \vz_n}) \leq \kappa\) to witness that the algorithm does not produce an approximation better than \(\widetilde{\Omega}(d \log \kappa)\).
While by definition \(\kappa = \frac{R}{r}\), our construction keeps \(r = 1\) (notice that any lower bound construction must be scale-invariant),
and for simplicity we use \(R = \kappa\).

Let \(\vz^\Delta_1, \vz^\Delta_2, \ldots, \vz^\Delta_{d+1} \in \R^{d}\) be the \(d+1\) vertices of a regular simplex \(\Delta_{d}\) that circumscribes \(B_2^d\).
Our adversary is described in \cref{alg:lower_bound}.
It uses a first phase that feeds \(\cA\) the vertices of \(\Delta_d\),
then a second phase that repeatedly feeds \(\cA\) points at a constant distance from the previous ellipsoid.
Specifically, every new point \(\vz_t\) in the second phase is in \(\vc_{t-1} + 2 \cdot \cE_{t-1}\), i.e. its distance is 2 from \(\vc_{t-1}\) in the norm that is the gauge of \(\cE_{t-1}\).

\begin{algorithm}[ht]
\caption{Lower bound adversary}\label{alg:lower_bound}
\textbf{Input}: Monotone algorithm \(\cA\), \(R \geq 1\)
\begin{algorithmic}[1]
\State \((\vc_0 + \cE_0, \alpha_0) = (0 + B_2^d, 1)\)
\Comment{Initialize to the unit ball}
\For{\(t \in \{1, 2, \ldots, d+1\}\)}
    \Comment{Phase I: feed \(\cA\) the vertices of a simplex}
    \State \((\vc_{t} + \cE_{t}, \alpha_{t}) = \cA(\vc_{t-1} + \cE_{t-1}, \alpha_{t-1}, \vz^\Delta_{t})\)
\EndFor
\State \(t \leftarrow d+2\)
\While{\(\vol(\cE_{t-1}) \leq \vol\p{\frac{R}{2} \cdot B_2^d}\)}\label{alg_line:adv_stopvol}
\Comment{Phase II: feed \(\cA\) points outside the previous ellipsoid}
    \State Let \(F_{t-1} = \partial(\vc_{t-1} + 2 \cE_{t-1}) \cap (R \cdot B_2^d)\)
    \If{\(F_{t-1} = \varnothing\)}
        \State \textbf{stop}\label{alg_line:adv_stopdisj}
    \EndIf
    \State Let arbitrary \(\vz_{t} \in F_{t-1}\)
    \State \((\vc_{t} + \cE_{t}, \alpha_{t}) = \cA(\vc_{t-1} + \cE_{t-1}, \alpha_{t-1}, \vz_{t})\)
    \State \(t \leftarrow t + 1\)
\EndWhile
\end{algorithmic}
\end{algorithm}

\begin{remark}
 This particular construction we give of the hard case is adaptive, meaning that the adversary's choice of points depend on the previous ellipsoids the algorithm outputs.
   However, this adversary can be made non-adaptive by taking an \(\varepsilon\)-net \(S\) of \(B_2^d\) for sufficiently small \(\varepsilon\),
   then feeding \(\cA\) the sequence of points in sets \(S, 2 \cdot S, 4 \cdot S, \ldots, 2^{\log_2 R - 1}, 2^{\log_2 R} \cdot S\).
   In consequence, this means that randomization on the part of the monotone algorithm does not help, unlike some other online settings.
\end{remark}

Let \(T\) be the largest value of \(t -1\) before the adversary halts.
We first show that the adversary only gives finitely many points before halting.
\begin{claim}
    \(T \leq O(d \log R)\)
\end{claim}
\begin{proof}
    We argue that the volume of \(\cE_{t}\) increases by at least a constant factor on each iteration.
    This is sufficient to bound the number of iterations by \(O(d \log R)\), 
    as \(\cE_0 = B_2^d\),
    and Line \ref{alg_line:adv_stopvol} is no longer true when the volume of \(\cE_t\) exceeds \(\p{\frac{R}{2}}^d \cdot \vol(B_2^d)\).

    We claim that for all \(t \geq d + 2\), \(\vol(\cE_{t}) \geq \frac{3}{2} \cdot \vol(\cE_{t-1})\).
    By applying a nonsingular affine transformation, we can assume without loss of generality that \(\cE_{t-1} = B_2^d\).
    With a further rotation, we can assume the newly received point is \(\vz_t = 2 \ve_1\).
    From monotonicity of \(\cA\) we must have that \(\vc_{t} + \cE_{t} \supseteq B_2^d \cup \{2 \ve_1\}\).
    Clearly every semi-axis of \(\cE_{t}\) must have length 
    at least 1 in order to contain \(\cE_{t-1}\).
    Observe that \(\cE_{t}\) must also contain the segment connecting
    \(-1 \ve_1\) and \(2 \ve_1\), and so at least one semi-axis must have length at least \(\frac{3}{2}\)
    (if not, the diameter of \(\cE_{t}\) would be strictly less than \(3\)).
    Hence as \(\frac{\vol(\cE_{t})}{\vol(B_2^d)}\) equals the product of the length of the semi-axes of \(\cE_{t}\), we have \(\vol(\cE_{t}) \geq \frac{3}{2} \vol(B_2^d)\).
\end{proof}

For the analysis we define quantities \(A_t, P_t\) associated with the sequence of ellipsoids for \(1 \leq t \leq T\):
\[A_t \defeq \frac{1}{\alpha_t}, \quad P_t \defeq \log \p{\frac{\vol(\cE_t)}{\vol(B_2^d)}}\]
By the monotonicity of \(\cA\), we have that \(A_t\) and \(P_t\) are both nondecreasing in \(t\).
We first observe that the adversary guarantees that the final volume of the ellipsoid output by \(\cA\) is large:
\begin{claim}\label{claim:lower_bound_vol}
    At the conclusion of \Cref{alg:lower_bound}'s execution, we have
   \[P_T  \geq d \log \frac{R}{2}\]
\end{claim}
\begin{proof}
    There are two ways that the adversary stops:
    if the condition in Line \ref{alg_line:adv_stopvol} is no longer true,
    or if Line \ref{alg_line:adv_stopdisj} is reached.
    If the former occurs, then we have \(\vol(\cE_{T}) > \vol(\frac{R}{2} \cdot B_2^d)\),
    and clearly then \(P_T \geq d \log\p{\frac{R}{2}}\).
    
    In the latter stopping condition, the algorithm halts at time \(T\) when \(\partial (c_{T} + 2 \cE_{T}) \cap R \cdot B_2^d = \varnothing\).
    The sets \(\partial (c_{T} + 2 \cE_{T})\) and \(R \cdot B_2^d\) can be disjoint in two cases:
    \(c_T + 2 \cE_T\) and \(R \cdot B_2^d\) are disjoint;
    or \(R \cdot B_2^d \subseteq c_T + 2 \cE_t\) with the boundaries of both ellipsoids disjoint.
    By the monotonicity of \(\cA\), we have \(1 \cdot B_2^d \subseteq c_T + 2 \cE_{T}\),
    and so eliminate the former case.
    But then \(\vol(2 \cdot \cE_T) \geq \vol(R \cdot B_2^d) = R^d \vol(B_2^d)\), and taking logarithms on both sides yields the claim.
\end{proof}

Now in contrast to the upper bound where we essentially gave an algorithm for which \(\frac{\Delta A}{\Delta P}\) was upper bounded by a constant,
here we will show a constant lower bound on the same quantity for any monotone algorithm.
\begin{claim}\label{claim:lb_step_main}
    There exists a constant \(C_{\ref{eqn:lb_step_main}} > 0\) such that if \(A_t \geq d\), we have
    \begin{equation}\label{eqn:lb_step_main}
    A_{t+1} - A_t \geq C_{\ref{eqn:lb_step_main}} (P_{t+1} - P_t)
    \end{equation}
\end{claim}

Observe that this lower bound requires \(A_t \geq d\),
hence necessitating a first phase using the simplex,
whose optimal roundings show tightness for John's theorem for general convex bodies.
In order to prove the lower bound we also need a second property, that \(A_{t}\) is large compared to \(P_t\).
\begin{claim}\label{claim:lb_simplex_symmetry}
    Let \(0 \leq \alpha \leq 1, \vc \in \R^d\), and \(\cE\) be an ellipsoid such that
    \[\vc + \alpha \cdot \cE \subseteq \Delta_d \subseteq \vc + \cE\]
    then we have: \begin{enumerate} 
        \item\label{item:lb_simplex_symmetry_1} \(\alpha \leq \frac{1}{d}\)
        \item \(\log\p{\frac{\vol(\cE)}{\vol(B_2^d)}} \leq O\p{\log(d) \cdot \frac{1}{\alpha}}\)
    \end{enumerate}
\end{claim}

With the statements of these claims in hand, we are ready to prove the lower bound.
\begin{proof}[Proof of \Cref{thm:main_four}]
 It is clear that \(\kappa(\conv\p{\vz_1, \ldots, \vz_T)} \leq R\), as for every \(1 \leq t \leq T\) the adversary guarantees \(1 \leq \|\vz_t\| \leq R\). So we focus on showing a lower bound on the quality of the approximation produced by \(\cA\).

As \(\cA\) is monotone, after the end of Phase I
we must have that \begin{equation}\label{eq:lb_ph1_approx}
\vc_{d+1} + \alpha_{d+1} \cdot \cE_{d+1} \subseteq \Delta_d \subseteq \vc_{d+1} + \cE_{d+1}
\end{equation}
Now because \(\cE_{d+1}\) satisfies the conditions of \Cref{claim:lb_simplex_symmetry},
we get using the definition \(A_{d+1} = \frac{1}{\alpha_{d+1}}\) that \(A_{t} \geq A_{d+1} \geq d\) for any \(t \geq d+1\).
Then we can apply \Cref{claim:lb_step_main} for every \(t \geq d + 1\) until termination of the algorithm:
\begin{align*}
A_{d+2} - A_{d+1} &\geq C_{\ref{eqn:lb_step_main}} \p{P_{d+2} - P_{d+1}} \\
A_{d+3} - A_{d+2} &\geq C_{\ref{eqn:lb_step_main}} \p{P_{d+3} - P_{d+2}} \\
&~\vdots \\
A_{T-1} - A_{T-2} &\geq C_{\ref{eqn:lb_step_main}} \p{P_{T-1} - P_{T-2}} \\
A_{T} - A_{T-1} &\geq C_{\ref{eqn:lb_step_main}} \p{P_{T} - P_{T-1}}
\end{align*}
Summing these inequalities, we have
\begin{align*}
    \sum_{t=d+1}^{T-1} A_{t+1} - A_{t} \geq C_{\ref{eqn:lb_step_main}} \p{\sum_{t=d+1}^{T-1} P_{t+1} - P_t}
\end{align*}
Both sides of this inequality are telescoping sums, so simplifying we get
\begin{equation}\label{eqn:thm_lower_bound_1}
A_T \geq A_{d+1} + C_{\ref{eqn:lb_step_main}} (P_T - P_{d+1})
\end{equation}

% Thus \(\vc_{d+1} + \cE_{d+1}\) is a rounding ellipsoid of the \(d\)-simplex, and it is well known that in this case the approximation factor is \(\alpha_{d+1} \leq \frac{1}{d}\) (see e.g. \cite{howard1997john}, Remark 1.1); i.e. \(A_{d+1} \geq d\).\maxcmt{fix flow}
Again because we can apply \Cref{claim:lb_simplex_symmetry} for \(\cE_{d+1}\), we have \(P_{d+1} \leq O(\log(d) \cdot A_{d+1})\), which along with (\ref{eqn:thm_lower_bound_1}) yields
\[A_T \geq A_{d+1} + \Omega(P_T - \log(d) \cdot A_{d+1}) \geq  \Omega (P_T - \log(d) \cdot A_{d+1})\]
Thus we have \[
A_T \geq \Omega(\max(A_{d+1}, P_T - \log(d) \cdot A_{d+1})) \geq \Omega\p{\frac{P_T}{\log(d)}}
\]
and we get the desired bound using \Cref{claim:lower_bound_vol}.
\end{proof}

Our proof of \Cref{claim:lb_step_main}
relies on a symmetrization argument to a reduced case (essentially two-dimensional, like for our algorithms). We now define this reduced case, and related quantities.
\begin{definition}\label{defn:lb_reduced}
In the reduced case, the previous outer and inner ellipsoids are given by \(B_2^d, \alpha \cdot B_2^d\), and the received point is \(\vz = 2 \ve_1\).
The next outer and inner ellipsoids are given by \(c \cdot \ve_1 + \cE_{\mM}, c \cdot \ve_1 + \alpha' \cdot \cE_{\mM}\)
for \(c \in \R\), and \(\mM = \Diag(a, b, b, \ldots, b, b)\) for \(a, b \geq 1\).
We let \(\Delta A = \frac{1}{\alpha'} - \frac{1}{\alpha}\) and \(\Delta P = \log \p{\frac{\vol(\cE_{\mM})}{\vol(B_2^d)}}\).
\end{definition}

Note that the update in this reduced case is monotone if
\(B_2^d \cup \{ 2 \ve_1 \} \subseteq c \cdot \ve_1 + \cE_{\mM}\)
and \(c \cdot \ve_1 + \alpha' \cdot \cE_{\mM} \subseteq \conv\p{(\alpha \cdot B_2^d) \cup \{ 2 \ve_1 \}}\).

Now we state the lower bound on \(\frac{\Delta A}{\Delta P}\) in this setting, which is established in \Cref{sec:lb_2d}.
It is exactly the bound of \Cref{claim:lb_step_main} in this special case.
\begin{claim}\label{claim:lb_reduced}
     In the reduced case, for any monotone update \(c \cdot \ve_1 + \cE_{\mM}, c \cdot \ve_1 + \alpha' \cdot \cE_{\mM}\) when \(\alpha' \leq \frac{1}{d}\) we have
    \begin{equation}
    \frac{\Delta A}{\Delta P} \geq C_{\ref{eqn:lb_step_main}}
     \end{equation}
\end{claim}

We now give the symmmetrization argument that shows that the above bound in the special case
implies the bound in the general case.
\begin{proof}[Proof of \Cref{claim:lb_step_main}]
    % We show this using a symmetrization argument.
    % Recall that an update \(\vc_t + \cE_t, \vc_{t+1} + \cE_{t+1}\) is monotone if 
    By the monotonicity of \(\cA\), we have \((\vc_t + \cE_t) \cup \{\vz_{t+1}\} \subseteq \vc_{t+1} + \cE_{t+1}\)
    and \(\vc_{t+1} + \alpha_{t+1} \cdot \cE_{t+1} \subseteq \conv\p{(\vc_t + \alpha_t \cdot \cE_t) \cup \{\vz_{t+1}\}}\).
    Without loss of generality we assume that \(\vc_t + \cE_t = B_2^d\) and \(\vz_{t+1} = 2 \cdot \ve_1\);
    we do this by applying a nonsingular affine transformation that maps \(\vc_t\) to the origin and \(\cE_t\) to \(B_2^d\), then apply a rotation
    that maps \(\vz_{t+1}\) to \(2 \cdot e_1\).
    Let \(\vc = \vc_{t+1}\), \(\cE = \cE_{t+1}\), and \(\alpha = \alpha_{t+1}\).
    Summarizing the conditions guaranteed by the monotonicity of \(\cA\), we have that \(B_2^d \cup \{ 2 \ve_1 \} \subseteq \vc + \cE\)
    and \(\vc + \alpha \cdot \cE \subseteq \conv\p{(\alpha_t \cdot B_2^d) \cup \{ 2 \ve_1 \}}\).

    To perform the reduction to the two-dimensional case, we apply a sequence of volume-preserving symmetrizations to the new inner and outer ellipsoids; these symmetrizations will also ensure that the update remains monotone.
    We will first apply two Steiner symmetrizations. The first of these Steiner symmetrizations transforms the ellipsoids so that their center lies on the \(\ve_1\)-axis. The second ensures that the ellipsoids have a semi-axis that is parallel to \(\ve_1\).
    Then, by a final symmetrization step we can transform the ellipsoids into bodies of revolution about \(\ve_1\).
    At that point it will suffice to consider the two-dimensional reduced case.

    Let \(\vc'\) be the projection of \(\vc\) onto the \(\ve_1\)-axis.
    The goal of the first symmetrization step is to transform \(\vc + \cE\) to \(\vc' + \cE'\) so that \(\vc'\) lies on the \(\ve_1\) axis. If \(\vc = \vc'\) then we do not need to do anything, otherwise we apply Steiner symmmetrization and consider \(S_{\vc - \vc'}(\vc+ \cE)\).
    By \Cref{claim:steiner_ellipsoid} this is still an ellipsoid,
    and we also have that the center of \(S_{\vc - \vc'}(\vc + \cE)\) is actually \(\vc'\);
    thus we may write \(S_{\vc - \vc'}(\vc + \cE) = \vc' + \cE'\) for some \(\cE'\).
    Further, we have that \(S_{\vc - \vc'}(\vc + \alpha \cdot \cE) = \vc' + \alpha \cdot \cE'\), as the Steiner symmetrization acts similarly on the scaled version of \(\cE\).
    To show that the update is still monotone, we observe that \(\vc' + \cE' = S_{\vc - \vc'}(\vc + \cE) \subseteq S_{\vc - \vc'}( \conv\p{(\alpha_t \cdot B_2^d) \cup \{ 2 \ve_1 \}})\).
    But by \Cref{claim:steiner_sym} and that \(\vc - \vc' \perp \ve_1\), \(\conv\p{(\alpha_t \cdot B_2^d) \cup \{ 2 \ve_1 \}}\) is invariant under the symmetrization \(S_{\vc - \vc'}\) and so we still have the inclusion \(\vc' + \alpha \cdot \cE' \subseteq \conv\p{(\alpha_t \cdot B_2^d) \cup \{ 2 \ve_1 \}}\).
    The `outer' inclusion
    \(B_2^d \cup \{ 2 \ve_1 \} \subseteq \vc' + \cE'\) follows in the same way.

    We now apply the second and final Steiner symmetrization. Let \(\vr\) be the rightmost point of \(\vc' + \cE'\) along \(\ve_1\); i.e. \(\vr = \arg\max_{\vr \in \vc' + \cE'} \pgo{\vr, \ve_1}\).
    Also let \(\vr'\) be its projection along the \(\ve_1\)-axis;
    if \(\vr = \vr'\) we again do not need to perform this symmetrization step, otherwise
    the Steiner symmetrization we apply is \(S_{\vr - \vr'}(\vc' + \cE')\).
    Since \(\vc'\) is at the midpoint of \(\vc' + \R (\vr - \vr')\) the center of the new ellipsoid is still \(\vc'\),
    so we may write \(S_{\vr - \vr'}(\vc' + \cE') = \vc' + \cE''\) and similarly
    \(S_{\vr - \vr'}(\vc' + \alpha \cdot \cE') = \vc' + \alpha \cdot \cE''\).
    Like for the previous symmetrization, the fact that \(\vr - \vr' \perp \ve_1\)
    means that both inclusions of the monotone update are preserved.
    Note finally that \(\vr'\) is the rightmost point of \(\vc' + \cE''\)
    and that the tangent plane of \(\vc + \cE''\) at \(\vr'\) is orthogonal to the line segment \(\overline{\vc' \vr'}\),
    so \(\ve_1\) is a semi-axis of \(\vc' + \cE''\).

    Our last transformation is a symmetrization of a different form, to turn \(\vc' + \cE''\) into a body of revolution.
    Let \(\sigma_1\) be the length of the semi-axis \(\ve_1\) of \(\cE''\), and \(\sigma_2, \ldots, \sigma_d\) be the lengths of the other semi-axes of \(\cE''\).
    We let \(\cE'''\) be the ellipsoid that has a \(\ve_1\) as a semi-axis of length \(\sigma_1\),
    and where every other semi-axis of \(\cE'''\) has length \(\sigma' \defeq \p{\prod_{i=2}^d \sigma_i}^{1/(d-1)}\).
    Clearly \(\vc' + \cE'''\) is now a body of revolution about \(\ve_1\) whose volume is the same as that of \(\vc' + \cE''\) (and hence also of \(\vc + \cE\)).
    Note that \(\vc' + \alpha \cE'''\) is also now a body of revolution.
    Since \(\sigma' \ge \min_{2 \le i \le d} \sigma_i\) we have \(B_2^d \cup \{ 2 \ve_1 \} \subseteq \vc' + \cE'''\), and correspondingly since \(\sigma' \le \max_{2 \le i \le d} \sigma_i\) we have that \(\vc' + \alpha \cdot \cE''' \subseteq \conv\p{(\alpha_t \cdot B_2^d) \cup \{ 2 \ve_1 \}}\).
    
    Clearly \(\vc' + \cE''', \vc + \alpha \cdot \cE'''\) now adhere to the reduced case of \Cref{defn:lb_reduced}.
    Since the update is monotone as well (and still \(\alpha \le 1/d\)) we can apply \Cref{claim:lb_reduced}.
    As \(\vol(\cE''') = \vol(\cE)\), this means we have
    \[
    A_{t + 1} - A_{t} \geq C_{\ref{eqn:lb_step_main}} \cdot (P_{t+1} - P_{t})
    \]
    as desired.
\end{proof}

\begin{proof}[Proof of \Cref{claim:lb_simplex_symmetry}]
For the first property, this is exactly the well-known fact that the best ellipsoidal rounding for the simplex \(\Delta_d\) (see e.g. \cite[Remark 1.1]{howard1997john}) has approximation factor \(d\).

Now we show the second property.
Again because the ball rounds the simplex \(\Delta_d\) with approximation factor \(d\), we have
\[\frac{1}{d} \cdot \Delta_d \subseteq B_2^d \subseteq \Delta_d\]
As a result of this, we have
\begin{align*}
    \log \p{\frac{\vol(\vc + \alpha \cdot \cE)}{\vol(B_2^d)}} &\le \log \p{\frac{\vol(\Delta)}{\vol(B_2^d)}} \\
    &\le \log \p{\frac{\vol(d \cdot B_2^d)}{\vol(B_2^d)}} \\
    &\le O(d \log d) 
\end{align*}
And so
\begin{align*}
\log \p{\frac{\vol(\cE)}{\vol(B_2^d)}} &= \log \p{\frac{\vol(\vc + \alpha \cdot \cE)}{\vol(B_2^d)}} + d \log\p{\frac{1}{\alpha}} \\
&\le O\p{d \log \p{\frac{1}{\alpha}}} & \text{as } \alpha \le \frac{1}{d}
\end{align*}
To establish the second property, it remains to show \(d \log(1/\alpha) \le O((1/\alpha) \log(d))\).
Observe that \(x \mapsto \frac{x}{\log x}\) is increasing for \(x \ge e\), so we have
\[\frac{d}{\log d} \le O\p{\frac{\sfrac{1}{\alpha}}{\log(\sfrac{1}{\alpha})}}\]
for all \(d \ge 2\) as \(d \le \sfrac{1}{\alpha}\).
Rearranging gives the desired inequality and thus the second property.
\end{proof}

\subsection{Analysis of the reduced case}
\label{sec:lb_2d}
In this section, we establish a lower bound on \(\frac{\Delta A}{\Delta P}\),
assuming we are in the `reduced case' defined in \Cref{defn:lb_reduced}.
Observe that in this case all relevant convex bodies \(\cE, \alpha \cE, c \cdot \ve_1 + \cE', c \cdot \ve_1 + \alpha' \cE', \conv\p{\alpha \cE \cup \{\vz\}}\) are all bodies of revolution about the \(x_1\)-axis, so to analyze the quantities involved we may instead look at any two-dimensional slice.
Accordingly we talk about the ellipses \(\cE, \alpha \cE, c + \cE', c + \alpha' \cE'\) in this two-dimensional slice,
where again \(\cE = B_2^2\), and \(c + \cE'\) and \(c + \alpha' \cE'\) are defined by
\begin{gather*}
c + \cE' = \pb{(x, y) \in \R^d \middle| \p{\frac{x-c}{a}}^2 + \p{\frac{y}{b}}^2 \leq 1}\\
c + \alpha' \cE' = \pb{(x, y) \in \R^d \middle| \p{\frac{x-c}{a}}^2 + \p{\frac{y}{b}}^2 \leq \alpha'^2}
\end{gather*}
for \(a, b > 0, c \in \R\).
We also use for convenience \(A = \frac{1}{\alpha}\) and \(A' = \frac{1}{\alpha'}\) so that \(\Delta A = A' - A\).
Also note in this reduced case we have by symmetry that
\[\Delta P = \log\p{\frac{\vol(c \cdot \ve_1 + \alpha' \cE')}{\vol(B_2^d)}} - \log\p{\frac{\vol(B_2^d)}{\vol(B_2^d)}} = \log(a \cdot b^{d-1})\]

Our lower bound in this reduced case is the following:
\begin{claim} There exists a fixed constant \(C_{\ref{eq:lb_cpct}} > 0\) such that 
    \[\frac{\Delta A}{\Delta P} \geq \min\p{C_{\ref{eq:lb_cpct}}, \frac{1}{10} \frac{A}{d}}\]    
\end{claim}

Clearly this claim yields \Cref{claim:lb_reduced} as a corollary, as by assumption in \Cref{claim:lb_reduced} we have \(A \geq d\) and so we get \(\frac{\Delta A}{\Delta P} \geq \Omega(1)\).

\begin{figure}[ht]
\centering
\includegraphics[height=6cm]{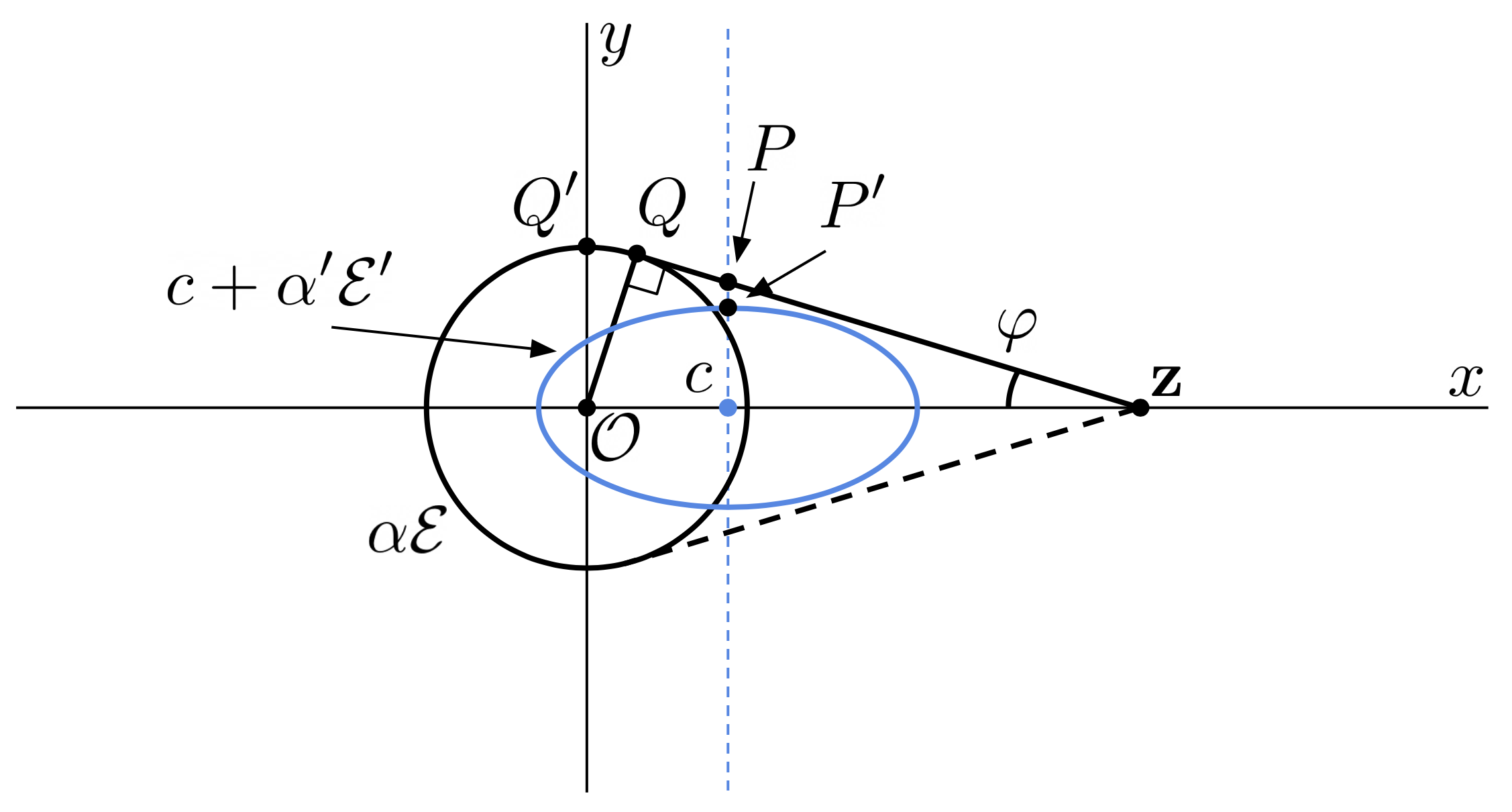}
\caption[The inner ellipses in the two-dimensional lower bound.]{The inner ellipses in the two-dimensional lower bound. \(\cO\) is the origin. The black solid circle is the previous inner ellipse \(\alpha \cE\),
and the blue solid circle is the next inner ellipse \(c + \alpha' \cE'\). The vertical dotted blue line \(x=c\) through the center \(c\) marks the location of the next inner ellipse on the \(x\)-axis. The new point is \(\vz = 2 \ve_1\),
and \(\overline{\vz Q}\) is one of the lines through \(\vz\) tangent to \(\alpha \cE\), with \(Q\) the point of tangency.
\(Q'\) is the intersection of \(\alpha \cE\) with the \(y\)-axis on the same side of the \(x\)-axis as \(Q\).
\(P'\) is the intersection of the line \(x=c\) with \(c + \alpha' \cE'\) on the same side as \(Q\),
and \(P\) is the intersection of this line with \(\overline{\vz Q}\).
We denote the angle \(\angle P \vz c\) with \(\varphi\).
}
\label{fig:lb_inner}
\end{figure}

The inner ellipses in this lower bound, and some relevant points used in the proof of this claim, are depicted in \Cref{fig:lb_inner}.
%GDrive Link: https://docs.google.com/drawings/d/1QNMgITYQrYlHKHLYSxmiCLNFNIxT9kxx4jiWoWz-PdI/edit?usp=sharing

\begin{figure}[ht]
\centering
\resizebox{\textwidth}{!}{%
\begin{tikzpicture}[
level 1/.style={sibling distance=75mm},
level 2/.style={sibling distance=55mm},
level 3/.style={sibling distance=25mm},
]
\tikzset{v/.append style={font=\bfseries}}
  \node {\eqref{eq:lb_5}}
    child {
        node[v] {I}
        child {
            node[v] {Ia}
            edge from parent
                node[left] {\(a, b \le 16\)}
        }
        child {
            node[v] {Ib}
            edge from parent
                node[right] {\(a > 16\) or \(b > 16\)}
        }
        edge from parent
            node[left] {New ellipse close to previous}
    }
    child {
        node[v] {II}
        child {
            node[v] {IIa}
            child {
                node[v] {IIa-i}
                edge from parent
                    node[left] {\(A > 2\)}
            }
            child {
                node[v] {IIa-ii}
                child {
                    node[v] {IIa-ii-A}
                    edge from parent
                        node[left] {\(b \le 100\)}
                }
                child {
                    node[v] {IIa-ii-B}
                    edge from parent
                        node[right] {\(b > 100\)}
                }
                edge from parent
                    node[right] {\(A \le 2\)}
            }
            edge from parent
                node[left] {\(a \le 16\)}
        }
        child {
            node[v] {IIb}
            child {
                node[v] {IIb-i}
                 edge from parent
                    node[left] {Term I smaller}
            }
            child {
                node[v] {IIb-ii}
                 edge from parent
                    node[right] {Term II smaller}
            }
            edge from parent
                    node[right] {\(a > 16\)}
        }
        edge from parent
            node[right] {New ellipse far from previous}
    };
\end{tikzpicture}
}
\caption{Tree of cases in the lower bound}
\label{fig:lb_cases}
\end{figure}

\begin{proof}
We establish this claim through a geometric argument that we break down by cases
(the logical tree of cases is visualized in \Cref{fig:lb_cases}).
First, as the new outer ellipse \(c + \cE'\) contains \(\cE = B_2^d\) we readily have that \(a, b \geq 1\).

As the rightmost point of the new outer ellipse must be to the right of \(\vz\), we have 
\begin{equation}\label{eq:lb_1}
c + a > 2
\end{equation}

As the leftmost point of the new inner ellipse must be to the right of the leftmost point of the previous inner ellipse, we have
\begin{equation}\label{eq:lb_4}
c \geq \alpha' a - \alpha
\end{equation}

\begin{claim}\label{claim:lb_1}
    We have \(\alpha' \cdot b \leq \alpha\), or equivalently \(A' \geq b \cdot A\).
\end{claim}
\begin{proof}
    The geometry of this fact is visualized in \Cref{fig:lb_inner}.
    We overload notation so that \(c\) will also denote the point \((c, 0)\), the center of the new inner ellipse.
We denote \(\cO\) as the origin.
Let \(\overline{\vz Q}\) be one of the lines through \(\vz\) and tangent to \(\alpha \cE\),
with \(Q\) the point of tangency  (the choice of which line is arbitrary, in the figure we choose the one whose intersection with \(\alpha \cE\) is above the \(x\)-axis).
We let \(P\) be the intersection of the vertical line through \((c, 0)\) with \(\overline{\vz Q}\),
and \(P'\) be the intersection of this line with the ellipse \(c + \alpha' \partial \cE'\)
on the same side of the \(x\)-axis as \(Q\).

    Observe that \(\alpha' b = \overline{c P'}\) as the vertical semi-axis of the ellipse \(c + \alpha' \cE\), and \(\alpha = \overline{\cO Q'}\).
    Due to the fact that \(c + \alpha' \cE' \subseteq \conv\p{\alpha \cE \cup \{\vz\}}\),
    the projection of both sets onto the \(y\)-axis satisfies the same inclusion, and this gives the desired inequality.
\end{proof}

Observe that as \(c + \cE'\) must contain both the points \((-1, 0)\) and \((0, 2)\), we have
\begin{equation}
\label{eq:lb_a_3_2}
a \geq \frac{3}{2}
\end{equation}

Observe that we can split \(\Delta P\) into two terms: \[\Delta P = \underbrace{(d-1) \log b}_{\text{I}} + \underbrace{\log a}_{\text{II}}\]
First we show that if term I is larger, then we have a constant lower bound on \(\frac{\Delta A}{\Delta P}\). 
\begin{claim}\label{claim:lb_term2}
    If \((d-1) \log b \geq \log a\), then \(\frac{\Delta A}{\Delta P} \geq \frac{1}{2} \frac{A} {d}\).
\end{claim}
\begin{proof}
    Under the assumption, we have \(\Delta P \leq 2 (d-1) \log b\).
    Combining this with \Cref{claim:lb_1}, we have
    \begin{align*}
    \frac{\Delta A}{\Delta P} &\geq \frac{A' - A}{2 (d-1) \log b} \\
    &\geq \frac{b \cdot A - A}{2 (d-1) \log b} \\
    &= \frac{b-1}{2 \log b} \frac{A}{d-1} \\
    &\geq \frac{1}{2} \frac{A}{d} \\
    \end{align*}
    where the last line uses that \(\frac{x-1}{2 \log (x)} > \frac{1}{2}\) when \(x > 1\).
\end{proof}

In light of \Cref{claim:lb_term2}, we can then assume in the sequel Term II is larger, meaning that
\begin{equation}\label{eq:lb_5}
\Delta P \leq 2 \log a
\end{equation}

\textbf{Case I }\textit{(New ellipse is close to the previous one)}. Assume that 
\begin{equation}\label{eq:lb_2}
    c + \alpha' \cdot a \leq \frac{11}{10} \alpha
\end{equation}
i.e. that the rightmost point of the new inner ellipse is to the left of \(\frac{11}{10}\).

\begin{claim}\label{claim:lb_close_1}
    In Case I, we have \(\Delta A \geq \frac{4}{11} A\).
\end{claim}
\begin{proof}
    We prove this by cases.
    First, if we assume that \(\alpha' \leq \frac{\alpha}{2}\),
    we get \(A' \geq 2 A\) and \(\Delta A \geq A\).

    In the second case, we have \(\alpha' > \frac{\alpha}{2}\).
    We first use this to show \(c > 0\).
    By (\ref{eq:lb_4}) and (\ref{eq:lb_1}) we have \(\frac{c + \alpha}{\alpha'} \geq a > 2 - c\), so
    \(c (1 + \alpha') > 2 \alpha' - a > 0\) and so \(c > 0\).

    Using (\ref{eq:lb_2}) and that \(c > 0\), we have \(\alpha' a \leq \frac{11}{10} \alpha\).
    Thus \(A' \geq \frac{10}{11} a A\).
    By (\ref{eq:lb_a_3_2}) we get \(A' \geq \frac{15}{11} A\), and finally
    \(\Delta A \geq \frac{4}{11} A\).
\end{proof}

\begin{claim}\label{claim:lb_2}
    In Case I, we have \(A' \geq \frac{10}{21} a \cdot A\).
\end{claim}
\begin{proof}
From (\ref{eq:lb_4}) we also get the weaker lower bound \(c \geq - \alpha\).
Combined with (\ref{eq:lb_2}), this gives \(\alpha' a \leq \frac{21}{10} \alpha\), which is equivalent to the desired inequality.
\end{proof}

We divide Case I into two sub-cases.

\textbf{Case Ia} \textit{(\(a, b \leq 16\))}.
First, assume that \(a, b \leq 16\).
Then \(\Delta P = \log(a \cdot b^{d-1}) \leq d \log(16)\),
and by \Cref{claim:lb_close_1} we get \[\frac{\Delta A}{\Delta P} \geq \frac{4}{11 \log(16)} \frac{A}{d} \geq \frac{1}{10} \frac{A}{d}\]

\textbf{Case Ib} \textit{(\(a > 16\) or \(b > 16\))}. Now assume that either \(a\) or \(b\) is greater than \(16\).

Combining \Cref{claim:lb_1} and \Cref{claim:lb_2} together, we get \(A'^2 \geq \frac{10}{21} ab A^2\), or \(A' \geq \sqrt{\frac{10}{21} a b} \cdot A\).
We have \(\Delta A = A' - A \geq \p{\sqrt{\frac{10}{21} a b} - 1} A\). As \(ab \geq 16\) we get \(\sqrt{ab} \geq 2 \sqrt{\frac{21}{10}}\), so
\(\Delta A = \p{\sqrt{\frac{10}{21} a b} - 1} A \geq \sqrt{\frac{5}{21} ab} \cdot A\).

Using the analytic inequality that \(\log x \leq \sqrt{x}\) for all \(x > 0\), we get
\(\Delta P = \log a + (d-1) \log b \leq \sqrt{a} + (d-1) \sqrt{b} \leq d \cdot \sqrt{ab}\).

Combining these inequalites for \(\Delta A\) and \(\Delta P\), we obtain
\[\frac{\Delta A}{\Delta P} \geq \sqrt{\frac{5}{21}} \frac{A}{d} \geq \frac{4}{10} \frac{A}{d}\]

\textbf{Case II }\textit{(New ellipse is far from the previous one)}. Assume that
\begin{equation}\label{eq:lb_3}
    c + \alpha' \cdot a > \frac{11}{10} \alpha
\end{equation}

\begin{claim}\label{claim:lb_3}
    We have 
    \begin{equation}\label{eq:lb_trig_b}
    \alpha' b \leq (2 - c) \cdot \frac{\frac{\alpha}{2}}{\sqrt{1 - \p{\frac{\alpha}{2}}^2}}
    \end{equation}
\end{claim}
\begin{proof}
Again, the proof of this claim is pictured in \Cref{fig:lb_inner},
where we construct the points in the same way as in the proof of \Cref{claim:lb_1}.
Let \(\angle P \vz c\) be denoted by \(\varphi\).
Note that the angle \(\angle P c \vz\) is a right angle, and so \(\tan \varphi = \frac{\overline{c P}}{\overline{c v}}\). The line \(\overline{c \vz}\) has length \(2 - c\), so we get \(\overline{c P} = (2 - c) \tan \varphi\).
The segment \(\overline{c P'}\), of length \(\alpha' b\), is contained within the segment \(\overline{c P}\), and so
\(\alpha' b \leq (2 - c ) \tan \varphi\).
    
Observe that the angle \(\angle \cO Q \vz\) is also a right angle.
Further, clearly the length of \(\overline{\cO Q}\) is \(\alpha\) and the length of \(\overline{0 v}\) is \(2\).
Since we have that \(\varphi\) is also the angle \(\angle Q \vz \cO\), we get \(\sin \varphi = \frac{\alpha}{2}\).
Now using the standard trigonometric identity that \(\tan \varphi = \frac{\sin \varphi}{\sqrt{1 - \sin^2 \varphi}}\) for $\varphi\in[-\pi/2,\pi/2]$,
we get the desired inequality.
\end{proof}

We split Case II into several sub-cases, as for Case I. 

\textbf{Case IIa} \textit{(\(a \leq 16\))}. First, we look at the case where \(a \leq 16\).

\textbf{Case IIa-i} \textit{(\(A \geq 2\))}. Assume \(A \geq 2\).
\begin{claim}\label{claim:lb_4}
    When \(A \geq 2\), we have \(\Delta A \geq \frac{1}{2}\).
\end{claim}
\begin{proof}
Adding (\ref{eq:lb_4}) and (\ref{eq:lb_3}) together gives \(c > \frac{1}{20} \alpha\).
Using this in (\ref{eq:lb_trig_b}) and rearranging using the definitions of \(A\) and \(A'\) yields
% \[\alpha' b \leq \p{1 - \frac{1}{40} \alpha} \frac{\alpha}{\sqrt{1 - \p{\frac{\alpha}{2}}^2}}\]
\[A' \geq b A \cdot \frac{1}{1-\frac{1}{40} \frac{1}{A}} \p{1 - \frac{1}{A^2}}\]
and therefore we get the inequality
\[\Delta A \geq b A \cdot \frac{1}{1-\frac{1}{40} \frac{1}{A}} \p{1 - \frac{1}{A^2}} - 1\]
and using \(b \geq 1\), we obtain
\[\Delta A \geq A \cdot \frac{1}{1-\frac{1}{40} \frac{1}{A}} \p{1 - \frac{1}{A^2}} - 1\]
To prove the claim, it suffices to show the right hand side exceeds \(\frac{1}{2}\) when \(A \geq 2\). Upon rearranging, this is equivalent to the inequality \(A - \frac{77}{80} \frac{1}{A} \geq \frac{3}{2}\) when \(A \geq 2\).
\end{proof}
Combining the assumption that \(a \leq 16\) with (\ref{eq:lb_5}) and \Cref{claim:lb_4} yields \(\frac{\Delta A}{\Delta P} \geq \frac{1}{4 \log 16} \geq \frac{1}{20}\).

\textbf{Case IIa-ii} \textit{(\(A \leq 2\))}. Next, we look at the other case where \(A \leq 2\).

\textbf{Case IIa-ii-A} \textit{(\(b \leq 100\))}. Now we look at the case where \(b \leq 100\).
\begin{claim}
    If \(a \leq 16, A \leq 2, b \leq 100\), then there is \(C_{\ref{eq:lb_cpct}} > 0\) such that
    \begin{equation}\label{eq:lb_cpct}
        \frac{\Delta A}{\Delta P} \geq C_{\ref{eq:lb_cpct}}
    \end{equation}
\end{claim}
\begin{proof}
    We show this by a compactness argument.
    By (\ref{eq:lb_5}) and the assumption that \(a \leq 16\) we have \(\frac{\Delta A}{\Delta P} \geq \frac{\Delta A}{2 \log 16}\).
    Now, observe that next outer and inner ellipsoids \(c + \cE'\) and \(c + \alpha' \cE'\) are fully determined by the parameters
    \(a, b, c, A, A'\).
    Further, we assume without loss of generality that \(A'\) is a function of the other parameters.
    This is because when \(A'\) is decreased as much as possible while preserving the monotonicity of the update, \(\Delta A = A' - A\) only decreases.
    To show a lower bound on \(\Delta A\) it then suffices to only do so in this hardest case.
    
    Note that we have \(1 \leq a \leq 16, 1 \leq b \leq 100, -1 \leq c \leq 2\), and \(1 \leq A \leq 2\); thus all the parameters defining the next inner and outer ellipsoids are bounded.
    Observe that \(\Delta A\) is a continuous function of these parameters, and as a continuous function of a compact set it attains its minimum. Finally, we argue that it is impossible for the minimum of \(\Delta A\) to be zero, and so the minimum is some strictly positive constant \(\frac{C_{\ref{eq:lb_cpct}}}{2 \log 16}\), which suffices to prove the claim.
    
    The following argument only concerns the inner ellipsoids, and can be pictured in \Cref{fig:lb_inner}. If \(c = 0\), then as the leftmost point of \(c + \alpha' \cE'\) must be to the right of the leftmost point of \(\alpha \cE'\),
    we have \(\alpha \geq \alpha' a\). But by (\ref{eq:lb_a_3_2}), we have \(\alpha \geq \frac{3}{2} \alpha'\), so \(\alpha' < \alpha\) and \(\Delta A > 0\).
    If \(c \neq 0\) then the vertical semi-axis of \(c + \alpha' \cE'\) must have length strictly less than \(\alpha\),
    and so \(\alpha > \alpha' b\). As \(b \geq 1\), this also gives \(\alpha > \alpha'\) and again \(\Delta A > 0\).
\end{proof}

\textbf{Case IIa-ii-B} \textit{(\(b > 100\))}.
Observe that the horizontal axis of the next inner ellipsoid \(c + \alpha' \cE'\) must be contained within the interval \([-\alpha, 2]\), thus we have that \(2 + \alpha \geq 2 b \alpha' > 200 \alpha'\).
Using the definitions of \(A, A'\) this is equivalent to \(2 + \frac{1}{A} > \frac{200}{A'}\), i.e. \(A' > \frac{200}{1 + \frac{1}{A}}\). As \(A \geq 1\) we get \(2 + \frac{1}{A} \leq 3\), and so \(A' \geq \frac{200}{3}\).

As \(A \leq 2\), we obtain \(\Delta A = A' - A \geq \frac{194}{3}\).
Now by \eqref{eq:lb_5} and that \(a \leq 16\) we have 
\[\frac{\Delta A}{\Delta P} \geq \frac{\Delta A}{2 \log 16} \geq \frac{194}{6 \log 16} \geq 11\]

\textbf{Case IIb} \textit{(\(a > 16\))}. Now, we examine the case where \(a > 16\).
Scaling (\ref{eq:lb_4}) by \(\frac{11}{10}\) and adding it to (\ref{eq:lb_3}), we have \(\frac{21}{10} c - \frac{1}{10} \alpha' a \geq 0\), i.e. \(c > \frac{1}{21} \alpha' a\).
Using this in (\ref{eq:lb_trig_b}), using \(b \geq 1\), using the definitions of \(A\) and \(A'\) and rearranging, we obtain
\[A' \geq A \cdot \frac{1}{1 - \frac{1}{42} \cdot \frac{a}{A'}} \cdot \sqrt{1 - \p{\frac{\alpha}{2}}^2}\]

Using the inequalities \(\sqrt{1-\p{\frac{x}{2}}^2} \geq 1 - \frac{x^2}{7}\) for \(0 \leq x \leq 1\) and \(\frac{1}{1-x} \geq 1 + x\) for \(0 \leq x \leq 1\), we have
\[A' \geq A \p{1- \frac{\alpha^2}{7}} \p{1 + \frac{1}{42} \frac{a}{A'}}\]
and thus
\(A'^2 - A \cdot A' \p{1 - \frac{\alpha^2}{7}} - \frac{1}{42}  \p{1 - \frac{\alpha^2}{7}} \geq 0\),
which implies by the quadratic formula that \begin{align*}
A' &\geq \frac{A \p{1 - \frac{\alpha^2}{7}} + \sqrt{A^2 \p{1 - \frac{\alpha^2}{7}} + \frac{4}{42} a A \p{1 - \frac{\alpha^2}{7}}}}{2} \\
&= A \p{1 - \frac{\alpha^2}{7}} \cdot \frac{1 + \sqrt{1 + \frac{2}{21} \frac{a}{A \p{1 - \frac{\alpha^2}{7}}}}}{2} \\
&= A \p{1 - \frac{\alpha^2}{7}} \cdot \p{1 + \frac{\sqrt{1 + \frac{2}{21} \frac{a}{A \p{1 - \frac{\alpha^2}{7}}}} - 1}{2}}
\end{align*}
Using the inequality \(\sqrt{1+x}-1 \geq \frac{2}{5} \min(x, \sqrt{x})\) for all \(x \geq 0\), we get that
\begin{equation}\label{eq:lb_2b}
A' \geq A \p{1 - \frac{\alpha^2}{7}} \p{1+ \frac{1}{5}\min\p{\smash[b]{\underbrace{\frac{2}{21} \frac{a}{A \p{1 - \frac{\alpha^2}{7}}}}_{\text{I}}, \underbrace{\sqrt{\frac{2}{21} \frac{a}{A \p{1 - \frac{\alpha^2}{7}}}}}_{\text{II}}}}}
\end{equation}
\\ % need this so the term I II in eqn don't collide with the paragraph below

To finish this case, we show the lower bound in the case where either term in the \(\min\) of (\ref{eq:lb_2b}) is the smaller term.

\textbf{Case IIb-i} \textit{(Term I in (\ref{eq:lb_2b}) is smaller)}. In this case, (\ref{eq:lb_2b}) is equivalent to
\begin{align*}
    A' &\geq A \p{1 - \frac{\alpha^2}{7}} + \frac{2}{105} a
\end{align*}
Using the definition of \(A\), we have 
\(A' \geq A - \frac{1}{7A} + \frac{2}{105} a\), and so \(\Delta A \geq - \frac{1}{7A} + \frac{2}{105} a\).
As \(A \geq 1\), we have \(\Delta A \geq \frac{2}{105} a - \frac{1}{7}\).

Now by \eqref{eq:lb_5}, we get
\[\frac{\Delta A}{\Delta P} \geq \frac{\frac{2}{105} a - \frac{1}{7}}{2 \log a}\]
Now, we complete this case by noticing the right hand side is at least \(\frac{1}{35}\) when \(a > 16\).

\textbf{Case IIb-ii} \textit{(Term II in (\ref{eq:lb_2b}) is smaller)}.  In this case, (\ref{eq:lb_2b}) is equivalent to
\begin{equation}\label{eq:lb_c2b2_1}
    A' \geq A \p{1 - \frac{\alpha^2}{7}} + \sqrt{\frac{2}{525} a A \p{1 - \frac{\alpha^2}{7}}}
\end{equation}
Using the definition of \(A\) and that \(A \geq 1\), we have \(A \p{1 - \frac{\alpha^2}{7}} = A - \frac{1}{7A} \geq - \frac{1}{7}\). Using this and the definition of \(\Delta A\) in \eqref{eq:lb_c2b2_1}, we have \[
\Delta A \geq - \frac{1}{7} + \sqrt{\frac{2}{525} a A \p{1 - \frac{\alpha^2}{7}}}
\]
Further, as \(0 \leq \alpha \leq 1\) we get \(A \p{1 - \frac{\alpha^2}{7}} = \frac{1}{\alpha} - \frac{\alpha}{7} \geq \frac{6}{7}\), so
\[\Delta A \geq - \frac{1}{7} + \sqrt{\frac{12}{3675} a}
\]
Now by \eqref{eq:lb_5}, we get
\[\frac{\Delta A}{\Delta P} \geq \frac{- \frac{1}{7} + \sqrt{\frac{12}{3675} a}}{2 \log a}\]
We finish with the fact that the right hand side is at least \(\frac{1}{65}\) when \(a > 16\).
\end{proof}

\section[Details of two-dimensional update analysis]{Details of Analysis in \texorpdfstring{\Cref{sec:two_dim_update}}{Section~\ref{sec:two_dim_update}}}
Here, we give the details for the outstanding claims in \Cref{sec:two_dim_update}. We stay in the context from that section and reuse notation (specifically, the definition of parameters in (\ref{eqn:update_params})). 

We begin with some well-known bounds on \(e^x\).

\begin{claim}\label{claim:wk_e} Recall the following well-known inequalities regarding \(e^x\).
\begin{enumerate}
\item \label{item:wk_1} \(1 + x \leq e^x\) for all \(x \in \R\);
\item \label{item:wk_2} \(1 + x + \frac{x^2}{2} \leq e^x\) for \(x \geq 0\).
\end{enumerate}
\end{claim}

We will also use a more specialized upper bound on \(e^x\).
\begin{claim}\label{claim:custom_exp_upper_bound} For \(0 \leq x \leq \frac{4}{3}\), we have
\(e^x \leq 1 + x + \frac{x^2}{2} + \frac{x^3}{4}\).
\end{claim}
\begin{proof}
Using the Taylor series for \(e^x\) about \(0\),
we get \[
\left(1 + x + \frac{x^2}{2} + \frac{x^3}{4}\right) - e^x = \frac{x^3}{12} - \sum_{k=4}^\infty \frac{x^k}{k!}
= x^3 \left(\frac{1}{12} - \sum_{k=4}^\infty \frac{x^{k-3}}{k!}\right).\]
Clearly \(x^3 \geq 0\) for \(x \geq 0\), so it remains to show \(\frac{1}{12} - \sum_{k=4}^\infty \frac{x^{k-3}}{k!} \geq 0\) for \(0 \leq x \leq \frac{4}{3}\). 
Note that \(\sum_{k=4}^\infty \frac{x^{k-3}}{k!}\) is increasing (the derivative is clearly positive when \(x \ge 0\)),
and we finish by noting \[\frac{1}{12} - \left.\sum_{k=4}^\infty \frac{x^{k-3}}{k!} \right|_{x = \frac{4}{3}} = \left.\frac{1 + x + \frac{x^2}{2}+\frac{x^3}{4} - e^x}{x^3} \right|_{x = \frac{4}{3}} > 0.\]
\end{proof}

Now we show some facts used in \Cref{claim:pty_outer}, which \Cref{claim:update_step_outer} reduces to.
The proof of \Cref{claim:pty_outer} will reduce to the following analytic inequality.
\begin{claim}\label{claim:outer_t_ineq}
  For all \(\gamma \geq 0\),
  \[\frac{(e^\gamma - 1)^2}{e^{2\gamma} - (1 + \frac{\gamma}{4})^2} \leq \frac{3}{2} \gamma.\]
\end{claim}
\begin{proof}
For the numerator of the left hand side, we have \((e^\gamma - 1)^2 = e^{2\gamma} - 2 e^\gamma + 1 \leq e^{2\gamma} - 2\gamma - 1\) using \Cref{claim:wk_e}-(\ref{item:wk_1}), \(1 + x \leq e^x\).
Further, \(e^\gamma \geq 1 + \frac{\gamma}{4}\) implies \(e^{2 \gamma} - (1 + \frac{\gamma}{4})^2 \geq 0\), so after multiplying both sides by \(e^{2 \gamma} - (1 + \frac{\gamma}{4})^2\) and rearranging, it suffices to show
\[\frac{3}{2} \gamma \left(1 + \frac{\gamma}{4}\right)^2 - 1 - 2\gamma \leq \left(\frac{3}{2} \gamma - 1\right) e^{2 \gamma}.\]
We split this into two cases, based on the value of \(\gamma\).
If \(\gamma \geq \frac{2}{3}\), then the right hand side is at least \(\left(\frac{3}{2} \gamma - 1\right) ( 1 + 2 \gamma + 2 \gamma^2)\) using \Cref{claim:wk_e}-(\ref{item:wk_2}), \(1 + x + \frac{x^2}{2} \leq e^x\),
so it is sufficient to show \(\left(\frac{3}{2} \gamma - 1\right) ( 1 + 2 \gamma + 2 \gamma^2) \geq \frac{3}{2} \gamma \left(1 + \frac{\gamma}{4}\right)^2 - 1 - 2 \gamma\).
Expanding both sides, this is equivalent to showing \(\frac{\gamma^2}{4} + \frac{93}{32} \gamma^3 \geq 0\), which is clearly true for \(\gamma \geq 0\).

If \(\gamma < \frac{2}{3}\), then we use \Cref{claim:custom_exp_upper_bound} to lower bound the right hand side with \(\left(\frac{3}{2} \gamma - 1 \right) (1 + 2 \gamma + 2 \gamma^2 + 2 \gamma ^3)\),
so it is sufficient to show  \(\left(\frac{3}{2} \gamma - 1\right) ( 1 + 2 \gamma + 2 \gamma^2 + 2 \gamma^3) \geq \frac{3}{2} \gamma \left(1 + \frac{\gamma}{4}\right)^2 - 1 - 2\gamma\).
Similar to before, after expanding both sides this is equivalent to showing \(\frac{\gamma^2}{4} + \frac{93}{32} \gamma^3 + \frac{3}{2} \gamma^4 \geq 0\), which is true for \(\gamma \geq 0\).
\end{proof}

We also show some relations between the parameters in the update step.
Recall that we assumed \(\alpha \leq \frac{1}{2}\).
\begin{claim}\label{claim:outer_int}
We have
\begin{enumerate}
\item \(b \leq 1 + \frac{\gamma}{4}\) \label{item:outer_int_1}
\item \(b \leq a\) \label{item:outer_int_2}
\item \(\frac{(a-1)^2}{a^2-b^2} \leq 1\) \label{item:outer_int_3}
\item \(b^2 \geq 1 + \alpha - \alpha'\) \label{item:outer_int_4}
\end{enumerate}
\end{claim}
\begin{proof}
We start by showing (\ref{item:outer_int_1}). As \(\frac{1}{\alpha'} = \frac{1}{\alpha} + 2\gamma\), we have \(\alpha = \alpha' + 2 \gamma \alpha \alpha'\).
Thus \(b = 1 + \gamma \alpha \alpha' \leq 1 + \gamma \alpha^2 \leq 1 + \gamma/4\) as \(\alpha' \leq \alpha \leq \frac{1}{2}\).

For (\ref{item:outer_int_2}), observe that \(b \leq 1 + \frac{\gamma}{4} \leq 1 + \gamma \leq e^\gamma = a\) using \Cref{claim:wk_e}-(\ref{item:wk_1}), \(1+x \leq e^x\), so \(a \geq b\).

To show (\ref{item:outer_int_3}), we first argue it is sufficient to show \(1 + b^2 \leq 2 a\). As a consequence \(-2a + 1 \leq - b^2\), so \((a-1)^2 \leq a^2 - b^2\).
Because \(b \geq 1\) by \Cref{claim:update_params}-(\ref{item:claim_update_params_2}), from (\ref{item:outer_int_2}) we can say that \(a^2 - b^2 \geq 0\), so that \(\frac{(a-1)^2}{a^2 - b^2} \leq 1\).

Now to show \(1 + b^2 \leq 2a\), we write as a series in terms of \(\gamma\).
On the left hand side using (\ref{item:outer_int_1}), we have \(1 + b^2 \leq 1 + \left(1 + \frac{\gamma}{4}\right)^2 = 2 + \frac{\gamma}{2} + \frac{\gamma^2}{4}\).
Further, by \Cref{claim:wk_e}-(\ref{item:wk_2}), \(e^x \geq 1 + x + \frac{x^2}{2}\), we have that \(2a \geq 2 + 2\gamma + \gamma^2\).
Clearly \(2 + \frac{\gamma}{2} + \frac{\gamma^2}{4} \leq 2 + 2\gamma + \gamma^2\) when \(\gamma \geq 0\), so we are finished.

For (\ref{item:outer_int_4}), we have by definition that \(b^2 = 1 + \alpha - \alpha' + \frac{(\alpha - \alpha')^2}{4}\), so \(b^2 \geq 1 + \alpha - \alpha'\).
\end{proof}

As the proof of \Cref{claim:update_step_outer} shows, that claim reduces to the following inequality.
\begin{claim}\label{claim:pty_outer}
We have \begin{equation}\label{eqn:pty_outer_eqn} % equation: property outer equation
        c^2 \leq \frac{b^2 - 1}{b^2} \cdot (a^2 - b^2).
    \end{equation}
\end{claim}
\begin{proof}
We first upper bound \(c\) to reduce the number of variables in (\ref{eqn:pty_outer_eqn}).
As \(b = 1 + \frac{\alpha - \alpha'}{2}\), we have \(2 (b - 1) = \alpha - \alpha'\) and so
\(\alpha = \alpha' + 2(b-1)\).
Thus \[c = - \alpha + \alpha' \cdot a = -(\alpha' + 2(b-1)) + \alpha' \cdot a = \alpha' \cdot (a - 1) + 2(1-b) a.\]
As \(b \geq 1\) by \Cref{claim:update_params}-(\ref{item:claim_update_params_2}), we have that \(2(1-b) \leq 0\) and therefore
\begin{equation}\label{eqn:pty_outer_uc} % equation: outer property upper bound on c
    c \leq \alpha' \cdot (a-1). \nonumber
\end{equation}
Using this in (\ref{eqn:pty_outer_eqn}), it suffices to show \(\alpha'^2 (a-1)^2 \leq \frac{b^2 - 1}{b^2} \cdot (a^2 - b^2)\),
which rearranges to
\begin{equation}\label{eqn:pty_outer_eqn1} % property outer equation 1
\frac{b^2 - 1}{\alpha'^2} \geq \frac{(a-1)^2 b^2}{a^2 - b^2}. \nonumber
\end{equation}

Using \Cref{claim:outer_int}-(\ref{item:outer_int_4}), this reduces to \begin{equation}\label{eqn:pty_outer_eqn2}
    \frac{\alpha - \alpha'}{\alpha'^2} \geq \frac{(a-1)^2 b^2}{a^2 - b^2}.
\end{equation}
The left hand side of (\ref{eqn:pty_outer_eqn2}) equals \(\frac{1}{\alpha'} \left(\frac{\alpha}{\alpha'} - 1\right)\).
    Because \(\frac{1}{\alpha'} = \frac{1}{\alpha} + 2 \gamma\), we have \(\frac{\alpha}{\alpha'} - 1 = 2 \gamma \alpha\),
so \(\frac{1}{\alpha'}\left(\frac{\alpha}{\alpha'} - 1\right) = 2\gamma \cdot \frac{\alpha}{\alpha'} = 2\gamma \cdot (1 + 2 \gamma \alpha)\).
So it is sufficient to show
\begin{equation}\label{eqn:pty_outer_eqn3}
2\gamma (1 + 2 \gamma \alpha)\geq \frac{(a-1)^2 b^2}{a^2 - b^2}.
\end{equation}

Now, we will eliminate the other variables in this inequality to transform it
into a statement involving only \(\gamma\).
We have
\begin{align*}
\frac{(a-1)^2 b^2}{a^2 - b^2} &= \frac{(a-1)^2}{a^2 - b^2} \left(1 + 2 \alpha \alpha' \gamma + \alpha'^2 \alpha^2 \gamma^2\right)\\
&\leq \frac{(a-1)^2}{a^2 - b^2} + \frac{\gamma}{2} + \alpha \frac{\gamma^2}{8},
\end{align*}
where the first line uses that \(b = 1 + \alpha \alpha' \gamma\), and
the second line inequality follows from \Cref{claim:outer_int}-(\ref{item:outer_int_3}) and the fact that \(\alpha' \leq \alpha \leq \frac{1}{2}\).
Thus we can reduce (\ref{eqn:pty_outer_eqn3}) to \(\frac{(a-1)^2}{a^2-b^2} \leq \frac{3}{2} \gamma + \frac{31}{8} \alpha \gamma^2\), or further to

\begin{equation}\label{eqn:pty_outer_eqn4}
\frac{(a-1)^2}{a^2 - b^2} \le \frac{3}{2} \gamma.
\end{equation}

Using \Cref{claim:outer_int}-(\ref{item:outer_int_1}) and that \(a = e^\gamma\), we have \(\frac{(a-1)^2}{a^2 - b^2} \leq \frac{(e^\gamma - 1)^2}{e^{2\gamma} - (1 + \frac{\gamma}{4})^2}\), so finally (\ref{eqn:pty_outer_eqn4}) reduces to
\[\frac{(e^\gamma - 1)^2}{e^{2\gamma} - (1 + \frac{\gamma}{4})^2} \le \frac{3}{2} \gamma,\]
which is proved in \Cref{claim:outer_t_ineq}.
\end{proof}

Recall in the proof of \Cref{claim:translate_ellipse_angle} we defined \(\ell_1 = \frac{1}{c+a}, \ell_2 = \sqrt{\frac{1}{\alpha^2} - \frac{1}{(c+a)^2}}, r = \frac{a^2 \ell_1^2}{b^2 \ell_2^2}\).
That claim reduces to the following.
\begin{claim}\label{claim:pty_inner}
We have
\[a - \alpha' \cdot a \sqrt{\frac{1+r}{r}} \geq 0.\]
\end{claim}
\begin{proof}
%To show that the translated center \(c_t\) is to the right of \(c + \alpha' \cE'\), we need that \(c \leq c_t\),
% or equivalently that \[a - \alpha' \cdot a \sqrt{\frac{1+r}{r}} = a \left(1 - \alpha' \sqrt{\frac{1+r}{r}}\right) \geq 0\]
As by definition \(a \geq 0\), it suffices to show
\begin{equation}\label{eqn:pty_inner_3}
    \alpha'^2 \cdot \left(\frac{1}{r}+1\right) \leq 1.
\end{equation}
Observe that \(\ell_2^2 = \frac{1}{\alpha^2} - \ell_1^2\),
so we can write \(\frac{1}{r} = \frac{b^2}{a^2} \left(\frac{1}{\alpha^2 \ell_1^2} - 1\right)\), and hence rewrite (\ref{eqn:pty_inner_3}) as
\[\alpha'^2 \left(1 + \frac{b^2}{a^2} \left(\left(\frac{c+a}{\alpha}\right)^2 - 1\right)\right) \leq 1.\]

Multiplying both sides by \(\frac{\alpha^2}{\alpha'^2}\) and rearranging, this is equivalent to
\begin{equation}\label{eqn:pty_inner_4}
\frac{b^2}{a^2} \left((c+a)^2 - \alpha^2\right) \leq \frac{\alpha^2}{\alpha'^2} - \alpha^2.
\end{equation}
Now, by definition of \(c\) we can write \(c + a = a(1+\alpha') - \alpha\), so that \((c+a)^2 - \alpha^2 = a^2 (1+\alpha')^2 - 2 \alpha a (1 + \alpha')\).
Thus, (\ref{eqn:pty_inner_4}) is equivalent to
\[\frac{b^2}{a^2} (a^2 ( 1 + \alpha')^2 - 2 \alpha a (1+\alpha')) \leq \frac{\alpha^2}{\alpha'^2} (1 + \alpha')(1 - \alpha').\]
Dividing by \(1 + \alpha'\) and simplifying the left hand side, this is equivalent to
\[
b^2 \left(1 + \alpha' - \frac{2 \alpha}{a}\right) \leq \frac{\alpha^2}{\alpha'^2}(1-\alpha'),
\]
which we show in \Cref{claim:pty_inner_main}.
\end{proof}

\begin{claim}\label{claim:pty_inner_main}
We have
    \[
b^2 \left(1 + \alpha' - \frac{2 \alpha}{a}\right) \leq \frac{\alpha^2}{\alpha'^2}(1-\alpha').
\]
\end{claim}
\begin{proof}
Using \Cref{claim:wk_e}-(\ref{item:wk_1}), \(e^{-x} \geq 1 - x\); and the fact that by definition \(\frac{1}{a} = e^{-\gamma}\), it suffices to show
\[
b^2 \left(1 + \alpha' - 2 \alpha (1-\gamma)\right) \leq \frac{\alpha^2}{\alpha'^2}(1-\alpha'). \nonumber
\]
Using \Cref{claim:outer_int}-(\ref{item:outer_int_4}), this reduces further to
\begin{equation}\label{eqn:pim_1}
    (1 + \alpha - \alpha')(1 + \alpha' - 2 \alpha ( 1 -  \gamma)) \leq \frac{\alpha^2}{\alpha'^2} (1 - \alpha').
\end{equation}
We expand both sides of this inequality into polynomials involving \(\gamma\) and \(\alpha\), and then analyze the resulting expression.
Using the definition of \(\alpha'\), we have \(\alpha ' = \frac{\alpha}{1+2\gamma \alpha}\), and thus
\(1 + \alpha' = \frac{1 + 2 \gamma \alpha + \alpha}{1 + 2 \gamma \alpha}\) and \(1 - \alpha' = \frac{1 + 2 \gamma \alpha - \alpha}{1 + 2 \gamma \alpha}\).
We also have \(\frac{\alpha}{\alpha'} = 1 + 2 \gamma \alpha\), and finally \(\alpha - \alpha' = \frac{2 \gamma \alpha^2}{1 + 2 \gamma \alpha}\).
Substituting these equalities into (\ref{eqn:pim_1}), we obtain the equivalent inequality
\[
\left(\frac{1+2 \gamma \alpha + \gamma \alpha^2}{1 + 2 \gamma \alpha}\right) \left(\frac{1 + 2 \gamma \alpha + \alpha}{1 + 2 \gamma \alpha} - 2 \alpha (1 - \gamma)\right) \leq (1 + 2 \gamma \alpha)^2 \left(\frac{1 + 2 \gamma \alpha - \alpha}{1 + 2 \gamma \alpha}\right).
\]
Multiplying both sides by \((1 + 2 \gamma \alpha)^2\) and rearranging the terms so that they are all on the same side, we get
\[(1 + 2 \gamma \alpha)^3 (1+2 \gamma \alpha - \alpha)-\left(1 + 2 \gamma \alpha + \gamma \alpha^2\right) (1+2 \gamma \alpha + \alpha - 2 \alpha (1- \gamma) (1 + 2 \gamma \alpha)) \geq 0.\]
Next, we expand this inequality:
\begin{align*}
    16 \alpha ^4 \gamma^4-16 \alpha ^4 \gamma^3+8 \alpha ^4 \gamma^2+24 \alpha ^3 \gamma^3-12 \alpha ^3 \gamma^2+12
   \alpha ^2 \gamma^2+2 \alpha ^3 \gamma-2 \alpha ^2 \gamma+2 \alpha  \gamma \geq 0
\end{align*}
As \(\gamma \alpha \geq 0\), we can divide both sides of this inequality by \(2 \gamma \alpha\).
Grouping by powers of \(\alpha\), we obtain
\[4 \alpha ^3 \gamma \left(2 \gamma^2-2 \gamma+1\right)+\alpha ^2 \left(12 \gamma^2-6 \gamma+1\right)+\alpha  (6
   \gamma-1)+1 \geq 0.\]
Upon inspection, both quadratics \(2 \gamma^2 - 2\gamma + 1\) and \(12 \gamma^2 - 6 \gamma + 1\) are positive for all \(\gamma\).
Thus we only need to show \(\alpha (6\gamma-1)+1 \geq 0\), but this is clear from writing it as \(1 - \alpha + 6 \gamma \alpha \geq 0\) and using that \(\alpha \leq 1\).
\end{proof}

\chapter[Sparsifying Matrix Block Norms]{The Change-of-Measure Method, Block Lewis Weights, and Approximating Matrix Block Norms}

\section{Introduction}

Suppose we are given a large dataset that is computationally inconvenient to work with in a downstream task. To alleviate this, we can try to randomly sample a small representative subset of the original dataset. The design and analysis of randomized sampling algorithms for this purpose is well-explored (for example, see \cite{ss08,mmwy21,wy22_oneshot,wy23_sens} for preserving $\ell_p$ objectives, \cite{fl11} for preserving objectives for $k$-median, projective clustering, subspace approximation, and more, \cite{ss08,kkty21a,jls22,lee22} for preserving graph and hypergraph $\ell_2$-energy, and \cite{jlls23} for sums (of powers) of general norms).

In order to design randomized sampling algorithms, we first need to understand the properties of the original dataset we want to preserve. To this end, we study the problem of preserving \textit{block $p$-norm objectives}. Let $\cG = (\mA \in \R^{n\times d}, S_1,\dots,S_m, p_1,\dots,p_m)$ be a dataset consisting of a matrix $\mA \in \R^{n\times d}$. Consider a partitioning of $[n]$ into groups $S_1,\dots,S_m$ and consider positive numbers $p_1,\dots,p_m$. Let $\mA$ have rows $\va_1,\dots,\va_n$ and denote by $\mA_{S_i}$ the matrix in $\R^{\abs{S_i} \times d}$ whose rows are the rows of $\mA$ indexed by $S_i$. Consider the function $\gnorm{\mA\vx}{p}$ on some input vector $\vx \in \R^d$:
\begin{align}
    \gnorm{\mA\vx}{p}^p \coloneqq \sum_{i=1}^m \norm{\mA_{S_i}\vx}_{p_i}^p \label{eq:energy}
\end{align}
We use the norm notation because we can easily verify that for $p \ge 1$ and $p_i \ge 1$ for all $i$, $\gnorm{\cdot}{p}$ is a norm. We remark that objectives of the form of \eqref{eq:energy} are widely studied in geometric functional analysis, theoretical computer science, and data science. In \Cref{sec:results}, we go over one important application of the objective \eqref{eq:energy}. We defer a broader discussion of more applications and connections to \Cref{sec:related_works}.

Our goal in this paper is to design and analyze randomized sampling algorithms to output a weighted subset that preserves \eqref{eq:energy} for all $\vx \in \R^d$. We give a formal problem statement for the general problem we study in \Cref{prob:sparsification_main_problem}. 

\begin{problem}[$\ell_p$ block norm sampling]
\label{prob:sparsification_main_problem}
We are given as input $\cG = \inparen{\mA \in \R^{n\times d}, S_1,\dots,S_m, p_1,\dots,p_m}$, $p > 0$, and an error parameter $\eps$. For all $i \in [m]$, we must output a probability distribution $\rho_1,\dots,\rho_m$ over $[m]$ such that if we choose a collection of groups $\cM = (i_1,\dots,i_{\mtilde})$ where each $i_h$ is independently distributed according to $\rho_i$, then the following holds with probability $\ge 1-\delta$:
\begin{align}
    \text{for all } \vx \in \R^d:\quad\quad\inparen{1-\eps}\gnorm{\mA\vx}{p}^p \le \frac{1}{\mtilde}\sum_{i\in \cM} \frac{1}{\rho_i}\cdot\norm{\mA_{S_i}\vx}_{p_i}^p \le \inparen{1+\eps}\gnorm{\mA\vx}{p}^p\label{eq:sparsification_main_objective}
\end{align}
We would like $\mtilde$ to be small with probability $1-\delta$ (for example, $\mtilde$ should not depend on $m$ and the dependence on $\delta^{-1}$ should be polylogarithmic).
\end{problem}
Observe that the formulation of \Cref{prob:sparsification_main_problem} is an instantiation of an \textit{importance sampling} framework. Specifically, we can think of the distribution $\cD$ as consisting of importance scores for each group. We form our sparse approximation by sampling group $i$ with probability $\rho_i$ and reweighting appropriately so that the function we return is an unbiased estimator of $\gnorm{\mA\vx}{p}$. We call $\mtilde$ the \textit{sparsity} of the procedure described in \Cref{prob:sparsification_main_problem}. Additionally, in the statement of our results, we will assume that $p$ is a constant (and thus any function solely of $p$ will treated as a constant in any $O\inparen{\cdot}$ or $\Omega\inparen{\cdot}$ terms). 

In this paper, we give new results for Problem \ref{prob:sparsification_main_problem} and show how these imply faster algorithms for commonly implemented optimization problems.

\subsection{Our results}
\label{sec:results}

For a quick summary of our existence results for the block norm sampling problem (\Cref{prob:sparsification_main_problem}), see \Cref{table:comparison}. 

We begin with stating our main result\footnote{In the statement of \Cref{thm:one_shot_lewis}, writing the lower bound $p \ge 1/\log d$ instead of $p > 0$ is somewhat arbitrary -- we choose this lower bound to make our calculations easier later on.}, \Cref{thm:one_shot_lewis}.

\begin{restatable}[Block Lewis weight sampling]{theorem}{oneshotlewis}
\label{thm:one_shot_lewis}
Let $\cG = (\mA \in \R^{n \times d}, S_1,\dots,S_m, p_1,\dots,p_m)$ where $S_1,\dots,S_m$ form a partition of $[k]$. Suppose at least one of the following holds:
\begin{itemize}
    \item $1 \le p < \infty$ and $p_1,\dots,p_m \ge 2$;
    \item $1/\log d \le p_1 = \dots = p_m = p < \infty$;
    \item $p_1 = \dots = p_m = 2$ and $1/\log d \le p < \infty$.
\end{itemize}
Let $P \coloneqq \max\inparen{1, \max_{i \in [m]} \min(p_i,\log\abs{S_i})}$. Then, there exists a probability distribution $\cD = \inparen{\rho_1,\dots,\rho_m}$ such that if
\begin{align*}
    \mtilde &= \Omega\inparen{\logv{\sfrac{1}{\delta}}\vbrho P \cdot  d^{\max(1,p/2)}},
\end{align*}
and if we sample $\cM \sim \cD^{\mtilde}$, then, with probability $\ge 1-\delta$,
\begin{align*}
    \text{for all } \vx \in \R^d,\quad (1-\eps)\gnorm{\mA\vx}{p}^p \le \frac{1}{\mtilde}\sum_{i \in \cM} \frac{1}{\rho_i} \cdot \norm{\mA_{S_i}\vx}_{p_i}^p \le (1+\eps)\gnorm{\mA\vx}{p}^p.
\end{align*}
\end{restatable}

We prove \Cref{thm:one_shot_lewis} in \Cref{sec:lewis}. It will follow from \Cref{thm:concentration} (stated and proven in \Cref{sec:generic_chaining}), which is a more general but more technical statement that also includes a description of the relevant distributions $\cD$.

We remark that when $p \ge 1$ and $p_1,\dots,p_m \ge 2$, the sampling probabilities \(\vrho\) mentioned in \Cref{thm:one_shot_lewis} can be found using the optimality conditions of a particular optimization problem that was stated and analyzed by \textcite[Section 4]{jlls23}. That problem itself can be viewed as the natural generalization of the determinant maximization problem that yields the existence of Lewis's measure (see \cite[Section 2]{sz01} for details). However, \cite{jlls23} did not address the question of whether sparsification guarantees could be obtained with these weights beyond the case where the ``outer norm'' satisfies $p=2$.

Additionally, although \cite{jlls23} study sparsification of sums of norms and sums of powers $p > 1$ of uniformly smooth norms, we obtain an improved sparsity in the case entailed by \Cref{prob:sparsification_main_problem} (by a factor of $\psi_d\logv{\sfrac{d}{\eps}}^{\min(p-1,2)}$, where $\psi_d$ is the KLS ``constant'' in $d$ dimensions). We defer a more detailed comparison of our existence results to \Cref{sec:related_works}.

Furthermore, it is well-known that the polynomial terms in the sparsities in \Cref{thm:one_shot_lewis} are optimal. In particular, \textcite[Corollary 1.6 and Theorem 1.7]{lww19} show that $\Omega(d^{\max(1,p/2)}+\eps^{-2}\mathrm{polylog}(\eps^{-1})d)$ rows must be chosen in order to satisfy the requirement imposed by (\ref{eq:sparsification_main_objective}).

Finally, the setting where $p = p_1 = \dots = p_m$ is a particularly important case of \Cref{prob:sparsification_main_problem}. Here, we see that $\norm{\mA\vx}_p = \gnorm{\mA\vx}{p}$, and so \Cref{prob:sparsification_main_problem} amounts to finding an $\ell_p$ subspace embedding under a group constraint (that certain rows must be kept together in the subsample). This might be a useful notion in practice, where the $S_i$ denote related observations that should be kept together for some downstream application. Moreover, this can be viewed as a higher-rank analog of $\ell_p$ row sampling, somewhat similarly to how the matrix Chernoff bound gives a higher-rank analog for the concentration of sums of bounded random matrices when compared to the rank-$1$ variant of \textcite{rudelson1996}.

\paragraph{Computing sampling probabilities.}
The previous results show the existence of sampling probabilities \(\rho_1, \ldots, \rho_m\) such that sampling using those probabilities gives a sparsifier in the setting of \Cref{prob:sparsification_main_problem}.
To get a sparsification \textit{algorithm}, we need to also compute (or approximate) the sampling probabilities.

We give efficient algorithms to do so in natural cases.

\begin{restatable}[Computation of block Lewis weights]{theorem}{computeblw}
\label{thm:computeblw}
Consider the setting of \Cref{thm:one_shot_lewis} and suppose at least one of the following holds:
\begin{itemize}
    \item $p = 2$ and $p_1,\dots,p_m \ge 2$;
    \item $1/\log n \le p_1=\dots=p_m=p < \infty$;
    \item $p_1 = \dots = p_m = 2$ and $1/\log n \le p < \infty$.
\end{itemize}
Let $P = \max\inparen{1, \max_{i \in [m]} \min(p_i,\log\abs{S_i})}$ and set
\begin{align*}
    \mtilde &= O\inparen{\logv{\sfrac{1}{\delta}}\vbrho P \cdot d^{\max(1,p/2)}}.
\end{align*}
Then, there is an algorithm that outputs a probability distribution $\cD = (\rho_1,\dots,\rho_m)$ such that sampling a multiset $\cM \sim \cD^{\mtilde}$ satisfies, with probability \(1 - \delta\),
\begin{align*}
    \text{for all } \vx \in \R^d,\quad (1-\eps)\gnorm{\mA\vx}{p}^p \le \frac{1}{\mtilde}\sum_{i \in \cM} \frac{1}{\rho_i} \cdot \norm{\mA_{S_i}\vx}_{p_i}^p \le (1+\eps)\gnorm{\mA\vx}{p}^p,
\end{align*}
Further, the algorithm to find $\cD$ performs at most \(\text{polylog}(k,n,m)\) leverage score overestimate computations or linear system solves.
\end{restatable}

We prove \Cref{thm:computeblw} in \Cref{sec:alg}.

We formally define a \textit{leverage score overestimate computation} in \Cref{def:lev_score_over}. Alternately, these can be implemented using linear system solvers that solve systems of the form $\mA^{\top}\mD\mA\vy=\vz$ for diagonal $\mD$ (see \cite{ls19} for details). Although the runtime of this primitive depends on the structure of the input, each such iteration runs in $\widetilde{O}(\mathsf{nnz}(\mA) + d^{\omega})$ time. Moreover, in the special case where the matrix $\mA$ is a graph edge-incidence matrix, the runtime improves to $\widetilde{O}(\mathsf{nnz}(\mA))$.

Finally, we note that our algorithms are faster than the log-concave sampling-based routines given in \cite{jlls23} for calculating sparse approximations to sums (of powers of) more general norms, when the outer norm $p$ satisfies $1 \le p \le 2$ (they do not give algorithms for the case where $p > 2$). In particular, while their algorithm applies to a more general setting, the runtime is $\widetilde{O}(m + d^5)$. In contrast, since our algorithms only depend on a polylogarithmic number of leverage score overestimate computations or linear system solves, we can obtain much faster runtimes (in particular improved powers of $n$). This means that we can apply our algorithms to downstream optimization tasks where the main computational primitive is a linear system solver (as is the case for many general frameworks for convex programming).

\paragraph{Applications to minimizing sums of Euclidean norms.} A well-studied regression task is the \textit{minimizing sums of Euclidean norms} (MSN) problem. We are given \(\mA \in \R^{n \times d}\) and \(\vb \in \R^n\), and a partition \(S_1, \ldots, S_m\) of \([n]\).
In this problem, we would like to find
\begin{align}
    \quad\quad\min_{\vx\in\R^d} \sum_{i=1}^m \norm{\mA_{S_i} \vx-\vb_{S_i}}_2. \label{eq:msn}
\end{align}
Solving the MSN objective \eqref{eq:msn} subsumes several widely implemented optimization problems such as variants of Euclidean single facility location, Euclidean multifacility location, Euclidean Steiner minimum tree under a given topology, and plastic collapse analysis. See the long line of work on this problem \cite{and96,xy97,acco00,qsz02} for a more detailed discussion. Additionally, observe that if all $\abs{S_i}=1$, then \eqref{eq:msn} is nothing but $\ell_1$ regression (i.e., $\min_{\vx\in\R^d} \norm{\mA\vx-\vb}_1$). Thus, \eqref{eq:msn} is a generalization of $\ell_1$ regression. Finally, notice that \eqref{eq:msn} subsumes the \textit{stochastic robust approximation problem} when the norm in question is the Euclidean norm and the design $\mA$ assumes a finite number of values -- see \cite[Section 6.4.1]{boyd2004convex} for further discussion.

In this paper, we will be interested in algorithms that return a $(1+\eps)$-multiplicative approximation to the objective -- namely, we desire a point $\widehat{\vx} \in \R^d$ such that
\begin{align*}
    \sum_{i=1}^m \norm{\mA_{S_i}\widehat{\vx}-\vb_{S_i}}_2 \le (1+\eps)\min_{\vx\in\R^d} \sum_{i=1}^m \norm{\mA_{S_i} \vx-\vb_{S_i}}_2.
\end{align*}
To our knowledge, the best known algorithms based on interior point methods output a $(1+\eps)$-approximate solution to \eqref{eq:msn} $\widetilde{O}(\sqrt{m}\logv{\sfrac{1}{\eps}})$ calls to a linear system solver \cite{and96,xy97} for matrices of the form $\mA^{\top}\mD\mA$ for block-diagonal matrices $\mD$.

By applying \Cref{thm:computeblw} on the matrices $\insquare{\mA_{S_i} \vert \vb_{S_i}} \in \R^{\abs{S_i} \times (d+1)}$ with $p_1=\dots=p_m=2$ and $p=1$, observe that within $\widetilde{O}(1)$ linear system solves in matrices $\mA^{\top}\mD\mA$ for nonnegative diagonal $\mD$, we obtain an objective with $\widetilde{O}(\eps^{-2} \cdot d)$ terms that approximates \eqref{eq:msn} up to a $(1\pm\eps)$ multiplicative factor on all vectors $\vx \in \R^{d+1}$ whose last coordinate is $1$. This immediately implies \Cref{thm:msnalg}.

\begin{restatable}[Minimizing sums of Euclidean norms]{theorem}{msnalg}
\label{thm:msnalg}
Let $\mA \in \R^{n \times d}$ and \(\vb \in \R^n\), and $S_1, \dots, S_m$ be a partition of $k$. There exists an algorithm that, with probability $\ge 1-\delta$, returns $\widehat{\vx}$ such that
\begin{align*}
    \sum_{i=1}^m \norm{\mA_{S_i}\widehat{\vx}-\vb_{S_i}}_2 \le (1+\eps)\min_{\vx\in\R^d} \sum_{i=1}^m \norm{\mA_{S_i} \vx-\vb_{S_i}}_2.
\end{align*}
The algorithm runs in $\widetilde{O}\inparen{\sfrac{\sqrt{d}}{\eps} \cdot \sqrt{\logv{\sfrac{1}{\delta}}}}$ calls to a linear system solver in matrices of the form $\mA^{\top}\mD\mA$ for block-diagonal matrices $\mD$, where each block has size $(\abs{S_i}+1) \times (\abs{S_i}+1)$.
\end{restatable}

We prove \Cref{thm:msnalg} in \Cref{sec:applications_msn}.

\Cref{thm:msnalg} improves over the best-known iteration complexities for solving \eqref{eq:msn} when the number of summands is much larger than the input dimension, i.e., $m \gg d$. Furthermore, the iteration complexity stated in \Cref{thm:msnalg} matches the iteration complexity for $\ell_1$ regression up to the $\eps^{-1}$ term \cite{bllssw21}. It is an interesting (but probably challenging) open problem to design and analyze an algorithm for \eqref{eq:msn} with iteration complexity $\widetilde{O}(\sqrt{d}\logv{\sfrac{1}{\eps}})$, which would exactly match what is known for $\ell_1$ regression.

Finally, we note that in the special case of the geometric median, where all the $\mA_{S_i} = \mI_r$ for some fixed dimension \(r\), an algorithm with runtime $\widetilde{O}(\mathsf{nnz}(\vb)\logv{\sfrac{1}{\eps}}^3)$ is known due to \textcite{clmps16}. The algorithm is a long-step interior point method with a custom analysis and follows from different techniques from ours.

\paragraph{Outline.} The rest of this paper is organized as follows. In the remainder of this section, we establish notation that we use throughout the rest of the paper (\Cref{sec:notation}), give an overview of our technical methods (\Cref{sec:sparsification_overview}), and discuss some prior and related works (\Cref{sec:related_works}). In \Cref{sec:background}, we give background from linear algebra, convex geometry, and probability that we rely on for the rest of the paper. In \Cref{sec:coverings}, we prove bounds on geometric quantities known as \textit{covering numbers}. These play a crucial role in our concentration arguments. In \Cref{sec:generic_chaining}, we prove that our general sampling scheme concentrates and therefore preserves the original objective on all $\vx \in \R^d$, with high probability. In \Cref{sec:applications}, we show how to apply our general sampling scheme to the problems we discuss in \Cref{sec:results}. Finally, in \Cref{sec:alg}, we describe our algorithmic results. %, and suggest future directions (\Cref{sec:future_work})

\subsection{Notation and definitions}
\label{sec:notation}

\paragraph{General notation.} For positive integer $N$, we let $[N]$ denote the set $\inbraces{i \in \Z \suchthat 1 \le i \le N}$. All $\log$s are base $2$; we use $\ln$ to denote the natural logarithm. We let $\ve_1,\dots,\ve_d$ denote the standard basis vectors in $\R^d$. When we write $a \lesssim b$, we mean that $a \le Cb$ for some universal constant $C > 0$.

\paragraph{Linear algebra notation.} In this paper, we work extensively with matrices and vectors. We always denote matrices with capital letters in boldface (e.g. $\mA$) and vectors with lowercase letters in boldface (e.g. $\vx$). With a few exceptions, we write the rows of a matrix using the lowercase boldface version of the same letter used to write the matrix along with a subscript denoting which index the row corresponds to. For example, $\va_i$ denotes the $i$th row of matrix $\mA$. In a slight abuse of notation, for a symmetric matrix $\mM$, we let $\mM^{-1} \coloneqq \sum_{i=1}^{\rank{\mM}} \lambda_i^{-1}\vu_i\vu_i^{\top}$, where $\vu_i$ is the $i$th eigenvector of $\mM$. In other words, we write $\mM^{-1}$ to denote the pseudoinverse of $\mM$ when $\mM$ is symmetric. We will never use the inverse notation $\mM^{-1}$ for a non-symmetric matrix $\mM$.

\subsection{Technical overview}
\label{sec:sparsification_overview}

In this subsection, we give a bird's eye view of the technical methods behind our proof of \Cref{thm:one_shot_lewis}.

\subsubsection{Concentration}
\label{sec:sparsification_overview_concentration}

We begin with an explanation of our concentration proof. This type of argument has become standard in the line of work on sparsification (particularly in \cite{lee22,jlls23}), but we include a description for completeness.

Let $B_p \coloneqq \inbraces{\vx\in\R^d\suchthat\gnorm{\mA\vx}{p}\le1}$. By a standard symmetrization reduction, it suffices to fix $i_1,\dots,i_{\mtilde}$ (not necessarily distinct) and argue that for independent $R_1,\dots,R_{\mtilde}$ where $R_h \sim \mathsf{Unif}\inparen{\pm 1}$ we have
\begin{align}
    \exvv{R_h}{\sup_{\vx \in B_{p}} \abs{\sum_{h=1}^{\mtilde} R_h \cdot \frac{\norm{\mA_{S_{i_h}}\vx}_{p_{i_h}}^p}{\rho_{i_h}}}} \le \mtilde\cdot\eps.\label{eq:intro_empirical_process}
\end{align}
Intuitively, satisfying \eqref{eq:intro_empirical_process} means that for the rebalancing of the groups given by the $\rho_{i_h}$, a Rademacher average of the groups evaluated on every point in $\vx$ is close to $0$. It is straightforward to check that the above instantiation is a subgaussian process under an appropriately chosen distance function $\myfunc{\dtwo}{\R^{d} \times \R^{d}}{\R}$. Thus, we will apply chaining \cite{tal21}, which can be thought of as simultaneously controlling \eqref{eq:intro_empirical_process} on $\eps$-nets of $B_p$ using the metric $\dtwo$, for all $\eps>0$.
% the distance
% \begin{align*}
%     \dtwo(\vx,\xhat) \coloneqq \inparen{\sum_{h=1}^{\mtilde} \inparen{\frac{\norm{\mA_{S_{i_h}}\vx}_{p_{i_h}}^p}{\rho_{i_h}} - \frac{\norm{\mA_{S_{i_h}}\xhat}_{p_{i_h}}^p}{\rho_{i_h}}}^2}^{1/2}.
% \end{align*}

To apply chaining, the main technical task is to understand the \textit{entropy numbers} $e_N(B_{p},\dtwo)$. The entropy numbers $e_N(B_p, \dtwo)$ are the values $\eta$ that answer the question, ``what is the smallest $\eta$ such that $B_p$ can be covered by at most $2^{2^N}$ balls of $\dtwo$-radius $\eta$?'' (or see \Cref{defn:entropy_numbers}).

\subsubsection{Covering numbers}
\label{sec:sparsification_overview_geometry}

In this subsection, we explain how to control the entropy numbers as required by \Cref{sec:sparsification_overview_concentration}. We first define the sampling body (\Cref{defn:sampling_polyhedron}).

\begin{definition}[Sampling body]
\label{defn:sampling_polyhedron}
Let $S$ be some subset of $[m]$ and $\rho_1,\dots,\rho_m$ be a probability distribution. Define the norm $\polynorm{\vx} \coloneqq \max_{i \in S} \rho_i^{-1/p}\norm{\mA_{S_i}\vx}_{p_i}$. We call the unit norm ball of $\polynorm{\vx}$ the \textit{sampling body}.
\end{definition}

Recall that the covering number $\cN(K_1,K_2)$ for two symmetric convex bodies $K_1$ and $K_2$ is the minimum number of translates of $K_2$ required to cover $K_1$. Additionally, recall from the previous subsection the notion of \textit{entropy numbers} (which we will define in \Cref{defn:entropy_numbers}). We will reduce controlling \eqref{eq:intro_empirical_process} to bounding the entropy numbers
\begin{align*}
    e_N\inparen{\inbraces{\vx\in\R^d\suchthat\gnorm{\mA\vx}{p}\le 1}, \inbraces{\vx\in\R^d\suchthat\polynorm{\vx}\le 1}}
\end{align*}
when $N$ is small. This places us in the setting where a simple volume-based argument becomes suboptimal. In this range, the dual Sudakov inequality (\Cref{fact:sudakov}) is the technical workhorse that allows us to get sharper bounds than what we would get if we applied just a volume-based bound. It states that if $B$ is the Euclidean ball in $d$ dimensions and $K$ is some symmetric convex body in $d$ dimensions, then we have
\begin{align*}
    \log\cN(B,\eta K) \lesssim \eta^{-2} \exvv{\vg\sim\cN(0,\mI_d)}{\norm{\vg}_K}^2,
\end{align*}
where $\norm{\cdot}_K$ is the \textit{gauge norm} for $K$, defined by $\norm{\vx}_K \coloneqq \inf\inbraces{t > 0 \suchthat \vx/t \in K}$.

However, applying the dual Sudakov inequality requires that we analyze covering numbers of the form $\log\cN(B, K)$ where $B$ is the Euclidean ball in $d$ dimensions and $K$ is some symmetric convex body. Denoting $\{\vx\in\R^d\suchthat\polynorm{\vx}\le 1\}$ by $K$, we see that we cannot immediately apply the dual Sudakov inequality to bound $\log\cN(B_p,\eta K)$. This is because $B_p$ is not a (linear transformation of a) Euclidean ball. The work of \cite{jlls23} resolve this by generalizing the dual Sudakov inequality to cover arbitrary symmetric convex bodies. Unfortunately, this approach is not optimal in every setting. One source of the loss arises from exploiting the concentration of Lipschitz functionals of isotropic log-concave random vectors -- improving the bounds on this concentration depends on further progress on the KLS conjecture. Another is that the one-dimensional conditionals of isotropic log-concave random variables, without any further assumptions, are only subexponential.

To escape these inefficiencies, we will want to try to find a way to apply the dual Sudakov inequality as-is. We may then exploit the concentration of Lipschitz functionals of Gaussian random vectors, which we do have a tight understanding of (for a precise statement, see \Cref{fact:lip_concentration}). A natural attempt is to first observe that for any $t > 0$,
\begin{align}
    \log\cN\inparen{B_p, \eta K} \le \log\cN\inparen{B_p, t\widehat{B_2}} + \log\cN\inparen{t\widehat{B_2}, \eta K} = \log\cN\inparen{B_p, t\widehat{B_2}} + \log\cN\inparen{\widehat{B_2}, \frac{\eta}{t}\cdot K}.\label{eq:intro_split_covering}
\end{align}
We will choose $\widehat{B_2}$ to be a linear transformation of a Euclidean ball so that we can control $\log\cN\inparen{\widehat{B_2}, \eta/t \cdot K}$ using the dual Sudakov inequality. 

Here, we will split our argument based on whether $p \ge 2$. When $p \ge 2$, it will become clear later on that it will be sufficient to choose $\widehat{B_2}$ so that $B_p \subseteq \widehat{B_2}$. Then, it is easy to see that when $t=1$, we get $\log\cN(B_p,t\widehat{B_2}) = 0$. Hence, we have $\log\cN(B_p, \eta K) \le \log\cN(\widehat{B_2}, \eta K)$, and the required bound will follow from exploiting the concentration of Lipschitz functionals of Gaussian random vectors and then applying the dual Sudakov inequality.

However, when $p < 2$, we are still left with a pesky $\log\cN(B_p,t \widehat{B_2})$ term. Loosely, this is almost dual to the statement of the dual Sudakov inequality. Now, because it is known that covering number duality does hold when one of the bodies in question is the Euclidean ball, it may be tempting to simply write $\log\cN(B_p, t\widehat{B_2}) = \log\cN(\widehat{B_2},t B_q)$ where $B_q$ is the dual ball to $B_p$ after applying some linear transformation to map $\widehat{B_2}$ to $B$. The challenge here is that we do not believe that the gauge of the resulting $B_q$ has a form that is amenable to analysis. We will therefore need to be more careful, and we describe our alternative approach in \Cref{sec:sparsification_overview_measure}.

\subsubsection{The change-of-measure principle and norm interpolation}
\label{sec:sparsification_overview_measure}

Recall from the previous part that our goal is to bound $\log\cN(B_p,t\widehat{B_2})$ when $p < 2$.

We are now ready to introduce our main conceptual message -- \textit{by changing the measure under which we take norms, we can almost automatically identify a linear transformation of a Euclidean ball $\widehat{B_2}$ that is a good approximation to $B_p$}. This sort of idea has already been used by \textcite{blm89} and \textcite{sz01} to obtain the required $\widehat{B_2}$ in the special case where all the $S_i$ are singletons. We will generalize this machinery to give similar results for the block norm sampling problem.

Let us describe this idea further. Let $\vlambda = [\lambda_1,\dots,\lambda_m]^{\top}$ denote a probability measure over the groups. Let $\mLambda \in \R^{n\times n}$ be the diagonal matrix such that if $j \in S_i$, then $\mLambda_{jj} = \lambda_i$. Finally, for any $r > 0$ and $\vy\in\R^n$, let $\gnorml{\vy}{r} = (\sum_{i \le m} \lambda_i\norm{\vy_{S_i}}_{p_i}^{r})^{1/r}$ and $B_r \coloneqq \{\vx\in\R^d \suchthat \gnorml{\mLambda^{-1/p}\mA\vx}{r} \le 1\}$. Notice that under this definition, we  still have \(B_p\) as before. We will first describe the argument when the groups are singletons, then explain how to move onto the general case. We will take $\widehat{B_2} = B_2$; it is easy to see that this is a linear transformation of a Euclidean ball.

Next, notice that by log-convexity of norms, if we choose $0 < \theta < p$ and $r > 2$ for which $1/2 = (\theta/2)/p + (1-\theta/2)/r$, we have
\begin{align}
    \gnorml{\vy}{2}^2 \le \gnorml{\vy}{p}^{\theta} \cdot \gnorml{\vy}{r}^{2-\theta}.\label{intro:compact_rounding_interpolation}
\end{align}
We will exploit this observation as follows. For all integers $h \ge 0$, we will show that there exists a set $\cL_h$ that is (a subset of) the unit ball of $B_2$ such that every pair of points in $\cL_h$ is $\delta_h$-separated according to $\gnorml{\cdot}{r}$. We will find $\delta_h$ according to the interpolation inequality \eqref{intro:compact_rounding_interpolation}. Furthermore, we will generate $\cL_h$ using a sort of compactness argument arising from a $B_2$-maximally separated subset of $B_p$. This means we get, for every $h \ge 0$,
\begin{align*}
    \log\cN(\widehat{B_2}, \delta_h B_r) \ge \log\abs{\cL_h} \ge \logv{\frac{\cN(B_p,8^h t \widehat{B_2})}{\cN(B_p, 8^{h+1} t \widehat{B_2})}}.
\end{align*}
Then, summing over $h \ge 0$ (noting that once $h$ is sufficiently large, $\cN(B_p, 8^ht \widehat{B_2}) = 1$), we have
\begin{align*}
    \log\cN(B_p, t\widehat{B_2}) \le \sum_{h \ge 0} \log\cN(\widehat{B_2},\delta_h B_r).
\end{align*}
Notice that the right hand side can be evaluated using the dual Sudakov inequality\footnote{For technical reasons that will be clearer in \Cref{sec:coverings}, we will have to do this after another interpolation step.} (recall the previous section), so it suffices to show that $\log\cN(\widehat{B_2},\eta B_r)$ is small.

This is where the choice of measure becomes crucial. Since both $\widehat{B_2}$ and $B_r$ are dependent on our choice of measure $\vlambda$, we will need to carefully choose the measure so that our covering numbers are well-behaved. A classical result of \textcite{Lewis1978} establishes the existence of a change-of-measure under which we simultaneously get:
\begin{equation}\label{intro:rounding}
  \begin{aligned}
    d^{1/2-1/r}B_r &\subset \widehat{B_2} \subset B_r & \text{for all } r < 2 \\
    B_r &\subset \widehat{B_2} \subset d^{1/2-1/r}B_r & \text{for all } r > 2
  \end{aligned}
\end{equation}
This change-of-measure corresponds to the ``$\ell_p$ Lewis weights'' of $\mA$ (in particular, if $w_i$ is the $i$th $\ell_p$ Lewis weight, then we set $\lambda_i = w_i/n$). It will turn out that this choice of $\vlambda$ is enough for us to ensure that $\cN(\widehat{B_2}, \eta B_r)$ is sufficiently small for our purposes, which eventually follows from \eqref{intro:rounding}.

\paragraph{Handling general $S_i$.} The main challenge with directly porting this argument to the block norm sampling problem is that $B_2$ is not a linear transformation of a Euclidean ball unless $p_1=\dots=p_m=2$. We will therefore have to choose $\widehat{B_2}$ to be a ``rounding'' of $B_2$ such that $B_2 \subseteq \widehat{B_2}$. Observe that the interpolation step \eqref{intro:compact_rounding_interpolation} will continue to hold here, as we will get $\norm{\vy}_{\widehat{B_2}} \le \gnorml{\vy}{2}$. However, if $\widehat{B_2}$ is chosen suboptimally, then there could a large loss in the interpolation step \eqref{intro:compact_rounding_interpolation}.

To understand what we need from our measure and rounding, let us try to derive a version of \eqref{intro:rounding} for general $S_i$. We show an example of this calculation for $r = p \le 2$; the other cases follow similarly. Let $\vlambda \in \R^m_{\ge 0}$ denote a probability measure. Let $\mW$ be a diagonal ``rounding matrix'' so that for all $\vx \in \R^d$, we have
\begin{align*}
    \norm{\mW^{1/2}\mLambda^{1/2-1/p}\mA\vx}_2 \le \gnorm{\mLambda^{1/2-1/p}\mA\vx}{2} = \gnorml{\mLambda^{-1/p}\mA\vx}{2}.
\end{align*}
Letting $\widehat{B_2} = \inbraces{\vx\in\R^d\suchthat \norm{\mW^{1/2}\mLambda^{1/2-1/p}\mA\vx}_2 \le 1}$, the above inequality gives $B_2 \subseteq \widehat{B_2}$, as desired. Next, observe that since $\vlambda$ is a probability measure, we get $B_2 \subseteq B_p$ for free. For the other direction, we write
\begin{align*}
    \gnorml{\mLambda^{-1/p}\mA\vx}{2}^2 &= \gnorm{\mLambda^{1/2-1/p}\mA\vx}{2}^2 = \sum_{i=1}^m \lambda_i\norm{\lambda_i^{-1/p}\mA\vx}_{p_i}^2 = \sum_{i=1}^m \lambda_i\norm{\lambda_i^{-1/p}\mA\vx}_{p_i}^{p}\norm{\lambda_i^{-1/p}\mA\vx}_{p_i}^{2-p} \\
    &\le \sum_{i=1}^m \lambda_i\norm{\lambda_i^{-1/p}\mA\vx}_{p_i}^{p} \cdot \max_{i \in [m]} \norm{\lambda_i^{-1/p}\mA\vx}_{p_i}^{2-p} = \gnorm{\mA\vx}{p}^p \cdot \max_{i \in [m]} \norm{\lambda_i^{-1/p}\mA\vx}_{p_i}^{2-p} \\
    &\le \gnorm{\mA\vx}{p}^p \cdot \max_{i \in [m]} \inparen{\max_{\vx \in \R^d} \frac{\norm{\lambda_i^{-1/p}\mA\vx}_{p_i}^{2}}{\gnorm{\mLambda^{1/2-1/p}\mA\vx}{2}^2}}^{1-p/2} \cdot \gnorm{\mLambda^{1/2-1/p}\mA\vx}{2}^{2-p}.
\end{align*}
We combine the $\gnorm{\mLambda^{1/2-1/p}\mA\vx}{2}$ terms and take the $p$th root of both sides, giving
\begin{align*}
    \gnorm{\mLambda^{1/2-1/p}\mA\vx}{2} &\le \gnorm{\mA\vx}{p} \cdot \max_{i \in [m]} \inparen{\max_{\vx \in \R^d} \frac{\norm{\lambda_i^{-1/p}\mA\vx}_{p_i}^{2}}{\gnorm{\mLambda^{1/2-1/p}\mA\vx}{2}^2}}^{1/p-1/2} \\
    &= \gnorm{\mA\vx}{p} \cdot \max_{i \in [m]} \inparen{\frac{1}{\lambda_i} \cdot \underbrace{\max_{\vx \in \R^d} \frac{\norm{\lambda_i^{1/2-1/p}\mA\vx}_{p_i}^{2}}{\gnorm{\mLambda^{1/2-1/p}\mA\vx}{2}^2}}_{\widehat{\tau}_i(\mLambda^{1/2-1/p}\mA)}}^{1/p-1/2}.
\end{align*}
We may think of the quantity $\widehat{\tau}_i$ as a generalized \textit{leverage score}. Specifically, it upper bounds the contribution of the term $\|\lambda_i^{1/2-1/p}\mA\vx\|_{p_i}^{2}$ to the objective $\gnorm{\mLambda^{1/2-1/p}\mA\vx}{2}^2$. The above calculation shows us that if we make $\widehat{\tau}_i/\lambda_i$ small for all $i$, then we can get a tight relationship between $B_2$ and $B_p$. A slight weakening of the definition of the $\widehat{\tau}_i$ motivates the notion of a \textit{block Lewis overestimate} that we use in the remainder of the paper.

\begin{restatable}[Block Lewis overestimate]{definition}{blocklewisoverestimate}
% \begin{definition}[Block Lewis overestimate]
\label{defn:block_lewis_overestimate}
Let $\tau_j(\mM)$ denote the leverage score of the $j$th row of $\mM$. Let \(F^{\star} > 0\). For $p > 0$ and $p_i > 0$, we say the probability measure $\vlambda$ and rounding $\mW$ form an $\Fstar$-block Lewis overestimate if for all $i \in [m]$, we have
\begin{align*}
    \frac{1}{\lambda_i}\inparen{\sum_{j \in S_i} \inparen{\frac{\tau_j(\mW^{1/2}\mLambda^{1/2-1/p}\mA)}{w_j}}^{p_i/2}}^{2/p_i} \le \Fstar.
\end{align*}
\end{restatable}
Following the above argument, establishing a probability measure $\vlambda$ and a rounding matrix $\mW$ that form an $\Fstar$-block Lewis overestimate will imply
\begin{equation}\label{intro:blw_rounding}
  B_2 \subseteq \widehat{B_2}\quad \text{ and }\quad
  \begin{aligned}
    (\Fstar)^{1/2-1/r}B_r &\subset B_2 \subset B_r & \text{for all } r < 2 \\
    B_r &\subset B_2 \subset (\Fstar)^{1/2-1/r}B_r & \text{for all } r > 2
  \end{aligned}.
\end{equation}
With \eqref{intro:blw_rounding} in hand, we at least have enough reason to believe that establishing $(\vlambda, \mW)$ that form an $\Fstar$-block Lewis overestimate may yield the requisite control over $\log\cN(\widehat{B_2},\eta B_r)$. To actually get this, by the dual Sudakov inequality, we estimate $\norm{\vg}_{B_r}$ for $\vg \sim \cN(0,\mI_d)$. Doing so is a matter of applying again the fact that the concentration of Lipschitz functionals of Gaussian vectors is determined entirely by the Lipschitz parameter of the functional. 

To actually find $\vlambda$ and $\mW$ with a small value of $\Fstar$, we split into cases. When $p \ge 1$ and $p_1=\dots=p_m \ge 2$, we extract the relevant $\vlambda$ and $\mW$ from the analysis in the proof of \cite[Lemma 4.2]{jlls23}, which yields $\Fstar = d$. When $p_1=\dots=p_m=p\ge1/\log d$ or $p_1=\dots=p_m=2$ and $p\ge1/\log d$, we separately prove that we can find $\vlambda$ and $\mW$ satisfying \Cref{defn:block_lewis_overestimate}, again with $\Fstar = d$. Hence, in all cases, the control we get over $\log\cN(\widehat{B_2},\eta B_r)$ is essentially as good as what we get in the case where all the $S_i$ are singletons.

\paragraph{Change-of-measures in functional analysis.} We note that other change-of-measure arguments are used throughout the study of finite-dimensional subspaces of $L_p$, as they are a very useful way to compare the $L_p$ norm to some other norm of interest. See the survey by \textcite[Section 1.2]{js01} for more information.

\subsection{Prior results, related works, and connections}
\label{sec:related_works}

\paragraph{Relevance of matrix block norms.} We discuss the importance of the matrix block norm objective \eqref{eq:energy} to functional analysis, theoretical computer science, and data science, beyond our previous discussion of the MSN problem \eqref{eq:msn}.

In the special case where all the $p_i$ are equal to one another (call this value $q$), the set of $\vx$ for which $\gnorm{\mA\vx}{p} < \infty$ yields a subspace of a mixed $p,q$ norm space (sometimes notated as $\ell_p(\ell_q)$). Observe that \Cref{thm:one_shot_lewis} implies a finite-dimensional subspace embedding result for finite-dimensional subspaces of infinite-dimensional $L_p(\ell_q^r)$ (where $r$ is finite).  Spaces of the form $\ell_p(\ell_q)$ are widely studied in the geometric functional analysis and approximation theory communities; see, e.g., \cite{ps12,kv15,mu19,jkb22} and the references therein. 

As mentioned in those works, a central motivation for studying $\ell_p(\ell_q)$ is that they are natural testbeds with which to evaluate and further our understanding of the geometry of symmetric convex bodies in high dimensions. Consequently, studying block norm subspace embedding problems (\Cref{prob:sparsification_main_problem}) is a fruitful direction through which to improve our geometric handle of subspaces of $\ell_p(\ell_q)$ and symmetric convex bodies in general. We note that understanding the correct $\mathrm{polylog}$ dependencies in $d$ for this problem typically requires new geometric insights. For instance, the necessity of additional polylogarithmic dependencies on the dimension $d$ is not even totally understood when $\abs{S_i} = 1$ and $p\neq 2$, and resolving them likely requires significant new geometric ideas \cite[Conjecture 2]{hrr22}. 

The matrix block norms are also ubiquitous in both theoretical computer science and data science. For example, the block norm objective has been studied in the context of hypergraph Laplacians. One recovers this by choosing $p=2$ and $p_1=\dots=p_m=\infty$; see the discussion in \cite[Section 1.2]{jls22} to see how to rewrite the hypergraph Laplacian in the form of \eqref{eq:energy}. Within data science, the block norms are used to encourage structured solutions to underdetermined linear systems (i.e., in a noiseless setting, we can set up and solve the convex optimization problem\footnote{This is similar to how basis pursuit can be seen as encouraging sparsity in a noiseless setting, while LASSO does so in the presence of noise \cite[Section 7.2]{Wainwright_2019}.}, ``find a vector $\vy$ in the affine space $\mB\vy = \vb$ minimizing $\gnorm{\vy}{p}^p$''); see, e.g., \cite{yy06,bach07,nie2010efficient,sfht13} and other applications mentioned in \cite{sra12}. As a concrete candidate application of our results to such settings, inspired by \cite{cd21,mmwy21}, we believe that our results can be used as subroutines to give runtime and query-efficient algorithms for active $\gnorm{\cdot}{p}$ regression when $p > 0$ and $p_1=\dots=p_m=2$ (generalizing the basis pursuit equivalent of the group Lasso objective) or when $p=2$ and $p_1,\dots,p_m \ge 2$. 

On a more conceptual level, we are optimistic that some of our results will be useful for designing faster algorithms for norm-constrained optimization problems. This is partly motivated by our discussion around the MSN problem \eqref{eq:msn} and is similar to how an improved geometric understanding of Lewis weights improved the iteration complexities for linear programming and $\ell_p$ regression \cite{ls19,jls21}.

\paragraph{Lewis weights for $\ell_p$ row sampling.} When each group $S_i$ has size $1$, notice that we have $\gnorm{\mA\vx}{p}^p = \sum_{i=1}^m \abs{\ip{\va_i,\vx}}^p = \norm{\mA\vx}_p^p$. Consequently, in this special case, satisfying (\ref{eq:sparsification_main_objective}) is exactly equivalent to computing an \textit{$\ell_p$ subspace embedding} for $\mA$. There is a long line of work studying computing $\ell_p$ subspace embeddings using Lewis weights, starting with that of \textcite{blm89}. For the details of this argument, see \cite[Section 7]{blm89} and \cite{sz01}. 

\paragraph{Sparsifying sums of norms.} The work perhaps most closely related to ours is \cite{jlls23}. There, the authors give existence results for sparse approximations to sums (of powers) of norms. It is easy to see that this is a more general problem than the one we study. However, this generality comes at a cost. In particular, the sparsity given by our \Cref{thm:one_shot_lewis} improves over theirs by a factor of $\psi_d\logv{\sfrac{d}{\eps}}^{\min(p-1,2)}$, where $\psi_d$ denotes the KLS ``constant'' in $d$ dimensions (which is currently $\sqrt{\log d}$, due to \textcite{kla23}). See their Theorem 1.3 for more details. And, as mentioned earlier, we believe understanding \Cref{prob:sparsification_main_problem} down to the correct polylogarithmic dependencies in $d$ is an important geometric question.

The authors also define the block Lewis weights as the natural generalization of the determinant-maximization program that \textcite{sz01} use to prove the existence of Lewis's measure for all $p > 0$. They use this to give results for sparsification of sums of certain powers of arbitrary norms (obtaining a sparsity of $\sim d^{2-1/p}$ when $1 \le p \le 2$) and for \Cref{prob:sparsification_main_problem} when the outer norm $p=2$ (obtaining a sparsity of $\sim d$). We note that the result they obtain when $p=2$ provides logarithmic-factor improvements over ours, as the main technical primitive they use is a chaining estimate developed by \textcite{lee22} that meaningfully exploits the fact that the space of events is a subset of a $2$-uniformly convexity set. However, they did not address whether the block Lewis weights could yield to sparsification guarantees for \Cref{prob:sparsification_main_problem}. Additionally, their construction of the block Lewis weights does not yield a change-of-measure that allows for sparsification when the inner norms $p_i \le 2$ or when the outer norm $p \le 1$. 

\paragraph{Summary for sparsifying sums of norms.} See \Cref{table:comparison} for a comparison between our new results and a selection of the most relevant prior work on sparsifying sums of norms. We focus on results concerning \(\ell_p\)-norms specifically, although \cite{jlls23} has results for more general classes of norms.
By ``sampling'', we mean the work provides an analysis that shows how sampling according to some sampling probabilities gives a sparsifier of size $\widetilde{O}(\eps^{-2}d^{\max(1,p/2)})$ with good probability.
By ``fast computation'', we mean the work provides an algorithm to compute sampling probabilities with \(\text{polylog}(k, m, d)\) leverage score computations or linear system solves (or some other primitive that can be implemented in time $\widetilde{O}(\mathsf{nnz}(\mA) + d^{\omega})$). For works that only explicitly handle \(\abs{S_i}=1\), we leave \(p_1, \ldots, p_m\) blank because the choice of inner norms does not affect
the objective.
\begin{table}[H]
\centering
\begin{tabular}{l c c c c r}
\hline
\textbf{Block size} & \(p\) & \(p_1, \ldots, p_m\) & \textbf{Sampling} & \textbf{Fast computation} & \\
\hline
\(1\) & \(1 \le p < \infty \) &  & \checkmark & & \cite{blm89} \\
\(1\) & \(0 < p < 1\) &  & \checkmark & & \cite{sz01} \\
\(1\) & \(0 < p < 4 \) & & & \checkmark & \cite{cp15} \\
\(\ge 1\) & \(p = 2\) & \(p_1 = \cdots = p_m = \infty\) & \checkmark & & \cite{lee22} \\
\(\ge 1\) & \(p=2\) & \(p_1=  \cdots = p_m = \infty\) & \checkmark & \checkmark & \cite{jls22} \\
\(\ge 1\) & \(1 \le p < \infty\) & \(p_1, \ldots, p_m \ge 2\) & \checkmark & & \cite{jlls23} \\
\rowcolor{blue!5} \(\ge 1\) & \(1 \le p < \infty\) & \(p_1, \ldots, p_m \ge 2\) & \checkmark & & This work \\
\rowcolor{blue!5} \(\ge 1\) & \(\sfrac{1}{\log d} \le p < \infty\) & \(p_1, \ldots, p_m = p\) & \checkmark & \checkmark & This work \\
\rowcolor{blue!5} \(\ge 1\) & \(\sfrac{1}{\log d} \le p < \infty\) & \(p_1, \ldots, p_m = 2\) & \checkmark & \checkmark & This work \\
\rowcolor{blue!5} \(\ge 1\) & \(p=2\) & \(p_1, \ldots, p_m \ge 2\) & \checkmark & \checkmark & This work \\
\hline
\end{tabular}
\caption{Comparison of sparsification results for sums of norms.}
\label{table:comparison}
\end{table}

\section{Preliminaries}
\label{sec:background}

In this section, we set up and review definitions and existing facts that will play crucial roles in our analyses. In Section \ref{sec:linalg_background}, we review material from linear algebra, and in Section \ref{sec:convex_geo_background}, we review material from convex geometry.

\subsection{Linear algebra background}
\label{sec:linalg_background}

We introduce a few definitions concerning \textit{leverage scores} (Definition \ref{defn:lev_score}). 
\begin{definition}
\label{defn:lev_score}
For a matrix $\mA \in \R^{n\times d}$, we let $\tau_i(\mA) \coloneqq \va_i^{\top}\inparen{\mA^{\top}\mA}^{-1}\va_i$ denote the \textit{leverage score} of row $\va_i$ with respect to the matrix $\mA$. When $\mA$ is clear from context, we omit it and simply write $\tau_i$ in place of $\tau_i(\mA)$.
\end{definition}

The following are well-known properties of leverage scores.

\begin{fact}
\label{fact:lev_score_facts}
For a matrix $\mA \in \R^{n\times d}$, we have:
\begin{itemize}
    \item $\sum_{i=1}^n \tau_i(\mA) = \rank{\mA}$;
    \item $\tau_i(\mA) = \max_{\vx\in\R^d\setminus\inbraces{0}} \frac{\abs{\ip{\va_i,\vx}}^2}{\norm{\mA\vx}_2^2}$ for all $i$;
    \item $0 \le \tau_i(\mA) \le 1$;
    \item For any positive constant $C$, we have $\tau_i(C\mA) = \tau_i(\mA)$ for all $i$.
\end{itemize}
\end{fact}

We will also need the following fact relating the leverage scores of a matrix $\mA$ to its singular value decomposition.

\begin{fact}
\label{fact:whitening}
Let $\mA\in\R^{n\times d}$ and $\mU\mSigma\mV^{\top}$ be a singular value decomposition for $\mA$, where $\mU\in\R^{n\times d}$ and $\mSigma,\mV\in\R^{d\times d}$. Then, $\tau_i(\mA) = \norm{\vu_i}_2^2$.
\end{fact}
\begin{proof}[Proof of \Cref{fact:whitening}]
To understand why this equality might hold, observe that we can think of $\mU$ as the resulting matrix from applying a statistical whitening transform to $\mA$. More precisely, recall that
\begin{align*}
    \tau_i(\mA) &= \va_i^{\top} \inparen{\mV\mSigma^2\mV^{\top}}^{-1}\va_i^{\top} = \va_i^{\top}\inparen{\sum_{j=1}^d \frac{1}{\sigma_i^2} \cdot \vv_j\vv_j^{\top}}\va_i^{\top} = \sum_{j=1}^d \frac{1}{\sigma_j^2}\ip{\va_i,\vv_j}^2.
\end{align*}
We now calculate $\ip{\va_i,\vv_j}$. Notice that
\begin{align*}
    \ip{\va_i,\vv_j} = \ip{\vv_j, \ve_i^{\top}\sum_{j'=1}^d \sigma_{j'}\mU\ve_{j'}\vv_{j'}^{\top}} = \ip{\vv_j, \sum_{j'=1}^d \sigma_{j'}\ip{\ve_i,\mU\ve_{j'}}\vv_{j'}^{\top}} = \sigma_{j}\ip{\ve_i,\mU\ve_j}.
\end{align*}
Substituting this back in gives
\begin{align*}
    \tau_i(\mA) = \sum_{j=1}^d \ip{\ve_i,\mU\ve_j}^2 = \norm{\vu_i}_2^2.
\end{align*}
This concludes the proof of \Cref{fact:whitening}.
\end{proof}

\subsection{Convex geometry background}
\label{sec:convex_geo_background}

In this subsection, we review foundational facts regarding convex geometry we use throughout the remainder of this paper. 

We will need the notions of covering and entropy numbers. 
\begin{definition}[Covering numbers {\cite[p. 69]{rothvoss}}]
\label{defn:covering_numbers}
Let $X, Y \subset \R^d$. The covering number $\cN(X,Y)$ is the minimum number of translates of $Y$ required to cover $X$. Formally, we have
\begin{align*}
    \cN(X,Y) \coloneqq \min\inbraces{N \in \N \suchthat \text{ there exists } \vx_1,\dots,\vx_N \in \R^d \text{ such that } X \subseteq \bigcup_{i=1}^N \inparen{\vx_i + Y}}.
\end{align*}
\end{definition}

\begin{definition}[Entropy numbers {\cite[Definition 2.1]{vh15}}]
\label{defn:entropy_numbers}
Let $X, Y \subset \R^d$. The entropy number $e_N(X,Y)$ is the minimum radius $\eta$ such that $\log\cN(X,\eta \cdot Y) \le 2^N$.
\end{definition}

Sometimes, when writing $e_N$, we will write $e_N(X,\norm{\cdot})$ for some quasi-norm $\norm{\cdot}$. Here, we take $Y$ to be the object formed by the unit ball of $\norm{\cdot}$.

Finally, we state the dual Sudakov inequality.
\begin{fact}[Dual Sudakov inequality, due to \textcite{ptj87}]
\label{fact:sudakov}
For a symmetric convex body $K \subset \R^d$, define
\begin{align*}
    \norm{\vx}_K \coloneqq \inf\inbraces{t > 0 \suchthat \vx/t \in K}.
\end{align*}
Let $K$ be a symmetric convex body in $\R^d$. We have the below.
\begin{align}
    \log\cN(B_2^d, \eta \cdot K) &\lesssim \eta^{-2} \cdot \exvv{\vg\sim\cN(0,\mI_d)}{\norm{\vg}_K}^2.\label{eq:sudakov_dual}
\end{align}
\end{fact}

\subsection{Probability background}

In this subsection, we review a few facts about subgaussian random variables. These are mostly derived from the presentation of \textcite{vershynin_2018}.

\begin{definition}[$\subgnorm{\cdot}$ and subgaussian random variable {\cite[Definition 2.5.6]{vershynin_2018}}]
\label{defn:subgaussian}
Let $X$ be a random variable. Define $\subgnorm{X} \coloneqq \inf\inbraces{t > 0 \suchthat \exv{\expv{\sfrac{X^2}{t^2}}} \le 2}$. If $\subgnorm{X} < \infty$, we say $X$ is subgaussian.
\end{definition}

\begin{fact}[Properties of subgaussian random variables {\cite[Proposition 2.5.2]{vershynin_2018}}]
\label{fact:subgaussian_properties}
The following properties equivalently characterize a subgaussian random variable $X$ up to constants:
\begin{itemize}
    \item For all $t \ge 0$, $\prv{\abs{X} \ge t} \le 2\expv{-\frac{t^2}{\subgnorm{X}^2K_1^2}}$;
    \item For all $r \ge 1$, $\exv{\abs{X}^r} \lesssim r^{r/2}$;
    \item For all $\lambda$ such that $\abs{\lambda} \lesssim \subgnorm{X}^{-1}$, we have $\exv{\expv{\lambda^2X^2}} \le \expv{\subgnorm{X}^2\lambda^2}$.
\end{itemize}
\end{fact}

\begin{fact}[Maximum of subgaussian random variables {\cite[Exercise 2.5.10]{vershynin_2018}}]
\label{fact:max_subgaussians}
Let $X_1,\dots,X_N$ be a sequence of (not necessarily independent) subgaussian random variables. Then
\begin{align*}
    \exv{\max_{i \in [N]} \abs{X_i}} \lesssim \max_{i \in [N]} \subgnorm{X_i}\sqrt{\log N}.
\end{align*}
\end{fact}

\begin{fact}[Decentering]
\label{fact:uncentering}
We have
\begin{align*}
    \subgnorm{X} \lesssim \subgnorm{X - \exv{X}} + \exv{\abs{X}}.
\end{align*}
\end{fact}
\begin{proof}[Proof of \Cref{fact:uncentering}]
By the triangle inequality, we get
\begin{align*}
    \subgnorm{X} \le \subgnorm{X - \exv{X}} + \subgnorm{\exv{X}} \lesssim \subgnorm{X-\exv{X}} + \exv{\abs{X}},
\end{align*}
which is exactly the statement of \Cref{fact:uncentering}.
\end{proof}

\begin{fact}[Lipschitz functionals of Gaussians are subgaussian {\cite[Theorem 5.2.2]{vershynin_2018}}]
\label{fact:lip_concentration}
If $\vg \sim \cN(0,\mI_d)$ and if $\myfunc{f}{\R^d}{\R}$, then
\begin{align*}
    \subgnorm{f(\vg) - \exv{f(\vg)}} \lesssim \norm{f}_{\mathsf{Lip}}.
\end{align*}
\end{fact}

%%%%%%%%%%%%%%%%%%%%%%%%%%%%%%%%%%%%%%%%%%%%%%%%%%%%%%%%%%%%%%%%%%%%%%%%%%%%%%%%%%%%%%%%%%%
% \newpage
\section{Covering number estimates}
\label{sec:coverings}

In this section, we develop our metric entropy estimates. It will be helpful to keep in mind the context and outline from \Cref{sec:sparsification_overview}.

\subsection{Notation and general formula}

We begin with some definitions that are necessary for our results.

\begin{definition}[Block-constant diagonal matrix]
\label{defn:block_constant}
We say that a vector $\vv \in \R^n$ and corresponding diagonal matrix $\mV\in\R^{n\times n}$ is block-constant or ``constant down the blocks'' if for every $j_1,j_2 \in S_i$, we have $v_{j_1}=v_{j_2}$.
\end{definition}

In particular, when we define a probability measure $\vlambda \in \R^{m}$ over $[m]$, we will find it useful to extend it to a block-constant diagonal matrix $\mLambda\in\R^{n \times n}$.

\begin{definition}[Rounding matrix]
\label{defn:rounding_matrix}
For a probability measure $\vlambda$ over $[m]$, we say that a positive diagonal matrix $\mW \in \R^{k \times k}$ rounds the measure matrix $\mLambda \in \R^{n\times n}$ if for all $\vx\in\R^d$ we have $\norm{\mW^{1/2}\mLambda^{1/2-1/p}\mA\vx}_2 \le \gnorm{\mLambda^{1/2-1/p}\mA\vx}{2}$. We also denote
\begin{align*}
    \widehat{B_2} \coloneqq \inbraces{\vx\in\R^d \suchthat \norm{\mW^{1/2}\mLambda^{1/2-1/p}\mA\vx}_2 \le 1}.
\end{align*}
\end{definition}

The $\widehat{B_2}$ defined in \Cref{defn:rounding_matrix} is the linear transformation of the Euclidean ball that we will ``pass through'' to get our covering number estimates (recall \eqref{eq:intro_split_covering}).

Next, recall our notion of measure overestimates. This is a generalization of prior definitions of Lewis measure overestimates (see e.g. \cite[Definition 2.4]{jls21}, \cite[Definition 2.3]{wy22}) and group leverage score overestimates (\cite[Definition 1.1]{jls22}).

\blocklewisoverestimate*

For example, observe that when all the $S_i$ have size $1$, then \Cref{defn:block_lewis_overestimate} corresponds to standard definitions of Lewis weight overestimates, and there exist weights such that $\Fstar \lesssim d$. Furthermore, we will see that there exist a $\vlambda$ and $\vw$ such that $\Fstar = d$ (it will follow from \Cref{lemma:block_instantiation_small}).

Next, we define the vector $\valpha$, whose entries capture a notion of group importance.

\begin{definition}
\label{defn:alpha}
Let $\vlambda$ be a probability measure over $[m]$ and let $\mW$ be a rounding matrix for $\mLambda$ (\Cref{defn:rounding_matrix}). If $p_1,\dots,p_m \ge 2$, then let $\valpha \in \R^{m}$ be the vector such that for all $i \in [m]$, we have
\begin{align*}
    \alpha_i^p \coloneqq \lambda_i^{1-p/2}\inparen{\sum_{j\in S_i}\inparen{\frac{\tau_j\inparen{\mW^{1/2}\mLambda^{1/2-1/p}\mA}}{w_j}}^{p_i/2}}^{p/p_i}.
\end{align*}
Equivalently, if $\mU$ is a matrix whose columns consist of the left singular vectors of $\mW^{1/2}\mLambda^{1/2-1/p}\mA$, and if we denote by $\vu_j$ the $j$th row of $\mU$ and let $\vf_j \coloneqq \lambda_{i}^{-1/2}w_j^{-1/2}\vu_j$, then by \Cref{fact:whitening}, we may also write
\begin{align*}
    \alpha_i^p &\coloneqq \lambda_i\inparen{\sum_{j \in S_i} \norm{\vf_j}_2^{p_i}}^{p/p_i}.
\end{align*}
On the other hand, if $p_1=\dots=p_m=p<2$, then let $\widehat{\vlambda}$ be a probability measure over $[n]$ and let $\widehat{\valpha}\in\R^n$ be defined as above accordingly. Finally, let $\valpha\in\R^m$ be such that
\begin{align*}
    \alpha_i^p \coloneqq \inparen{\sum_{j \in S_i}\widehat{\alpha}_j^p}^{1/p}.
\end{align*}
\end{definition}

To help ground \Cref{defn:alpha}, notice that combining \Cref{defn:block_lewis_overestimate} with \Cref{defn:alpha} gives us, for $p_1,\dots,p_m\ge 2$ and $p_1,\dots,p_m=p < 2$, respectively,
\begin{equation}
\begin{aligned}
    \alpha_i^p &\le \lambda_i^{1-p/2}\inparen{\lambda_i\Fstar}^{p/2} = \lambda_i\inparen{\Fstar}^{p/2} \\
    \alpha_i^p &= \sum_{j \in S_i} \widehat{\alpha}_j^p = \sum_{j \in S_i} \widehat{\lambda}_j^{1-p/2}\tau_j\inparen{\widehat{\mLambda}^{1/2-1/p}\mA}^{p/2} \le \sum_{j \in S_i}\widehat{\lambda}_j\inparen{\Fstar}^{p/2} = \inparen{\sum_{j\in S_i} \widehat{\lambda_j}}\inparen{\Fstar}^{p/2},
\end{aligned}
\label{eq:alpha_to_fstar}
\end{equation}
and that when $\Fstar \le 2d$ (say), we get $\norm{\valpha}_p^p \lesssim d^{p/2}$. Thus, at least when $p\ge 2$, we can think of $\norm{\valpha}_p^p$ as giving the sparsity we should expect when we sample with probabilities proportional to the $\alpha_i$. Although this does not quite work when $p < 2$, a minor modification of it will.

\begin{definition}[Notation for unit balls and norms under change-of-measure]
\label{defn:unit_ball}
Let $\vlambda$ be a probability measure over $[m]$ and $\mLambda\in\R^{n\times n}$ be its corresponding block-constant diagonal matrix (\Cref{defn:block_constant}). For any $r > 0$, and $\vy\in\R^n$ we define
\begin{align*}
    \gnorml{\vy}{r} \coloneqq \inparen{\sum_{i=1}^m \lambda_i\inparen{\sum_{j \in S_i} \abs{\vy_j}^{p_i}}^{r/p_i}}^{1/r}.
\end{align*}
We also define
\begin{align*}
    B_r \coloneqq\inbraces{\vx \in\R^d \suchthat \gnorml{\mLambda^{-1/p}\mA\vx}{r} \le 1}.
\end{align*}
\end{definition}

From \Cref{defn:unit_ball}, it is easy to verify that $\gnorml{\mLambda^{-1/p}\mA\vx}{p} = \gnorm{\mA\vx}{p}$. Indeed, we have
\begin{align*}
    \gnorml{\mLambda^{-1/p}\mA\vx}{p}^p = \sum_{i=1}^m \lambda_i\norm{\mLambda_{S_i}^{-1/p}\mA_{S_i}\vx}_{p_i}^{p} = \sum_{i=1}^m \norm{\mA_{S_i}\vx}_{p_i}^p = \gnorm{\mA\vx}{p}^p.
\end{align*}
We will also require the crucial property that $\gnorml{\vy}{r}$ is log-convex in $1/r$. To see this, note that the vector in $\R^m$ formed by calculating all the inner norms $p_1,\dots,p_m$ is constant regardless of the outer norm, and then we can use the fact that for a fixed measure $\vmu$, the $\ell_r^m(\vmu)$ norms are log-convex in $1/r$.

We now have the language to state the main result of this section, \Cref{thm:general_covering}.

\begin{theorem}
\label{thm:general_covering}
Let $\vlambda$ be a probability measure and $\mW$ be a rounding matrix (\Cref{defn:rounding_matrix}) so that $\vlambda$ and $\mW$ form an $\Fstar$-block Lewis overestimate (\Cref{defn:block_lewis_overestimate}). Suppose at least one of the following holds:
\begin{itemize}
    \item $p > \frac{1}{\log d}$ and $\abs{S_1}=\dots=\abs{S_m}=1$;
    \item $p = p_1 = \dots = p_m$ and $p < 2$;
    \item $p \ge 1$ and $p_1,\dots,p_m \ge 2$;
    \item $p_1=\dots=p_m=2$ and $1/\log d \le p < \infty$.
\end{itemize}
If $H \ge 1$ is such that the sampling probabilities $\rho_i$ satisfy $H\rho_i \ge \alpha_i^p/\norm{\valpha}_p^p$ for all $i \in [m]$, and if we write $p^{\star}\coloneqq\max\inbraces{1,\max_{i}\min\inbraces{p_i,\log\abs{S_i}}}$, then (recall \Cref{defn:sampling_polyhedron} for the definition of $\polynorm{\cdot}$)
\begin{align*}
    \log\cN\inparen{B_p, \eta\inbraces{\vx\in\R^d \suchthat \polynorm{\vx} \le 1}} \lesssim \eta^{-\min(2,p)} \cdot H^{2/\max(p,2)}\cdot C(p)p^{\star}\Fstar\log\max\inbraces{\mtilde,\Fstar},
\end{align*}
where $C(p)$ is a constant that only depends on $p$.
\end{theorem}

Although \Cref{thm:general_covering} is stated abstractly, we will see that there exists a convenient instantiation for all the parameters stated.

\begin{corollary}
\label{corollary:specific_cover}
In the same cases as in \Cref{thm:general_covering}, there exists a probability measure $\vlambda$ over $[m]$ and a rounding $\mW$ for which in the same setting as \Cref{thm:general_covering}, we have
\begin{align*}
    \log\cN\inparen{B_p, \eta\inbraces{\vx\in\R^d \suchthat \polynorm{\vx} \le 1}} \lesssim_p \eta^{-\min(2,p)}\cdot{\underset{i \in S}{\max}\ \min\inbraces{p_i,\log\abs{S_i}}d\log\mtilde}.
\end{align*}
\end{corollary}

\subsection{Block Lewis weights}
\label{sec:blw_defs_covering}

For the sake of motivation, let us first prove \Cref{corollary:specific_cover} given \Cref{thm:general_covering}. We first need \Cref{lemma:block_lewis}, which is derived from the \textit{block Lewis weights} of \textcite{jlls23}.

For a nonnegative diagonal matrix $\mV$, let $\beta_i(\mV) \coloneqq \inparen{\sum_{j \in S_i} \inparen{\va_j^{\top}(\mA^{\top}\mV\mA)^{-1}\va_j}^{p_i/2}}^{1/p_i}$. We call the $\beta_i(\mV)^p$ the \textit{block Lewis weights}.

\begin{lemma}
\label{lemma:block_lewis}
If $p_i \in [2, \infty]$ and $p \in [1, \infty)$, then there exist diagonal $\mV, \mLambda \in \R^{n \times n}$ such that $\vlambda$ is a probability measure over $[m]$ and the corresponding \(\mLambda\in\R^{n\times n}\) is constant on the blocks, then
\(
\sum_{i=1}^m \beta_i(\mV)^p = d
\) and for all $\vx\in\R^d$,
\begin{align*}
    \norm{\mV^{1/2}\mA\vx}_2 \le d^{1/2-1/p}\gnorm{\mLambda^{1/2-1/p}\mA\vx}{2} \le d^{\max(0,1/2-1/p)}\gnorm{\mA\vx}{p}
\end{align*}
\end{lemma}

\begin{proof}[Proof of \Cref{lemma:block_lewis}]
The reader familiar with the work of \textcite{jlls23} will notice that \Cref{lemma:block_lewis} is a strengthened variant of Lemma 4.2 from that work. 

Indeed, consider the context of the proof of Lemma 4.2 from \cite{jlls23}. There, notice that $\mW$ is initially chosen so that $\sum_{i=1}^m \beta_i(\mW)^p = d$ and \(\mU = (\mA^\top \mW \mA)^{-1/2}\). We choose $\mV$ in the same way. Next, using their choice of $\vu$, we have $\norm{\vu_{S_i}}_{p_i}^{p-2} = \inparen{\beta_i(\mV)^{p}}^{1-2/p}$. 

Restating (4.8) from \cite{jlls23} in our notation, we have for all $\vx\in\R^d$ that
\begin{align*}
    \norm{\mV^{1/2} \mA \vx}_2^2 \le \sum_{i=1}^m \norm{\vu_{S_i}}_{p_i}^{p-2} \norm{\mA_{S_i} \vx}_{p_i}^2.
\end{align*}
For each \(i \in [m]\) let \(\lambda_i = \frac{\beta_i(\mV)^p}{d}\), so that \(\vlambda\) is a probability measure.
Then
\begin{align*}
    \norm{\mV^{1/2} \mA \vx}_2^2 &\le d^{1-2/p} \sum_{i=1}^m \lambda_i^{1-2/p} \norm{\mA_{S_i} \vx}_{p_i}^2.
\end{align*}
Let \(\mLambda\) be a \(n \times n\) diagonal matrix, where for every \(i \in [m]\) and \(j \in S_i\),
we define \(\mLambda_{jj} = \lambda_i\).
Because \(\lambda_i^{1-2/p} \norm{\mA_{S_i} \vx}_{p_i}^2 = \norm{(\mLambda^{1/2-1/p} \mA)_{S_i} \vx}_{p_i}^2\), we obtain
\begin{equation}\label{eq:vax_atmost_nlax}
    \norm{\mV^{1/2}\mA\vx}_2 \le d^{1/2-1/p}\gnorm{\mLambda^{1/2-1/p}\mA\vx}{2}.
\end{equation}
Since $p$-norms taken with respect to a probability measure are increasing in $p$ we immediately get for all $p \ge 2$ that
\begin{align*}
    \norm{\mV^{1/2}\mA\vx}_2 &\stackrel{\eqref{eq:vax_atmost_nlax}}{\le} d^{1/2-1/p}\gnorm{\mLambda^{1/2-1/p}\mA\vx}{2} = d^{1/2-1/p}\gnorml{\mLambda^{-1/p}\mA\vx}{2} \\
    &\le d^{1/2-1/p}\gnorml{\mLambda^{-1/p}\mA\vx}{p} = d^{1/2-1/p}\gnorm{\mA\vx}{p}.
\end{align*}
The case where $p \le 2$ follows from the ``$1 \le q \le 2$'' subcase of the proof of Lemma 4.2 from \cite{jlls23}, which yields
\begin{align*}
    \norm{\mV^{1/2}\mA\vx}_2 \stackrel{\eqref{eq:vax_atmost_nlax}}{\le} d^{1/2-1/p}\gnorm{\mLambda^{1/2-1/p}\mA\vx}{2} \le \gnorm{\mA\vx}{p}.
\end{align*}
We therefore conclude the proof of \Cref{lemma:block_lewis}.
\end{proof}

We use \Cref{lemma:block_lewis} to give an instantiation for the parameters in \Cref{thm:general_covering}.

\begin{lemma}
\label{lemma:block_instantiation_small}
Let \(\mV, \mLambda\) be the matrices from \Cref{lemma:block_lewis} and let $\mU$ and $\vf_j$ be as defined in \Cref{defn:alpha}. Let $p \ge 1$ and $p_i \ge 2$ for all $i \in [m]$. If we choose \(\mW\) such that
\begin{align*}
    \frac{\mV^{1/2}}{d^{1/2-1/p}} = \mW^{1/2}\mLambda^{1/2-1/p},
\end{align*}
then:
\begin{itemize}
    \item $\norm{\mW^{1/2}\mLambda^{1/2-1/p}\mA\vx}_2 \le \gnorm{\mLambda^{1/2-1/p}\mA\vx}{2}$;
    \item for all $i$, $\frac{\alpha_i}{d^{1/2-1/p}} = \beta_i(\mV)$;
    \item $\norm{\valpha}_p^p = d^{p/2}$;
    \item for all $i$, $\inparen{\sum_{j \in S_i} \norm{\vf_j}_2^{p_i}}^{1/p_i} = \norm{\valpha}_p = d^{1/2}$.
    \item The rounding matrix $\mW$ and measure $\vlambda$ are an $\Fstar$-block Lewis overestimate (\Cref{defn:block_lewis_overestimate}) with $\Fstar = d$.
\end{itemize}
\end{lemma}
\begin{proof}[Proof of \Cref{lemma:block_instantiation_small}]
The first property follows immediately from \Cref{lemma:block_lewis}. Using \Cref{fact:whitening}, notice that
\begin{align*}
    \va_j^{\top}(\mA^{\top}\mV\mA)^{-1}\va_j = \frac{\tau_j(\mV^{1/2}\mA)}{v_j} = \frac{\tau_j(\mW^{1/2}\mLambda^{1/2-1/p}\mA)}{d^{1-2/p}w_j\lambda_i^{1-2/p}} = \frac{\norm{\vu_j}_2^2}{d^{1-2/p}w_j\lambda_i^{1-2/p}} = \frac{\lambda_i^{2/p}\norm{\vf_j}_2^2}{d^{1-2/p}},
\end{align*}
so after substituting into the formula for $\beta_i(\mV)$,
\begin{align*}
    \beta_i(\mV) &= \inparen{\sum_{j \in S_i} \inparen{\va_j^{\top}(\mA^{\top}\mV\mA)^{-1}\va_j}^{p_i/2}}^{1/p_i} = \inparen{\sum_{j \in S_i} \inparen{\frac{\lambda_i^{2/p}\norm{\vf_j}_2^2}{d^{1-2/p}}}^{p_i/2}}^{1/p_i} \\
    &= \frac{\lambda_i^{1/p}\inparen{\sum_{j \in S_i} \norm{\vf_j}_2^{p_i}}^{1/p_i}}{d^{1/2-1/p}} = \frac{\alpha_i}{d^{1/2-1/p}},
\end{align*}
where the last equality follows from the formula for $\valpha$ stated in \Cref{thm:general_covering}. This also implies that
\begin{align*}
    \norm{\valpha}_p^p = \sum_{i=1}^m \alpha_i^p = \sum_{i=1}^p \beta_i(\mV)^pd^{p/2-1} = d^{p/2}.
\end{align*}
Finally, observe that the above calculation shows that $\lambda_i \propto \alpha_i^p$, since we have defined $\lambda_i \propto \beta_i(\mV)^p$ and we have just seen that $\beta_i(\mV)^p \propto \alpha_i^p$. This means we can write $\lambda_i = \alpha_i^p/\norm{\valpha}_p^p$. Using this, we have
\begin{align*}
    \alpha_i = \lambda_i^{1/p}\inparen{\sum_{j \in S_i} \norm{\vf_j}_2^{p_i}}^{1/p_i} = \frac{\alpha_i}{\norm{\valpha}_p} \cdot \inparen{\sum_{j \in S_i} \norm{\vf_j}_2^{p_i}}^{1/p_i}.
\end{align*}
After rearranging, we have
\begin{align*}
    \inparen{\sum_{j \in S_i} \norm{\vf_j}_2^{p_i}}^{1/p_i} = \norm{\valpha}_p = d^{1/2},
\end{align*}
and so we may take $\Fstar = d$. This concludes the proof of \Cref{lemma:block_instantiation_small}.
\end{proof}

We now handle the cases that are not covered by the block Lewis weight construction of \cite{jlls23}.

\begin{lemma}
\label{lemma:block_instantiation_alleq}
If $0 < p_1=\dots=p_m=p < 2$ or if $p_1=\dots=p_m=2$ and $1/\log d \le p < \infty$, then there exists a probability measure $\widehat{\vlambda}$ over $[k]$ and corresponding $\widehat{\valpha} \in \R^n$ such that $\widehat{\vlambda}$ is an $\Fstar$-block Lewis overestimate for $\Fstar=n$.
\end{lemma}
\begin{proof}
For the case where $0 < p_1=\dots=p_m=p < 2$, we simply use the fact that Lewis's measure tells us that there exists a measure $\widehat{\vlambda}$ such that
\begin{align*}
    \frac{\tau_j\inparen{\widehat{\mLambda}^{1/2-1/p}\mA}}{\widehat{\lambda}_j} \le d.
\end{align*}
In the other case, we will see later that the guarantee of a natural contraction mapping (\Cref{alg:blw_lewismap} and \Cref{lemma:contract_alg}) imply that $\mW = \mI_n$ and the resulting $\vlambda$ form an $d$-block Lewis overestimate, thereby concluding the proof of \Cref{lemma:block_instantiation_alleq}.
\end{proof}

\Cref{lemma:block_instantiation_small} and \Cref{lemma:block_instantiation_alleq} easily imply \Cref{corollary:specific_cover}.

\begin{proof}[Proof of \Cref{corollary:specific_cover}]
We combine \Cref{thm:general_covering} with the instantiations in \Cref{lemma:block_instantiation_small} and \Cref{lemma:block_instantiation_alleq}, directly yielding \Cref{corollary:specific_cover}.
\end{proof}

In light of \Cref{corollary:specific_cover}, the goal of the remainder of this section is to prove \Cref{thm:general_covering}. 

It will be useful to consider a corresponding change-of-basis that arises from our setting of $\vlambda$. Let $\mU\mSigma\mV^{\top}$ be a singular value decomposition of $\mW^{1/2}\mLambda^{1/2-1/p}\mA$ where $\mU \in \R^{m \times d}$ and $\mSigma, \mV \in \R^{d \times d}$. Let $\mR$ be the invertible matrix $\mV\mSigma^{-1}$ (we assume without loss of generality that $\rank{\mA} = d$, and it is easy to extend the results of this section to the case where $\rank{\mA} < d$). We take $\mR$ as our change-of-basis matrix. Using this, it is easy to see that $\mW^{1/2}\mLambda^{1/2-1/p}\mA\mR = \mU$ consists of orthonormal columns. Furthermore, we have $\mLambda^{-1/p}\mA\mR = \mW^{-1/2}\mLambda^{-1/2}\mU$.
 
\subsection{Covering numbers for \texorpdfstring{$0 < p < 2$}{0 < p < 2}}

The goal of this section is to prove \Cref{lemma:covering_two_to_one_clean} under the notion of overestimate given by \Cref{defn:block_lewis_overestimate}. 

We are now ready to state the main result of this subsection.

\begin{lemma}
\label{lemma:covering_two_to_one_clean}
Let $\vlambda$ and $\vw$ be such that they form an $\Fstar$-block Lewis overestimate. Then,
\begin{align*}
    \log\cN(B_p,\eta \widehat{B_2}) \lesssim \eta^{-\frac{2p}{2-p}} \cdot C(p)\max_{i}\min\inparen{p_i,\log\abs{S_i}}\Fstar\log\Fstar,
\end{align*}
where $C(p)$ is a constant that only depends on $p$.
\end{lemma}

The goal of the rest of this subsection is to prove \Cref{lemma:covering_two_to_one_clean}. We follow the outline detailed in \Cref{sec:sparsification_overview_measure}. In short, our plan is the following:
\begin{enumerate}
    \item We first reduce bounding $\log\cN(B_p, \eta\widehat{B_2})$ to bounding $\log\cN(\widehat{B_2}, \delta_h B_r)$ for all $h \ge 0$ and appropriate choices of $r$ and $\delta_h$.
    \item We then control each term $\log\cN(\widehat{B_2}, \delta_h B_r)$. To do so, we will apply the dual Sudakov inequality (\Cref{fact:sudakov}, \eqref{eq:sudakov_dual}). To actually estimate $\E\norm{\vg}_{B_r}$ where $\vg \sim \cN(0,\mI_d)$, we need to prove that every resulting summand of the form $\norm{\cdot}_{p_i}$ is subgaussian with a parameter that only depends on $p_i$. To do so, we exploit the fact that these summands are Lipschitz and then apply \Cref{fact:lip_concentration}.
    \item We finally assemble all the previous pieces together to get the desired handle on $\log\cN(B_p, \eta\widehat{B_2})$. 
\end{enumerate}

\subsubsection{Reduction to bounding \texorpdfstring{$\log\cN(\widehat{B_2},\eta B_r)$}{log N(\^{B2}, n Br)}}

As stated in \Cref{sec:sparsification_overview_measure}, we begin with reducing the calculation of $\log\cN(B_p, \eta \widehat{B_2})$ to calculating $\log\cN(\widehat{B_2}, \eta B_r)$ (for a different $\eta$).

\begin{lemma}
\label{lemma:poor_mans_dualization}
Let $\theta$ and $r$ be such that $r=(2-\theta)p/(p-\theta)$. Define
\begin{align*}
    \delta_h \coloneqq \inparen{\frac{8^{h+1}\eta}{2\cdot 8^{2/\theta}}}^{\frac{\theta}{2-\theta}} = \eta^{\frac{\theta}{2-\theta}}\cdot 8^{(h+1)\cdot\frac{\theta}{2-\theta}} \cdot \inparen{2\cdot 8^{2/\theta}}^{-\frac{\theta}{2-\theta}}
\end{align*}
Then, we have
\begin{align*}
     \log\cN(B_p,\eta \widehat{B_2}) &\le \sum_{h\ge 0} \log\cN(\widehat{B_2},\delta_hB_r).
\end{align*}
\end{lemma}
\begin{proof}[Proof of \Cref{lemma:poor_mans_dualization}]
For $h \in \N_{\ge 0}$, let $\cN_h$ be a maximal subset of $B_p$ such that for any two distinct elements $\vz_1, \vz_2 \in B_p$, we have $\norm{\mW^{1/2}\mLambda^{-1/p}\mA(\vz_1-\vz_2)}_{2(\vlambda)} \ge 8^h\eta$ (where by $\norm{\cdot}_{p(\vlambda)}$ we mean the $\ell_p$ norm taken with respect to the measure given by $\vlambda$). This yields $\abs{\cN_h} \ge \cN(B_p, 8^h \eta \widehat{B_2})$.

Next, since for every $h$ there are $\vz_i \in B_p$ for which $B_p \subseteq \bigcup_{i=1}^{\cN(B_p,8^{h+1}\eta \widehat{B_2})} \inbraces{\vz_i + 8^{h+1}\eta \widehat{B_2}}$, for every $h$ there must exist a $\zstar_h \in B_p$ for which
\begin{align*}
    \abs{\inbraces{\zstar_h + 8^{h+1} \eta \widehat{B_2}} \cap \cN_h} \ge \frac{\abs{\cN_h}}{\cN(B_p,8^{h+1} \eta \widehat{B_2})} \ge \frac{\cN(B_p,8^h \eta \widehat{B_2})}{\cN(B_p,8^{h+1} \eta \widehat{B_2})}.
\end{align*}
Let
\begin{align*}
    \cL_h \coloneqq \inbraces{\frac{\vz-\zstar_h}{8^{h+1} \eta} \suchthat \vz \in \inbraces{\zstar_h + 8^{h+1} \eta \widehat{B_2}} \cap \cN_h}
\end{align*}
from which we get by the sub-triangle inequality that
\begin{align*}
    \norm{\mW^{1/2}\mLambda^{-1/p}\mA\vz}_{2(\vlambda)} \le 1 \text{ and } \gnorml{\mLambda^{-1/p}\mA\vz}{p} \le \frac{\max\inbraces{2^{1/p},2}}{8^{h+1} \eta} \text{ for any } \vz \in \cL_h
\end{align*}
and
\begin{align*}
    \norm{\mW^{1/2}\mLambda^{-1/p}\mA\inparen{\vz_1-\vz_2}}_{2(\vlambda)} \ge \frac{1}{8} \text{ for any distinct } \vz_1,\vz_2 \in \cL_h.
\end{align*}
We now apply an interpolation estimate. Let $\vz_1$ and $\vz_2$ be distinct elements from $\cL_h$, set $0<\theta<2$ and $r=(2-\theta)p/(p-\theta)$, and observe that $\theta = p(r-2)/(r-p)$ and 
\begin{align*}
    \frac{1}{8^2} &\le \norm{\mW^{1/2}\mLambda^{-1/p}\mA\inparen{\vz_1-\vz_2}}_{2(\vlambda)}^2 \\
    &\le \gnorml{\mLambda^{-1/p}\mA\inparen{\vz_1-\vz_2}}{2}^2 \\
    &\le \gnorml{\mLambda^{-1/p}\mA\inparen{\vz_1-\vz_2}}{p}^{\theta}\cdot\gnorml{\mLambda^{-1/p}\mA\inparen{\vz_1-\vz_2}}{r}^{2-\theta} \\
    &\le \inparen{\frac{\max\inbraces{2^{1/p},2}}{8^{h+1}\eta}}^{\theta}\gnorml{\mLambda^{-1/p}\mA\inparen{\vz_1-\vz_2}}{r}^{2-\theta} 
\end{align*}
which means that after rearranging we have
\begin{align*}
    \gnorml{\mLambda^{-1/p}\mA\inparen{\vz_1-\vz_2}}{r} \ge \inparen{\inparen{\frac{8^{h+1}\eta}{\max\inbraces{2^{1/p},2}}}^{\theta} \cdot \frac{1}{8^2}}^{\frac{1}{2-\theta}} \ge \delta_h.
\end{align*}
The above argument gives
\begin{align*}
    \log\cN(\widehat{B_2},\delta_h B_r) \ge \log\abs{\cL_h} \ge \log\cN\inparen{B_p,8^h\eta\widehat{B_2}}-\log\cN\inparen{B_p,8^{h+1}\eta\widehat{B_2}}.
\end{align*}
We sum these inequalities over all $h \ge 0$ (noting that when $h$ is sufficiently large, we have $\log\cN(B_p,8^{h+1} \eta \widehat{B_2})=0$), and get
\begin{align*}
    \log\cN(B_p,\eta \widehat{B_2}) &\le \sum_{h\ge 0} \log\cN(\widehat{B_2},\delta_hB_r).
\end{align*}
This concludes the proof of \Cref{lemma:poor_mans_dualization}.
\end{proof}

\subsubsection{Bounding \texorpdfstring{$\log\cN(\widehat{B_2},\eta B_r)$}{log N(\^{B2}, n Br)}}

As we saw in \Cref{lemma:poor_mans_dualization}, it will be enough to understand the behavior of $\log\cN(\widehat{B_2},\eta B_r)$. Since $\widehat{B_2}$ is a linear transformation of a Euclidean ball, we will be able to apply the dual Sudakov inequality (\Cref{fact:sudakov}, \eqref{eq:sudakov_dual}).

To prepare for an application of the dual Sudakov inequality, we bound the Gaussian width of the ball $\inbraces{\vx \in \R^d \suchthat \gnorml{\mLambda^{-1/p}\mA\mR\vx}{r} \le 1}$. As we will see in a moment, the relevance of this ball arises from the fact that it is the $r$-ball with respect to the $\vlambda$ measure after a suitable linear transformation of the underlying space. In particular, it is under the invertible mapping $\vx \mapsto \mR\vx$ that we get $\mW^{1/2}\mLambda^{1/2-1/p}\mA\vx \mapsto \mW^{1/2}\mLambda^{1/2-1/p}\mA\mR\vx = \mU\vx$.

\begin{lemma}
\label{lemma:one_group_subgaussian}
Let $\vg \sim \cN(0,\mI_d)$. We have
\begin{align*}
    \subgnorm{\inparen{\sum_{j \in S_i} \abs{\ip{\vf_j,\vg}}^{p_i}}^{1/p_i}} \lesssim \inparen{1+\sqrt{p_i}}\inparen{\sum_{j \in S_i} \norm{\vf_j}_{2}^{p_i}}^{1/p_i}.
\end{align*}
\end{lemma}
\begin{proof}[Proof of \Cref{lemma:one_group_subgaussian}]
Observe the following Lipschitzness bound, i.e., for any $\vx$, by Cauchy-Schwarz, we have
\begin{align*}
   \inparen{\sum_{j \in S_i} \abs{\ip{\vf_j,\vx}}^{p_i}}^{1/p_i} \le \inparen{\sum_{j \in S_i} \norm{\vf_j}_{2}^{p_i}}^{1/p_i}\norm{\vx}_2
\end{align*}
which means by \Cref{fact:lip_concentration}, we get
\begin{align*}
    \subgnorm{\inparen{\sum_{j \in S_i} \abs{\ip{\vf_j,\vg}}^{p_i}}^{1/p_i}-\exvv{\vg\sim\cN(0,\mI_d)}{\inparen{\sum_{j \in S_i} \abs{\ip{\vf_j,\vg}}^{p_i}}^{1/p_i}}} \lesssim \inparen{\sum_{j \in S_i} \norm{\vf_j}_{2}^{p_i}}^{1/p_i}.
\end{align*}
Now, observe that
\begin{align*}
    \exvv{\vg\sim\cN(0,\mI_d}{\inparen{\sum_{j \in S_i} \abs{\ip{\vf_j,\vg}}^{p_i}}^{1/p_i}} \le \inparen{\sum_{j \in S_i} \norm{\vf_j}_2^{p_i}\exvv{\vg\sim\cN(0,\mI_d)}{\abs{\ip{\frac{\vf_j}{\norm{\vf_j}_2},\vg}}^{p_i}}}^{1/p_i} \asymp p_i^{1/2}\inparen{\sum_{j \in S_i} \norm{\vf_j}_{2}^{p_i}}^{1/p_i},
\end{align*}
and by \Cref{fact:uncentering},
\begin{align*}
    \subgnorm{\inparen{\sum_{j \in S_i} \abs{\ip{\vf_j,\vg}}^{p_i}}^{1/p_i}} \lesssim \inparen{1+\sqrt{p_i}}\inparen{\sum_{j \in S_i} \norm{\vf_j}_{2}^{p_i}}^{1/p_i},
\end{align*}
completing the proof of \Cref{lemma:one_group_subgaussian}.
\end{proof}

Next, we estimate the expected norm of a Gaussian random vector under the norm given by $\inbraces{\vx\in\R^d\suchthat \gnorml{\mLambda^{-1/p}\mA\mR\vx}{r} \le 1}$.

\begin{lemma}
\label{lemma:khintchine}
For $r \ge 2$, we have
\begin{align*}
    \exvv{\vg\sim\cN(0,\mI_d)}{\gnorml{\mLambda^{-1/p}\mA\mR\vg}{r}} \lesssim \begin{cases} r^{1/2}\inparen{1+\sqrt{\max_i p_i}}\inparen{\sum_{i=1}^m \lambda_i \inparen{\sum_{j \in S_i} \norm{\vf_j}_{2}^{p_i}}^{r/p_i}}^{1/r} & \text{ if } r \le \log m \\ \sqrt{\log m} \cdot \max_i \inparen{1+\sqrt{p_i}}\inparen{\sum_{j \in S_i} \norm{\vf_j}_{2}^{p_i}}^{1/p_i} & \text{ otherwise} \end{cases}
\end{align*}
\end{lemma}
\begin{proof}[Proof of \Cref{lemma:khintchine}]
Let $\vu_j$ denote the rows of $\mU$. Note that by \Cref{fact:whitening}, we have
\begin{align*}
    \norm{\vu_j}_2^2 = \tau_j(\mW^{1/2}\mLambda^{1/2-1/p}\mA).
\end{align*}
Now, observe that $\mLambda^{-1/p}\mA\mR = \mW^{-1/2}\mLambda^{-1/2}\mU$. By \Cref{lemma:one_group_subgaussian}, we know that
\begin{align*}
    \exvv{\vg\sim\cN(0,\mI_d)}{\inparen{\sum_{j \in S_i} \abs{\ip{\vf_j,\vg}}^{p_i}}^{r/p_i}} \lesssim r^{r/2}\inparen{1+\sqrt{p_i}}^r\inparen{\sum_{j \in S_i} \norm{\vf_j}_{2}^{p_i}}^{r/p_i}
\end{align*}
We first handle the case where $r \lesssim \log m$. Notice that
\begin{align*}
    \exvv{\vg\sim\cN(0,\mI_d)}{\gnorml{\mLambda^{-1/p}\mA\mR\vg}{r}} &= \exvv{\vg\sim\cN(0,\mI_d))}{\inparen{\sum_{i=1}^m \lambda_i\inparen{\sum_{j \in S_i}\abs{\ip{w_j^{-1/2}\lambda_i^{-1/2}\vu_j,\vg}}^{p_i}}^{r/p_i}}^{1/r}} \\
    &\le \inparen{\sum_{i=1}^m \lambda_i\exvv{\vg\sim\cN(0,\mI_d))}{\inparen{\sum_{j \in S_i}\abs{\ip{w_j^{-1/2}\lambda_i^{-1/2}\vu_j,\vg}}^{p_i}}^{r/p_i}}}^{1/r} \\
    &\lesssim \inparen{\sum_{i=1}^m \lambda_i\inparen{r^{r/2}\inparen{1+\sqrt{p_i}}^r\inparen{\sum_{j \in S_i} \norm{\vf_j}_{2}^{p_i}}^{r/p_i}}}^{1/r} \\
    &\le r^{1/2}\inparen{1+\sqrt{\max_i p_i}}\inparen{\sum_{i=1}^m \lambda_i \inparen{\sum_{j \in S_i} \norm{\vf_j}_{2}^{p_i}}^{r/p_i}}^{1/r}.
\end{align*}
We now handle the case where $r \gtrsim \log m$. We have
\begin{align*}
    \exvv{\vg\sim\cN(0,\mI_d)}{\gnorml{\mLambda^{-1/p}\mA\mR\vg}{r}} &= \exvv{\vg\sim\cN(0,\mI_d))}{\inparen{\sum_{i=1}^m \lambda_i\inparen{\sum_{j \in S_i}\abs{\ip{w_j^{-1/2}\lambda_i^{-1/2}\vu_j,\vg}}^{p_i}}^{r/p_i}}^{1/r}} \\
    &\lesssim \exvv{\vg\sim\cN(0,\mI_d))}{\max_i \inparen{\sum_{j \in S_i}\abs{\ip{w_j^{-1/2}\lambda_i^{-1/2}\vu_j,\vg}}^{p_i}}^{1/p_i}} \\
    &\lesssim \sqrt{\log m} \cdot \max_i \inparen{1+\sqrt{p_i}}\inparen{\sum_{j \in S_i} \norm{\vf_j}_{2}^{p_i}}^{1/p_i}
\end{align*}
and conclude the proof of \Cref{lemma:khintchine} (the last line follows from \Cref{fact:max_subgaussians}).
\end{proof}

Now, we show how to relate $(\sum_{j \in S_i} \norm{\vf_j}_2^{p_i})^{1/p_i}$ to $\Fstar$.

\begin{lemma}
\label{lemma:convert_f_to_fstar}
For all $i \in [m]$, we have
\begin{align*}
    \inparen{\sum_{j \in S_i} \norm{\vf_j}_2^{p_i}}^{2/p_i} \le \Fstar.
\end{align*}
\end{lemma}
\begin{proof}[Proof of \Cref{lemma:convert_f_to_fstar}]
Recall \Cref{fact:whitening}; this gives us
\begin{align*}
    \inparen{\sum_{j \in S_i} \norm{\vf_j}_{2}^{p_i}}^{2/p_i}
    &= \inparen{\sum_{j \in S_i} \inparen{\frac{\tau_j\inparen{\mW^{1/2}\mLambda^{1/2-1/p}\mA}}{w_j\lambda_i}}^{p_i/2}}^{2/p_i} \\
    &= \frac{1}{\lambda_i}\inparen{\sum_{j \in S_i} \inparen{\frac{\tau_j\inparen{\mW^{1/2}\mLambda^{1/2-1/p}\mA}}{w_j}}^{p_i/2}}^{2/p_i}.
\end{align*}
We recall that $\Fstar$ satisfies \Cref{defn:block_lewis_overestimate} and conclude the proof of \Cref{lemma:convert_f_to_fstar}.
\end{proof}

We now have enough tools to build a na\"ive estimate of $\log\cN\inparen{\widehat{B_2},\eta B_r}$ via directly applying the dual Sudakov inequality.

\begin{lemma}
\label{lemma:r_to_two}
We have
\begin{align*}
    \log\cN\inparen{\widehat{B_2},\eta B_r} \lesssim \eta^{-2} \cdot  \begin{cases} r\inparen{1+\sqrt{\max_i p_i}}^2\inparen{\sum_{i=1}^m \lambda_i \inparen{\sum_{j \in S_i} \norm{\vf_j}_{2}^{p_i}}^{r/p_i}}^{2/r} & \text{ if } r \le \log m \\ \log m \cdot \max_i \inparen{1+\sqrt{p_i}}^2\inparen{\sum_{j \in S_i} \norm{\vf_j}_{2}^{p_i}}^{2/p_i} & \text{ otherwise} \end{cases}
\end{align*}
Simply put, we may also write
\begin{align*}
    \log\cN\inparen{\widehat{B_2},\eta B_r} \lesssim \eta^{-2} \cdot r\max_{i}\min\inparen{p_i, \log\abs{S_i}}\Fstar.
\end{align*}
\end{lemma}
\begin{proof}[Proof of \Cref{lemma:r_to_two}]
Since $\mR$ is invertible, it will be enough to bound the covering number
\begin{align*}
    \cN \coloneqq \cN\inparen{\inbraces{\vx\in\R^d \suchthat \norm{\mU\vx}_2 \le 1}, \eta\inbraces{\vx\in\R^d \suchthat \gnorml{\mLambda^{-1/p}\mA\mR\vx}{r} \le 1}}.
\end{align*}
Because $\norm{\mU\vx}_2 = \norm{\vx}_2$, we can apply the dual Sudakov Inequality (\Cref{fact:sudakov}, \eqref{eq:sudakov_dual}). This means we get
\begin{align*}
    \log\cN \lesssim \eta^{-2} \inparen{\exvv{\vg\sim\cN(0,\mI_d)}{\gnorml{\mLambda^{-1/p}\mA\mR\vg}{r}}}^2.
\end{align*}
We plug in the result from \Cref{lemma:khintchine} and conclude the proof of \Cref{lemma:r_to_two}. The statement after the ``simply put'' follows from \Cref{lemma:convert_f_to_fstar}.
\end{proof}

Although the calculation in \Cref{lemma:r_to_two} works pretty well for small $r$, this degrades quite rapidly once $r$ is large (say, larger than $\log d$).

To resolve this, we build another estimate for $\log\cN\inparen{\widehat{B_2},\eta B_r}$ that performs better when $r$ is larger than $\log d$ or so. We will be able to do this after an interpolation step and a simple geometric observation relating $\widehat{B_2}$ and $B_r$.

\begin{lemma}
\label{lemma:r_ellipsoid}
Let $\Delta_i$ be defined such that 
\begin{align*}
    \Delta_i^{1/2} \coloneqq \max_{\vx\in\R^d\setminus\inbraces{0}} \frac{\lambda_i^{-1/p}\norm{\mA_{S_i}\vx}_{p_i}}{\norm{\mW^{1/2}\mLambda^{1/2-1/p}\mA\vx}_2},
\end{align*}
and let $\Delta \coloneqq \max_{i \in [m]} \Delta_i$.

For all $\vx\in\R^d$ and $r > 2$, if $\abs{S_i} = 1$ for all $i$, then we have
\begin{align*}
    \norm{\mLambda^{-1/2}\mU\vx}_{r(\vlambda)} \le \Delta^{1/2-1/r} \cdot \norm{\mU\vx}_2 \le \Delta^{1/2-1/r} \cdot \norm{\mLambda^{-1/2}\mU\vx}_{r(\vlambda)}.
\end{align*}
Moreover, if there exists at least one $S_i$ for which $\abs{S_i} > 1$, then we have
\begin{align*}
    \gnorml{\mLambda^{-1/p}\mA\vx}{r} \le \Delta^{1/2}\norm{\mW^{1/2}\mLambda^{1/2-1/p}\mA\vx}_2.
\end{align*}
\end{lemma}
\begin{proof}[Proof of \Cref{lemma:r_ellipsoid}]
For the sake of intuition and an interpretation, the reader may think $\Delta \approx d$.

Note that for the case where all the $S_i$ are singletons, we may assume $\mW = \mI_m$.

Since $\vlambda$ is a probability measure, we have for any $r \ge 2$ and for all $\vx\in\R^d$ that
\begin{align*}
    \norm{\mU\vx}_2 = \norm{\mLambda^{-1/2}\mU\vx}_{2(\vlambda)} \le \norm{\mLambda^{-1/2}\mU\vx}_{r(\vlambda)}.
\end{align*}
We now prove the lower bound. We have
\begin{align*}
    \norm{\mLambda^{-1/2}\mU\vx}_{r(\vlambda)} &= \inparen{\sum_{i=1}^m \lambda_i\abs{\ip{\lambda_i^{-1/2}\vu_i,\vx}}^{r}}^{1/r} =  \inparen{\sum_{i=1}^m \lambda_i\abs{\ip{\lambda_i^{-1/2}\vu_i,\vx}}^{2}\cdot\abs{\ip{\lambda_i^{-1/2}\vu_i,\vx}}^{r-2}}^{1/r} \\
    &\le \inparen{\norm{\mU\vx}_2^2 \cdot \max_i \abs{\ip{\lambda_i^{-1/2}\vu_i,\vx}}^{r-2}}^{1/r} \le \inparen{\norm{\mU\vx}_2^2 \cdot \inparen{\Delta^{1/2}\norm{\vx}_2}^{r-2}}^{1/r} \\
    &= \Delta^{1/2-1/r} \norm{\vx}_2.
\end{align*}
We now move onto the more general case where the $S_i$ are allowed to have multiple elements. We write
\begin{align*}
    \gnorml{\mLambda^{-1/p}\mA\vx}{r} &= \inparen{\sum_{i=1}^m \lambda_i\norm{\mLambda_{S_i}^{-1/p}\mA_{S_i}\vx}_{p_i}^r}^{1/r} \\
    &\le \inparen{\sum_{i=1}^m \lambda_i\Delta_i^{r/2}\norm{\mW^{1/2}\mLambda^{1/2-1/p}\mA\vx}_{2}^{r}}^{1/r} \le \Delta^{1/2}\norm{\mW^{1/2}\mLambda^{1/2-1/p}\mA\vx}_2,
\end{align*}
which concludes the proof of \Cref{lemma:r_ellipsoid}.
\end{proof}

At last, we have the tools we need to give a characterization of $\log\cN\inparen{\widehat{B_2},\eta B_r}$ when $r \gtrsim \log d$.

\begin{lemma}
\label{lemma:r_to_two_complicated}
If all the $S_i$ have size $1$, then
\begin{align*}
    \log\cN\inparen{\widehat{B_2}, \eta B_r} \lesssim \inparen{\frac{\eta}{2}}^{-\frac{2r}{r-2}} \cdot r\Fstar\log\Fstar,
\end{align*}
and if there is at least one $S_i$ larger than $1$, then
\begin{align*}
    \log\cN\inparen{\widehat{B_2}, \eta B_r} \lesssim \inparen{\frac{\eta}{2}}^{-\frac{2r}{r-2}} \cdot \frac{2r^2}{r-2}\max_i \min(p_i+1, \log\abs{S_i})\Fstar\log\Fstar.
\end{align*}
\end{lemma}
\begin{proof}[Proof of \Cref{lemma:r_to_two_complicated}]
The reader familiar with the work of \textcite{blm89} can think of the present Lemma as a generalization of (7.13) of Proposition 7.2 of that work.

Define
\begin{align*}
    M(\mF,r) &\coloneqq \begin{cases} r\inparen{1+\sqrt{\max_i p_i}}^2\inparen{\sum_{i=1}^m \lambda_i \inparen{\sum_{j \in S_i} \norm{\vf_j}_{2}^{p_i}}^{r/p_i}}^{2/r} & \text{ if } r \le \log m \\ \log m \cdot \max_i \inparen{1+\sqrt{p_i}}^2\inparen{\sum_{j \in S_i} \norm{\vf_j}_{2}^{p_i}}^{2/p_i} & \text{ otherwise} \end{cases}.
\end{align*}

Let $q > r$ and $0 < \theta < 1$ be such that
\begin{align*}
    \frac{1}{r} = \frac{1-\theta}{2} + \frac{\theta}{q}.
\end{align*}
By interpolation, observe that we have
\begin{align*}
    \gnorml{\mLambda^{-1/p}\mA(\vx_1-\vx_2)}{r} &\le \gnorml{\mLambda^{-1/p}\mA(\vx_1-\vx_2)}{2}^{1-\theta} \cdot \gnorml{\mLambda^{-1/p}\mA(\vx_1-\vx_2)}{q}^{\theta} \\
    &\le 2\gnorml{\mLambda^{-1/p}\mA(\vx_1-\vx_2)}{q}^{\theta}
\end{align*}
which means that
\begin{align*}
    \log\cN\inparen{B_2,\eta B_r} \le \log\cN\inparen{B_2,(\eta/2)^{1/\theta}B_q} \le \inparen{\frac{\eta}{2}}^{-2/\theta}M(\mF,q).
\end{align*}
Let us set $q = r\log D$, where we will choose $D$ in a moment. Then, notice that
\begin{align*}
    \inparen{\frac{\eta}{2}}^{-2/\theta} = \inparen{\frac{\eta}{2}}^{-2r(q-2)/(q(r-2))} = \inparen{\frac{\eta}{2}}^{-\frac{2r}{r-2}\inparen{1-\frac{2}{r\log D}}} = \inparen{\frac{\eta}{2}}^{-\frac{2r}{r-2} + \frac{4}{(r-2)\log D}}.
\end{align*}
It is now sufficient to identify $D$ such that whenever $\eta$ is small enough to have $\log\cN > 0$, we have
\begin{align*}
    \inparen{\frac{\eta}{2}}^{\frac{4}{(r-2)\log D}} \lesssim 1.
\end{align*}
To identify this $D$, notice that \Cref{lemma:r_ellipsoid} implies that if all the $S_i$ have size $1$, then only values of $\eta$ such that $\eta \le \Delta^{1/2-1/r}$ contribute to $\log\cN$. In the more general setting, observe only $\eta \le \Delta^{1/2}$ counts.

Hence, if all the $S_i$s are singletons, we choose $D = \Delta$. For any $\eta \le 2\Delta^{1/2-1/r} = 2D^{1/2-1/r}$, we see that
\begin{align*}
    \inparen{\frac{\eta}{2}}^{\frac{4}{(r-2)\log D}} \le \Delta^{\frac{r-2}{2r} \cdot \frac{4}{(r-2)\logv{\Delta}}} = \Delta^{\frac{2}{r\log\Delta}} = 2^{2/r} \le 2.
\end{align*}
Similarly, for the case where the $S_i$ are more generally sized, we choose $D = \Delta^{2r/(r-2)}$. Now, for any $\eta \le 2\Delta^{1/2}$, we get
\begin{align*}
    \inparen{\frac{\eta}{2}}^{\frac{4}{(r-2)\log D}} \le \Delta^{\frac{4}{(r-2)\logv{\Delta^{(2r/(r-2))}}}} = \Delta^{\frac{2}{r\log\Delta}} = 2^{2/r} \le 2.
\end{align*}
Putting everything together, if all the $S_i$s are singletons, we get
\begin{align*}
    \log\cN\inparen{\widehat{B_2}, \eta B_r} \lesssim \inparen{\frac{\eta}{2}}^{-\frac{2r}{r-2}} \cdot M(\mF,q) \lesssim \inparen{\frac{\eta}{2}}^{-\frac{2r}{r-2}} \cdot r\Fstar\log\Fstar,
\end{align*}
and in the more general case,
\begin{align*}
    \log\cN\inparen{\widehat{B_2}, \eta B_r} \lesssim \inparen{\frac{\eta}{2}}^{-\frac{2r}{r-2}} \cdot M(\mF,q) \lesssim \inparen{\frac{\eta}{2}}^{-\frac{2r}{r-2}} \cdot \frac{2r^2}{r-2}\max_i \min(p_i+1, \log\abs{S_i})\Fstar\log\Fstar.
\end{align*}
This concludes the proof of \Cref{lemma:r_to_two_complicated}.
\end{proof}

\subsubsection{Putting everything together}

We are finally ready to combine all the tools we have built in the last few subsections to prove our entropy estimate when $0 < p < 2$.

Below, we state \Cref{lemma:covering_two_to_one}, which more precisely characterizes the behavior of the dependence on $p$ referred to by \Cref{lemma:covering_two_to_one_clean}. 

\begin{lemma}
\label{lemma:covering_two_to_one}
We have
\begin{align*}
    \log\cN(B_p,\eta \widehat{B_2}) \lesssim \eta^{-\frac{2p}{2-p}} \cdot C(p)\max_{i}\min\inparen{p_i+1,\log\abs{S_i}}\Fstar\log\Fstar,
\end{align*}
where $C(p)$ is a constant that only depends on $p$. The constant $C(p)$ is defined as follows. For $0 < \theta < p$, let $r=(2-\theta)p/(p-\theta)$. Then, we define
\begin{align}
    \widehat{C}(p,\theta) &\coloneqq \inparen{\frac{\inparen{2\cdot 8^{2/\theta}}^{-\frac{\theta}{2-\theta}}}{2}}^{-\frac{2r}{r-2}}\nonumber \\
    \widehat{\widehat{C}}(p) &\coloneqq \begin{cases} \min\inbraces{\widehat{C}(p,p/2), \widehat{C}(p,1)} & \text{ if } 1 \le p < 2 \\ \widehat{C}(p,p/2) & \text{ if } 0 < p < 1 \end{cases}\label{eq:hathat} \\
    C(p) &\coloneqq \widehat{\widehat{C}}(p)\begin{cases} r & \text{ if } \abs{S_i} = 1 \text{ for all } i \\ 2r + \frac{4r}{r-2} & \text{ otherwise} \end{cases}\label{eq:choosing_c_from_hathat},
\end{align}
where, in an abuse of notation, $r$ in \eqref{eq:choosing_c_from_hathat} is chosen according to the value of $\theta$ that is selected by $\widehat{\widehat{C}}(p)$ in \eqref{eq:hathat}.
\end{lemma}
\begin{proof}[Proof of \Cref{lemma:covering_two_to_one} and \Cref{lemma:covering_two_to_one_clean}]
This time, following \Cref{lemma:r_to_two_complicated}, we define
\begin{align*}
    M(\mF,r) &\coloneqq \begin{cases} r\Fstar\log\Fstar & \text{ if } \abs{S_i}=1 \\ (2r + \frac{4r}{r-2})\inparen{\max_i \min(p_i+1, \log\abs{S_i})}\Fstar\log\Fstar & \text{ otherwise} \end{cases}.
\end{align*}
Use \Cref{lemma:poor_mans_dualization} and \Cref{lemma:r_to_two} to write
\begin{align}
    \log\cN(B_p,\eta \widehat{B_2}) &\le \sum_{h\ge 0} \log\cN(\widehat{B_2},\delta_hB_r) \le M(\mF,r)\sum_{h \ge 0} \inparen{\frac{\delta_h}{2}}^{-\frac{2r}{r-2}}\nonumber \\
    &= M(\mF,r) \sum_{h \ge 0} \inparen{\eta^{\frac{\theta}{2-\theta}}\cdot 8^{(h+1)\cdot\inparen{\frac{\theta}{2-\theta}}} \cdot \frac{\inparen{2\cdot 8^{2/\theta}}^{-\frac{\theta}{2-\theta}}}{2}}^{-\frac{2r}{r-2}} \nonumber \\
    &= \eta^{-\frac{\theta}{2-\theta}\cdot\frac{2r}{r-2}} \cdot \underbrace{\inparen{\frac{\inparen{2\cdot 8^{2/\theta}}^{-\frac{\theta}{2-\theta}}}{2}}^{-\frac{2r}{r-2}}}_{= \widehat{C}(p,\theta)} M(\mF,r)\sum_{h \ge 0} 8^{(h+1) \cdot \inparen{-\frac{\theta}{2-\theta}}\cdot\frac{2r}{r-2}} \label{eq:substitute_theta}.
\end{align}
We now make the substitution $\theta=p/2$. By the formula in \Cref{lemma:poor_mans_dualization}, this means that $r = 4-p$ and $-\theta/(2-\theta) \cdot 2r/(r-2) = -2p/(2-p)$. We continue.
\begin{align*}
    \log\cN(B_p,\eta B_2) &\le \eta^{-\frac{\theta}{2-\theta}\cdot\frac{2r}{r-2}} \cdot \widehat{C}(p,\theta) M(\mF,r)\sum_{h \ge 0} 8^{(h+1) \cdot -\frac{\theta}{2-\theta}\cdot\frac{2r}{r-2}} \\
    &= \eta^{-\frac{2p}{2-p}} \cdot \widehat{C}(p,p/2) M(\mF,4-p)\sum_{h \ge 0} 8^{(h+1) \cdot -\frac{2p}{2-p}} \\
    &\lesssim \eta^{-\frac{2p}{2-p}} \cdot \widehat{C}(p,p/2) M(\mF,4-p).
\end{align*}
A regrettable consequence of the above calculation is that the ``constant'' $C(p,\theta)=C(p,p/2)$ explodes as $p\rightarrow 2$, as observed by \textcite{sz01}. To fix this, we perform a slightly different variant of this calculation in the regime where $1 < p < 2$. We resume from \eqref{eq:substitute_theta} except we use $\theta = 1$. Here, again using \Cref{lemma:poor_mans_dualization}, we check that $r=p/(p-1)$ and $-\theta/(2-\theta) \cdot 2r/(r-2) = -2p/(2-p)$ (note that now $r$ is the conjugate exponent of $p$). This means that
\begin{align*}
    \log\cN(B_p,\eta \widehat{B_2}) &\le \eta^{-\frac{\theta}{2-\theta}\cdot\frac{2r}{r-2}} \cdot \widehat{C}(p,\theta) M(\mF,r)\sum_{h \ge 0} 8^{(h+1) \cdot -\frac{\theta}{2-\theta}\cdot\frac{2r}{r-2}} \\
    &\lesssim \eta^{-\frac{2p}{2-p}} \cdot \widehat{C}(p,1)\sum_{h \ge 0} 8^{(h+1) \cdot -\frac{2p}{2-p}} \lesssim \eta^{-\frac{2p}{2-p}}M\inparen{\mF,\frac{p}{p-1}}.
\end{align*}
Taking the minimum over all the cases (of course where applicable) and expanding out the definition of $M(\mF,r)$ concludes the proof of \Cref{lemma:covering_two_to_one} and \Cref{lemma:covering_two_to_one_clean}.
\end{proof}

\subsection{Covering numbers for \texorpdfstring{$p \ge 2$}{p > 2}}

We will see that compared to the previous section, our task when $p\ge 2$ is far easier. The main technical lemma we need is \Cref{lemma:covering_poly_to_two}, which we need for both regimes of $p$.

We first state and prove \Cref{lemma:covering_poly_to_two}.

\begin{lemma}
\label{lemma:covering_poly_to_two}
Let $\mW$ and $\mLambda$ be chosen according to \Cref{thm:general_covering}. Suppose that $H \ge 1$ is such that $H\rho_i \ge \sfrac{\alpha_i^p}{\norm{\valpha}_p^p}$ for all $i \in [m]$. Then,
\begin{align*}
    \log\cN\inparen{\widehat{B_2}, \eta\inbraces{\vx\in\R^d \suchthat \polynorm{\vx} \le 1}} \lesssim \eta^{-2}\cdot{\underset{i \in S}{\max}\ \min\inbraces{p_i,\log\abs{S_i}}H^{2/p}\norm{\valpha}_p^2\log\mtilde}.
\end{align*}
\end{lemma}
\begin{proof}[Proof of \Cref{lemma:covering_poly_to_two}]
Following the proof of \Cref{lemma:r_to_two} and the references therein, our goal here is to analyze the quantity
\begin{align*}
    \log\cN\inparen{\inbraces{\vx\in\R^d \suchthat \norm{\mW^{1/2}\mLambda^{1/2-1/p}\mA\vx}_2 \le 1}, \eta\inbraces{\vx\in\R^d \suchthat \polynorm{\vx} \le 1}}.
\end{align*}
Using the same type of linear transformation argument as in \Cref{lemma:r_to_two} (so, replacing every $\vx$ above with $\mR\vx$), we find that it is in fact sufficient to analyze
\begin{align*}
    \log\cN \coloneqq \log\cN\inparen{\inbraces{\vx \suchthat \norm{\mU\vx}_2 \le 1}, \eta\inbraces{\vx \suchthat \max_{i \in S}\ \rho_i^{-1/p}\norm{\mW_{S_i}^{-1/2}\mLambda^{1/p-1/2}_{S_i}\mU\vx}_{p_i} \le 1}}.
\end{align*}
Recall that $\norm{\mU\vx}_2 = \norm{\vx}_2$, so a natural plan is to apply the dual Sudakov inequality (\Cref{fact:sudakov}, \eqref{eq:sudakov_dual}). We first consider the quantity (when $2 \le p_i \le \log\abs{S_i}$)
\begin{align*}
    \exvv{\vg\sim\cN(0,\mI_d)}{\norm{\mW_{S_i}^{-1/2}\mLambda^{1/p-1/2}_{S_i}\mU\vg}_{p_i}} &= \exvv{\vg\sim\cN(0,\mI_d)}{\inparen{\sum_{j \in S_i} \abs{\lambda_i^{1/p}\ip{w_{j}^{-1/2}\lambda_i^{-1/2}\vu_j,\vg}}^{p_i}}^{1/p_i}} \\
    &\le \inparen{\sum_{j\in S_i} \inparen{\lambda_i^{1/p}\norm{w_j^{-1/2}\lambda_i^{-1/2}\vu_j}_2}^{p_i}\exvv{g\sim\cN(0,1)}{\abs{g}^{p_i}}}^{1/p_i} \\
    &\lesssim p_i^{1/2} \inparen{\sum_{j\in S_i} \inparen{\lambda_i^{1/p}\norm{w_j^{-1/2}\lambda_i^{-1/2}\vu_j}_2}^{p_i}}^{1/p_i} \\
    &= p_i^{1/2}\inparen{\sum_{j \in S_i} \inparen{\lambda_i^{1/p}\norm{\vf_j}_2}^{p_i}}^{1/p_i} \\
    &= p_i^{1/2}\lambda_i^{1/p}\inparen{\sum_{j \in S_i} \norm{\vf_j}_2^{p_i}}^{1/p_i} \\
    &= p_i^{1/2}\alpha_i.
\end{align*}
On the other hand, if $p_i \ge \log\abs{S_i}$, then we get
\begin{align*}
    \exvv{\vg\sim\cN(0,\mI_d)}{\norm{\mW^{-1/2}_{S_i}\mLambda^{1/p-1/2}_{S_i}\mU\vg}_{p_i}} &\lesssim \exvv{\vg\sim\cN(0,\mI_d)}{\norm{\mW^{-1/2}_{S_i}\mLambda^{1/p-1/2}_{S_i}\mU\vg}_{\infty}} \\
    &\lesssim  \lambda_i^{1/p}\max_{j\in S_i}\norm{\vf_j}_2\sqrt{\log\abs{S_i}} \asymp\alpha_i\sqrt{\log\abs{S_i}}.
\end{align*}
Finally, when $p_1=\dots=p_m=p < 2$, we have
\begin{align*}
    \exvv{\vg\sim\cN(0,\mI_d)}{\norm{\widehat{\mLambda}_{S_i}^{1/p-1/2}\mU\vg}_p} &= \exvv{\vg\sim\cN(0,\mI_d)}{\inparen{\sum_{j \in S_i}\widehat{\lambda}_j\abs{\ip{\widehat{\lambda}_j^{-1/2}\vu_j,\vg}}^{p}}^{1/p}} \\
    &\le \inparen{\sum_{j\in S_i} \widehat{\lambda}_j\norm{\widehat{\lambda}_j^{-1/2}\vu_j}_2^p\exvv{g\sim\cN(0,1)}{\abs{g}^{p}}}^{1/p} \lesssim p^{1/2}\inparen{\sum_{j \in S_i} \widehat{\alpha}_j^p}^{1/p} = p^{1/2}\alpha_i.
\end{align*}
This means that throughout the rest of the proof, we assume without loss of generality that $p_i \le \log\abs{S_i}$. Next, observe that for any $\vx\in\R^d$ and when $p_1,\dots,p_m \ge 2$,
\begin{align*}
    \norm{\mW^{-1/2}_{S_i}\mLambda^{1/p-1/2}_{S_i}\mU\vx}_{p_i} = \inparen{\sum_{j \in S_i} \abs{\lambda_i^{1/p}\ip{\vf_j,\vx}}^{p_i}}^{1/p_i} \le \inparen{\sum_{j \in S_i} \abs{\lambda_i^{1/p}\norm{\vf_j}_2\norm{\vx}_2}^{p_i}}^{1/p_i} = \alpha_i\norm{\vx}_2.
\end{align*}
Similarly, when $p_1=\dots=p_m=p$, we first observe
\begin{align*}
    \abs{\ip{\widehat{\lambda}_j^{1/p-1/2}\vu_j,\vx}} \le \widehat{\alpha}_j\norm{\vx}_2.
\end{align*}
We therefore get
\begin{align}
    \norm{\widehat{\mLambda}_{S_i}^{1/p-1/2}\mU\vx}_p \le \inparen{\sum_{j \in S_i} \widehat{\alpha}_j^p}^{1/p}\norm{\vx}_2 = \alpha_i\norm{\vx}_2.\label{eq:alpha_alleq_lipschitz}
\end{align}

Hence, after applying \Cref{fact:lip_concentration} and \Cref{fact:uncentering}, we notice that $\alpha_i(H\rho_i)^{-1/p}\le\norm{\valpha}_p$ and get
\begin{align*}
    \subgnorm{\rho_i^{-1/p}\norm{\mLambda^{1/p-1/2}_{S_i}\mU\vg}_{p_i}} \lesssim \inparen{1 + \sqrt{p_i}}H^{1/p}\norm{\valpha}_p.
\end{align*}
All of this implies that for any subset $S$ of size $\mtilde$ (see Exercise 2.5.10 of \cite{vershynin_2018}),
\begin{align*}
    \exvv{\vg\sim\cN(0,\mI_d)}{\max_{i \in S} \rho_i^{-1/p}\norm{\mLambda^{1/p-1/2}_{S_i}\mU\vg}_{p_i}} \lesssim \max_{i \in S} p_i^{1/2}H^{1/p}\norm{\valpha}_p\sqrt{\log\mtilde}.
\end{align*}
Thus,
\begin{align*}
    \log\cN \lesssim \eta^{-2}\inparen{\inparen{\max_{i \in S} p_i+1}H^{2/p}\norm{\valpha}_p^2\log\mtilde}.
\end{align*}
This concludes the proof of \Cref{lemma:covering_poly_to_two}.
\end{proof}

We are finally ready to prove \Cref{thm:general_covering}.

\begin{proof}[Proof of \Cref{thm:general_covering}]
For notational simplicity in this proof, write
\begin{align*}
    K \coloneqq \inbraces{\vx\in\R^d\suchthat \polynorm{\vx} \le 1}.
\end{align*}
Let us first handle the case where $p \ge 2$. By our choice of $\mW$, we have for all $\vx\in\R^d$ that
\begin{align*}
     \norm{\mW^{1/2}\mLambda^{1/2-1/p}\mA\vx}_{2} \le \gnorml{\mLambda^{-1/p}\mA\vx}{p} = \gnorm{\mA\vx}{p}.
\end{align*}
This implies the containment $B_p \subseteq \widehat{B_2}$, and thus
\begin{align*}
    \log\cN\inparen{B_p, \eta K} \le \log\cN\inparen{\widehat{B_2}, \eta K} \le \eta^{-2}\max_{i}\min\inbraces{p_i,\log\abs{S_i}}H^{2/p}\Fstar\log\mtilde.
\end{align*}
The desired result now follows immediately from \Cref{lemma:covering_poly_to_two}.

For the case where $p < 2$ and $p_i \ge 2$ for all $i$, we require a bit more work. By our choice of $\mW$, we have
\begin{align*}
    \norm{\mW^{1/2}\mLambda^{1/2-1/p}\mA\vx}_{2} \le \gnorml{\mLambda^{-1/p}\mA\vx}{2} = \norm{\mLambda^{1/2-1/p}\mA\vx}_2,
\end{align*}
and for any $t > 0$ (after remembering \eqref{eq:alpha_to_fstar} which tells us $\norm{\valpha}_p \le \sqrt{\Fstar}$),
\begin{align*}
    \log\cN\inparen{B_p, \eta K} &\le \log\cN\inparen{B_p, t\widehat{B_2}} + \log\cN\inparen{t\widehat{B_2}, \eta K} = \log\cN\inparen{B_p, t\widehat{B_2}} + \log\cN\inparen{\widehat{B_2}, \frac{\eta}{t}\cdot K} \\
    &\le t^{-\frac{2p}{2-p}} \cdot C(p)\max_{i}\min\inbraces{p_i+1,\log\abs{S_i}}\Fstar\log\Fstar \\
    &\quad+\inparen{\frac{\eta}{t}}^{-2}\max_{i}\min\inbraces{p_i+1,\log\abs{S_i}}H^{2/p}\Fstar\log\mtilde,
\end{align*}
where the last line follows from \Cref{lemma:covering_two_to_one_clean} and \Cref{lemma:covering_poly_to_two}.
Choose $t=\eta^{1-p/2} \cdot H^{1/2-1/p}$, and for simplicity let $p^{\star} \coloneqq \max_i\min\inbraces{p_i+1,\log\abs{S_i}}$. We write
\begin{align*}
    \log\cN\inparen{B_p, \eta K} &\le t^{-\frac{2p}{2-p}} \cdot C(p)p^{\star}\Fstar\log\Fstar +\inparen{\frac{\eta}{t}}^{-2}p^{\star}H^{2/p}\Fstar\log\mtilde \\
    &\lesssim \eta^{-p}\cdot H \cdot C(p)\inparen{p^{\star}\Fstar\log\max\inbraces{\mtilde,\Fstar}}.
\end{align*}
Finally, we need to address the case where $p_1=\dots=p_m=p < 2$, as this is not covered by the construction of the block Lewis weights (\Cref{lemma:block_lewis}).

To do so, notice that $\gnorm{\mA\vx}{p} = \norm{\mA\vx}_p$ for all $\vx\in\R^d$. So, we reduce to the case where all the $S_i$ have size $1$. In particular, let $\widehat{\vlambda}$ be a probability measure over $[k]$ such that for all $j\in [k]$,
\begin{align*}
    \Fstar \ge \frac{\tau_j\inparen{\widehat{\mLambda}^{1/2-1/p}\mA}}{\widehat{\lambda}_j},
\end{align*}
and let
\begin{align*}
    \widehat{B_2} = \inbraces{\vx\in\R^d\suchthat \norm{\widehat{\mLambda}^{1/2-1/p}\mA\vx}_2 \le 1}.
\end{align*}
Notice that this setup is in accordance with \Cref{defn:alpha}. By \Cref{lemma:covering_two_to_one_clean}, we get
\begin{align*}
    \log\cN(B_p,t\widehat{B_2}) \lesssim t^{-\frac{2p}{2-p}} \cdot C(p)\Fstar\log\Fstar.
\end{align*}
It remains to bound $\log\cN(\widehat{B_2},(\eta/t)K)$. Let $\vlambda$ be a probability measure over $[m]$, where $\lambda_i = \sum_{j \in S_i} \widehat{\lambda}_j$. Define $\widehat{\valpha} \in \R^n$ analogously to $\widehat{\vlambda}$. Notice that, for these choices of $\widehat{\valpha}$ and $\widehat{\vlambda}$, we see that the conclusion of \Cref{lemma:covering_poly_to_two} still holds, and we have
\begin{align*}
    \log\cN\inparen{\widehat{B_2},\inparen{\frac{\eta}{t}} \cdot K} \lesssim \inparen{\frac{\eta}{t}}^{-2}H^{2/p}\Fstar\log\mtilde.
\end{align*}
Now, the calculation is the same as before, and we conclude the proof of \Cref{thm:general_covering}.
\end{proof}

\subsection{Volume-based metric entropy}

In this subsection, we prove \Cref{lemma:covering_easy}, which is an easy consequence of a volume-based argument to obtain a covering number guarantee.

We start with \Cref{lemma:containments}.

\begin{lemma}
\label{lemma:containments}
Let $S \subseteq [m]$ have size $\mtilde$. Let $H \ge 1$ be such that $H\rho_i \ge \sfrac{\alpha_i^p}{\norm{\valpha}_p^p}$ for all $i \in [m]$. For all $\vx\in\R^d$, we have
\begin{align*}
    \max_{i \in S} \frac{\norm{\mA_{S_i}\vx}_{p_i}}{\inparen{\Fstar}^{\max(1/2,1/p)}\rho_i^{1/p}H^{1/p}} &\le \gnorm{\mA\vx}{p}.
\end{align*}
\end{lemma}
\begin{proof}[Proof of \Cref{lemma:containments}]
Consider the invertible mapping $\vx \mapsto \mR\vx$, and write
\begin{align}
    \norm{\mW^{-1/2}_{S_i}\mLambda^{1/p-1/2}_{S_i}\mU\vx}_{p_i} = \inparen{\sum_{j \in S_i} \abs{\lambda_i^{1/p}\ip{\vf_j,\vx}}^{p_i}}^{1/p_i} \le \inparen{\sum_{j \in S_i} \abs{\lambda_i^{1/p}\norm{\vf_j}_2\norm{\vx}_2}^{p_i}}^{1/p_i} = \alpha_i\norm{\vx}_2.\label{eq:alpha_importance_proof}
\end{align}
This means that when $p \ge 2$,
\begin{align*}
    \norm{\mA_{S_i}\vx}_{p_i} \le \alpha_i\norm{\mW^{1/2}\mLambda^{1/2-1/p}\mA\vx}_2 \le \alpha_i\gnorm{\mLambda^{1/2-1/p}\mA\vx}{2} \le \alpha_i\gnorm{\mA\vx}{p} \le \norm{\valpha}_p\rho_i^{1/p}H^{1/p}\gnorm{\mA\vx}{p}.
\end{align*}
Dividing both sides by $\norm{\valpha}_p\rho_i^{1/p}H^{1/p}$ and then recalling \eqref{eq:alpha_to_fstar} yields the desired conclusion (in particular, we see that $\norm{\valpha}_p \le \inparen{\Fstar}^{1/2}$).

We now analyze what happens when $p \le 2$ and $p_1,\dots,p_m \ge 2$. Let
\begin{align*}
    \Delta_i^{1/2} \coloneqq \max_{\vx\in\R^d\setminus\inbraces{0}}\frac{\lambda_i^{-1/p}\norm{\mA_{S_i}\vx}_{p_i}}{\gnorm{\mLambda^{1/2-1/p}\mA\vx}{2}}.
\end{align*}
Let $\Delta \coloneqq \max_{i \in [m]} \Delta_i$. We now see that
\begin{align*}
    &\ \gnorm{\mLambda^{1/2-1/p}\mA\vx}{2} = \inparen{\sum_{i=1}^m \lambda_i\norm{\mLambda_{S_i}^{-1/p}\mA_{S_i}\vx}_{p_i}^2}^{1/2} = \inparen{\sum_{i=1}^m \lambda_i\norm{\mLambda_{S_i}^{-1/p}\mA_{S_i}\vx}_{p_i}^{p}\cdot\norm{\mLambda_{S_i}^{-1/p}\mA_{S_i}\vx}_{p_i}^{2-p}}^{1/2} \\
    &\le \inparen{\gnorm{\mA\vx}{p}^p\inparen{\Delta^{1/2} \cdot \gnorm{\mLambda^{1/2-1/p}\mA\vx}{2}}^{2-p}}^{1/2} = \gnorm{\mA\vx}{p}^{p/2}\inparen{\Delta^{1/2} \cdot \gnorm{\mLambda^{1/2-1/p}\mA\vx}{2}}^{1-p/2}.
\end{align*}
Rearranging and taking the $2/p$-power gives
\begin{align*}
    \gnorm{\mLambda^{1/2-1/p}\mA\vx}{2} \le \gnorm{\mA\vx}{p} \Delta^{1/p-1/2}.
\end{align*}
Observe that this yields the inequalities
\begin{align*}
     \frac{\norm{\mA_{S_i}\vx}_{p_i}}{\Delta^{1/p-1/2}} &\le \frac{\alpha_i}{\Delta^{1/p-1/2}}\norm{\mW^{1/2}\mLambda^{1/2-1/p}\mA\vx}_2 \le \frac{\alpha_i}{\Delta^{1/p-1/2}}\gnorm{\mLambda^{1/2-1/p}\mA\vx}{2} \\
     &\le \alpha_i\gnorm{\mA\vx}{p} \le \norm{\valpha}_p\rho_i^{1/p}H^{1/p}\gnorm{\mA\vx}{p}.
\end{align*}
It remains to bound the $\Delta_i$. By the same sort of argument from the $p\ge 2$ case (i.e., \eqref{eq:alpha_importance_proof}), we get
\begin{align*}
    \lambda_i^{1/p}\Delta_i^{1/2} \le \alpha_i = \lambda_i^{1/p}\inparen{\sum_{j \in S_i} \norm{\vf_j}_2^{p_i}}^{1/p_i}
\end{align*}
which means that
\begin{align*}
    \Delta \le \max_{i \in [m]} \inparen{\sum_{j \in S_i} \norm{\vf_j}_2^{p_i}}^{2/p_i} \le \Fstar.
\end{align*}
Now, again using the fact that $\norm{\valpha}_p \le \inparen{\Fstar}^{1/2}$, we see that $\Delta^{1/p-1/2}\norm{\valpha}_p \le \inparen{\Fstar}^{1/p}$. 

Finally, we analyze the case where $p_1=\dots=p_m=p < 2$. Using \eqref{eq:alpha_alleq_lipschitz}, and the same sort of argument from above, we have
\begin{align*}
    \norm{\mA_{S_i}\vx}_p \le \alpha_i\norm{\widehat{\mLambda}^{1/2-1/p}\mA\vx}_2 \le \alpha_i\inparen{\Fstar}^{1/p-1/2}\gnorm{\mA\vx}{p} \le \norm{\valpha}_pH^{1/p}\rho_i^{1/p}\inparen{\Fstar}^{1/p-1/2}\gnorm{\mA\vx}{p}.
\end{align*}
Once again, we use $\norm{\valpha}_p \le \inparen{\Fstar}^{1/2}$.

We have covered all our cases and may conclude the proof of \Cref{lemma:containments}.
\end{proof}

\Cref{lemma:containments} also suggests a useful sanity check, as the denominator points to a sparsity of $\sum_{i \le m} \inparen{\inparen{\Fstar}^{\max(1/2,1/p)}\rho_i^{1/p}H^{1/p}}^p = H\inparen{\Fstar}^{\max(1,p/2)}$. And, recall that we should be able to set $\Fstar \sim d$, which indeed gives us the dependence on $d$ we see in \Cref{thm:one_shot_lewis}. 

\begin{lemma}
\label{lemma:covering_easy}
Let $S \subseteq [m]$ have size $\mtilde$. We have
\begin{align*}
    \log\cN\inparen{B_p,\eta\inbraces{\vx\in\R^d\suchthat\polynorm{\vx}\le 1}} \le n\logv{\frac{4H^{1/p}\inparen{\Fstar}^{\max(1/p,1/2)}}{\eta}}
\end{align*}
\end{lemma}
\begin{proof}[Proof of \Cref{lemma:covering_easy}]
Define
\begin{align*}
    K &\coloneqq \inbraces{\vx\in\R^d\suchthat\polynorm{\vx}\le 1}.
\end{align*}
Let $C$ be a value such that for all $i \in S$ and $\vx\in\R^d$, we have $(C\rho_i)^{-1/p}\norm{\mA_{S_i}\vx}_{p_i} \le \gnorm{\mA\vx}{p}$. This means that $B_p \subseteq C^{1/p} K$. Then,
\begin{align*}
    \log\cN\inparen{B_p,\eta K} \le \log\cN\inparen{C^{1/p} K, \eta K} = \log\cN\inparen{K, \frac{\eta}{C^{1/p}} \cdot K} \le n\logv{\frac{4C^{1/p}}{\eta}}.
\end{align*}
By \Cref{lemma:containments}, when $p\ge 2$, we can choose $C = H\inparen{\Fstar}^{\max(1/p,1/2)}$.
This concludes the proof of \Cref{lemma:covering_easy}. 
\end{proof}

%%%%%%%%%%%%%%%%%%%%%%%%%%%%%%%%%%%%%%%%%%%%%%%%%%%%%%%%%%%%%%%%%%%%%%%%%%%%%%%%%%%%%%%%%%%
% \newpage
\section{Concentration analysis}
\label{sec:generic_chaining}

In this section, we prove Theorem \ref{thm:concentration}. Theorem \ref{thm:concentration} states our main result in its fullest generality. \Cref{thm:one_shot_lewis} follows easily from this, as we show in \Cref{sec:applications}.

We first state \Cref{thm:concentration}.

\begin{restatable}[General concentration result]{theorem}{concentration}
\label{thm:concentration}
Let $\cG = (\mA \in \R^{n \times d}, S_1,\dots,S_m, p_1,\dots,p_m)$ where $S_1,\dots,S_m$ form a partition of $[n]$. Suppose at least one of the following holds:
\begin{itemize}
    \item $1 \le p < \infty$ and $p_1,\dots,p_m \ge 2$;
    \item $1/\log d \le p_1 = \dots = p_m = p < \infty$;
    \item $p_1 = \dots = p_m = 2$ and $1/\log n \le p < \infty$.
\end{itemize}
Let $P \coloneqq \max\inparen{1, \max_{i \in [m]} \min(p_i,\log\abs{S_i})}$. Suppose that $\vlambda \in \R^m$ is a probability measure over $[m]$, let $\mW$ be a rounding matrix (\Cref{defn:rounding_matrix}) such that we get an $\Fstar$-block Lewis overestimate (\Cref{defn:block_lewis_overestimate}), and define $\valpha$ according to \Cref{defn:alpha}. Let $\cD = \inparen{\rho_1,\dots,\rho_m}$ be a probability distribution over $[m]$ and $H \ge 1$ be such that $H\rho_i \ge \alpha_i^p/\norm{\valpha}_p^p$.

If
\begin{align*}
    \mtilde &= \Omega\inparen{\logv{\sfrac{1}{\delta}}\vbrho \cdot H \cdot P \inparen{\Fstar}^{\max(1,p/2)}},
\end{align*}
and if we sample $\cM \sim \cD^{\mtilde}$, then, with probability $\ge 1-\delta$, we have:
\begin{align*}
    \text{for all } \vx \in \R^d,\quad (1-\eps)\gnorm{\mA\vx}{p}^p \le \frac{1}{\mtilde}\sum_{i \in \cM} \frac{1}{\rho_i} \cdot \norm{\mA_{S_i}\vx}_{p_i}^p \le (1+\eps)\gnorm{\mA\vx}{p}^p
\end{align*}
\end{restatable}
The goal of the rest of this section is to prove \Cref{thm:concentration}. It may be helpful to recall the argument sketch given in \Cref{sec:sparsification_overview_concentration}. 

To formalize the idea given there, we first introduce the following notation (recall that $\rho_i$ is the probability that we choose group $i$ in a round of sampling and that \Cref{defn:entropy_numbers} defines the $e_N$).
\begin{align}
    g_i(\vx) &\coloneqq \norm{\mA_{S_i}\vx}_{p_i}\label{eq:chaining_g_defn} \\
    \dtwo(\vx, \xhat) &\coloneqq \inparen{\sum_{h=1}^{\mtilde} \inparen{\frac{g_{i_h}(\vx)^p}{\rho_{i_h}} - \frac{g_{i_h}(\xhat)^p}{\rho_{i_h}}}^2}^{1/2}\label{eq:chaining_dtwo_defn} \\
    \gamma_2(B_{p},\dtwo) &\coloneqq \inf_{\substack{\abs{T_N} \le 2^{2^N};\\ T_N  \subset B_{p}}} \sup_{\vx \in B_{p}} \sum_{N \ge 0} 2^{N/2} \cdot d_2(\vx,T_N)\label{eq:chaining_gamma2_defn}
\end{align}
The goal is to control $\gamma_2(B_p, \dtwo)$. This quantity represents the worst-case approximation error that one incurs by using the discretization scheme given by the $T_N$, where the discretization is taken with respect to the $\dtwo$ metric.

Towards this goal, we first apply a standard symmetrization reduction. Informally, this reduction (\Cref{lemma:symmetrization}) states that it is enough to analyze the average fluctuations of a Rademacher average of any set of $\mtilde$ (not necessarily distinct) reweighted groups.
\begin{lemma}[Symmetrization reduction]
\label{lemma:symmetrization} 
Let $R_1,\dots,R_{\mtilde}$ be independent Rademacher random variables (i.e., $\mathsf{Unif}\inparen{\pm 1}$). We have
\begin{align*}
    \exvv{\cG'}{\abs{\gnorm{\mA\vx}{p}^p - \norm{\vx}_{\cG'}^p}} \le 2 \underset{\cG'}{\E} \exvv{R_1, \ldots, R_{\mtilde}}{\frac{1}{\mtilde}\abs{\sum_{h=1}^{\mtilde} R_h \frac{g_{i'_h}(\vx)^p}{\rho_{i'_h}}}}.
\end{align*}
\end{lemma}
\begin{proof}[Proof of \Cref{lemma:symmetrization}]
We follow the proof of Lemma 3.1 due to \textcite{lee22}. Let \(\tilde{\cG}\) be an independent copy of \(\cG'\). Fixing \(\cG'\), we have by Jensen's inequality that
\begin{align*}
    \abs{\gnorm{\mA\vx}{p}^p - \norm{\vx}_{\cG'}^p} &= \abs{\exvv{\tilde{\cG}}{\norm{\vx}_{\tilde{\cG}}^p} - \norm{\vx}_{\cG'}^p}
    \leq \exvv{\tilde{\cG}}{\abs{\norm{\vx}_{\tilde{\cG}}^p - \norm{\vx}_{\cG'}^p}}.
\end{align*}
Thus, taking expectation over \(\cG'\),
\begin{align*}
    \exvv{\cG'}{\abs{\gnorm{\mA\vx}{p}^p - \norm{\vx}_{\cG'}^p}} &\leq \exvv{\cG', \tilde{\cG}}{\abs{\norm{\vx}_{\tilde{\cG}}^p - \norm{\vx}_{\cG'}^p}}
    = \exvv{\cG', \tilde{\cG}}{\frac{1}{\mtilde}\abs{\sum_{h=1}^{\mtilde} \frac{g_{\tilde{i}_h}(\vx)^p}{\rho_{\widetilde{i}_h}} -\frac{g_{i'_h}(\vx)^p}{\rho_{i'_h}}}}.
\end{align*}
Observe that \(\frac{g_{\tilde{i}_h}(\vx)^p}{\rho_{\widetilde{i}_h}} -\frac{g_{i'_h}(\vx)^p}{\rho_{i'_h}}\) is symmetric,
and so is distributed the same as \(R_h \inparen{\frac{g_{\tilde{i}_h}(\vx)^p}{\rho_{\widetilde{i}_h}} -\frac{g_{i'_h}(\vx)^p}{\rho_{i'_h}}}\)
where \(R_h\) is an independent Rademacher variable. Then,
\begin{align*}
    \exvv{\cG', \tilde{\cG}}{\abs{\frac{1}{\mtilde}\sum_{h=1}^{\mtilde} \inparen{\frac{g_{\tilde{i}_h}(\vx)^p}{p_{\tilde{i}_h}} -\frac{g_{i'_h}(\vx)^p}{p_{i'_h}}}}} &= \underset{R_1, \ldots, R_{\mtilde}}{\E} \exvv{\cG', \tilde{\cG}}{\frac{1}{\mtilde}\abs{\sum_{h=1}^{\mtilde} R_h \inparen{\frac{g_{\tilde{i}_h}(\vx)^p}{\rho_{\widetilde{i}_h}} -\frac{g_{i'_h}(\vx)^p}{\rho_{i'_h}}}}} \\
    &\leq 2 \underset{\cG'}{\E} \exvv{R_1, \ldots, R_M}{\frac{1}{\mtilde}\abs{\sum_{h=1}^{\mtilde} R_h \frac{g_{i'_h}(\vx)^p}{\rho_{i'_h}}}}.
\end{align*}
This concludes the proof of \Cref{lemma:symmetrization}.
\end{proof}
With \Cref{lemma:symmetrization} in hand, we set up our chaining argument in \Cref{lemma:chaining_setup}. We first confirm that our random process is subgaussian with respect to our choice of $\dtwo$. 
\begin{lemma}[Choosing the distance]
\label{lemma:chaining_setup}
The random process $\sum_{h=1}^{\mtilde} R_{i_h}\cdot\sfrac{g_{i_h}(\vx)^p}{\rho_{i_h}}$  is subgaussian with respect to $\dtwo$ as defined in  \eqref{eq:chaining_dtwo_defn}.
\end{lemma}
\begin{proof}[Proof of \Cref{lemma:chaining_setup}]
Let
\begin{align*}
    P &\coloneqq \abs{\sum_{h=1}^{\mtilde} R_{h}\inparen{\frac{g_{i_h}(\vx)^p}{\rho_{i_h}} - \frac{g_{i_h}(\xhat)^p}{\rho_{i_h}}}}.
\end{align*}
Let us first calculate $\subgnorm{P}$. Using the fact that every term in this sum is independent and \Cref{fact:subgaussian_properties}, we get
\begin{align*}
    \subgnorm{P}^2 = \sum_{h=1}^{\mtilde} \subgnorm{R_{h}\inparen{\frac{g_{i_h}(\vx)^p}{\rho_{i_h}} - \frac{g_{i_h}(\xhat)^p}{\rho_{i_h}}}}^2 \le \sum_{h=1}^{\mtilde} 2\inparen{\frac{g_{i_h}(\vx)^p}{\rho_{i_h}} - \frac{g_{i_h}(\xhat)^p}{\rho_{i_h}}}^2
\end{align*}
By \Cref{fact:subgaussian_properties}, we have
\begin{align*}
    \prvv{R_{h}}{P \ge v\dtwo(\vx,\xhat)} &\le 2\expv{-\frac{v^2\dtwo(\vx,\xhat)^2}{4\sum_{h=1}^{\mtilde} \inparen{\frac{g_{i_h}(\vx)^p}{\rho_{i_h}} - \frac{g_{i_h}(\xhat)^p}{\rho_{i_h}}}^2}} = 2\expv{-\frac{v^2}{2}}.
\end{align*}
This concludes the proof of \Cref{lemma:chaining_setup}.
\end{proof}
\Cref{lemma:chaining_setup} tells us that $\dtwo$ is a choice of distance on $B_{p}$ that allows us to use the subgaussian form of chaining to analyze our random process. Along with the way we have set up our sampling process, we have enough to apply Theorem \ref{thm:generic_chaining_dirksen}. This is simply a restatement of Lemma 2.6 of \cite{jlls23} for our setting.

\begin{theorem}[Restatement of Lemma 2.6 from \cite{jlls23}, $\alpha=2$]
\label{thm:generic_chaining_dirksen}
Recall $\dtwo$ (\ref{eq:chaining_dtwo_defn}) and $\gamma_2$ (\ref{eq:chaining_gamma2_defn}). Suppose that for some $D$ and for every choice of $i_1,\dots,i_{\mtilde}$, we have
\begin{align*}
    \gamma_2\inparen{B_p,\frac{\dtwo}{\mtilde}} \lesssim D\inparen{\max_{\vx\in B_p} \norm{\vx}_{\cG'}^p}^{1/2},
\end{align*}
where
\begin{align*}
    \norm{\vx}_{\cG'}^p = \frac{1}{\mtilde}\sum_{h=1}^{\mtilde} \frac{g_{i_h}(\vx)^p}{\rho_{i_h}}.
\end{align*}
Then, we have the following.
\begin{align*}
    \exvv{\cD}{\sup_{\vx\in B_p} \abs{\gnorm{\mA\vx}{p}-\norm{\vx}_{\cG'}^p}} &\lesssim D.
\end{align*}
If we also have for all choices $i_1,\dots,i_{\mtilde}$ and for some $\widehat{D}$ that
\begin{align*}
    \mathsf{diam}\inparen{B_p, \frac{\dtwo}{\mtilde}} \lesssim \widehat{D}\inparen{\max_{\vx\in B_p} \norm{\vx}_{\cG'}^p}^{1/2},
\end{align*}
then there exists a universal constant $C > 0$ such that for all $0 \le t \le \sfrac{1}{2K\widehat{D}}$,
\begin{align*}
    \prvv{\cD}{\sup_{\vx\in B_p} \abs{\gnorm{\mA\vx}{p}^p-\norm{\vx}_{\cG'}^p} \ge C(D + t\widehat{D})} \le \expv{-\frac{Kt^2}{4}}.
\end{align*}
\end{theorem}

With Theorem \ref{thm:generic_chaining_dirksen} in our arsenal, our task becomes to compute $\gamma_2(B_p,\dtwo)$ (which we will then divide by $\mtilde$ so that we can apply \Cref{thm:generic_chaining_dirksen}). Hereafter, we will simply abbreviate $\gamma_2(B_p,\dtwo)$ as $\gamma_2$. We will first weaken the definition of $\gamma_2$, which is essentially equivalent to Dudley's integral. Recall the definition of the entropy numbers $e_N$ (\Cref{defn:entropy_numbers}) and notice that
\begin{align*}
    \gamma_2 \le \sum_{N \ge 0} 2^{N/2}e_N(B_{p},\dtwo). 
\end{align*}
We now rewrite $\dtwo$ in a form that will be more convenient for us.

\begin{lemma}
\label{lemma:concentration_g2_rewrite_d}
We have
\begin{align*}
    \dtwo(\vx, \xhat) \le \max(p,2) \inparen{\Fstar}^{\max(0,p/4-1/2)}{\mtilde}^{1/2} \cdot \polynorm{\vx-\xhat}^{\min(p/2,1)} \cdot \inparen{\max_{\vx \in B_{p}}\norm{\vx}_{\cG'}^p}^{1/2}
\end{align*}
and therefore
\begin{align*}
    \gamma_2 &\le {\mtilde}^{1/2}\max(p,2) \inparen{H^{1/p}\inparen{\Fstar}^{1/2}}^{\max(0,p/2-1)}\inparen{\max_{\vx \in B_{p}}\norm{\vx}_{\cG'}^p}^{1/2}\sum_{N \ge 0} 2^{N/2}e_N\inparen{B_p, \polynorm{\cdot}}^{\min(p/2,1)}.
\end{align*}
\end{lemma}
\begin{proof}[Proof of \Cref{lemma:concentration_g2_rewrite_d}]
We have two cases. We first address the case $0 < p < 2$. Recall that in this regime, we have $\abs{a^{p/2}-b^{p/2}} \le \abs{a-b}^{p/2}$. Since $g_{i_h}(\vx)$ is a norm, the triangle inequality tells us that $\abs{g_{i_h}(\vx) - g_{i_h}(\xhat)} \le g_{i_h}(\vx-\xhat)$. We use these and write
\begin{align*}
    \dtwo(\vx, \xhat) &= \inparen{\sum_{h=1}^{\mtilde} \inparen{\frac{g_{i_h}(\vx)^{p/2}}{\sqrt{\rho_{i_h}}} - \frac{g_{i_h}(\xhat)^{p/2}}{\sqrt{\rho_{i_h}}}}^2\inparen{\frac{g_{i_h}(\vx)^{p/2}}{\sqrt{\rho_{i_h}}} + \frac{g_{i_h}(\xhat)^{p/2}}{\sqrt{\rho_{i_h}}}}^2}^{1/2} \\
    &\le \inparen{\sum_{h=1}^{\mtilde} \inparen{\frac{g_{i_h}(\vx-\xhat)^{p/2}}{\sqrt{\rho_{i_h}}}}^2\inparen{\frac{g_{i_h}(\vx)^{p/2}}{\sqrt{\rho_{i_h}}} + \frac{g_{i_h}(\xhat)^{p/2}}{\sqrt{\rho_{i_h}}}}^2}^{1/2} \\
    &\le \polynorm{\vx-\xhat}^{p/2}\cdot\inparen{\sum_{h=1}^{\mtilde} \inparen{\frac{g_{i_h}(\vx)^{p/2}}{\sqrt{\rho_{i_h}}} + \frac{g_{i_h}(\xhat)^{p/2}}{\sqrt{\rho_{i_h}}}}^2}^{1/2} \\
    &\le 2{\mtilde}^{1/2} \cdot \polynorm{\vx-\xhat}^{p/2}\inparen{\max_{\vx \in B_{p}}\norm{\vx}_{\cG'}^p}^{1/2}
\end{align*}
which concludes the proof in the range $0 < p < 2$.

We now move onto the case where $p \ge 2$. Recall that by Lipschitzness, we get
\begin{align*}
    \abs{a^{p/2}-b^{p/2}} \le \frac{p}{2} \cdot \max(a,b)^{p/2-1}\abs{a-b}.
\end{align*}
Next, by \Cref{lemma:containments}, we know that for all $i$, 
\begin{align*}
    \norm{\mA_{S_i}\vx}_{p_i} \le \inparen{\Fstar}^{1/2}\rho_i^{1/p}H^{1/p}\gnorm{\mA\vx}{p} \le \inparen{\Fstar}^{1/2}\rho_i^{1/p}H^{1/p}.
\end{align*}
We use these to rewrite $\dtwo(\vx,\xhat)$.
\begin{align*}
    \dtwo(\vx, \xhat) &= \inparen{\sum_{h=1}^{\mtilde} \inparen{\frac{g_{i_h}(\vx)^{p/2}}{\sqrt{\rho_{i_h}}} - \frac{g_{i_h}(\xhat)^{p/2}}{\sqrt{\rho_{i_h}}}}^2\inparen{\frac{g_{i_h}(\vx)^{p/2}}{\sqrt{\rho_{i_h}}} + \frac{g_{i_h}(\xhat)^{p/2}}{\sqrt{\rho_{i_h}}}}^2}^{1/2} \\
    &\le \frac{p}{2}\cdot H^{1/2-1/p}\inparen{\Fstar}^{p/4-1/2}\inparen{\sum_{h=1}^{\mtilde} \inparen{\rho_{i_h}^{-1/p}g_{i_h}(\vx-\xhat)}^2\inparen{\frac{g_{i_h}(\vx)^{p/2}}{\sqrt{\rho_{i_h}}} + \frac{g_{i_h}(\xhat)^{p/2}}{\sqrt{\rho_{i_h}}}}^2}^{1/2} \\
    &\le \frac{p}{2}\cdot H^{1/2-1/p}\inparen{\Fstar}^{p/4-1/2}\polynorm{\vx-\xhat}\cdot\inparen{\sum_{h=1}^{\mtilde} \inparen{\frac{g_{i_h}(\vx)^{p/2}}{\sqrt{\rho_{i_h}}} + \frac{g_{i_h}(\xhat)^{p/2}}{\sqrt{\rho_{i_h}}}}^2}^{1/2} \\
    &\le p \cdot H^{1/2-1/p}\inparen{\Fstar}^{p/4-1/2}{\mtilde}^{1/2} \cdot \polynorm{\vx-\xhat}\inparen{\max_{\vx \in B_{p}}\norm{\vx}_{\cG'}^p}^{1/2}
\end{align*}
We conclude the proof of \Cref{lemma:concentration_g2_rewrite_d}.
\end{proof}

In light of \Cref{lemma:concentration_g2_rewrite_d}, observe that it is enough to calculate each term of the sum
\[
\sum_{N \ge 0} 2^{N/2}e_N(B_p, \polynorm{\cdot})^{\min(p/2,1)}.
\]
We begin this analysis with \Cref{lemma:concentration_g2_two_one_volume}.

\begin{lemma}
\label{lemma:concentration_g2_two_one_volume}
For all $N \ge 0$, we have
\begin{align*}
    2^{N/2}e_N\inparen{B_p, \polynorm{\cdot}}^{\min(p/2,1)} \le 2^{N/2 + \min(p/2,1)(2+(1/p)\log C - 2^N/d)},
\end{align*}
where $C$ is such that for all $i \in S$ and $\vx\in\R^d$, we have $\inparen{C\rho_i}^{-1/p}\norm{\mA_{S_i}\vx}_{p_i} \le \gnorm{\mA\vx}{p}$.
\end{lemma}
\begin{proof}[Proof of \Cref{lemma:concentration_g2_two_one_volume}]
Recall that by \Cref{lemma:covering_easy}, we have
\begin{align*}
    \log\cN\inparen{B_p,\eta\inbraces{\vx\in\R^d\suchthat\polynorm{\vx}\le 1}} \le d\logv{\frac{4C^{1/p}}{\eta}}.
\end{align*}
We set $\eta = 2^{2+(1/p)\log C - 2^N/d}$ so that
\begin{align*}
    \log\cN\inparen{B_p,\eta\inbraces{\vx\in\R^d\suchthat\polynorm{\vx}\le 1}} \le 2^N.
\end{align*}
Then,
\begin{align*}
    2^{N/2}e_N\inparen{B_p, \polynorm{\cdot}}^{\min(p/2,1)} &\le 2^{N/2 + \min(p/2,1)(2+(1/p)\log C - 2^N/d)},
\end{align*}
concluding the proof of Lemma \ref{lemma:concentration_g2_two_one_volume}.
\end{proof}

Using \Cref{lemma:concentration_g2_two_one_volume}, we get a rapidly converging tail in our summation for large values of $N$. See \Cref{lemma:g2_tail}.

\begin{lemma}
\label{lemma:g2_tail}
Let $N_S \coloneqq \logv{6d\log d}$. We have
\begin{align*}
    \sum_{N \ge 0} 2^{N/2}e_N\inparen{B_p, \polynorm{\cdot}}^{\min(p/2,1)} &\lesssim \sqrt{nH^{1/\max(2,p)}\inparen{\Fstar}^{1/2}\log d}\\
    &\quad+\sum_{N \le N_S} 2^{N/2}e_N\inparen{B_p, \polynorm{\cdot}}^{\min(p/2,1)}.
\end{align*}
\end{lemma}
\begin{proof}[Proof of \Cref{lemma:g2_tail}]
For now, let $C$ be such that for all $i \in S$ and $\vx\in\R^d$, we have $\inparen{C\rho_i}^{-1/p}\norm{\mA_{S_i}\vx}_{p_i} \le \gnorm{\mA\vx}{p}$.

Let $N_S$ be a threshold such that for all $N \ge N_S$, we use the entropy number bound given by \Cref{lemma:concentration_g2_two_one_volume} (as the volume-based covering number bound is much better for small values of $e_N$). Let us enforce the constraint $N_S \ge \ceil{\logv{\sfrac{3d}{\min(p,2)}}}$, so that the entropy number bound in \Cref{lemma:concentration_g2_two_one_volume} is decreasing in $N$ and is dominated from above by a geometric series with common ratio $\sfrac{1}{2}$.

We now set $N_S = \logv{6d\log d}$. Since $p \ge \sfrac{1}{\log d}$, we know that $N_S \ge \ceil{\logv{\sfrac{3d}{\min(p,2)}}}$. Now, since $2^{N/2}e_N$ is bounded above by a geometric series with common ratio $\sfrac{1}{2}$, the summation for all $N \ge N_S$ is dominated by the first term. Let us evaluate this. We first observe that
\begin{align*}
    \frac{N_S}{2} + \min\inparen{\frac{p}{2},1}\inparen{-\frac{2^{N_S}}{d} + 2 + \frac{\log C}{p}} &= \frac{\logv{6d\log d}}{2} + \min\inparen{\frac{p}{2},1}\inparen{-\frac{2^{\logv{6d\log d}}}{d} + 2 +\frac{\log C}{p}} \\
    &= \frac{\logv{6d\log d}}{2} + \min\inparen{\frac{p}{2},1}\inparen{-6\log d + 2 +\frac{\log C}{p}} \\
    &\le \frac{\logv{6d\log d}}{2} + \min\inparen{\frac{p}{2},1}\inparen{-6\log d + 2 +\frac{\log C}{p}} \\
    &\le \frac{\logv{6d\log d}}{2} + \frac{\log C}{\max(2,p)} \le \frac{6Cd\log d}{2}
\end{align*}
By \Cref{lemma:concentration_g2_two_one_volume}, we see that
\begin{align*}
    2^{N_S/2}e_{N_S}\inparen{B_p, \polynorm{\cdot}}^{\min(p/2,1)} \lesssim \sqrt{dC^{1/\max(2,p)}\log d},
\end{align*}
and by \Cref{lemma:containments}, we can choose
\begin{align*}
    C = H\inparen{\Fstar}^{\max(1,p/2)}.
\end{align*}
We plug this in, account for the remaining terms in the summation, and conclude the proof of \Cref{lemma:g2_tail}.
\end{proof}

We now give another way to evaluate the terms of our summation when the indices $N$ are such that the entropy numbers are rather large. See \Cref{lemma:concentration_g2_two_one_sudakov}.

\begin{lemma}
\label{lemma:concentration_g2_two_one_sudakov}
For all $N \ge 0$, we have
\begin{align*}
    2^{N/2}e_N\inparen{B_p, \polynorm{\cdot}}^{\min(p/2,1)} \le \inparen{C(p) \cdot \max_i\min\inbraces{p_i,\log\abs{S_i}}H^{2/\max(2,p)}\Fstar\logv{\Fstar+\mtilde}}^{1/2},
\end{align*}
where $C(p)$ is a constant that only depends on $p$.
\end{lemma}
\begin{proof}[Proof of \Cref{lemma:concentration_g2_two_one_sudakov}]
For now, let
\begin{align*}
    f(\Fstar,\cG) \coloneqq C(p) \cdot \max_i\min\inbraces{p_i,\log\abs{S_i}}H^{2/\max(2,p)}\Fstar\logv{\Fstar+\mtilde}.
\end{align*}
By \Cref{thm:general_covering}, we have
\begin{align*}
    \log \cN\inparen{B_{p}, \eta \cdot \inbraces{\vy \in \R^d \suchthat \polynorm{\vx} \le 1}} \lesssim \eta^{-\min(2,p)} \cdot f(\Fstar,\cG).
\end{align*}
so if we choose, for some universal constant $C_0$,
\begin{align*}
    \eta = C_0 \cdot 2^{-N/\min(p,2)}\inparen{f(\Fstar,\cG)}^{\max(1/2,1/p)},
\end{align*}
then we get
\begin{align*}
    \log \cN\inparen{B_{p}, \eta \cdot \inbraces{\vy \in \R^d \suchthat \polynorm{\vx} \le 1}} \le 2^N.
\end{align*}
Thus, $e_N\inparen{B_p, \polynorm{\cdot}} \le \eta$. Exponentiating and substituting the definition of $f(\Fstar,\cG)$ concludes the proof of \Cref{lemma:concentration_g2_two_one_sudakov}.
\end{proof}

We now show how to complete the sum by combining \Cref{lemma:concentration_g2_two_one_sudakov} and \Cref{lemma:g2_tail}.

\begin{lemma}
\label{lemma:concentration_subgaussian}
We have
\begin{align*}
    \sum_{N \ge 0} 2^{N/2}e_N\inparen{B_p, \polynorm{\cdot}}^{\min(p/2,1)} &\lesssim \sqrt{nH^{1/\max(2,p)}\inparen{\Fstar}^{1/2}\log n} \\
    &\quad +\log n\inparen{C(p) \cdot \max_i\min\inbraces{p_i,\log\abs{S_i}}H\Fstar\logv{\Fstar+\mtilde}}^{1/2}.
\end{align*}
\end{lemma}
\begin{proof}[Proof of \Cref{lemma:concentration_subgaussian}]
Noting that $N_S = \logv{6d\log d} \lesssim \log d$, we combine the conclusions of \Cref{lemma:concentration_g2_two_one_sudakov} and \Cref{lemma:g2_tail} to obtain the statement of \Cref{lemma:concentration_subgaussian}.
\end{proof}

Finally, we translate Lemma \ref{lemma:concentration_subgaussian} into an upper bound on the process we started with by using \Cref{thm:generic_chaining_dirksen}. We then use this to complete the proof of Theorem \ref{thm:concentration}.

\begin{proof}[Proof of Theorem \ref{thm:concentration}]
As we have done in previous proofs, as a shorthand, we define
\begin{align*}
    p^{\star} \coloneqq \max_{i}\min\inbraces{p_i,\log\abs{S_i}}.
\end{align*}
We first weaken the statement of \Cref{lemma:concentration_subgaussian} to read
\begin{align*}
    &\quad\sum_{N \ge 0} 2^{N/2}e_N\inparen{B_p, \polynorm{\cdot}}^{\min(p/2,1)} \\
    &\lesssim \max\inbraces{d,\Fstar}H^{1/2\max(2,p)}\sqrt{\log d} \\
    &\quad+ \log d \inparen{C(p) \cdot H^{2/\max(2,p)}p^{\star}\max\inbraces{d,\Fstar}\logv{d + \Fstar + \mtilde}}^{1/2} \\
    &\lesssim \log d \inparen{C(p) \cdot H^{2/\max(2,p)}p^{\star}\max\inbraces{d,\Fstar}\logv{d + \Fstar + \mtilde}}^{1/2}.
\end{align*}
Let $V$ denote the right hand side of the above. Combining this rewrite with \Cref{lemma:concentration_g2_rewrite_d}, we get
\begin{align*}
    \gamma_2 \lesssim {\mtilde}^{1/2}\max(p,2) \inparen{H^{1/p}\inparen{\Fstar}^{1/2}}^{\max(0,p/2-1)}V\inparen{\max_{\vx \in B_{p}}\norm{\vx}_{\cG'}^p}^{1/2}.
\end{align*}
By the symmetrization reduction (\Cref{lemma:symmetrization}) and \Cref{thm:generic_chaining_dirksen}, we have
\begin{align*}
    \exv{\sup_{\vx\in B_p} \abs{\gnorm{\mA\vx}{p}-\norm{\vx}_{\cG'}^p}} \lesssim \frac{\max(p,2) \inparen{H^{1/p}\norm{\valpha}_p}^{\max(0,p/2-1)}V}{\mtilde^{1/2}}
\end{align*}
and so to make the RHS upper-bounded by $\eps$, it is sufficient to set $\mtilde$ according to
\begin{align*}
    \mtilde &\asymp \frac{\max(p,2)^2 \inparen{H^{1/p}\inparen{\Fstar}^{1/2}}^{\max(0,p-2)}V^2}{\eps^2}.
\end{align*}
This means that when $p < 2$, we have
\begin{align*}
    \mtilde &\asymp \frac{(\log d)^2\cdot C(p) \cdot Hp^{\star}\max\inbraces{d,\Fstar}\logv{d + \Fstar + \mtilde}}{\eps^2} \\
    &\asymp \frac{(\log d)^2\logv{\max\inbraces{d,\Fstar}/\eps}\cdot C(p) \cdot Hp^{\star}\max\inbraces{d,\Fstar}}{\eps^2},
\end{align*}
which is what we desired.

For $p \ge 2$, we have
\begin{align*}
    \mtilde &\asymp \frac{p^2H^{1-2/p}\inparen{\Fstar}^{p/2-1}\inparen{(\log d)^2H^{2/p}p^{\star}\max\inbraces{d,\Fstar}\logv{d+\Fstar+\mtilde}}}{\eps^2} \\
    &\asymp \frac{(\log d)^2\logv{\max\inbraces{d,\Fstar}/\eps}\cdot p^2 \cdot Hp^{\star}\max\inbraces{d,\Fstar}^{p/2}}{\eps^2}
\end{align*}
To bound $\mathsf{diam}(B_p,\dtwo)$, by the triangle inequality, it is enough to estimate $\dtwo(\vx,0)$ for all $\vx \in B_p$. Recalling \Cref{lemma:containments}, we have
\begin{align*}
    \dtwo(\vx,0) &= \inparen{\sum_{h=1}^{\mtilde} \inparen{\frac{g_{i_h}(\vx)^p}{\rho_{i_h}} - \frac{g_{i_h}(0)^p}{\rho_{i_h}}}^2}^{1/2} = \inparen{\sum_{h=1}^{\mtilde} \inparen{\frac{g_{i_h}(\vx)^p}{\rho_{i_h}}}\cdot\inparen{\frac{g_{i_h}(\vx)^p}{\rho_{i_h}}}}^{1/2} \\
    &\le \inparen{\sum_{h=1}^{\mtilde} H\inparen{\Fstar}^{\max(1,p/2)}\cdot\inparen{\frac{g_{i_h}(\vx)^p}{\rho_{i_h}}}}^{1/2} \le \mtilde^{1/2}H^{1/2}\inparen{\Fstar}^{\max(1/2,p/4)}\max_{\vx\in B_p}\inparen{\norm{\vx}_{\cG'}^{p}}^{1/2},
\end{align*}
which means that we may set $\widehat{D}$ in \Cref{thm:generic_chaining_dirksen} according to
\begin{align*}
    \widehat{D} \asymp \frac{H^{1/2}\inparen{\Fstar}^{\max(1/2,p/4)}\max_{\vx\in B_p}\inparen{\norm{\vx}_{\cG'}^{p}}^{1/2}}{\mtilde^{1/2}}.
\end{align*}
We now verify that if we choose
\begin{align*}
    \mtilde \asymp \vbrho \cdot H\max_{i \in [m]} \min(p_i,\log\abs{S_i}) \inparen{\Fstar}^{\max(1,p/2)}\logv{\sfrac{1}{\delta}},
\end{align*}
that we indeed get for some universal constant $C$ that
\begin{align*}
    \prvv{\cD}{\max_{\vx\in B_p} \abs{\gnorm{\mA\vx}{p}^p-\norm{\vx}_{\cG'}^p} \ge C\eps} \lesssim \delta.
\end{align*}
We rescale $\eps$ appropriately and conclude the proof of \Cref{thm:concentration}.
\end{proof}

%%%%%%%%%%%%%%%%%%%%%%%%%%%%%%%%%%%%%%%%%%%%%%%%%%%%%%%%%%%%%%%%%%%%%%%%%%%%%%%%%%%%%%%%%%%
% \newpage
\section{Applications and algorithms}
\label{sec:applications}
% , \Cref{corollary:hypergraph_lp}
At this point in the paper, we are ready to prove our main results (\Cref{thm:one_shot_lewis} and \Cref{thm:computeblw}).

\subsection{Block norm approximations via block Lewis weights (Proof of Theorem \ref{thm:one_shot_lewis})}
\label{sec:lewis}

We restate and prove the main result of the paper.
\oneshotlewis*
\begin{proof}[Proof of Theorem \ref{thm:one_shot_lewis}]
Observe that \Cref{lemma:block_instantiation_small} proves the existence of a probability measure $\vlambda$ over $[m]$ and a rounding $\mW$ that are $\Fstar$-block Lewis overestimates for $\Fstar = d$ if $p_i \ge 2$, and \Cref{lemma:block_instantiation_alleq} proves the existence of a probability measure $\widehat{\vlambda}$ over $[n]$ and corresponding $\widehat{\valpha} \in \R^n$ such that we get an $\Fstar = d$-Lewis overestimate.

We now apply \Cref{thm:concentration} and conclude the proof of Theorem \ref{thm:one_shot_lewis}.
\end{proof}

\subsection{Efficient computation of block Lewis weight overestimates (Proof of \texorpdfstring{\Cref{thm:computeblw}}{Theorem 2})}
\label{sec:alg}

In this subsection, we restate and prove \Cref{thm:computeblw}.
\computeblw*
We break up the proof into two sections -- one where $p_1=\dots=p_m=2$ and $p > 0$ and another where $p=2$ and $p_1,\dots,p_m \ge 2$.

\subsubsection{Special case -- \texorpdfstring{$p > 0, p_1=\dots=p_m=2$}{p>0,p1=...=pm=2}}

Following \Cref{defn:block_lewis_overestimate} and the discussion in \Cref{sec:blw_defs_covering}, observe that when $p_1=\dots=p_m=2$, we have \[\beta_i(\mV) \coloneqq \inparen{\sum_{j \in S_i} \va_j^{\top}(\mA^{\top}\mV\mA)^{-1}\va_j}^{1/2}.\]
As before, we call \(\beta_i(\mV)^p\) block Lewis weights.
Given \(\vb \in \R^m\), we let \(\mB \in \R^{n}\) be a diagonal matrix given by
\(\mB_{jj} = \vb_i\) for all \(i \in [m]\), \(j \in S_i\).
First, let us specialize the definition of block Lewis weight overestimates (recall \Cref{defn:block_lewis_overestimate}) to this special case.
\begin{definition}\label{def:lev_score_over}
    For \(\nu \ge 0\), we say \(\vb \in \R_{\geq 0}^m\) is a vector of \(\nu\)-bounded block Lewis weight overestimates for \(\mA\) if
    \[\|\vb\|_1 \le \nu,\]
    and for all \(i \in [m]\),
    \[b_i \ge \beta_i(\mB^{1-2/p})^p.\]
\end{definition}

We can think of the definition of block Lewis weight overestimates as being a relaxation of the fixed point condition for block Lewis weights that is described in \cite[Page 31, Proof of Lemma 4.2]{jlls23}. 

As a primitive, our algorithm will use \textit{leverage score overestimates} (see \cite[Definition 2.2]{jls22}). They are approximate forms of leverage scores \(\tau_j(\mM)\).
\begin{definition}
    For \(\nu \ge 0\), we say \(\widetilde{\tau} \in \R^m_{\ge 0}\) is a vector of \(\nu\)-bounded leverage score overestimates for \(\mM \in \R^{m \times d}\) if
    \[\|\widetilde{\tau}\|_1 \le \nu\]
    and for all \(i \in [k]\),
    \[\widetilde{\tau}_i \ge \tau_i(\mM).\]
\end{definition}

There are known efficient algorithms for computing leverage score overestimates or reducing leverage score computations to linear system solves.

\begin{theorem}[\protect{\cite[Theorem 3]{jls22}}]\label{thm:lev_ove}
There is an algorithm \textnormal{\textsc{OverLev}} that, given \(\mM \in \R^{n \times d}\), produces \(O(n)\)-bounded leverage score overestimates for \(\mM\) in \(\widetilde{O}(\nnz{\mM} + d^\omega)\) time, where \(\omega\) is the matrix multiplication exponent.
\end{theorem}

We will use two different algorithms depending on the value of $p$. If $p \le 2$, then we present a contractive scheme reminiscent of the algorithm of \textcite{cp15}. If $p > 2$, we present an algorithm similar to those of \textcite{ccly19} and \textcite{jls21}.

We begin with the case where $0 < p \le 2$ (in fact, we will see that this algorithm yields guarantees where $p < 4$). The main objects of interest here are \Cref{alg:blw_lewismap} and \Cref{lemma:contract_alg}.

\begin{algorithm}
\caption{Algorithm to compute block Lewis weight overestimates, $0 < p < 4$ and $p_1=\dots=p_m=2$.}
\begin{algorithmic}[1]\label{alg:blw_lewismap}
\State \textbf{Input:} \(\mA \in \R^{n \times d}\), outer norm $0 < p < 4$, group structure \((S_1,\dots,S_m,2,\dots,2)\).
\State Initialize \(\vb^{(0)} = \frac{d}{m} \cdot \onev_m\).
\State Define $\psi$ such that\label{line:psi_def}
\begin{align*}
    \psi_i(\vb) \coloneqq \inparen{\inparen{\sum_{j \in S_i} \inparen{\va_j^{\top}\inparen{\mA^{\top}\mB^{1-2/p}\mA}^{-1}\va_j}^{p_i/2}}^{2/p_i}}^{p/2}.
\end{align*}
\For{\(t = 1, \ldots, T\)}
    \State \(\vb^{(t)} = \max(\psi(\vb^{(t-1)}),\onev_{m} \cdot 1/m)\) \Comment{The $\max$ is taken elementwise here.}
\EndFor
\State \Return \(1.1\vb^{(T)}\)
\end{algorithmic}
\end{algorithm}

\begin{lemma}
\label{lemma:contract_alg}
Let
\begin{align*}
    T \ge \frac{\ln\inparen{\frac{\ln\inparen{\frac{m}{d}}}{\ln\inparen{1+\eps}}}}{\ln\inparen{\abs{\frac{2}{2-p}}}}.
\end{align*}
Then, the weights $1.1\vb^{(T)}$ output by \Cref{alg:blw_lewismap}  are a $1.1n$-block Lewis overestimate (\cref{defn:block_lewis_overestimate}). Furthermore, computing $\vb^{(T)}$ requires at most $T$ computations of the vector whose entries are the $\va_j^{\top}\inparen{\mA^{\top}\mD\mA}^{-1}\va_j$ for all $j$, where $\mD$ is a diagonal matrix.
\end{lemma}

To prove \Cref{lemma:contract_alg}, we first state and prove \Cref{lemma:lewis_contract_easy}.

\begin{lemma}
\label{lemma:lewis_contract_easy}
Let $\psi$ be as defined in Line \ref{line:psi_def} in \Cref{alg:blw_lewismap}. For all $\vu \in \R^{m}_{\ge 0}$ and $\vv \in \R^{m}_{\ge 0}$, we have
\begin{align*}
    \max_{1 \le i \le m} \abs{\lnv{\frac{\psi_i(\vu)}{\psi_i(\vv)}}} \le \abs{\frac{p}{2}-1}\max_{1 \le i \le m} \abs{\lnv{\frac{u_i}{v_i}}},
\end{align*}
and therefore $\psi(\vu)$ is a contraction whenever $0 < p < 4$.
\end{lemma}
\begin{proof}[Proof of \Cref{lemma:lewis_contract_easy}]
Fix some index $i \le m$. For notational simplicity in this proof, let $\alpha$ be such that $\lnv{\alpha} \coloneqq \triangle(\vu,\vv)$.

This easily implies that
\begin{align*}
    \frac{1}{\alpha^{\abs{1-2/p}}} \cdot \va_j^{\top}\inparen{\mA^{\top}\mV^{1-2/p}\mA}^{-1}\va_j \le \va_j^{\top}\inparen{\mA^{\top}\mU^{1-2/p}\mA}^{-1}\va_j \le \alpha^{\abs{1-2/p}} \cdot \va_j^{\top}\inparen{\mA^{\top}\mV^{1-2/p}\mA}^{-1}\va_j.
\end{align*}
We take the $p_i/2$-norm and take the $p/2$ power, which tells us that for all $1 \le i \le m$,
\begin{align*}
    \frac{1}{\alpha^{\abs{p/2-1}}} \cdot \psi_i(\vv) \le \psi_i(\vu) \le \alpha^{\abs{p/2-1}} \cdot \psi_i(\vv).
\end{align*}
Hence,
\begin{align*}
    \max_{1 \le i \le m} \abs{\lnv{\frac{\psi_i(\vu)}{\psi_i(\vv)}}} \le \abs{\frac{p}{2}-1}\lnv{\alpha} = \abs{\frac{p}{2}-1}\max_{1 \le i \le m} \abs{\lnv{\frac{u_i}{v_i}}},
\end{align*}
completing the proof of \Cref{lemma:lewis_contract_easy}.
\end{proof}

We are now ready to complete the proof of \Cref{lemma:contract_alg}.

\begin{proof}[Proof of \Cref{lemma:contract_alg}]
The computational complexity guarantee is immediate, so we focus on the approximation guarantee.

By \Cref{lemma:lewis_contract_easy} and the Banach fixed point theorem, we know that $\psi$ has a unique fixed point. 
Denote this fixed point by $\vb^{\star}$. W
e would like to argue that since the convergence in the $\ln$-metric is linear, it takes roughly $\log\log(1+\eps)/\log\abs{2/(2-p)}$ applications of $\psi$ to reach a $(1+\eps)$-multiplicative approximation to $\vb^{\star}$. 
An annoying technicality is that if $\vb^{\star}$ has some elements arbitrarily close to $0$, then the convergence rate could be very slow. 
To fix this, we simply enforce that the coordinates of the iterates never drop below $1/m$. 
It is easy to see that this only overestimates the true weights and therefore does not affect our sampling guarantees (in particular, $\norm{\max(\vb^{\star}, \onev_m \cdot 1/m)}_1 \le d + 1$).

More precisely, let $\vb \coloneqq \max(\vb^{\star}, \onev_m \cdot 1/m)$. Notice that for all $1 \le i \le m$, $\abs{\ln\inparen{\sfrac{b_i^{(0)}}{b_i}}} \le \ln\inparen{\sfrac{m}{d}}$. This means that after $T$ iterations, we have
\begin{align*}
    \max_{1 \le i \le m} \abs{\ln\inparen{\frac{b^{(T)}_i}{b_i}}} \le \abs{\frac{p}{2}-1}^T\max_{1 \le i \le m} \abs{\ln\inparen{\frac{b_i^{(0)}}{b_i}}} \le  \abs{\frac{p}{2}-1}^T\ln\inparen{\frac{m}{d}}.
\end{align*}
Choosing
\begin{align*}
    T \ge \frac{\ln\inparen{\frac{\ln\inparen{\frac{m}{d}}}{\ln\inparen{1+\eps}}}}{\ln\inparen{\abs{\frac{2}{2-p}}}}
\end{align*}
and observing that for sampling that it is sufficient to choose $\eps = 0.1$ implies that $\vb^{(T)}$ is an entrywise $1.1$-approximation to $\vb$. As this is sufficient to get the concentration in the setting of \Cref{thm:concentration}, we may  complete the proof of \Cref{lemma:contract_alg}.
\end{proof}

Now, we move onto the case where $p \ge 2$. This covers the cases of $p$ where \Cref{alg:blw_lewismap} is not a contraction (whenever $p \ge 4$). In this setting, we have \Cref{alg:blw}. At a high level, observe that Line \ref{line:alg_iter_next} of \Cref{alg:blw} performs a fixed point iteration on the stationary condition \(b_i = \beta_i(\mB^{1-2/p})^p\) that holds for the optimal choice of block Lewis weights (see \cite[proof of Lemma 4.2 and (4.5)]{jlls23}).

\begin{algorithm}[H]
\caption{Algorithm to compute block Lewis weight overestimates, $p \ge 2$}
\begin{algorithmic}[1]\label{alg:blw}
\State \textbf{Input:} \(\mA \in \R^{n \times d}\)
\State \textbf{Output:}
\State Initialize \(\vb^{(1)} = \frac{d}{m} \cdot \onev\) \label{line:alg_init}
\For{\(t = 1, \ldots, T - 1\)}
    \State \(\mB^{(t)}_{jj} = b^{(t)}_i\) for all \(i \in [m]\), \(j \in S_i\)
    \State \(\widetilde{\tau}^{(t)} = \textsc{OverLev}((\mB^{(t)})^{1/2 - 1/p} \mA)\)\label{line:alg_iter_lev}
    \State \(b^{(t+1)}_i = \sum_{j \in S_i} \widetilde{\tau_j}\) for all \(i \in [m]\)\label{line:alg_iter_next}
\EndFor
\State \(\overline{\vb} = \frac{1}{T} \sum_{t=1}^T \vb^{(t)}\)
\State \Return \(\vb = \frac{3}{2} \overline{\vb}\)
\end{algorithmic}
\end{algorithm}

The guarantee we obtain for \Cref{alg:blw} is captured by \Cref{thm:blw}.

\begin{lemma}
\label{thm:blw}
The return value \(\vb\) of  \Cref{alg:blw} is a vector of \(O(d)\)-bounded block Lewis weight overestimates.
Further, this vector of overestimates is found in \(\text{polylog}(k, m, d)\) leverage score overestimate computations.
\end{lemma}

The goal of the rest of this section is to analyze \Cref{alg:blw} and to prove \Cref{thm:blw}. Let us briefly describe the analysis of \Cref{alg:blw}. We first give a collection of potential functions \(\phi_i(\vb)\), with the goal of showing the potential of \(\phi_i(\overline{\vb})\) decreases with \(T\). A low potential will also imply that \(\overline{\vb}\) is nearly a vector of block Lewis weight overestimates.
For each \(i \in [m]\) define \(\phi_i \colon \R^m \to \R\) by
\[\phi_i(\vb) \coloneqq \ln\inparen{\frac{1}{\vb_i} \sum_{j \in S_i} \tau_j(\mB^{1/2 - 1/p} \mA)}.\]

The key property of this potential is convexity, which we now show.
\begin{lemma}
\label{lemma:alg_phi_convex}
    For each \(i \in [m]\), \(\phi_i\) is convex.
\end{lemma}
\begin{proof}[Proof of \Cref{lemma:alg_phi_convex}]
     Our argument for the convexity of this function is similar to the one given in \cite[Lemma A.2]{jls21}.
     First, notice that by the definition of \(\tau_j\), \(\phi(\vb)\) is equal to
     \begin{align*}
         \phi_i(\vb) = \ln\inparen{\frac{1}{\vb_i^{2/p}} \sum_{j \in S_i} \va_j^\top \inparen{\mA^\top \mB^{1 - 2/p} \mA}^{-1} \va_j}.
     \end{align*}
     Since \(- \frac{2}{p} \ln(b_i)\) is convex, it suffices to show the convexity of
     \[f(\vb) \coloneqq \ln\inparen{\sum_{j \in S_i} \va_j^\top \inparen{\mA^\top \mB^{1 - 2/p} \mA}^{-1} \va_j)}.\]
     Now we define for each \(j \in [k]\) the function
     \[h_j(\vb) \coloneqq \ln \inparen{\va_j^\top (\mA^\top \mB^{1-2/p} \mA)^{-1} \va_j}.\]
     A version of this function without repeated entries in \(\mB\) was shown to be convex in \cite[Lemma A.2]{jls21}, but \(h_j(\vb)\) is still convex.
     Next, notice that we may write \[
     f(\vb) = \ln\inparen{\sum_{j \in S_i} \exp(h_j(\vb))}.
     \]
     Now taking \(\vb, \vb' \in \R^m\) and \(\lambda \in [0, 1]\) we have
     \begin{align}
        f(\lambda \vb + (1-\lambda) \vb') &= \ln\inparen{\sum_{j \in S_i} \exp(h_j(\lambda \vb + (1-\lambda) \vb'))} \nonumber
        \\
     &\le \ln\inparen{\sum_{j \in S_i} \exp(\lambda h_j(\vb) + (1-\lambda) h_j(\vb'))} \label{eq:alg_ph_cvx_1}\\
     &\le \lambda \ln\inparen{\sum_{j \in S_i} \exp(h_j(\vb))} + (1-\lambda) \ln\inparen{\sum_{j \in S_i} \exp(h_j(\vb'))} \label{eq:alg_ph_cvx_2} \\
     &= \lambda f(\vb) + (1-\lambda) f(\vb'), \nonumber
     \end{align}
     where \eqref{eq:alg_ph_cvx_1} follows from the convexity of \(h_j\) and the monotonicity of log-sum-exp, and \eqref{eq:alg_ph_cvx_2} is due to the convexity of log-sum-exp (see e.g. \cite[Section 3.1.5]{boyd2004convex}). Hence, we may conclude the proof of \Cref{lemma:alg_phi_convex}.
\end{proof}

We now give an argument that \(\phi_i(\overline{\vb}) = \widetilde{O}(1/T)\) using the convexity of \(\phi_i\).

\begin{lemma}\label{lemma:alg_analysis1}
    Assume that \textnormal{\textsc{OverLev}} returns \(\nu\)-bounded leverage score overestimates.
   Then after \Cref{alg:blw} runs for \(T\) iterations, we have for all \(i \in [m]\) that
    \[\phi_i(\overline{\vb}) \le \frac{1}{T} \ln\inparen{\frac{m \nu}{d}}.\]
\end{lemma}
\begin{proof}[Proof of \Cref{lemma:alg_analysis1}]
    We have
    \begin{align*}
        \phi_i(\overline{\vb}) &\le \frac{1}{T} \sum_{t=1}^T \phi_i(\vb^{(t)}) &\text{Jensen's inequality} \\
        &= \frac{1}{T} \sum_{t=1}^T \ln \inparen{\frac{1}{b_i^{(t)}} \sum_{j \in S_i} \tau_j((\mB^{(t)})^{1/2 - 1/p} \mA)} &\text{Definition of \(\phi_i\)}\\
        &\le \frac{1}{T} \sum_{t=1}^T \ln \inparen{\frac{1}{b_i^{(t)}} \sum_{j \in S_i} \widetilde{\tau}_j^{(t)}} &\text{Definition of \(\tau_j\)}\\
        &= \frac{1}{T} \sum_{t=1}^T \ln \inparen{\frac{b_i^{(t+1)}}{b_i^{(t)}}} &\text{Line \ref{line:alg_iter_next}} \\
        &= \frac{1}{T} \ln\inparen{\frac{b_i^{(T+1)}}{b_i^{(1)}}} \\
        &= \frac{1}{T} \ln\inparen{m/n} + \frac{1}{T} \ln(b_i^{(T+1)}). &\text{Line \ref{line:alg_init}}
    \end{align*}
    In the above, for the sake of analysis we define \(\vb_i^{(T+1)}\) to be as if the algorithm executed \(T+1\) iterations. Now, \(b_i^{(T+1)} \le \|\widetilde{\tau}\|_1 \le \nu\) by the fact that \(\textsc{OverLev}\) returns \(\nu\)-bounded leverage score overestimates. We therefore conclude the proof of \Cref{lemma:alg_analysis1}.
\end{proof}

Now, we show that \(\text{polylog}(k,m,d)\) leverage score overestimate computations suffice to find the \((\sfrac{3}{2}\cdot n\))-bounded block Lewis weight overestimates. This proves \Cref{thm:blw}.

\begin{proof}[Proof of \Cref{thm:blw}]
By \Cref{thm:lev_ove}, \textsc{OverLev} returns \(O(d)\)-bounded leverage score overestimates.
We take \(T = O\inparen{\ln\inparen{m}}\); clearly this satisfies the desired runtime guarantee. Further, using \Cref{lemma:alg_analysis1} and taking the constant in \(T\) large enough, we have for each \(i\) that \(\phi_i(\overline{\vb}) \le \ln\inparen{\frac{3}{2}}\).
Thus we have
\begin{equation}\label{eq:alg_b_guar}
    \frac{3}{2} \cdot \overline{b_i} \ge \sum_{j \in S_i} \tau_j(\overline{\mB}^{1/2 - 1/p} \mA).
\end{equation}
And so
\begin{align*}
    b_i &\ge \frac{3}{2} \overline{b_i} \\
    &\ge \sum_{j \in S_i} \tau_j(\overline{\mB}^{1/2 - 1/p} \mA) &\text{by \eqref{eq:alg_b_guar}} \\
    &= \sum_{j \in S_i} \tau_j(\mB^{1/2 - 1/p} \mA). &\text{\Cref{fact:lev_score_facts}}
\end{align*}
We now manipulate this guarantee into the desired guarantee of block Lewis weight overestimates
by some algebra.
% From the above, we get
% \[
% \vb_i^{2/p} \vb_i^{1-2/p} \ge \sum_{j \in S_i} \tau_j(\mB^{1/2 - 1/p} \mA)
% \]
Splitting powers on the left hand side of the above, we get
\[
b_i^{2/p} \ge \frac{1}{b_i^{1-2/p}} \sum_{j \in S_i} \tau_j(\mB^{1/2 - 1/p} \mA).
\]
Taking this to the \(p/2\)th power, we obtain
\begin{align*}
    \vb_i \ge \inparen{\frac{\sum_{j \in S_i} \tau_j(\mB^{1/2-1/p} \mA)}{b_i^{1-2/p}}}^{p/2} = \beta_i(\mB^{1-2/p})^p,
\end{align*}
as desired. To bound \(\norm{\vb}_1\), notice
\begin{align*}
    \norm{\vb}_1 = \frac{3}{2} \|\overline{\vb}\|_1 \le \frac{3}{2} \frac{1}{T} \sum_{t=1}^T \|\vb^{(t)}\|_1 \le \frac{3}{2} \nu \le O(n).
\end{align*}
Finally, to obtain the concentration statement, we observe that the above implies that we get a measure $\vlambda$ and a rounding $\mW$ that are $\Fstar$-block Lewis estimates for $\Fstar=O(d)$. This concludes the proof of \Cref{thm:blw}.
\end{proof}

\subsubsection{Special case -- \texorpdfstring{$p = 2, p_1,\dots,p_m\ge2$}{p=2,p1,...,pm>=2}}

Finally, we are ready to introduce and analyze the algorithm for the case where $p=2$ and $p_1,\dots,p_m \ge 2$. See \Cref{alg:blw_base}. The main property of \Cref{alg:blw_base} is given in \Cref{thm:blw_base}.

\begin{algorithm}[ht]
\caption{Algorithm to compute block Lewis weight overestimates, $p=2$ and $p_1,\dots,p_m \ge 2$}
\begin{algorithmic}[1]\label{alg:blw_base}
\State \textbf{Input:} \(\mA \in \R^{n \times d}\), group structure \((S_1,\dots,S_m,p_1,\dots,p_m)\).
\State Initialize \(\vb^{(0)} = \frac{d}{m} \cdot \onev\) and \(\vu^{(0)}\) such that for all $1 \le i \le m$, \(u^{(0)}_j = 1/\abs{S_i}\) for all \(j \in S_i\)
\For{\(t = 1, \ldots, T - 1\)}
    \State \(\widetilde{\tau}^{(t)} = \textsc{OverLev}(\mV(\vu^{(t-1)})^{1/2}\mA)\) \Comment{$\mV(\vu)$ is such that $v_j = u_j^{1-2/p_i}$ for all $1 \le i \le m$ and $j \in S_i$}
    \State \(b_{i}^{(t)} = \sum_{j \in S_i} \widetilde{\tau}_j^{(t)}\) for all $1 \le i \le m$
    \State \(u^{(t)}_j = \widetilde{\tau}_j^{(t)}/(\sum_{j' \in S_i} \widetilde{\tau}_{j'}^{(t)})\) for all $1 \le i \le m$ and all $j \in S_i$
\EndFor
\State \(\overline{\vb} = \abs{S_i}^{1/T}\cdot\frac{1}{T}\sum_{t=1}^T \vb^{(t)}\)
\State \(\overline{\vu} = \frac{1}{T}\sum_{t=0}^{T-1} \vu^{(t)}\)
\State \Return \((\overline{\vb},\overline{\vu})\)
\end{algorithmic}
\end{algorithm}

\begin{lemma}
\label{thm:blw_base}
If \textnormal{\textsc{OverLev}} is a routine that returns leverage score overestimates whose sum is at most \(\nu\), then \Cref{alg:blw_base} returns \(\Fstar\)-block Lewis overestimates (\Cref{defn:block_lewis_overestimate}) with \(\Fstar = \max_{1 \le i \le m} \abs{S_i}^{1/T} \cdot \nu\).

In particular, if $T = \max_{1 \le i \le m} \log\abs{S_i}$ and $\nu \le (4/e)d$, then we get $\Fstar = 4d$.
\end{lemma}

As with the analysis of \Cref{alg:blw}, we first prove that a relevant potential is convex and then show that we can control it effectively.

\begin{lemma}
\label{lemma:blw_convex_potential_general}
Let $\mU \in \R^{k \times k}$ be a nonnegative diagonal matrix. Let $\mV(\vu)$ denote the matrix such that for all $i$ and $j \in S_i$, we have $v_j = \vu_j^{1-2/p_i}$. Then, $\phi$ as defined below is log-convex in $\vu$.
\begin{align*}
    \text{for all } j \in S_1 \cup \dots \cup S_m:\quad\quad\phi_j(\vu) = \lnv{\frac{\tau_j\inparen{(\mV(\vu))^{1/2}\mA}}{u_j}}
\end{align*}
\end{lemma}
\begin{proof}[Proof of \Cref{lemma:blw_convex_potential_general}]
We expand the above definition after taking the $\ln$ of both sides.
\begin{align*}
    \phi_j(\vu) &= \lnv{\frac{\tau_j\inparen{(\mV(\vu))^{1/2}\mA}}{u_j}} = \lnv{\va_j^{\top}\inparen{\mA^{\top}\mV(\vu)\mA}^{-1}\va_j} + \lnv{\frac{v_j}{u_j}} \\
    &= \lnv{\va_j^{\top}\inparen{\mA^{\top}\mV(\vu)\mA}^{-1}\va_j} - \frac{2}{p_i}\lnv{u_j}.
\end{align*}
The last term is convex, so it is sufficient to argue that the first term is convex. To see this, first observe that by concavity of $x^{1-2/p_i}$, we have
\begin{align*}
    \lambda\mV(\vu^{(1)}) + (1-\lambda)\mV(\vu^{(2)}) \preceq \mV(\lambda\vu^{(1)} + (1-\lambda)\vu^{(2)}),
\end{align*}
which implies
\begin{align*}
    \lnv{\va_j^{\top}\inparen{\mA^{\top}\mV(\lambda\vu^{(1)} + (1-\lambda)\vu^{(2)})\mA}^{-1}\va_j} &\le \lnv{\va_j^{\top}\inparen{\lambda\mA^{\top}\mV(\vu^{(1)})\mA + (1-\lambda)\mA^{\top}\mV(\vu^{(2)})\mA}^{-1}\va_j} \\
    &\le \lambda\lnv{\va_j^{\top}\inparen{\mA^{\top}\mV(\vu^{(1)})\mA}^{-1}\va_j} \\
    &\quad + (1-\lambda)\lnv{\va_j^{\top}\inparen{\mA^{\top}\mV(\vu^{(2)})\mA}^{-1}\va_j}.
\end{align*}
Above, the last line follows from the well-known (see, e.g., \cite[Lemma 3.4]{ccly19}) fact that $\lnv{\va_j^{\top}\mM^{-1}\va_j}$ is convex in $\mM$ where $\mM$ is symmetric positive-semidefinite. This completes the proof of \Cref{lemma:blw_convex_potential_general}.
\end{proof}

Next, we will show that any choice of nonnegative $\vu$ such that $\sum_{j \in S_i} u_j = b_i^{\inparen{1-\frac{2}{p}}\cdot\inparen{\frac{p_i}{p_i-2}}}$ can be used to find a rounding matrix $\mW$.

\begin{lemma}
\label{lemma:blw_rounding_matrix}
For all $\vx\in\R^d$ and all nonnegative $\vu$ such that $\sum_{j \in S_i} u_j = b_i^{\inparen{1-\frac{2}{p}}\cdot\inparen{\frac{p_i}{p_i-2}}}$, we have
\begin{align*}
    \frac{\norm{\mV(\vu)^{1/2}\mA\vx}_2}{\inparen{\sum_{i' \le m} b_{i'}}^{1/2-1/p}} \le \gnorm{\mLambda^{1/2-1/p}\mA\vx}{2}.
\end{align*}
\end{lemma}
\begin{proof}[Proof of \Cref{lemma:blw_rounding_matrix}]
We start with the LHS.
\begin{align*}
    \norm{\mV(\vu)^{1/2}\mA\vx}_2^2 &= \sum_{i=1}^m \sum_{j \in S_i} u_j^{1-2/p_i}\abs{\ip{\va_j,\vx}}^2 \\
    &\le \sum_{i=1}^m b_i^{1-2/p}\norm{\mA_{S_i}\vx}_{p_i}^2 & \text{ by H\"older's Inequality with powers } \frac{p_i}{p_i-2}, \frac{p_i}{2}
\end{align*}
Normalizing concludes the proof of \Cref{lemma:blw_rounding_matrix}.
\end{proof}

Now, we show that finding uniform overestimates $\tau_j/u_j \le b_i^{\inparen{\frac{2}{p}-\frac{2}{p_i}}\cdot\frac{p_i}{p_i-2}}$ is enough to satisfy \Cref{defn:block_lewis_overestimate} with $\Fstar = \sum_{i \le m} b_i$.

\begin{lemma}
\label{lemma:blw_overestimates_conversion}
If we have $\vu$ such that $\tau_j(\mV(\vu)^{1/2}\mA)/u_j \le b_i^{\inparen{\frac{2}{p}-\frac{2}{p_i}}\cdot\frac{p_i}{p_i-2}}$ for all $i$, then the rounding matrix $\mW = \mV(\vu)\mB^{2/p-1}$ and measure $\lambda_i = b_i/(\sum_{i'\le m} b_{i'})$ is an $\Fstar$-block Lewis overestimate (\Cref{defn:block_lewis_overestimate}) with $\Fstar = \sum_{i'\le m} b_{i'}$.
\end{lemma}
\begin{proof}[Proof of \Cref{lemma:blw_overestimates_conversion}]
Observe that we must have
\begin{align*}
    \frac{\tau_j\inparen{\mV^{1/2}\mA}}{v_j} \le v_j^{\frac{2}{p_i-2}} \cdot b_i^{\inparen{\frac{2}{p}-\frac{2}{p_i}}\cdot\frac{p_i}{p_i-2}}.
\end{align*}
Let $T = \sum_{i' \le m} b_{i'}$. Following \Cref{lemma:block_lewis}, let $\lambda_i = b_i/T$ and $\mW$ be such that
\begin{align*}
    \frac{\mV^{1/2}}{T^{1/2-1/p}} = \mW^{1/2}\mLambda^{1/2-1/p}.
\end{align*}
In particular, this means that
\begin{align*}
    w_j = \frac{v_j}{(T\lambda_i)^{1-2/p}} = \frac{v_j}{b_i^{1-2/p}}.
\end{align*}
Hence,
\begin{align*}
    \inparen{\sum_{j \in S_i} \inparen{\frac{\tau_j\inparen{\mV^{1/2}\mA}}{w_j}}^{p_i/2}}^{2/p_i} &= b_i^{1-2/p}\inparen{\sum_{j \in S_i} \inparen{\frac{\tau_j\inparen{\mV^{1/2}\mA}}{v_j}}^{p_i/2}}^{2/p_i} \\
    &\le b_i^{1-2/p} \cdot b_i^{\inparen{\frac{2}{p}-\frac{2}{p_i}}\cdot\frac{p_i}{p_i-2}} \cdot \inparen{\sum_{j \in S_i} v_{j}^{\frac{2}{p_i-2} \cdot \frac{p_i}{2}}}^{2/p_i} \\
    &= b_i^{1-2/p}\cdot b_i^{\inparen{\frac{2}{p}-\frac{2}{p_i}}\cdot\frac{p_i}{p_i-2}} \cdot \inparen{\sum_{j \in S_i} u_j}^{2/p_i} \\
    &= b_i^{1-2/p}\cdot b_i^{\inparen{\frac{2}{p}-\frac{2}{p_i}}\cdot\frac{p_i}{p_i-2}} \cdot \inparen{\sum_{j \in S_i} u_j}^{2/p_i} \\
    &= b_i^{1-2/p}\cdot b_i^{\inparen{\frac{2}{p}-\frac{2}{p_i}}\cdot\frac{p_i}{p_i-2}} \cdot \inparen{b_i^{\inparen{1-\frac{2}{p}}\cdot\inparen{\frac{p_i}{p_i-2}}}}^{2/p_i} = b_i = \lambda_i\inparen{\sum_{i'\le m} b_{i'}},
\end{align*}
which is exactly the statement of \Cref{lemma:blw_overestimates_conversion}.
\end{proof}

We now have the tools we need to prove \Cref{thm:blw_base}.

\begin{proof}[Proof of \Cref{thm:blw_base}]
Note that \Cref{alg:blw_base} and \Cref{thm:blw_base} are reminiscent of \cite[Algorithm 2 and Theorem 4]{jls22}.

Using \Cref{lemma:blw_convex_potential_general}, we begin with using the convexity of the potential $\phi$. For all $j \in S_i$, we have
\begin{align*}
    \phi_j(\overline{\vu}) &\le \frac{1}{T}\sum_{t=0}^{T-1} \phi\inparen{\vu^{(t)}} = \frac{1}{T}\sum_{t=0}^{T-1} \lnv{\frac{\tau_j\inparen{\mV(\vu^{(t)})^{1/2}\mA}}{u_j^{(t)}}} \le \frac{1}{T}\sum_{t=0}^{T-1}\lnv{\frac{\widetilde{\tau}_j^{(t+1)}}{u_j^{(t)}}} \\
    &= \frac{1}{T}\sum_{t=0}^{T-1}\inparen{\lnv{\frac{u_j^{(t+1)}}{u_j^{(t)}}} + \lnv{\sum_{j' \in S_i} \widetilde{\tau}_{j'}^{(t+1)}}} = \frac{1}{T}\lnv{\frac{u_j^{(T)}}{u_j^{(0)}}} + \frac{1}{T}\sum_{t=0}^{T-1}\lnv{\sum_{j' \in S_i} \widetilde{\tau}_{j'}^{(t+1)}} \\
    &\le \frac{1}{T}\lnv{\frac{u_j^{(T)}}{u_j^{(0)}}} + \lnv{\frac{1}{T}\sum_{t=0}^{T-1}\sum_{j' \in S_i} \widetilde{\tau}_{j'}^{(t+1)}} = \frac{1}{T}\lnv{\frac{u_j^{(T)}}{u_j^{(0)}}} + \lnv{\abs{S_i}^{-1/T}b_{i}} \le \lnv{b_i}.
\end{align*}
We now apply \Cref{lemma:blw_overestimates_conversion} and see that the measure $\lambda_i = b_i/\norm{\overline{\vb}}_1$ is a $\Fstar$-block Lewis overestimate (\Cref{defn:block_lewis_overestimate}) with $\Fstar = \norm{\overline{\vb}}_1$.

Now, observe that
\begin{align*}
    \norm{\overline{\vb}}_1 \le \max_{1 \le i \le m} \abs{S_i}^{1/T} \cdot \frac{1}{T}\sum_{t=1}^T\norm{\vb^{(t)}}_1 \le \max_{1 \le i \le m} \abs{S_i}^{1/T} \cdot \nu \le 4d,
\end{align*}
where we use our setting of $T$ and the fact that $\textsc{OverLev}$ returns leverage score estimates whose sum is at most $(4/e)d$. This completes the proof of \Cref{thm:blw_base}.
\end{proof}

We are finally ready to give the proof of \Cref{thm:computeblw}.

\begin{proof}[Proof of \Cref{thm:computeblw}]
We have three cases.
\begin{itemize}
    \item If $0 < p < 4$ and $p_1=\dots=p_m=2$, then the algorithm and guarantee on the weights follow from \Cref{alg:blw_lewismap} and \Cref{lemma:contract_alg}.
    \item If $p \ge 2$ and $p_1=\dots=p_m=2$, then the algorithm and guarantee on the weights follow from \Cref{alg:blw} and \Cref{thm:blw}.
    \item If $p = 2$ and $p_1,\dots,p_m\ge 2$, then the algorithm and guarantee on the weights follow from \Cref{alg:blw_base} and \Cref{thm:blw_base}.
\end{itemize}
We plug these guarantees into \Cref{thm:concentration} and conclude the proof of \Cref{thm:computeblw}.
\end{proof}

\subsection{Minimizing sums of Euclidean norms (Proof of \texorpdfstring{\Cref{thm:msnalg}}{Theorem 3})}
\label{sec:applications_msn}

Recall the minimizing sums of Euclidean norms (MSN) problem \eqref{eq:msn}.
Given \(\mA \in \R^{n \times d}\), a partition \(S_1, \ldots, S_m\) of \([n]\), and \(\vb_1 \in \R^{|S_1|}, \ldots, \vb_m \in \R^{|S_m|}\), we would like to find $\widehat{\vx}$ such that
\[\sum_{i=1}^m \|\mA_{S_i} \widehat{\vx} - \vb_i\|_2\le (1+\eps)\min_{\vx \in \R^d} \sum_{i=1}^m \|\mA_{S_i} \vx - \vb_i\|_2.\]
\textcite{xy97} give an algorithm with iteration complexity $\widetilde{O}(\sqrt{m}\logv{\sfrac{1}{\eps}})$ for the above problem (though for an additive approximation guarantee instead of a multiplicative one), where each iteration reduces to solving linear systems in matrices $\mA^{\top}\mD\mA$ for block-diagonal matrices, where each block has size $(\abs{S_i}+1) \times (\abs{S_i}+1)$. Their algorithm is based on the primal-dual interior point method framework.
% Xue and Ye \cite{xy97} describe an interior-point algorithm to solve this problem. They use a potential primal-dual reduction method, which on each iteration determines the search direction by solving a linear system. In particular, finding one component of the search direction requires a system solve, and finding the other two can be done after in time \(O(k(k+n))\).
% Combined with setting up the system, the total time complexity per iteration is bounded by \(O((k+n)^2)\), not including solving the system.
% Under appropriate normalization, the number of iterations to obtain an \(\eps\)-optimal solution is \(\widetilde{O}(\sqrt{m} \log(\sfrac{1}{\eps}))\).

% For the dual-scaling scheme of \cite{xy97}, we remark that the linear system solved in each iteration is a generalization of the type handled by \Cref{thm:lev_ove}, and designing specialized solvers for this class of systems like those for computing leverage scores is an interesting problem.

Each system solve takes the following form. Let \(\widetilde{\mA}\) be a block matrix with \(- \mI_m\) in one block, and \(\mA\) in another.
The goal is to find \(\vy\) in the system
\[\widetilde{\mA}^\top \mD \widetilde{\mA} \vy = \vz\]
where \(\mD\) is a block matrix, with one \((1 + |S_i|) \times (1 + |S_i|)\) sized block for each group
(see \cite[equation (4.13)]{xy97}).
% Alternatively, using some algebra, \cite{xy97}
% describe a reformulation of the system that can be set up and solved in time \(O(n^3 + m d^2 \max_i |S_i|)\).

Our main result of this section is \Cref{thm:msnalg}, which gives an improved iteration complexity for \eqref{eq:msn} when $m \gg n$.

\msnalg*

We prove \Cref{thm:msnalg} by sparsifying the objective \eqref{eq:msn} using \Cref{thm:computeblw} and then applying the primal-dual interior point method from \cite{xy97}. We state the guarantee of this algorithm in \Cref{lemma:ipm_log_barrier}.

\begin{lemma}
\label{lemma:ipm_log_barrier}
Let $\mA \in \R^{n \times d}$ and \(\vb \in \R^n\), and  $S_1, \dots, S_m$ be a partition of $k$. There exists an algorithm that returns $\widehat{\vx}$ such that
\begin{align*}
    \sum_{i=1}^m \norm{\mA_{S_i}\widehat{\vx}-\vb_{S_i}}_2 \le (1+\eps)\min_{\vx\in\R^d} \sum_{i=1}^m \norm{\mA_{S_i} \vx-\vb_{S_i}}_2.
\end{align*}
The algorithm runs in $\widetilde{O}\inparen{\sqrt{m}\logv{\sfrac{1}{\eps}}}$ calls to a linear system solver in matrices of the form $\mA^{\top}\mD\mA$ for block-diagonal matrices $\mD$, where each block has size $(\abs{S_i}+1) \times (\abs{S_i}+1)$.
\end{lemma}
\begin{proof}[Proof of \Cref{lemma:ipm_log_barrier}]
The guarantee we will reduce to is \cite[Theorem 5.2]{xy97}. However, the guarantee there is stated for an additive approximation, and we desire a multiplicative approximation. We therefore apply a few transformations to our problem so that we can apply this guarantee.

Let
\begin{align*}
    \vx_0 &\coloneqq \argmin{\vx\in\R^d} \norm{\mA\vx-\vb}_2^2.
\end{align*}
This can be found in one linear system solve. Let $V \coloneqq \norm{\mA\vx_0-\vb}_2$ and consider the following modified optimization problem:
\begin{align}
    \min_{\vx-\vx_0\in\R^d} \frac{1}{V}\sum_{i=1}^m \norm{\mA_{S_i}(\vx-\vx_0)-(\vb_{S_i}-\mA_{S_i} \vx_0)}_2.\label{eq:modified_msn}
\end{align}
We will invoke \cite[Theorem 5.2]{xy97} on the above problem \eqref{eq:modified_msn}, folding the \(\frac{1}{V}\) factor into \(\mA\) and \(\vb\). Clearly, this problem is equivalent to the problem we started with. Next, let $\xstar$ be given by
\begin{align*}
    \xstar &\coloneqq \argmin{\vx\in\R^d} \sum_{i=1}^m \norm{\mA_{S_i}\vx-\vb_{S_i}}_2,
\end{align*}
where we use \(\vb \in \R^n\) for the vector formed by stacking \(\vb_{S_1}, \ldots, \vb_{S_m}\).
Let $\opt \coloneqq \sfrac{1}{V} \cdot \sum_{i=1}^m \norm{\mA_{S_i}\xstar-\vb_{S_i}}_2$. 
% We therefore have
% by the optimality of \(\xstar\)
% and  that
% \begin{align*}
%     \opt &= \frac{1}{V}\sum_{i=1}^m \norm{\mA_{S_i}(\xstar-\vx_0)-(\vb_{S_i}-\mA\vx_0)}_2 \le \frac{1}{V}\sum_{i=1}^m \norm{\mA_{S_i}(\vx_0-\vx_0)-(\vb_{S_i}-\mA_{S_i}\vx_0)}_2 \\
%     &\le \frac{\sqrt{m}}{V} \norm{\mA\vx_0-\vb}_2 = \sqrt{m},
% \end{align*}
% and by \(\|\cdot\|_2 \le \| \cdot\|_1\),
Because \(\|\cdot\|_2 \le \|\cdot\|_1\), we have
\begin{align*}
    1=\frac{\norm{\mA\vx_0-\vb}_2}{V} \le \frac{\norm{\mA\xstar-\vb}_2}{V} \le \frac{1}{V}\sum_{i=1}^m \norm{\mA_{S_i}\xstar-\vb_{S_i}}_2 = \opt.
\end{align*}
% Thus, $1\le \opt \le \sqrt{m}$. 
Furthermore, we have by \(\|\cdot\|_1 \le \sqrt{m}\|\cdot\|_2\) (applied by considering the summation over \(i\) as an \(\ell_1\) norm) that
\begin{align*}
    \max_{1 \le i \le m} \frac{1}{V}\norm{\vb_{S_i}-\mA_{S_i}\vx_0}_2 \le \frac{1}{V}\sum_{i=1}^m \norm{\vb_{S_i}-\mA_{S_i}\vx_0}_2 \le \sqrt{m} \cdot \frac{\norm{\mA\vx_0-\vb}_2}{V} = \sqrt{m}.
\end{align*}
This implies that all the offset vectors of our transformed problem \eqref{eq:modified_msn} have polynomially bounded norm. This suffices for our iteration complexity bound, because the iteration complexity of the algorithm in \cite{xy97} depends logarithmically on the maximum norm of these vectors. Now, using their method we can solve \eqref{eq:modified_msn} up to $\eps$ additive error. This means we find $\widehat{\vx}$ such that
\begin{align*}
    \sum_{i=1}^m \norm{\mA_{S_i}(\widehat{\vx}-\vx_0)-(\vb_{S_i}-\mA_{S_i} \vx_0)}_2 &\le \sum_{i=1}^m \norm{\mA_{S_i}(\xstar-\vx_0)-(\vb_{S_i}-\mA_{S_i} \vx_0)}_2 + V\eps \\
    &= V \cdot \opt\inparen{1+\frac{\eps}{\opt}} \le V \cdot \opt\inparen{1+\eps}.
\end{align*}
where in the last inequality, we use that \(\opt \ge 1\).
Since $V \cdot \opt$ is the original optimal objective value, this completes the proof of \Cref{lemma:ipm_log_barrier}.
\end{proof}

We remark that instead of the $\sqrt{m}$ above, one can get a $\sqrt{d}$-factor relationship between the optimal objective for a least squares relaxation of our problem by using the block Lewis weights with $p_1=\dots=p_m=2$ and $p=1$. However, this will only impact lower order terms.

We are now ready to prove \Cref{thm:msnalg}.

\begin{proof}[Proof of \Cref{thm:msnalg}]
We apply \Cref{thm:computeblw} for \(p_1 = \cdots = p_m = 2\) and \(p = 1\) with approximation $\eps/3$ to the group matrices $\insquare{\mA_{S_i} \vert \vb_{S_i}}$ to find a sparsified objective
with \(\widetilde{m} = O(\eps^{-2} \cdot d(\log d)^2 \log(\sfrac{d}{\eps}))\) terms.
This requires \(\widetilde{O}(1)\) linear system solves.
This implies a \((1+\eps)\) approximation to an un-sparsified objective over any vector in \(\R^{d+1}\),
and the approximation we need comes by only considering vectors in \(\R^{d+1}\) whose last entry is \(-1\).
We plug this into the guarantee of \Cref{lemma:ipm_log_barrier} with approximation $\eps/3$. Since $(1+\eps/3)^2 \le 1+\eps$ and $(1+\eps/3)/(1-\eps/3) \le 1+\eps$, this returns a $(1+\eps)$-approximate minimizer to \eqref{eq:msn}, completing the proof of \Cref{thm:msnalg}.
\end{proof}

\chapter[
  \texorpdfstring{Sparse \(\ell_2\)-Configuration}{Sparse l2-Configuration}
]{
  \texorpdfstring{
    First-Order Algorithms for \(\ell_2\)-Embedding and Sparse Euclidean Distance Feasibility
  }{
    First-Order Algorithms for \(\ell_2\)-Embedding and Sparse Euclidean Distance Feasibility
  }
}

\section{Introduction}

In this work, we study the algorithmic problem of finding configurations of vectors in \(\ell_2\) whose pairwise distances abide a given set of constraints; we call this the \emph{graph-constrained \(\ell_2\) configuration problem}.
In particular, let \(V\) be a ground set with \(n = |V|\), and \(E \subseteq \binom{V}{2}\) a set of `edges' on \(V\) with \(m = |E|\).
We are given distance constraints \(u_{ij} \ge l_{ij} \ge 0\) for every \((i, j) \in E\), and wish to solve the following feasibility problem (identifying \(V\) with \([n]\)):
\begin{align}\label{eq:main_prob}
  \left.\begin{aligned}\text{find}~ & v_1, \ldots, v_n \in \ell_2                                \\
               \text{s.t.}~ & l_{ij} \le \|v_i - v_j\| \le u_{ij} & \forall~(i, j) \in E
        \end{aligned}
  \qquad\right\}
\end{align}
In words, we wish to find a configuration of vectors \(v_1, \ldots, v_n \in \ell_2\)
such that the distance between \(v_i\) and \(v_j\) lies in the provided bounds for every \((i, j)\) for which a constraint is given.

A useful way to view \eqref{eq:main_prob} is that it interpolates between three traditions:
multidimensional scaling and similarity data, metric embeddings and cut rounding, and Euclidean distance geometry.
In multidimensional scaling, the input is often a matrix of similarities or dissimilarities rather than a complete metric, and the goal is to recover a geometric representation that exposes the structure of the data.
In Euclidean distance geometry and distance-matrix completion, one similarly seeks a point configuration realizing a partially observed set of distances \cite{ding2010sensor}.
The \(\ell_2\)-configuration formulation is a sparse feasibility version of these viewpoints: it allows arbitrary upper and lower distance bounds on a specified set of pairs, thereby accommodating missing, partially specified, or nonmetric dissimilarity data while retaining an SDP formulation.

On the theoretical computer science side, a central special case is the minimum-distortion embedding of a finite metric space into \(\ell_2\);
see, e.g., \cite{matousek2013lectures,naor2015metric} for background on low-distortion embeddings.
Here, \((V, d)\) is a metric space, and the configuration \(v_1, \ldots, v_n\) is called an \emph{embedding} of \(V\) into \(\ell_2\).
The \emph{distortion} of an embedding is defined as the infimum over \(D \ge 1\) such that there exists \(\lambda > 0\) such that for all \(i, j \in V\),
\[\lambda \cdot d(i, j) \le \|v_i - v_j\| \le D \cdot \lambda \cdot d(i, j).\]
Here, \(D\) measures how faithfully the pairwise distances between points in the
configuration \(v_1, \ldots, v_n\) represent the distances between the corresponding pairs in \(V\).
The parameter \(\lambda\) exists to make the distortion scale-invariant when the target space is not necessarily a vector space;
here we can always take \(\lambda = 1\) by scaling the configuration appropriately.

It is well-known in metric geometry that by a theorem of Bourgain \cite{bourgain1985lipschitz},
an embedding of distortion \(\lesssim \log n\) into \(\ell_2\) always exists and may be constructed explicitly.
In many contexts, however, it is crucial to construct as faithful an embedding into \(\ell_2\) as possible.
Low-distortion embeddings are algorithmic primitives rather than merely visualization tools.
For example, embedding a given metric space into \(\ell_2\) with minimal distortion is closely related to SDP rounding procedures for generalized sparsest cut \cite{arora2005euclidean}.

A standard `optimization-to-feasibility' reduction shows that minimum distortion embedding into \(\ell_2\) reduces to solving a sequence of \(\ell_2\)-configuration problems.
Thus the sparse \(\ell_2\)-configuration problem considered here generalizes metric embedding while exposing the dependence on the number \(m\) of constrained pairs.
A key question, then, is what the true algorithmic dependence is on the number of constraints when finding an \(\ell_2\)-configuration.
In this work we make progress on this question by giving a first-order SDP solver whose runtime scales with the number of constraint edges \(m\), rather than only with \(n^2\).

We will use the following multiplicative notion of approximation.
\begin{definition}[Approximately feasible \(\ell_2\)-configuration]\label{def:5_approx_feasible}
  A configuration \(v_1, \ldots, v_n \in \ell_2\) is \(\eps\)-approximately feasible for an instance \((V,E,u,l)\) if for every \((i,j)\in E\),
  \[
    (1 - \eps) l_{ij}^2 \le \|v_i - v_j\|^2 \le (1 + \eps) u_{ij}^2.
  \]
\end{definition}
It is clear that if we simply take the square of the constraints, then the feasibility problem \eqref{eq:main_prob} becomes a semidefinite program (SDP).
While there exist `black-box' solvers that solve generic SDPs that apply to this problem,
we use its structure to obtain a more efficient first-order method.
Our main algorithmic guarantee for the \(\ell_2\)-configuration problem is the following:
\begin{restatable}[Graph-constrained \(\ell_2\)-configuration]{theorem}{ltwoconfigmain}
\label{thm:l2_config_main}
Let \((V, E, u, l)\) be an instance of \eqref{eq:main_prob} with \(n = |V|\), \(m = |E|\), and \(\alpha = \max_{ij \in E} \frac{u_{ij}}{l_{ij}}\).
For every \(0 < \eps, \delta < 1\), there is a randomized first-order algorithm running in time
\[
  \widetilde{O}\p{n m \alpha^5 \eps^{-7.5}\log\frac{1}{\delta}}
\]
with the following guarantee. With probability at least \(1-\delta\), if the instance is feasible, the algorithm returns an \(\eps\)-approximately feasible solution in the sense of \Cref{def:5_approx_feasible}; if the algorithm reports \textnormal{\textsc{Infeasible}}, the instance must be infeasible.
\end{restatable}

The parameter \(\alpha = \max_{ij \in E} \frac{u_{ij}}{l_{ij}}\) denotes an `accuracy' parameter,
measuring the width of the allowed interval for each constrained pair.
In the all-pairs minimum-distortion embedding setting, \(\alpha\) is the candidate distortion.
For general sparse or nonmetric \(\ell_2\)-configuration instances, \(\alpha\) is an input condition parameter that can vary potentially unboundedly.

The way that the configuration \(v_1, \ldots, v_n\) is generally understood in metric geometry is
as a map \(f \colon V \to \ell_2\) where the distortion \(D(f)\) is the quantity defined above.
Now, we can define \(e_{\ell_2}(V)\) to be the minimum distortion of any map from \(V\) to \(\ell_2\):
\[e_{\ell_2}(V) \coloneqq \min_{f \colon V \to \ell_2} \mathrm{Distortion}(f)
\]
As a key application, we provide a fast algorithm for this problem:
\begin{restatable}[Minimum-distortion embedding into \(\ell_2\)]{corollary}{ltwodistortionmain}
\label{thm:l2_distortion_main}
Let \((V, d)\) be a finite metric space with \(n = |V|\).
For every \(0 < \eps < 1/2\), there is an algorithm running in time \(\widetilde{O}\p{n^3 \eps^{-7.5}}\)
that returns an embedding \(f \colon V \to \ell_2\) satisfying:
\[D(f) \le (1+\eps) e_{\ell_2}(V)\]
provided the \(\ell_2\)-configuration routine is implemented with certified infeasibility for failed feasibility probes.
\end{restatable}

\subsection{Related Work}

We first put \Cref{thm:l2_config_main}'s performance in the context of generic solvers for SDPs.
We use \(\omega\) to denote the matrix multiplication exponent; the current best known bound is \(\omega < 2.371339\) \cite{alman2025more}.

\begin{table}[htbp]
  \centering
  {\renewcommand{\arraystretch}{1.8}%
  \begin{tabular}{l l r}
    \hline
    \textbf{Runtime}                                                      & \textbf{Parameters}                         &                                                                                               \\
    \hline
    \(\widetilde{O}(n^4 \log(1/\eps) L)\)                                   &                                             & \cite{alizadeh1995interior}                                                \\
    \(\widetilde{O}(\sqrt{n} (mn^2 + m^\omega + n^\omega) \log(1/\eps))\) & & \cite{jiang2020faster}                                                    \\
    \(\widetilde{O}\p{n^{o(1)} \p{\sqrt{n}(m^2 + n^4) + m^\omega + n^{2 \omega}} \log^{O(1)}(1/\eps)}\) & & \cite{huang2022solving} \\
    \rowcolor{blue!5} \(\widetilde{O}\p{nm \cdot \alpha^5 \eps^{-7.5}}\) &   \(\alpha = \max_{ij} \frac{u_{ij}}{l_{ij}}\)   & \textbf{This work}                                                                           \\
    \hline
  \end{tabular}
  }
  \caption{Comparison of runtimes for general semidefinite programming solvers and this work.}
  \label{table:sdp-runtime-comparison}
\end{table}
In the rows citing \cite{alizadeh1995interior}, the parameter \(L\) comes from the bit-complexity analysis of the method
and cannot be easily bounded by the other parameters,
like other classical interior point methods.
\cite{huang2022solving} is best for \(m \gg n^{\sim 1.79}\).

\paragraph{Metric embeddings and cut rounding.}
For background on low-distortion embeddings, see \cite{matousek2013lectures,naor2015metric}.
In graph partitioning, low-distortion embeddings are fundamental primitives: Linial--London--Rabinovich connected metric embeddings with cuts and multicommodity flow \cite{linial1995geometry},
and a sequence of works starting with the seminal work of \cite{arora2009expander}
improved these partitioning primitives by using special classes of metric spaces that were especially
amenable to optimization by SDPs \cite{arora2005euclidean,chawla2005embeddings}.
For generalized sparsest cut, \cite{arora2005euclidean} gave an \(O(\sqrt{\log n}\log\log n)\)-approximation via a generalization of the same approach for uniform sparsest cut.
Very recently, \cite{chang2024random} showed that the Goemans--Linial SDP has integrality gap \(O(\sqrt{\log n})\), matching the \(\Omega(\sqrt{\log n})\) lower bound of Naor and Young \cite{naor2017integrality}.
% This SDP-and-rounding viewpoint is complementary to fast end-to-end graph algorithms for generalized sparsest cut.
% For example, one tradeoff of Madry's algorithm gives an \(O(\log^{2+o(1)} n)\)-approximation in \(\widetilde{O}(m+|E_D|+n^{4/3}\log U)\) time, where \(E_D\) is the demand edge set and \(U\) is the capacity ratio \cite{madry2010fast}.
% This approximation factor is worse than the ALN and CNR SDP-rounding guarantees, but it comes with a much faster graph-algorithmic runtime.
% Our results instead target the Euclidean embedding subproblem that appears in SDP-based rounding; they should therefore be viewed as improving a geometric optimization primitive, not as subsuming such end-to-end cut algorithms.

For minimum distortion embedding into \(\ell_2\), our algorithm gives the \(n^3\)-time bound stated above, up to logarithmic factors.
%Why this? \Cref{thm:l2_distortion_main}.
To our knowledge, this is the first first-order algorithm that improves on the dense all-pairs dependence of \cite{arora2005fast} for this particular problem.
\begin{table}[htbp]
  \centering
  {\renewcommand{\arraystretch}{1.8}%
  \setlength{\tabcolsep}{4pt}%
  \begin{tabular}{l l l r}
    \hline
    \textbf{Runtime} & \textbf{With \(\omega \sim 2.37\)} & \textbf{Parameters} & \\
    \hline
    \(\widetilde{O}(n^4 \log(1/\eps) L)\) & & & \cite{alizadeh1995interior} \\
    \(\widetilde{O}\p{ \min \p{ n^3 \kappa^{2.5} \eps^{-3.5},
    n^{3.5} \kappa^{1.5} \eps^{-2.5}}}\) & & \(\kappa = \frac{\sum_{ij} d_{ij}}{n^2 \min_{ij} d_{ij}}\) & \cite[Theorem 5]{arora2005fast} \\
    \(\widetilde{O}(n^{0.5+2\omega} \log(1/\eps))\) & \(\widetilde{O}(n^{\sim5.25} \log(1/\eps))\) & & \cite{jiang2020faster} \\
    \(\widetilde{O}\p{n^{2 \omega + o(1)} \log^{O(1)}(1/\eps)}\) & \(\widetilde{O}\p{n^{\sim4.74 + o(1)} \log^{O(1)}(1/\eps)}\) & & \cite{huang2022solving} \\
    \rowcolor{blue!5} \(\widetilde{O}\p{n^3 \eps^{-7.5}}\) & &  & \textbf{This work} \\
    \hline
  \end{tabular}
  }
  \caption{Comparison of runtimes for minimum-distortion embedding into \(\ell_2\).}
  \label{table:l2-embedding-runtime-comparison}
\end{table}

\cite{jiang2020faster} note that their method improves over previous methods, including the cutting-plane method of \cite{lee2015faster} and \cite{jiang2020improved}, in the regime where \(m \ge n\).
This is essentially the regime we care about: if the constraint graph has \(m < n-1\), then it is disconnected, and the problem splits into independent instances on each connected component.

Clearly, the \(\kappa\) parameter of \cite{arora2005fast}, 
an average-case form of the usual metric space aspect ratio \(\frac{\max_{ij} d_{ij}}{ \min_{ij} d_{ij}}\),
always satisfies \(\kappa \ge 1\).
Thus, the adaptive local-norm runtime is at least as fast as that of \cite{arora2005fast} in terms of pure dependence on the number of
points, at the expense of a worse dependence on the accuracy \(\eps^{-1}\).
Further, it removes the geometric condition parameter \(\kappa\) from the all-pairs minimum-distortion embedding runtime.
In applications where only moderate accuracy is required, such as finding embeddings up to a constant factor,
our algorithm avoids the expensive per-iteration cost of interior point methods and achieves a faster overall runtime.
We note that a more complete exposition of the algorithm of \cite{arora2005fast} can be
found in Kale's thesis \cite[\S 6.6.3]{kale2007efficient}.

\paragraph{Multidimensional scaling and fixed dimension.}
The problem of representing dissimilarity data by points in Euclidean space is traditionally studied under the name multidimensional scaling.
Classical MDS, also called Torgerson or Torgerson--Gower scaling, gives a spectral construction when the dissimilarities are treated as squared Euclidean distances
\cite{torgerson1952multidimensional,torgerson1958theory,gower1966some,borg2005modern}.
Nonmetric MDS and stress-based MDS instead fit more general similarity or dissimilarity data, leading to objectives such as Kruskal stress and to stress-majorization algorithms such as SMACOF
\cite{shepard1962analysisi,shepard1962analysisii,kruskal1964multidimensional,deleeuw2009multidimensional,borg2005modern}.
When the target dimension is fixed and the objective is distortion or stress, the corresponding optimization problems are generally nonconvex and can be computationally hard
\cite{matouvsek2010inapproximability,demaine2021multidimensional}.

\paragraph{Euclidean distance geometry and sparse distances.}
Euclidean distance geometry and Euclidean distance matrix completion ask for point configurations realizing partially observed distances
\cite{schoenberg1935remarks,young1938discussion,gower1966some,liberti2014euclidean}.
These problems arise in sensor network localization, molecular conformation, and statics, and have been studied via SDP relaxations and low-rank matrix-completion methods
\cite{alfakih1999solving,biswas2004semidefinite,so2007theory,tasissa2018exact}.
Ding, Krislock, Qian, and Wolkowicz explicitly formulate sensor network localization as an EDM completion problem and use SDP relaxations to handle noisy distance information \cite{ding2010sensor}.
Sparse distance constraints also arise in graph drawing and manifold learning, where graph distances or neighborhood distances are used as proxies for latent geometric structure
\cite{tenenbaum2000global,gansner2004graph}.

\paragraph{Positive SDP solvers.}
If the constraints in the SDP \eqref{eq:main_prob} are interpreted as Laplacian
matrices taking the inner product with \(X\), \(\pgo{L_e, X}\),
then we can see that every constraint involves a positive semidefinite \(L_e \succeq 0\).
An SDP where all constraints and objectives are PSD generally falls into a class
called positive SDPs, with a wide literature of fast algorithms, including some width-independent ones
\cite{allenzhu2016using,peng2012faster,jambulapati2020positive,jambulapati2023structured}.
While the graph-constrained \(\ell_2\)-configuration problems are not not within a class of SDPs
where we have good width-independent algorithms, we show that
they can be reduced to solving a sequence of positive SDPs that are potentially much more amenable
to speedup with custom solvers.

\section{Preliminaries}

We use \(\|\cdot\|_p\) for the \(p\)-norm,
and use \(\|\cdot\|\) as shorthand for the \(2\)-norm.
Eigenvalues of matrix are \(\lambda_1(M) \ge \lambda_2(M) \ge \cdots \ge \lambda_n(M)\).

\subsection{Graphs}

Let \(G = (V, E, w)\) be a graph with weights \(w \in \R^E_{\ge 0}\).
Let \(A \in \R^{m \times n}\) be the weighted incidence matrix of the graph,
i.e. \(A \defeq \Diag(w) B\) for \(B \in \R^{m \times n}\) the signed edge-vertex incidence matrix of \(G\).
Recall \(L = B^\top \Diag(w) B\) is the Laplacian for \(G\).
We let \(L_{ij}\) denote the Laplacian for just the edge \((ij)\), i.e. \(L_{ij} = b_{ij} b_{ij}^\top\)
for \(b_{ij} = e_i - e_j\).
We will occasionally interchange describing a graph directly by its Laplacian \(L\) or the weights of the Laplacian \(w\)
with the graph itself;
the underlying edge set and representation should be clear from context.

%
%\begin{definition}
%  \(\tilde{\sigma}\) are \textnormal{\(\nu\)-leverage score overestimates} of \(A\) if \(\|\tilde{\sigma}\|_1 \le \nu\),
%  and \(\tilde{\sigma} \ge \sigma(A)\).
%\end{definition}
%
%\begin{theorem}[\preserve{\cite[Theorem 3]{jambulapati2023chaining}}]
%  Let \(A \in \R^{m \times n}\) be the weighted incidence matrix of a graph.
%  There is an algorithm that computes \(O(n)\)-leverage score overestimates of \(A\) in time \(\widetilde{O}(m)\).
%\end{theorem}

\subsection{Numerical Linear Algebra}

In this paper we will depend heavily on Laplacian system solvers, i.e. 
fast nearly-linear algorithms solving systems of the form \(L x = b\).

\begin{theorem}[Laplacian system solvers, \protect{\cite{cohen2014solving}}]\label{thm:laplacian_solver}
  Let \(G = (V, E, w)\) be a weighted graph with \(n\) vertices and \(m\) edges, and let \(L\) be its Laplacian.
  For every \(b \in \operatorname{im}(L)\) and \(\eps, \delta \in (0, 1)\),
  there is a randomized algorithm \(\textsc{LapSolve}(L, b, \eps, \delta)\) that returns \(x \in \R^n\) satisfying
  \[
    \|x - L^+ b\|_L \le \eps \|L^+ b\|_L,
    \qquad
    \|y\|_L \defeq \sqrt{y^\top L y},
  \]
  with probability at least \(1-\delta\).
  The algorithm runs in time \(\widetilde{O}(m \log(1/\eps)\log(1/\delta))\).
\end{theorem}

\cite[Lemma 31]{kale2007efficient}
\cite[Theorem 4.2(a)]{kuczynski1992estimating}

Lanczos iteration with random start.
\(\textsc{LanczosRand}(M, \eps, \delta, \Lambda)\).
The only access that this algorithm needs to the matrix \(M\) is the product of \(M\) with any vector \(z\),
i.e. \(z \mapsto M z\).
Perhaps this theorem citation requires more rigorous explanation (differs from AHK).

\begin{theorem}\label{thm:lanczos_rand}
  Let \(M \in \R^{n \times n}\) be symmetric with eigenvalues \(\lambda_1 \ge \cdots \ge \lambda_n\).
  Let \(\eps > 0\) and \(1 > \delta > 0\) be an error tolerance and failure probability, respectively.
  Let \(\Lambda > 0\) be an estimate on the magnitude of the smallest eigenvalue,
  and \(\gamma\) be an accuracy parameter.

  There is an algorithm \(\textsc{LanczosRand}(M, \eps, \delta, \Lambda, \gamma)\) that either returns a unit vector \(x\) or reports \(\textsc{Nsd}\).
  If \(\Lambda \ge -\lambda_n\) and \(\gamma \le \min\p{1, \frac{\lambda_1 + \eps}{\lambda_1 + \Lambda}}\)
%   then the followig hold with probabiity at least \(1 - \delta\):
%   \begin{enumerate}
%     \item If \(\lambda_1 \ge 0\), then the algorithm returns unit vector \(x\) satisfying \(x^\top M x \ge - \eps\).
%     \item If the algorithm reports \(\textsc{Nsd}\), then \(\lambda_1 < 0\).
%   \end{enumerate}
    then with probability at least \(1-\delta\), if \(\lambda_1 \ge 0\), the algorithm returns a unit vector \(x\) satisfying \(x^\top M x \ge - \eps\).

  \textsc{LanczosRand} requires \(O\p{\frac{1}{\sqrt{\gamma}} \log\p{\frac{n}{\delta}}}\)
  matrix-vector products involving \(M\) and additional time \(O\p{\frac{n}{\sqrt{\gamma}} \log\p{\frac{n}{\delta}}}\).
\end{theorem}

The above theorem gives us a procedure for using Lanczos to solve eigenvalue problems, i.e. those corresponding
to Rayleigh quotients \(\max \frac{x^\top M x}{x^\top x}\).
However, later we will need to solve generalized eigenvalue problems corresponding to generalized Rayleigh quotients
\(\max \frac{x^\top M x}{x^\top H x}\).
First, we define these generalized eigenvalues:

\begin{definition}[Generalized eigenvalues of a pair]\label{def:generalized_eigenvalues_pair}
  Let \(M, H \in \R^{n \times n}\) be symmetric, with \(H \succeq 0\), and let \(r = \operatorname{rank}(H)\).
  The finite generalized eigenvalues of the pair \((M, H)\) are the eigenvalues
  \(\lambda_1 \ge \cdots \ge \lambda_r\) of the symmetric matrix \(H^{+/2} M H^{+/2}\) restricted to \(\operatorname{im}(H)\).
  Equivalently, they are the stationary values of the generalized Rayleigh quotient
  \[
    \frac{x^\top M x}{x^\top H x}
  \]
  over nonzero \(x \in \operatorname{im}(H)\).
  We will use the notation \(\lambda_i(M, H) \defeq \lambda_i\).
\end{definition}

Lanczos methods that solve these eigenvalue problems have been called generalized Lanczos schemes in the past \cite{kalamboukis1984lanczos} \cite{van1982generalized}.
Below, we demonstrate that the Lanczos iteration with random start method of \Cref{thm:lanczos_rand} can be modified
to find generalized eigenvalues.
\begin{lemma}\label{lem:generalized_lanczos_rand}
  Let \(M \in \R^{n \times n}\) be symmetric and let \(H \in \R^{n \times n}\) be a Laplacian matrix with \(r = \operatorname{rank}(H)\).
  Let \(\lambda_1 \ge \cdots \ge \lambda_r\) be the finite generalized eigenvalues of the pair \((M, H)\).
  Let \(\eps > 0\) and \(1 > \delta > 0\) be an error tolerance and failure probability, respectively.
  Let \(\Lambda > 0\) be an estimate on the magnitude of the smallest finite generalized eigenvalue,
  and \(\gamma\) be an accuracy parameter.

  There is an algorithm \(\textsc{GenLanczosRand}(M, H, \eps, \delta, \Lambda, \gamma)\) that either returns a vector \(x\) satisfying \(x^\top H x = 1\) or reports \textsc{Npos}. If \(\Lambda \ge -\lambda_r\), and \(\gamma \le \min\p{1, \frac{\lambda_1 + \eps}{\lambda_1 + \Lambda}}\), then with probability at least \(1-\delta\),
  if \(\lambda_1 \ge 0\), the algorithm returns \(x\) satisfying \(x^\top H x = 1\) and \(x^\top M x \ge - \eps\).
  \textsc{GenLanczosRand} requires \(O\p{\frac{1}{\sqrt{\gamma}} \log\p{\frac{n}{\delta}}}\)
  matrix-vector products involving \(M\)
  and additional time \(\widetilde{O}\p{\frac{n}{\sqrt{\gamma}} \log\p{\frac{1}{\delta}}}\).
\end{lemma}
\begin{proof}
Our plan is to simulate a call to \(\textsc{LanczosRand}(H^{+/2} M H^{+/2}, \eps, \delta, \Lambda)\),
while never actually materializing the matrix \(\tilde{M} \defeq H^{+/2} M H^{+/2}\) or any of the intermediate vectors, and instead keeping track of the intermediates in a different coordinate frame.
By the definition above, the eigenvalues of \(\tilde{M}\) on \(\operatorname{im}(H)\) are exactly the finite generalized eigenvalues of the pair \((M, H)\), so the parameter conditions in \Cref{lem:generalized_lanczos_rand} imply those required in the statement of \Cref{thm:lanczos_rand}.

The Lanczos algorithm with random start as described in \cite{kuczynski1992estimating}
(also see e.g. \cite[\S 3.3]{tropp2021acm}) involves a sequence of a restricted class of operations:
\begin{enumerate}
    \item An initial \(\tilde{x}_0\) that is distributed uniformly on the sphere.
    \item The next \(\tilde{x}_j\) can be \(\tilde{x}_j = \tilde{M} x_{i}\) for some \(i < j\).
    \item The next \(\tilde{x}_j\) can be \(\tilde{x}_j = \alpha_1 \tilde{x}_1 + \ldots + \alpha_{j-1} \tilde{x}_{j-1}\), where each coefficient
    \(\alpha_i\) for \(i < j\) is either \(0\), \(1/\|\tilde{x}_i \|\), or \(\pgo{\tilde{M} \tilde{x}_{i'}, \tilde{x}_{i''}}\).
\end{enumerate}
Instead of materializing each of the \(\tilde{x}_i\)s, we store a computed \(x_i\) and maintain the invariant \(\tilde{x}_i = H^{1/2} x_i\).
We argue that each \(\tilde{x}_i\) can be simulated efficiently using an inductive argument, arguing this for each possible operation.
\begin{enumerate}
    \item We must simulate \(\tilde{x}_0\) with \(x_0\) such that \(\tilde{x}_0 = H^{1/2} x_0\) is distributed uniformly on the sphere.
    Since \(H\) is Laplacian, let \(H = B^\top W B\) be its decomposition.
    Then set:
    \[x_0 = \frac{z}{\sqrt{z^\top H z}}, \qquad z \sim H^+ B^\top W^{1/2} \xi, \qquad \xi \sim \cN(0, I)\]
    To confirm that the simulated \(\tilde{x}_0\) has the right distribution,
    define \(\tilde{z} = H^{1/2} z\).
    We have that: \[\tilde{x}_0 = \frac{H^{1/2} z}{\sqrt{\tilde{z} H^{\sfrac{+}{2}} H H^{\sfrac{+}{2}} \tilde{z}}} = \frac{\tilde{z}}{\|\tilde{z}\|}\]
    where \(\tilde{z}\) is Gaussian with mean zero and has covariance \(H^{1/2} H^+ B^\top W^{1/2} W^{1/2} B H^+ H^{1/2} = I\), showing that \(\tilde{x}_0\)
    is indeed distributed uniformly on the sphere.
    To compute \(x_0\) requires only two matrix-vector multiplications with \(H\) and \(B^\top W^{1/2}\), and one Laplacian system solve with \(H\).
    \item We have \(\tilde{x}_j = \tilde{M} \tilde{x}_i = H^{\sfrac{+}{2}} M H^{\sfrac{+}{2}} \tilde{x}_i\), and so \(x_j = H^+ M x_i\).
    Thus applying the matrix-vector product with \(\tilde{M}\) can be simulated with one matrix-vector product for \(M\) and one Laplacian system solve with \(H\).
    \item Notice \(\| \tilde{x}_i\| = \sqrt{x_i^\top H x_i}\), which can be simulated with one matrix-vector product for \(H\). Next, observe \(\pgo{\tilde{M} \tilde{x}_{i'}, \tilde{x}_{i''}} = \pgo{H^{+/2} M H^{+/2} H^{1/2} x_{i'}, H^{1/2} x_{i''}} = \pgo{M x_{i'}, x_{i''}}\),
    so this inner product can be simulated with one matrix-vector product for \(M\).
    Finally, by linearity, we can clearly simulate \(\tilde{x}_j = \sum_{i < j} \alpha_i \tilde{x}_j\) with \(x_j = \sum_{i < j} \alpha_i x_i\).
\end{enumerate}
There is a final step of Lanczos by which the eigenvalues of a tridiagonal matrix formed from the scalars in operation type (3) are computed in \(O(n)\) time.
We have shown that all the values of all these scalars are equal to their corresponding values in the simulated call to \textsc{LanczosRand}, and so we can treat the final step of Lanczos as a black box to extract an extremal generalized finite eigenvalue of the pair \((M, H)\).

Finally, we return the last \(x_j\) instead of \(\tilde{x}_j\).
Observe that \(\tilde{x}^\top \tilde{x} = 1\) implies \(x^\top H x = 1\),
and \(\tilde{x}^\top \tilde{M} \tilde{x} \ge - \eps\) implies \(x^\top M x \ge - \eps\),
and the other case of \(\tilde{M}\) being negative semidefinite corresponds to all finite generalized eigenvalues
of the pair \((M, H)\) being nonpositive. Thus, the returned value satisfies the desired conditions.

For the final runtime, Lanczos performs at most \(O\p{\frac{1}{\sqrt{\gamma}} \log\p{\frac{n}{\delta}}}\) operations
of the three types we outlined above; each of these performs a constant number of \(O(n)\) time vector operations, matrix-vector products involving \(M\), and Laplacian system solves for \(H\).
By \Cref{thm:laplacian_solver}, each of those system solves takes time at most \(\widetilde{O}(n)\).
The final step of Lanczos takes \(O(n)\) time, giving the desired time complexity bounds.
\end{proof}

\subsection{Algorithms for Ellipsoidal Rounding}
Let \(A \in \R^{m \times n}\), and define \(K = \{x \in \R^n \colon \|Ax\|_\infty \le 1\}\).
As a key primitive we use an algorithm that computes a `rounding' ellipsoid of this convex body \(K\).
\begin{theorem}[\protect{\cite[Theorem 11]{cohen2019near}}]
  There exists an algorithm that \(\textsc{ApproxJohnEllipsoid}(A, \sigma, \eps, \delta)\), that for all \(\eps, \delta \in (0, 1)\),
  returns a vector \(\sigma \in \R^m_{\ge 0}\) such that for \(\cE = \{ x \in \R^n \colon x^\top A^\top \Diag(\sigma) A x \le 1\}\), we have with probability at least \(1-\delta\):
  \begin{equation}\label{eq:john_approx} \frac{1}{\sqrt{1+\eps}} \cdot \cE \subseteq K \subseteq \sqrt{n} \cdot \cE
  \end{equation}
  This algorithm requires solving at most \(O(\eps^{-2} \log(m/\delta))\) linear systems of the form
  \(A^\top \Sigma A x = b\), where \(\Sigma\) is a non-negative diagonal matrix.
\end{theorem}

The guarantee \eqref{eq:john_approx} ensures that \(\cE\) approximates the set \(K\) almost as well as possible.

Although this algorithm works for an arbitrary \(A\),
as observed in \cite{cohen2019near}, if \(A\) is the weighted incidence matrix of a graph,
then \(A^\top \Sigma A x = b\) is a Laplacian linear system.
This will be a key special case for us, so we define a specialized version of this algorithm where we interpret
the returned weights as modifying the weights of the input graph.
Given a weighted graph \(G = (V, E, w)\) with Laplacian \(L_G\), we define the weighted graph's Laplacian ellipsoid \(\cE_{L_G}\) (or sometimes \(\cE_{G}\)) as
\[\{x \in \R^n \colon x^\top L_G x \le 1\}\]
and the weighted graph's Lipschitz body \(K_{L_G}\) as
\[K_{L_G} \defeq \pc{x \colon \|\Diag(w) B\|_\infty \le 1}\]
Now we may state the guarantees of the algorithm when specialized to these graph-related convex bodies:
\begin{corollary}\label{corr:approx_lap_ellipsoid}
    There exists an algorithm \(\textsc{ApproxLapEllipsoid}(G)\) with a universal constant \(C_{\ref{eq:approx_lap_ellipsoid_guarantee}} > 0\) that, given a weighted graph \(G\), returns another graph Laplacian \(H\) on the same set of edges such that with high probability,
    \begin{equation}\label{eq:approx_lap_ellipsoid_guarantee}
    \frac{1}{C_{\ref{eq:approx_lap_ellipsoid_guarantee}} \sqrt{n}} \cE_{H} \subseteq K_{L_G} \subseteq \cE_{H}
    \end{equation}
    The algorithm takes time \(\widetilde{O}(m)\).
\end{corollary}

\subsection{Multiplicative Weights}

Recall that in the online convex optimization framework,
a `regret-minimizing' player players iterates against an adversary that
plays loss functions.
Formally, the regret-minimizing player plays iterates \(x_1, x_2, \ldots \in K\) for some convex set \(K\).
After each iterate \(x_t\) the adversary plays a convex loss function \(\ell_t \colon K \to \R\) upon which the player sees the entire loss function and suffers loss \(\ell_t(x_t)\). The goal of the player is to minimize regret,
i.e. to play a sequence of iterates such that the total loss is not much worse than the best a fixed solution could achieve in hindsight.
Formally, regret relative to a fixed solution \(u \in K\) is.
\[\regret_T(u) \defeq \sum_{t=1}^T\p{\ell_t(x_t)-\ell_t(u)}\]

An important special case of this framework where the set \(K = \Delta_N\), the set of distributions on \(N\)
elements.
In this case, there is a well known regret guarantee achieved by a simple regret minimizer called the multiplicative weight update (MWU), or sometimes exponentiated gradient (EG) algorithm.

\begin{theorem}[e.g. \protect{\cite[pg. 56]{orabona2019modern}}]
  There is an algorithm that plays iterates \(x_1, \ldots, x_T \in \Delta_N\) such that for any convex losses \(\ell_1, \ldots, \ell_T \colon \Delta_N \to \R\) and any \(u \in \Delta_N\),
  \[\regret_T \le \frac{\ln N}{\eta} + \frac{\eta}{2} \sum_{t=1}^T \|g_t\|_\infty^2\]
  where for each \(t = 1, \ldots, T\), \(g_t = \nabla \ell_t(x_t)\).
\end{theorem}

The quantity \(W = \max_t \|g_t\|_\infty\) is commonly called the `width' of the algorithm.

After some simple calculation and an optimal choice of step size,
the above regret bound yields the following iteration complexity bound.
\begin{corollary}\label{corr:mwu_regret}
  In the above setting, let \(W \ge \|g_t\|_\infty\) for all \(t\).
  Then after
  \begin{equation}\label{eq:mwu_regret_iteration_bound}
    T = O\p{\ln(N) \cdot \frac{W^2}{\eps^2}}
  \end{equation}
  many iterations we have \(\frac{1}{T} \regret_T \le \eps\).
\end{corollary}

We will also use a `local-norm' version of the same regret analysis.
From a high level, the above regret bound comes from using the strong
convexity of the entropy regularizer 
(the regularizer choice for mirror descent that leads to MWU/EG), 
a global second-order condition. 
ocal-norm regret bounds instead use the local second-order information for the regularizer at the current iterate.
Orabona derives the following exponentiated-gradient specialization in \cite[\S 6.6, after Remark~6.34]{orabona2019modern} by applying the local-norm OMD bound of \cite[Lemma~6.32]{orabona2019modern} to the entropy regularizer.
The bound we cite is the `signed-gradient' version.
\begin{theorem}[Exponentiated gradient local-norm regret bound, \protect{\cite[pg. 81]{orabona2019modern}}]\label{thm:orabona_local_norm_eg}
    For every \(W > 0\) and \(\eta \le \frac{1}{W}\), there is an algorithm (exponentiated gradient) with the following regret guarantee.
    Let \(\ell_1, \ldots, \ell_T \colon \Delta_N \to \R\) be losses such that for all \(t\), \(\max_{x \in \Delta_N} \|\nabla \ell_t(x)\|_\infty \le W\).
    For every \(u \in \Delta_N\), we have
\[
        \regret_T
        \le
        \frac{\ln N}{\eta}
        +
        \frac{e}{2} \eta \sum_{t=1}^T \pgo{g_t^2, x_t}.
\]
    where \(g_t = \nabla \ell_t(x_t)\), and \(g_t^2\) is the element-wise square of \(g_t\).
\end{theorem}

A bound like this for multiplicative weights appeared in \cite[Theorem 2.3]{arora2012multiplicative},
where it was also discussed in the context of Lagrangian-relaxation type solvers like ours.
However, this bound was specialized to multiplicative weights (specifically to \textsc{Hedge}), and was not using the local norm language that is more general and optimization-grounded.
Closely related regret bounds based on local norm appear in the positive-LP multiplicative-update analysis of \cite{allenzhu2015break} and for matrix multiplicative weights/FTRL in \cite{allenzhu2015spectral}.

The following lemma packages the local-norm regret bound into the form we will 
use later: once the per-round local quadratic term is uniformly bounded, it 
gives an immediate iteration bound for achieving small average regret.
\begin{lemma}\label{lem:local_norm_regret_iteration_bound}
    Fix \(\nu, \rho, \eps > 0\). Let \(\ell_1, \ldots, \ell_T \colon \Delta_N \to \R\) be a sequence of losses,
    and let \(x_1, \ldots, x_T \in \Delta_N\) be the responses from exponentiated gradient.
    Finally, let \(g_t = \nabla \ell_t(x_t)\) for all \(t\).
    If for all \(t\), we have \(\pgo{g_t^2, x_t} \le \nu\) and \(\|g_t\|_\infty \le W\),
    then after
    \begin{equation}\label{eq:local_norm_regret_iteration_bound}
        T = O\p{\ln( N) \cdot \max\p{\frac{\nu }{\eps^2}, \frac{W}{\eps}}}
    \end{equation}
    many iterations we have \(\frac{1}{T} \regret_T \le \eps\).
\end{lemma}
\begin{proof}
    By applying \Cref{thm:orabona_local_norm_eg} and the \(\nu\) bound, we have for any \(0 < \eta \le 1/W\) that:
    \[\frac{1}{T} \regret_T \le \frac{\ln N}{\eta T} + \eta \frac{e}{2} \nu \]
    In order for the average regret to match our target \(\eps\),
    we must have \(T \ge \frac{\ln N}{\eta \p{\eps - \eta \frac{e}{2} \nu}}\).
    Thus, the problem reduces to maximizing \(\eta\p{\eps - \eta \frac{e}{2} \nu}\) subject to the constraint that \(0 < \eta \le 1/W\).
    This objective is a concave quadratic with maximum at \(\eta = \Theta(\eps/\nu)\),
    and so the optimal choice of \(\eta\) is \(\Theta(\min(\eps/\nu, 1/W))\).
    Plugging this choice of \(\eta\) into the lower bound for \(T\) gives the desired bound.
\end{proof}

\section[Algorithm for Graph-Constrained \texorpdfstring{\(\ell_2\)}{l2}-Configuration]{Algorithm for Graph-Constrained \texorpdfstring{\(\ell_2\)}{l2}-Configuration}
Let \((V, E, u, l)\) be an instance of the \(\ell_2\) configuration problem.
As previously described, taking the squares of the constraints in \eqref{eq:main_prob} gives us an SDP.
\begin{align}\label{eq:main_sdp}
  \begin{aligned}\text{find}~ & v_1, \ldots, v_n \in \ell_2                                      \\
               \text{s.t.}~ & l_{ij}^2 \le \|v_i - v_j\|^2 \le u_{ij}^2 & \forall~(i, j) \in E
  \end{aligned}
\end{align}

We now describe how our algorithm works to solve this SDP.
First, we can conceptually rewrite the SDP into an optimization problem where the constraints are normalized.
\begin{align}\label{eq:main_sdp_trans}
  \begin{aligned}\text{find}~ & v_1, \ldots, v_n \in \ell_2                                       \\
               \text{s.t.}~ & 1 - \frac{\|v_i - v_j\|^2}{l_{ij}^2} \le 0 & \forall~(i, j) \in E \\
                            & \frac{\|v_i - v_j\|^2}{u_{ij}^2} - 1 \le 0 & \forall~(i, j) \in E
  \end{aligned}
\end{align}

Consider the Lagrangian of the problem \eqref{eq:main_sdp_trans}. Associated to the constraints we have dual variables \(y, z \in \R^E_{\ge 0}\):
\begin{equation}\label{eq:lang}\sL(v_1, \ldots, v_n, y, z) = \sum_{(i, j) \in E} y_{ij} \p{1 - \frac{\|v_i - v_j\|^2}{l_{ij}^2
    }} + \sum_{(i, j) \in E} z_{ij} \p{\frac{\|v_i - v_j\|^2}{u_{ij}^2} - 1}
\end{equation}

One further conceptual step is that on each iteration of the algorithm, we will consider an instance of the one dimensional version of the problem, where we have embedding variables \(x_1, \ldots, x_n \in \R\)
rather than \(v_1, \ldots, v_n \in \ell_2\).
With variables \(x \in \R^V\) and \(y, z \in \R^E_{\ge 0}\),
we define the Lagrangian of this problem:
\begin{equation}\label{eq:lang_1d}\sL_1(x, y, z) = \sum_{(i, j) \in E} y_{ij} \p{1 - \frac{(x_i - x_j)^2}{l_{ij}^2
    }} + \sum_{(i, j) \in E} z_{ij} \p{\frac{(x_i - x_j)^2}{u_{ij}^2} - 1}
\end{equation}

%Notice that \(g_1(x) \defeq \inf \{\sL_1(x, y, z) \mid y, z \ge 0\}\) is \(0\) if the one-dimensional embedding \(i \mapsto x_i\) satisfies the constraints, and is \( \infty\) otherwise.

Our algorithm follows the template of other primal-dual algorithms for feasibility problems (also called `Lagrangian relaxation' algorithms).
Our algorithm is set up as a game between a regret-minimizing \((y,z)\) `dual' player, and a `primal' \((x)\) player given by an \textsc{Oracle}.
This game proceeds via some number \(T\) of rounds,
and returns the concatenation of \(x\) values produced by the \textsc{Oracle}.

\subsection{Existence of oracle solution}

Now, we discuss the design and analysis of the \textsc{Oracle}.
We need to establish that if the problem instance is feasible, then on each iteration there is exists
a one-dimensional solution \(x\) that satisfies an `average case' version of the desired guarantees.
% While we did argue in the previous lemma that a `feasible' one dimensional solution exists when the problem instance is feasible,
% we were not able to bound the width of this solution.
The key idea of this nxt lemma is there must be a coordinate of the feasible solution that is at least as good as the feasible solution,
with the width of this solution depending on the number of dimensions.
By using the Johnson-Lindenstrauss lemma, we can make this dependence logarithmic in \(n\).
\begin{lemma}\label{lemma:struct_lemma}
    There exists a universal constant \(C_{\ref{eq:struct_lemma_width}}\) such that the following holds.
  If \(\{v\}\) is feasible for \eqref{eq:main_sdp}, then for any \((y, z) \in \Delta_{2m}\) and \(\eps > 0\), there is \(x \in \R^V\) such that \(\sL_1(x, y, z) \le \eps\)  and 
  \begin{equation}\label{eq:struct_lemma_width}
    \max_{ij} \frac{(x_i - x_j)^2}{l_{ij}^2} \le C_{\ref{eq:struct_lemma_width}} \cdot \alpha^2 \log n \cdot \frac{1+\eps}{\eps^2}
    \end{equation}
\end{lemma}

\begin{proof}
  We first establish a version of this bound where we guarantee \(\sL_1(x, y, z) \le 0\), but the width bound can be as large as \(\alpha^2 n\).
  As above, assume without loss of generality that \(v_1, \ldots, v_n \in \R^d\) for \(d \le n\).
  Observe that for any \(y, z \ge 0\), we have: \[0 = \sL(\{v\}, y, z) = \sum_{k=1}^d \sL_1((v_1)_k, \ldots, (v_n)_k, y, z) = \frac{1}{d} \sum_{k=1}^d \sL_1( \sqrt{d} \cdot (v_1)_k, \ldots, \sqrt{d} \cdot (v_n)_k, y, z)\]
  Thus, there must exist a coordinate \(k\) for which \(\sL_1(\sqrt{d} \cdot (v_1)_k, \ldots, \sqrt{d} \cdot (v_n)_k, y, z) \le 0\),
  and we may take \(x_i = \sqrt{d} \cdot (v_i)_k\).

  Again by feasibility, we have \(\sum_{k=1}^d ((v_i)_k - (v_i)_k)^2 \le u_{ij}^2 \le \alpha^2 \cdot \ell_{ij}^2\) for all \(i, j \in E\).
  Thus clearly we have \(\frac{(x_i - x_j)^2}{l_{ij}^2} \le \alpha^2 \cdot d\).

  In order to improve the width bound, we will use the Johnson-Lindenstrauss lemma (see e.g. \cite{dasgupta2003elementary}).
  By applying this lemma to \(v_1, \ldots, v_n\) with some fixed \(\eps > 0\), we obtain an embedding \(\tilde{v}_1, \ldots, \tilde{v}_n \in \R^{\tilde{d}}\)
  for \(\tilde{d} \lesssim \eps^{-2} \log n\),
  for which \((1-\eps) \|v_i - v_j\|^2 \le \|\tilde{v}_i - \tilde{v}_j\|^2 \le (1+\eps) \|v_i - v_j\|^2\) for all pairs \(i, j \in V\).
  Using this embedding and the fact that \(v_1, \ldots, v_n\) is feasible, we have
  \begin{align*}
    \sL(\{\tilde{v}\}, y, z) & \le \sum_{(i, j) \in E} y_{ij} \p{1 - \frac{\|v_i - v_j\|^2}{l_{ij}^2} + \eps} + z_{ij} \p{\frac{\|v_i - v_j\|^2}{u_{ij}^2} - 1 + \eps} \\
                             & \le \eps 
  \end{align*}
  By a similar argument as before, there is a coordinate \(k\) for which taking \(x_i = \sqrt{\tilde{d}} \cdot (\tilde{v}_i)_k\)
  yields \(\sL_1(x_1, \ldots, x_n, y, z) \le \eps\).
  To establish the width bound, observe that \(\sum_{k=1}^{\tilde{d}} ( (\tilde{v}_i)_k - (\tilde{v}_j)_k)^2 \le (1+\eps) u_{ij}^2 \le (1+\eps) \alpha^2 \cdot l_{ij}^2\). Plugging in \(x\) yields the desired width bound.
\end{proof}

\subsection{Local norm-bounded oracle}
In the previous section we gave an existence result for a `good' one-dimensional solution \(x\);
in this section we describe how to implement an oracle that approximately finds this solution.
In this section, pick \begin{equation}\label{eq:width_choice}
    \rho = C_{\ref{eq:struct_lemma_width}} \alpha^2 \log n \frac{4+2\eps}{\eps^2}
\end{equation}
i.e. the width guaranteed by \Cref{lemma:struct_lemma} for \(\eps/2\).
Recall that:
\[\cL_1(x, y, z) = \|y\|_1 - \|z\|_1 + x^\top \p{\sum_{ij \in E} \p{\frac{z_{ij}}{u_{ij}^2} - \frac{y_{ij}}{l_{ij}^2}} L_{ij}} x\]
We define the local-norm width of a value \(x\): \[\textrm{locn}_{y, z}(x) \defeq \sum_{ij \in E} y_{ij}\p{1 - \frac{(x_i - x_j)^2}{l_{ij}^2}}^2 + z_{ij} \p{\frac{(x_i - x_j)^2}{u_{ij}^2} - 1}^2\]
and the regular width:
\[\textrm{width}(x) \defeq \max_{ij} \frac{(x_i - x_j)^2}{l_{ij}^2}\]

Let \(w = \p{\frac{1}{l_{ij} \sqrt{\rho}}}_{ij \in E} \in \R^E\), then we define the low-width body:
\[K_{(V, E, w)} \defeq \{x \colon |x_i - x_j| \le l_{ij} \sqrt{\rho} \qquad\forall ij \in E\}\]

For concreteness, we define exactly what guarantees the \textsc{Oracle} must satisfy.
\begin{definition}\label{def:oracle_local_norm}
  Consider a procedure that takes \((y, z) \in \Delta_{2m}\) and constructs a weighted graph Laplacian \(H\) supported on edge set \(E\), and either returns \(\textsc{Witness}(x)\)
  where \(x \in \R^n\), or reports \(\textsc{NoWitness}\).
  We say this is a  \((W, \nu)\)-bounded \(\eps\)-approximate \(\delta\)-sure \textnormal{\textsc{Oracle}} if:
  \begin{enumerate}
    \item\label{item:def_oracle_local_norm_containment_guarantee} \(H\) always satisfies \(K_{(V, E, w)} \subseteq \cE_H\).
    \item\label{item:def_oracle_local_norm_witness_guarantee} Whenever it returns \(\textsc{Witness}(H, x)\), we have \(\cL_1(x, y, z) \le \eps\), \(x^\top H x \le 1\), \(\textnormal{width}(x) \le W\), and \(\textnormal{locn}_{y, z}(x) \le \nu\).
    \item\label{item:def_oracle_local_norm_high_prob_guarantee} With probability at least \(1-\delta\), 
    if \(\min_{x \in \cE_H} \cL_{1}(x, y, z) \le \eps/2\), the oracle returns \(\textsc{Witness}(x)\).
    % \begin{enumerate}
    %     \item\label{item:def_oracle_local_norm_completeness} If \(\min_{x \in \cE_H} \cL_{1}(x, y, z) \le \eps/2\), it returns \(\textsc{Witness}(x)\).
    %     \item\label{item:def_oracle_local_norm_soundness} If it reports \textsc{NoWitness}, we have \(\min_{x \in \cE_H} \sL_1(x,y,z) > \eps/2\).
    % \end{enumerate}
  \end{enumerate}
\end{definition}

Let \(H_K = \textsc{ApproxLapEllipsoid}((V, E, w))\), and recall that the guarantee \eqref{eq:approx_lap_ellipsoid_guarantee}
says that \begin{equation}\label{eq:local_norm_oracle_lap_ellipsoid_guarantee}
    \frac{1}{C_{\ref{eq:approx_lap_ellipsoid_guarantee}} \sqrt{n}} \cE_{H_K} \subseteq K_{(V, E, w)} \subseteq \cE_{H_K}
\end{equation}

Our implementation of the oracle is in \Cref{alg:local_norm_bd_oracle}.

\begin{algorithm}[ht]
  \caption{Local-norm bounded oracle \(\textsc{LnbOracle}(y, z, \eps, \delta)\)}
  \label{alg:local_norm_bd_oracle}
  \begin{algorithmic}[1]
    \Require \((V, E, u, l)\), \(W, \rho, \nu, H_K\) (context), \(y, z, \eps, \delta\)
    \Ensure \(H\), and (\(\textsc{Witness}(x)\) or \(\textsc{NoWitness}\))
    % \State \(y^{(0)}, z^{(0)} \leftarrow \frac{1}{2m} (1, \ldots, 1) \in \Delta_{2m}\)
    % \For{\(t = 0, \ldots, T-1\)}
    % \State \(x^{(t)} \leftarrow \textsc{Oracle}(y^{(t)}, z^{(t)})\)
    % \If{\textsc{Oracle} reports \textsc{Fail}}
    % \State \Return \textsc{Infeasible}
    % \EndIf
    % \State \(\ell^{(t)}(y, z) = \sL_1(x^{(t)}, y, z)\)
    % \State \(y^{(t+1)}, z^{(t+1)} \leftarrow \textsc{MWU}(y^{(t)}, z^{(t)}, -\ell^{(t)})\)
    % \EndFor
    % \State \(v_i = \frac{1}{\sqrt{T}} (x^{(0)}_i, \ldots, x^{(T-1)}_i)\) for all \(i = 1, \ldots, n\)
    % \State \Return \(\{v_i\}_{i=1}^n\)
    \State \(H_{\text{loc}} = \sum_{ij \in E} (y_{ij} + z_{ij}) \frac{1}{l_{ij}^2 \rho} L_{ij}\)
    \State \(H = \frac{1}{2} H_{\text{loc}} + \frac{1}{2} H_K\)
    \State \(c = \|y\|_1 - \|z\|_1\)
    \If{\(c \le \eps\)}
        \State \Return \(H, \textsc{Witness}(0)\) \label{alg_line:local_norm_bd_oracle_earlyreturn}
    \EndIf
    \State \(Q = \sum_{ij \in E} \p{\frac{z_{ij}}{u_{ij}^2} - \frac{y_{ij}}{l_{ij}^2}} L_{ij}\)
    \State \(M = \p{\frac{3}{4} \eps - c} H - Q\)
    \State \(\Lambda = 2 + 2 \rho\)
    \State \(\gamma = \eps/(4\Lambda)\)
    \State \(x = \textsc{GenLanczosRand}(M, H, \eps/4, \delta, \Lambda, \gamma)\)
    \If{\(x = \textsc{Npos}\)}
        \State \Return \(H, \textsc{NoWitness}\)
    \ElsIf{\(\cL_1(x, y, z) \le \eps\) and \(x^\top H x \le 1\) and \(\textrm{locn}_{y, z}(x) \le \nu\) and \(\textrm{width}(x) \le W\)} \label{alg_line:local_norm_bd_oracle_finalif}
        \State \Return \(H, \textsc{Witness}(x)\) \label{alg_line:local_norm_bd_oracle_witness_return}
    \Else
        \State \Return \(H, \textsc{NoWitness}\) \label{alg_line:local_norm_bd_oracle_nowitness_return}\
    \EndIf
  \end{algorithmic}
\end{algorithm}

\begin{lemma}\label{lem:local_norm_oracle_gen_eigenvalue_bounds}
    The following generalized eigenvalue bounds hold:
    \begin{enumerate}
        \item\label{item:local_norm_oracle_completeness_eigenvalue_bound}
        If \(\min_{x \in \cE_H} \cL_1(x, y, z) \le \eps/2\), then the largest finite generalized eigenvalue of the pair \((M, H)\)
        is at least \(\eps/4\).
        \item\label{item:local_norm_oracle_min_gen_eigenvalue_bound}
        The magnitude of any finite generalized eigenvalue of the pair \((M, H)\) is at
        most \(\Lambda = 2 + 2\rho\).
    \end{enumerate}
\end{lemma}
\begin{proof}
    For \eqref{item:local_norm_oracle_completeness_eigenvalue_bound},
    %  observe how the generalized eigenvalues of the pair \((M, H)\) are related to the pair \((Q, H)\): we have the relation \(\lambda_i(M, H) = \frac{3}{4} \eps - c - \lambda_i(Q, H)\).
    note first that
    \[\cL_1(x, y, z)  = c + x^\top Q x\]
    and that for any quadratic form \(F\), any extremal value of \(x^\top F x\) on \(\cE_H\)
    must lie on the boundary of \(\cE_H\). This means that the extremal values of \(x^\top F x + \beta \cdot x^\top H x\)
    and \(x^\top F x + \beta\) must be the same on \(\cE_H\).
    And so by assumption we have that \[\min_{x \in \cE_H} x^\top (c H + Q) x \le \eps/2\]
    and consequently \[\frac{1}{4} \eps \le \max_{x \in \cE_H} x^\top \p{\frac{3}{4} \eps \cdot H - (c H + Q) } x = \max_{x \in \cE_H} x^\top M x \]
    as desired.

    For \eqref{item:local_norm_oracle_min_gen_eigenvalue_bound}, we will argue this by first showing that \(\pa{x^\top Q x} \le 2 \rho \cdot x^\top H x\) for all \(x \in \R^n\).
    Now, we have
    \begin{align*}
        \pa{x^\top Q x} &\le \sum_{ij \in E} \p{\frac{y_{ij}}{l_{ij}^2} + \frac{z_{ij}}{u_{ij}^2}} (x_i - x_j)^2 \\
        &\le \sum_{ij \in E} \p{y_{ij} + z_{ij}}\frac{(x_i - x_j)^2}{l_{ij}^2} \\
        &= \rho \sum_{ij \in E} \p{y_{ij} + z_{ij}}\frac{(x_i - x_j)^2}{\rho l_{ij}^2} \\
        &\le 2 \rho \cdot x^\top H x
    \end{align*}

    Now, we can simply bound the generalized eigenvalues of the pair \((M, H)\) in the same way:
    \begin{align*}
        \pa{x^\top M x} &\le \pa{\frac{3}{4} \eps - c} x^\top H x + \pa{x^\top Q x} \\
        &\le \p{\pa{\frac{3}{4} \eps - c} + 2 \rho} x^\top H x
    \end{align*}
    Finally, because \(\eps \le 1\) and \(|c| \le 1\), after dividing through
    we have the desired bound \(\Lambda = 2 \rho + 2\).
\end{proof}

\begin{lemma}\label{lem:local_norm_oracle_K_contained_in_EH}
    \(K_{(V, E, w)} \subseteq \cE_{H}\).
\end{lemma}
\begin{proof}
    Fix any \(x \in K_{(V, E, w)}\). Then for all \(ij \in E\), we have
    \(|x_i - x_j| \le l_{ij} \sqrt{\rho}\).
    Hence \[x^\top H_{\text{loc}} x = \sum_{ij \in E} (y_{ij} + z_{ij}) \frac{(x_i - x_j)^2}{l_{ij}^2 \rho} \le 1\]
    In addition, the right inequality on \eqref{eq:local_norm_oracle_lap_ellipsoid_guarantee} implies that
    \(x^\top H_K x \le 1\).
    Putting these together, we have \(x^\top H x = \frac{1}{2} x^\top H_{\text{loc}} x + \frac{1}{2} x^\top H_K x \le 1\),
    implying the desired conclusion.
\end{proof}

Now, we establish the width bounds.
\begin{lemma}\label{lem:local_norm_oracle_ellipsoid_size_bounds}
    For every \(x \in \cE_{H}\), we have:
    \begin{enumerate}
        \item\label{item:local_norm_oracle_weighted_stretch_bound} \(\sum_{ij \in E} (y_{ij} + z_{ij}) \frac{(x_i - x_j)^2}{l_{ij}^2} \le 2 \rho\)
        \item\label{item:local_norm_oracle_width_bound} \(\textnormal{width}(x) \le O(n \rho)\)
        \item\label{item:local_norm_oracle_locn_bound} \(\textnormal{locn}_{y, z}(x) \le O(n \rho^2)\)
    \end{enumerate}
\end{lemma}
\begin{proof}
    For \eqref{item:local_norm_oracle_weighted_stretch_bound}, since \(x^\top H x \le 1\)
    and \(H_{\text{loc}} \preceq 2 H\), we have
    \[\sum_{ij \in E} (y_{ij} + z_{ij}) \frac{(x_i - x_j)^2}{l_{ij}^2 \rho} = x^\top H_{\text{loc}} x \le 2 x^\top H x \le 2\]
    
    For \eqref{item:local_norm_oracle_width_bound}, we can do something similar.
    We have \(x^\top H x \le 1\) and \(H_K \preceq 2 H\), so that \(x^\top H_K x \le 2\), i.e. \(\cE_{H} \in \sqrt{2} \cE_{H_K}\).
    The left inequality in \eqref{eq:local_norm_oracle_lap_ellipsoid_guarantee} can be rearranged as
    \(\cE_{H_K} \subseteq C_{\ref{eq:approx_lap_ellipsoid_guarantee}} \sqrt{n} K_{(V, E, w)}\).
    Combining these, we have \(\cE_{H} \subseteq C_{\ref{eq:approx_lap_ellipsoid_guarantee}} \sqrt{2n} K_{(V, E, w)}\).
    Thus for any \(ij \in E\), we have
    \begin{equation}\label{eq:local_norm_oracle_edge_width_bound}
        \frac{(x_i - x_j)^2}{l_{ij}^2 \rho} \le 2 C^2_{\ref{eq:approx_lap_ellipsoid_guarantee}} n \rho = O(n \rho)
    \end{equation}

    For \eqref{item:local_norm_oracle_locn_bound}, we first have by \eqref{item:local_norm_oracle_weighted_stretch_bound} and \eqref{item:local_norm_oracle_width_bound} that
    \[\sum_{ij \in E} (y_{ij} + z_{ij}) \frac{(x_i - x_j)^4}{l_{ij}^4} \le \text{width}(x) \sum_{ij \in E} (y_{ij} + z_{ij}) \frac{(x_i - x_j)^2}{l_{ij}^2} = O(n \rho^2)\]
    Now, we have \(0 \le \frac{(x_i - x_j)^2}{u_{ij}^2} \le \frac{(x_i - x_j)^2}{l_{ij}^2}\), and so we have
    \begin{align*}
        \text{locn}_{y, z}(x) &= \sum_{ij \in E} y_{ij} \p{1 - \frac{(x_i - x_j)^2}{l_{ij}^2}}^2 + z_{ij} \p{1 - \frac{(x_i - x_j)^2}{l_{ij}^2}}^2 \\
        &\le \sum_{ij \in E} (y_{ij} + z_{ij}) + \sum_{ij \in E} (y_{ij} + z_{ij}) \frac{(x_i - x_j)^4}{l_{ij}^4} \\
        &\stackrel{\eqref{eq:local_norm_oracle_edge_width_bound}}{\le} O(n \rho^2)
    \end{align*}

\end{proof}

Now, we combine these lemmata to prove the guarantees required for the oracle, including the width bounds and runtime.
\begin{theorem}\label{thm:local_norm_bounded_oracle}
    \Cref{alg:local_norm_bd_oracle} is a \((W=O(n \rho), \nu=O(n \rho^2))\)-bounded, \(\eps\)-approximate \(\delta\)-sure oracle.
    Further, its runtime is \(\widetilde{O}\p{(n+m) \alpha \eps^{-1.5}}\).
\end{theorem}
\begin{proof}
    Here, we combine all the previous lemmata from this section to show the correctness and bound the runtime of \textsc{LnbOracle}.

    \paragraph{Correctness}
    The first property \Cref{def:oracle_local_norm}-\eqref{item:def_oracle_local_norm_containment_guarantee}
    follows directly from \Cref{lem:local_norm_oracle_K_contained_in_EH}.
    
    Next, let us establish \Cref{def:oracle_local_norm}-\eqref{item:def_oracle_local_norm_witness_guarantee}.
    If we return witness \(x = 0\) in Line \ref{alg_line:local_norm_bd_oracle_earlyreturn},
     then clearly \(x^\top H X = 0\), \(\text{width}(x) = 0\), and \(\text{locn}_{y, z}(x) \le 1\).
     We also have \(\cL_1(x, y, z) = c \le \eps\) as required.
     The condition in Line \ref{alg_line:local_norm_bd_oracle_finalif} ensures that any returned \(x\) in Line \ref{alg_line:local_norm_bd_oracle_witness_return} also adheres to the requirements of \Cref{def:oracle_local_norm}-\eqref{item:def_oracle_local_norm_witness_guarantee}.

     Finally, we prove \Cref{def:oracle_local_norm}-\eqref{item:def_oracle_local_norm_high_prob_guarantee}.
     First, assume that \(\min_{x \in \cE_H} \cL_1(x, y, z) \le \eps/2\).
    We need to first establish that the input the algorithm gives to \textsc{GenLanczosRand} is valid.
    Recall the two requirements from \Cref{lem:generalized_lanczos_rand}: that the estimate \(\Lambda = 2 + 2 \rho\) satisfies \(\Lambda \ge -\lambda_n \)  which is immediately handled by \Cref{lem:local_norm_oracle_gen_eigenvalue_bounds}-\eqref{item:local_norm_oracle_min_gen_eigenvalue_bound};
    and that we must have \(\gamma \le \frac{\lambda_1 + \eps}{\lambda_1 + \Lambda}\).
    With our choice of \(\gamma = \eps/(4\Lambda)\), the latter condition is equivalent to
    \(\lambda_1 \ge \frac{- 3 \eps}{\Lambda - \eps}\), and is implied when \(\lambda_1 \ge 0\).
    Due to
    \Cref{lem:local_norm_oracle_gen_eigenvalue_bounds}-\eqref{item:local_norm_oracle_completeness_eigenvalue_bound},
    \(\min_{x \in \cE_H} \cL_1(x, y, z) \le \eps/2\) implies \(\lambda_1(M, H) \ge \eps/4 \ge 0\).
    Thus we may apply \Cref{lem:generalized_lanczos_rand},
    and so with probability \(1-\delta\) the algorithm returns an \(x\) satisfying \(x^\top H x = 1\)
    and \(x^\top M x \ge - \eps/4\).
    Using similar reasoning as for \Cref{lem:local_norm_oracle_gen_eigenvalue_bounds}-\eqref{item:local_norm_oracle_completeness_eigenvalue_bound}, the latter property implies that \( \cL_1(x, y, z) \le \eps\).
    Since \(x \in \cE_H\), we may apply 
    \Cref{lem:local_norm_oracle_ellipsoid_size_bounds}-\eqref{item:local_norm_oracle_width_bound}
    and \Cref{lem:local_norm_oracle_ellipsoid_size_bounds}-\eqref{item:local_norm_oracle_locn_bound}
    to get the bounds \(W = O(n \rho)\) and \(\nu = O(n \rho^2)\) respectively.
    Thus the returned \(x\) will satisfy all the conditions in Line \ref{alg_line:local_norm_bd_oracle_finalif},
    and so with probability \(1-\delta\) a \textsc{Witness} will be returned.

    \paragraph{Runtime} We can treat all matrices in \Cref{alg:local_norm_bd_oracle} as implicitly represented, and only used
    in \textsc{GenLanczosRand}, which takes \(\widetilde{O}\p{\frac{1}{\sqrt{\gamma}}}\) matrix-vector products involving
    \(M\). Since \(H\) and \(Q\) are both Laplacian matrices, each of these takes \(O(m)\) time.
    Combined with the operations requiring additional time, our total time is \(\widetilde{O}\p{(n+m)\frac{1}{\sqrt{\gamma}}}\).
    Finally, we have \(\gamma = \frac{\eps}{4 \Lambda} = \Omega\p{\frac{\eps}{1 + \rho}}\),
    giving a total time bound of 
    \[\widetilde{O}\p{(n+m) \sqrt{\frac{1+\rho}{\eps}}} = \widetilde{O}\p{(n+m) \alpha \eps^{-1.5}}\]
\end{proof}

\subsection[Full algorithm for l2-configuration]{Full algorithm for \(\ell_2\)-configuration}

We are now ready to describe the full algorithm, see \Cref{alg:l2_main}.

\begin{algorithm}[H]
  \caption{\(\textsc{L2GraphConfig}((V, E, u, l), \eps, \delta, T)\)}
  \label{alg:l2_main}
  \begin{algorithmic}[1]
    \Require \((V, E, u, l)\), oracle accuracy \(\eps\), oracle failure probability \(\delta\), number of rounds \(T\)
    \Ensure \(v_1, \ldots, v_n\)
    \State \(w = \p{\frac{1}{l_{ij} \sqrt{\rho}}}_{ij \in E}, \rho \) from \eqref{eq:width_choice}
    \State \(H_k = \textsc{ApproxLapEllipsoid}((V, E, w))\)
    \State \(y^{(0)}, z^{(0)} \leftarrow \frac{1}{2m} (1, \ldots, 1) \in \Delta_{2m}\)
    \For{\(t = 0, \ldots, T-1\)}
    \State \(x^{(t)} \leftarrow \textsc{Oracle}(y^{(t)}, z^{(t)}, \eps, \delta/T)\)
    \If{\textsc{Oracle} reports \textsc{Fail}}
    \State \Return \textsc{Infeasible}
    \EndIf
    \State \(\ell^{(t)}(y, z) = \sL_1(x^{(t)}, y, z)\)
    \State \(y^{(t+1)}, z^{(t+1)} \leftarrow \textsc{MWU}(y^{(t)}, z^{(t)}, -\ell^{(t)})\)
    \EndFor
    \State \(v_i = \frac{1}{\sqrt{T}} (x^{(0)}_i, \ldots, x^{(T-1)}_i)\) for all \(i = 1, \ldots, n\)
    \State \Return \(\{v_i\}_{i=1}^n\)
  \end{algorithmic}
\end{algorithm}

Our analysis of this algorithm template follows that of \cite[Theorem 3.4]{arora2012multiplicative}.

We show how the correctness and approximation factor of our algorithm follows from the \textnormal{\textsc{Oracle}}'s guarantees.
\begin{theorem}\label{thm:l2_main_full}
  Given any feasible instance \((V, E, u, l)\), with probability at least \(1-\delta\),
  \Cref{alg:l2_main} returns a \(2\eps\)-approximately feasible solution.
\end{theorem}

% \ltwoconfigmain*

Each oracle call errs with probability at most \(1/T\). By the union bound, all oracle
calls are simultaneously correct with probability at least \(1-\delta\), so in the following lemmata
we assume this event holds.

\begin{lemma}[Successful oracle calls imply approximate feasibility]\label{lem:l2_main_oracle_success_feasibility}
    If in a call to \textsc{L2GraphConfig} the oracle never fails and the procedure 
    returns \(v_1, \ldots, v_n\), that solution is \(2\eps\)-approximately feasible.
\end{lemma}

\begin{proof}
  Observe that
  \[g^{(t)} \defeq \nabla_{y, z} \ell^{(t)}(y^{(t)}, z^{(t)}) = \nabla_{y, z} \cL_1(x^{(t)}, y^{t}, z^{t}) = \pb{
    \p{1 - \frac{(x_i^{(t)} - x_j^{(t)})^2}{l_{ij}^2}}_{ij \in E} \middle| \p{\frac{(x_i^{(t)} - x_j^{(t)})^2}{u_{ij}^2} - 1}_{ij \in E}
    }\]
  Now, since again \(\frac{(x_i^{(t)} - x_j^{(t)})^2}{u_{ij}^2} \le \frac{(x_i^{(t)} - x_j^{(t)})^2}{l_{ij}^2}\), we have:
  \[\pn{g^{(t)}}_\infty \le 1 + \text{width}(x^{(t)}) \le O(n \rho)\]
  using the width bound \(W \le O(n \rho)\) from \Cref{thm:local_norm_bounded_oracle}.
  We also have
  \[\pg{\p{g^{(t)}}^2, (y^{(t)}, z^{(t)})} = \text{locn}_{y^{(t)}, z^{(t)}}(x^{(t)}) \le O(n\rho^2)\]
  using the local norm bound \(\nu \le O(n \rho^2)\) from \Cref{thm:local_norm_bounded_oracle}.
  Now we can apply \Cref{lem:local_norm_regret_iteration_bound} and our definition \eqref{eq:width_choice};
  we need
  \begin{equation}\label{eq:l2_main_iteration_bound}
    T = O\p{\ln(2m) \cdot \max\p{\frac{n \rho^2}{\eps^2}, \frac{n \rho}{\eps}}} \le \widetilde{O} \p{n\alpha^4\eps^{-6}}
  \end{equation}
  number of iterations to guarantee that \(\frac{1}{T} \regret_T \le \eps\).

  Now fix \((y^*, z^*) \in \Delta_{2m}\).
  Observe that:
  \[\sL(v_1, \ldots, v_n, y^*, z^*) = \frac{1}{T} \sum_{t=0}^{T-1} \sL_1(x^{(t)}, y^*, z^*)\]
  Applying our regret bound from \Cref{lem:local_norm_regret_iteration_bound}, we have:
  \begin{align*}
    \sL(v_1, \ldots, v_n, y^*, z^*) \le \frac{1}{T} \sum_{t=0}^{T-1} \sL_1(x^{(t)}, y^{(t)}, z^{(t)}) + \eps
  \end{align*}
  By \Cref{def:oracle_local_norm}-\eqref{item:def_oracle_local_norm_witness_guarantee},
  we have \(\sL_1(x^{(t)}, y^{(t)}, z^{(t)}) \le \eps\) for all \(t\), and so
  \(\sL(v_1, \ldots, v_n, y^*, z^*) \le 2 \eps\).
  As we have shown this for arbitrary \((y^*, z^*) \in \Delta_{2m}\), we have that
  \[\max_{ij \in E} \max\p{1 - \frac{\|v_i - v_j\|^2}{l_{ij}^2}, \frac{\|v_i - v_j\|^2}{u_{ij}^2} - 1} \le 2\eps\]
  which one can clearly see is the definition of approximately feasible from \Cref{def:5_approx_feasible}.
\end{proof}

\begin{lemma}[Feasible instances return vectors]\label{lem:l2_main_feasible_instance_returns}
    If the instance \((V, E, u, l)\) is feasible, then \Cref{alg:l2_main} always returns a set of vectors.
\end{lemma}
\begin{proof}
    Fix any round \(t\).
    By \Cref{lemma:struct_lemma}, there exists a point \(x \in \R^V\) such that \(\cL_1(x, y, z) \leq \eps/2\)
    such that \(\text{width}(x) \le \rho\) for \(\rho\) defined in \eqref{eq:width_choice}.
    But then because of \Cref{def:oracle_local_norm}-\eqref{item:def_oracle_local_norm_containment_guarantee},
    we have \(x \in \cE_H\), and so \(\min_{x \in \cE_H} \cL_1(x, y^{(t)}, z^{(t)}) \le \eps/2\).
    By \Cref{def:oracle_local_norm}-\eqref{item:def_oracle_local_norm_high_prob_guarantee}, the oracle must return a
    \(\textsc{Witness}(x)\).

    As every round had a witness, the algorithm must return a set of vectors.
\end{proof}

\begin{proof}[Proof of \Cref{thm:l2_main_full}]
    We prove the correctness and analyze the runtime of the algorithm.

    \paragraph{Correctness} This follows from chaining \Cref{lem:l2_main_feasible_instance_returns} with
    \Cref{lem:l2_main_oracle_success_feasibility}.

    \paragraph{Runtime}
    \textsc{ApproxLapEllipsoid} is called once for initialization, but as per \Cref{corr:approx_lap_ellipsoid} the total runtime is \(\widetilde{O}(m)\), so is dominated by other factors.

    Recall that the oracle from \Cref{thm:local_norm_bounded_oracle} had a runtime per call of \(\widetilde{O}((n+m) \alpha \eps^{-1.5})\).
    In \eqref{eq:l2_main_iteration_bound}, we showed that \(\widetilde{O}(n \alpha^4 \eps^{-6})\) iterations sufficed
    to get a solution with our desired feasibility.
    Thus, the total runtime is \(\widetilde{O}(n(n+m) \alpha^5 \eps^{-7.5})\).
    Note that if \(m \ll n\)
    then the algorithm can be run separately on disconnected components
    in the constraint graph; thus our final runtime is
    \(\widetilde{O}(nm \alpha^5 \eps^{-7.5})\).
\end{proof}

\section{Applications}

\subsection[Approximating Minimum Distortion Embedding into l2]{Approximating Minimum Distortion Embedding into \(\ell_2\)}

For completeness, we discuss the `optimization-to-feasibility' reduction that shows a minimum
distortion embedding from a finite metric space can be found by solving a sequence of
graph-constrained \(\ell_2\)-configuration
problems.
For a candidate distortion \(D \ge 1\), define the \(\ell_2\)-configuration instance
\(\cC(D)\) : \(d_{ij} \le \|v_i - v_j\| \le D d_{ij}\).
Then \(\cC(D)\) is feasible exactly when \(e_{\ell_2}(V) \le D\).
By rescaling the approximate feasibility guarantee of \Cref{def:5_approx_feasible},
we have for all \(\eps \le \frac{1}{\sqrt{2}}\) that a certified implementation either reports that
the problem is at least \(\eps\)-feasible, providing a solution such that \(e_{\ell_2}(V) \le (1 + 2 \eps) D\),
or that the problem is infeasible, meaning that \(e_{\ell_2}(V) > D\).

By definition, \(e_{\ell_2}(V) \ge 1\).
Further, Bourgain's theorem \cite{bourgain1985lipschitz} shows that there is a universal constant \(C > 0\) such that
\(e_{\ell_2}(V) \le C \cdot \log n\).
Thus, we know that \(e_{\ell_2}(V)\) must lie somewhere in the interval \([1, C \log n]\).
We can perform a binary search: maintain a current interval \(I = [a, b]\),
and repeatedly solve an instance of \(\ell_2\)-configuration \(\cC(\sqrt{ab})\) (the geometric midpoint) problems.
If we stop after our interval has (geometric) width at most \(\frac{b}{a} \le (1+ O(\eps))\),
we guarantee that the embedding produced by the instance \(\cC(D)\)
with the smallest \(D\) that succeeded during the search satisfies \(D \le (1+O(\eps)) e_{\ell_2}(V)\).
The total number of \(\ell_2\)-configurations we need to reduce the interval to a width of \((1+ O(\eps))\)
is \(O\p{\log \frac{\log(C \log n)}{\eps}}\).
Thus, the runtime to find a \(1 + O(\eps)\) approximate minimum distortion embedding
is at most \(O\p{\log \frac{\log(C \log n)}{\eps}}\) times the runtime to \((1+\eps)\)-approximately
solve the \(\ell_2\)-configuration problem, meaning the final runtimes are equivalent up to log factors in all parameters.
Rescaling \(\eps\) by a universal constant gives the guarantee in \Cref{thm:l2_distortion_main}:

\ltwodistortionmain*

\subsection{Generalized sparsest-cut rounding}

The same configuration primitive can be used for the Euclidean embedding step in SDP-based generalized sparsest-cut rounding.
Let \(C,D\in\R_{\ge0}^{n\times n}\) be the capacity and demand matrices, and let
\[
        m_{\rm round}=\abs{\operatorname{supp}(C)\cup\operatorname{supp}(D)}.
\]
The Goemans--Linial/ALN-type relaxation produces a negative-type metric \(d_{ij}=\|w_i-w_j\|^2\).
For a target distortion \(\Lambda\), the rounding step asks for an embedding that does not expand capacity pairs too much and does not contract demand pairs too much.
This can be written as an \(\ell_2\)-configuration instance on the support pairs with
\[
        l_{ij}=d_{ij}/\Lambda,
        \qquad
        u_{ij}=d_{ij},
        \qquad
        \alpha=\Lambda.
\]
Equivalently, one can use the one-sided version with upper constraints on capacity pairs and lower constraints on demand pairs; the same local-norm calculation only changes \(m\) to \(m_{\rm round}\).
Applying \Cref{thm:l2_config_main} gives rounding-stage time
\[
        \widetilde{O}\p{m_{\rm round}n\Lambda^5\eps^{-15/2}}.
\]
For the ALN rounding theorem, \(\Lambda=O(\sqrt{\log n}\log\log n)\) \cite{arora2005euclidean}; for the current optimal Goemans--Linial rounding theorem, \(\Lambda=O(\sqrt{\log n})\) \cite{chang2024random}.
Thus for constant accuracy this Euclidean rounding stage runs in \(\widetilde{O}(m_{\rm round}n)\) time.
Note that this is not an end-to-end \(\widetilde{O}(m_{\rm round}n)\) sparsest-cut algorithm, because the SDP relaxation must still be solved.

\section{Open Problems and Future Work}

\subsection{Improving dependence on accuracy}

The runtime dependence on accuracy \(\eps\) in our algorithms for \(\ell_2\)-configuration
and minimum distortion embedding into \(\ell_2\) is \(\eps^{-7.5}\).
One clear area this dependence could be improved is in the optimization algorithm used.
We use the local-norm regret bound for multiplicative weights which produces an \(\eps^{-2}\)
dependence; using an accelerated method (integrated with the \textsc{Oracle} implementation)
could likely improve this dependence by at least an \(\eps\) factor.

\subsection{Relation to Covering SDPs}

Terminology for classes of positive SDPs tends to be inconsistent in the literature,
so for clarity here we define our own.
First we define a packing SDP and its dual, a linear covering program over the PSD cone, as:
\[
        \max_{x\ge0}\pc{\ones^\top x:\sum_i x_iA_i\preceq I},
        \qquad
        \min_{X\succeq0}\pc{\langle I,X\rangle:\langle A_i,X\rangle\ge1\quad\forall i}.
\]

Now we define a covering SDP and its dual, a linear packing program over the PSD cone, as:
\[
        \min_{x\ge0}\pc{\ones^\top x:\sum_i x_iA_i\succeq I},
        \qquad
        \max_{X\succeq0}\pc{\langle I,X\rangle:\langle A_i,X\rangle\le1\quad\forall i}.
\]

In \cite{jambulapati2020positive}'s language, the packing SDP is a `pure' packing SDP
and the  covering SDP a `pure' covering SDP, as opposed to a `mixed' packing-covering SDP
which may have both packing and covering SDP constraints.
\cite{jambulapati2020positive} attempted to establish a width-independent solver for mixed packing-covering SDPs, but the ArXiv version
has detailed the issues in their analysis, which have not yet beeen resolved.

Fast width-independent algorithms for packing SDPs are broadly well understood,
while much less can be said about algorithms for covering SDPs.
Unlike the case of LPs, packing and covering SDPs are not dual to each other,
and form different families of positive SDPs.

By dualizing only the upper bound constraint (rather than both the upper and lower bounds
as in our current algorithm), we can reduce the \(\ell_2\) configuration problem
to solving a sequence of covering SDPs. This provides additional
motivation for finding width-independent solvers specifically for the class of covering SDPs.

% \subsection{Faster Generalized Sparsest Cut with Restricted Demand Support}

% The existential embedding results for negative-type metrics into \(\ell_2\)
% have a refinement when the demand support has size \(k\): \cite{arora2005euclidean} with \(O(\sqrt{\log k}\log\log k)\), and \cite{chang2024random} with \(O(\sqrt{\log k})\).

% Meanwhile, our method requires both sides of the inequalities to be present,
% and in fact its complexity depends strongly on \(\alpha = \max_{ij} \frac{u_{ij}}{l_{ij}}\).
% This makes it unclear how to get fast algorithms for generalized sparsest cut when the
% support of the demand is restricted to some number \(k\) vertices.

\printbibliography[heading=bibintoc,title={References}]

\end{document}